\documentclass[elsarticle,amsmath,amssymb,floatfix]{revtex4}
\usepackage{graphicx}
\usepackage{bm}
\usepackage{dsfont}
\usepackage{cancel}
\usepackage{comment}
\usepackage{ulem}
\usepackage{subfigure}

\DeclareMathAlphabet{\mathpzc}{OT1}{pzc}{m}{it}
\newcommand{\bem}{\begin{multline}}
\newcommand{\eem}{\end{multline}}

\newcommand{\gae}{\lower 2pt \hbox{$\,
\buildrel{\scriptstyle >}\over {\scriptstyle \sim}\,$}}
\newcommand{\lae}{\lower 2pt \hbox{$\,
\buildrel{\scriptstyle <}\over {\scriptstyle \sim}\,$}}

\newcommand{\rcite}[1]{Ref.~\cite{#1}}
\newcommand{\eref}[1]{Eq.~(\ref{#1})}
\newcommand{\exref}[1]{Exercise~(\ref{#1})}
\newcommand{\sref}[1]{Sec.~\ref{#1}}
\newcommand{\fref}[1]{Fig.~\ref{#1}}

\newcommand{\cref}[1]{Chp.~\ref{#1}}
\newcommand{\aref}[1]{Appendix~\ref{#1}}

\def \be{\begin{equation}}
\def \ee{\end{equation}}
\def \ba{\begin{array}}
\def \ea{\end{array}}
\def \beq{\begin{eqnarray}}
\def \eeq{\end{eqnarray}}
\def \nn{\nonumber}

\def \a{{\alpha}}

\def \b{{\beta}}

\usepackage{amsthm}
\theoremstyle{remark}
\newtheorem{ExInternal}{Exercise}[section]

\makeatletter
\let\@exercises\@empty%
\newcommand\exercise[2][]{%
    \g@addto@macro\@exercises{%
        \begin{ExInternal}[#1]%
            #2%
        \end{ExInternal}%
    }%
}

\newcommand\exerciseshere{%
    \@exercises%
    \global\let\@exercises\@empty%
}
\makeatother

\newcommand{\vect}[1]{\boldsymbol{\mathbf{#1}}}

\newcommand{\add}[1]{#1}

\begin{document}
\title{Geometry and non-adiabatic response in quantum and classical systems}

\author{Michael Kolodrubetz\footnote{mkolodru@gmail.com}$^{1,2,3}$, Dries Sels$^{1,4}$, Pankaj Mehta$^1$, Anatoli Polkovnikov$^1$}
\affiliation{$^1$Department of Physics, Boston University, Boston, MA 02215, USA\\
  $^2$Department of Physics, University of California, Berkeley, CA 94720, USA\\
  $^3$Materials Sciences Division, Lawrence Berkeley National Laboratory, Berkeley, CA 94720, USA\\
  $^4$Theory of Quantum and Complex Systems, Universiteit Antwerpen, B-2610 Antwerpen, Belgium}

\begin{abstract}
In these lecture notes, partly based on a course taught at the Karpacz Winter School in
March 2014, we explore the close connections between non-adiabatic response of a system
with respect to macroscopic parameters and the geometry of quantum and classical states.
We center our discussion around adiabatic gauge potentials, which are the generators of
unitary basis transformations in quantum systems and generators of special canonical
transformations in classical systems. In quantum systems, eigenstate expectation values of these
potentials are the Berry connections and the covariance matrix of these
gauge potentials is the geometric tensor, whose antisymmetric part defines the Berry 
curvature and whose symmetric part is the Fubini-Study metric tensor. In classical systems one
simply replaces the eigenstate expectation value by an average over the micro-canonical
shell. For complicated interacting systems, we show that a variational principle may be used to
derive approximate gauge potentials. We then express the non-adiabatic response of the physical observables of the system
through these gauge potentials, specifically demonstrating the close connection of the geometric
tensor to the notions of Lorentz force and renormalized mass. We highlight applications of this formalism
to deriving counter-diabatic (dissipationless) driving protocols in various systems, as well as
to finding equations of motion for slow macroscopic parameters coupled
to fast microscopic degrees of freedom that go beyond macroscopic
Hamiltonian dynamics. Finally, we illustrate these ideas with a number of simple examples
and highlight a few more complicated ones drawn from recent literature.
\end{abstract}

\maketitle


\tableofcontents

Geometry plays an important role in many aspects of modern physics. In these lecture notes, we will highlight one situation where geometry plays a crucial role, namely the dynamics of closed systems. We will do so by introducing the concept of {\em gauge potentials}, which are infinitesimal generators of unitary transformations. Specifically, gauge potentials are defined as $\mathcal A_\lambda = i\hbar \partial_\lambda$, where the derivative is understood as acting on a smooth manifold of basis states parameterized by the (potentially multi-component) parameter $\lambda$. Among gauge potentials, a very important role in these notes will be played by {\em adiabatic gauge potentials} where the family of basis states are chosen as the eigenstates of some Hamiltonian $\mathcal H(\lambda)$\footnote{\add{If we associate adiabatic transformations with parallel transport, the adiabatic gauge potentials are nothing but connections. We, however, prefer to avoid using this terminology to highlight the physical meaning of $\mathcal A_\lambda$.}}. These adiabatic gauge potentials are the fundamental objects of both adiabatic perturbation theory and geometry of quantum or classical states. We will see, for instance, that in a moving frame the Hamiltonian picks up an effective Galilean term $\mathcal H \to \mathcal H - \dot \lambda \mathcal A_\lambda$, which yields important non-adiabatic corrections to the dynamics. These corrections can be measured through standard linear response techniques or by their back action on $\lambda$ if it is treated as a dynamical degree of freedom. In classical systems, gauge potentials correspond to generators of infinitesimal canonical transformations parameterized by $\lambda$. The adiabatic gauge potentials are in turn generators of special canonical transformations $\vect q(\lambda)$ and $\vect p(\lambda)$ which leave the Hamiltonian invariant. In particular, the adiabatic gauge potentials ensure that the Hamiltonian \add{corresponding to coupling $\lambda'=\lambda+\delta \lambda$ in new coordinates $q', p'$ ``commutes,'' i.e., has vanishing Poisson bracket, with the Hamiltonian corresponding to coupling $\lambda$ in the original coordinates $q,p$}. Such special canonical transformations in classical systems are analogous to special unitary transformations in quantum systems which diagonalize the instantaneous Hamiltonian. \add{The vanishing Poisson brackets in classical systems then correspond to vanishing commutator between two diagonal matrices representing the instantaneous diagonalized Hamiltonian in quantum systems.}

We begin the lecture notes in \sref{sec:invitation} by introducing the major concepts in more detail using two simple examples: the quantum spin-1/2 and the simple harmonic oscillator (both quantum and classical). Next, in \sref{sec:gauge_potentials}, we introduce the concept of \add{adiabatic} gauge potentials for quantum and classical systems in full generality. \add{For simple integrable or highly symmetric systems the adiabatic gauge potentials can be found exactly. For more complex systems, we show in \sref{sec:approx_agp} that adiabatic gauge potentials can be found approximately by either perturbative or variational methods. We further discuss applications of exact and approximate adiabatic gauge potentials to designing counter-diabatic (transitionless) driving protocols, allowing fast state preparation. } In \sref{sec:geometric_tensor} we introduce the geometric tensor, an object which describes the geometric properties of quantum ground state manifolds via the Fubini-Study metric and Berry curvature tensors. We show how these ideas can be generalized to quantum systems that are far from their ground states as well as to classical systems. In \sref{sec:measuring_geometric_tensor}, we connect adiabatic gauge potentials to the geometric tensor by showing how the geometric tensor can be measured via dynamical response. Finally, in \sref{sec:emergent_newtonian_dynamics} we show one important consequence of these ideas, namely the emergence of effective Newtonian dynamics for the classical parameter $\lambda$ due to excitation of the quantum or classical system to which it is coupled. To help understand this general concept, in \sref{sec:newtonian_dynamics_examples} we show explicit examples of this emergent dynamics, ranging from relatively simple (particle in a box) to more complicated (dynamics of the order parameter in a quenched superconductor).

\section{Invitation: Quantum spins and classical oscillators out of equilibrium}
\label{sec:invitation}

We begin by considering two simple examples where geometry enters into dynamics. The goal of this section is to introduce the main ingredients of the formalism such as gauge potentials, geometric tensor, \add{counter-diabatic driving}, generalized Coriolis force and mass renormalization. In the following sections we will rigorously introduce these concepts, derive general statements and illustrate them with additional examples. As a first example, consider one of the staple problems in quantum mechanics, a single spin-1/2 particle in a time-dependent magnetic field \add{with Hamiltonian $\mathcal H=-\mu \vect{B} \cdot \vect{s}$}. We will restrict ourselves to considering the case where the field strength is fixed but the field's direction can vary with time. If we parameterize the magnetic field direction by the spherical angles $\theta$ and $\phi$, \add{express the magnitude as $h=\hbar \mu B / 2$, and represent the spin in terms of Pauli matrices $\vect \sigma=2 \vect s / \hbar$}, then the Hamiltonian for this problem takes the form
\be
\mathcal H =-h \left[ \cos (\theta) \sigma^z + \sin(\theta) \cos (\phi) \sigma^x + \sin(\theta) \sin (\phi) \sigma^y \right] = -h \left( \begin{array}{cc} \cos \theta & \mathrm{e}^{i \phi} \sin \theta  \\ \mathrm{e}^{- i \phi} \sin \theta  & -\cos \theta \end{array} \right) 
\ee
with ground ($|g\rangle$) and excited ($|e\rangle$) eigenstates 
\be
|g\rangle = \left( \begin{array}{c}  \cos (\theta / 2) \\ \mathrm{e}^{i \phi} \sin (\theta / 2)  \end{array} \right) ~,~|e\rangle = \left( \begin{array}{c}  \sin (\theta / 2) \\ -\mathrm{e}^{ i \phi} \cos (\theta / 2)\end{array} \right) ~.
\label{eq:gs_es_qubit}
\ee
The simple dynamical problem that we want to consider is the case where the field rotates around the z-axis in the lab frame.  This problem can be solved exactly by going to the moving (rotating) frame, i.e., by diagonalizing the Hamiltonian by a unitary rotation $U$ to give a diagonal matrix $\tilde{\mathcal H} = U^\dagger \mathcal H U$, where 
\be
U(\theta,\phi)=\left( \begin{array}{cc}  \cos (\theta / 2) &  \sin (\theta / 2) \\ \mathrm{e}^{i \phi} \sin (\theta / 2) & -\mathrm{e}^{i \phi} \cos (\theta / 2)\end{array} \right)
\ee
and $\tilde{\mathcal H} = -h \tilde \sigma^z$ Going to a rotating frame corresponds to a time-dependent unitary transformation on the wave function 
\be
|\tilde \psi \rangle = U^\dagger(\theta,\phi) |\psi \rangle ~.
\ee
This unitary transformation can be equivalently thought of as expanding the wave function in the rotated basis.
For rotations around the $z$-axis, only the angle $\phi$ changes and $|\tilde \psi \rangle$ satisfies a new Schr\"odinger equation given by
\beq
\nn i \hbar \frac{d |\tilde \psi \rangle}{dt} &=& i \hbar \frac{d (U^\dagger |\psi \rangle)}{dt} = i \hbar \frac{d U^\dagger}{dt} |\psi \rangle + i \hbar U^\dagger\frac{d |\psi \rangle}{dt}= i \hbar \frac{d\phi}{dt} \frac{\partial U^\dagger}{\partial \phi} |\psi \rangle + U^\dagger \mathcal H |\psi \rangle
\\  &=& \frac{d\phi}{dt} \left( i \hbar \frac{\partial U^\dagger}{\partial \phi} U \right) |\tilde \psi \rangle + U^\dagger \mathcal H U |\tilde\psi \rangle= \underbrace{\left( \tilde{\mathcal H} - \dot \phi \tilde{\mathcal{A}}_\phi \right)}_{\tilde{\mathcal H}_m} |\tilde\psi \rangle ~,
\label{eq:H_eff_rot_spin_half}
\eeq
where 
\be
\tilde{\mathcal{A}}_\phi = -i \hbar \left( \partial_\phi U^\dagger \right) U = -i \hbar \left[ \partial_\phi \left(U^\dagger U\right) - U^\dagger \partial_\phi U \right] = i \hbar U^\dagger \partial_\phi U 
\label{eq:A_phi_intro}
\ee
is a very important operator that we will refer to as the \emph{adiabatic gauge potential} with respect to the parameter $\phi$ (here in the moving frame)\footnote{As we explain later the word {\em adiabatic} highlights the fact that $U$ in \eref{eq:A_phi_intro} is a particular unitary operator connecting eigenstates of the Hamiltonian $\mathcal H$ at different parameters $\phi$. Since in these notes we are mostly dealing with adiabatic gauge potentials we will often omit the word ``adiabatic'' and simply use gauge potential.}.  In writing these equations, we have introduced the notation, which we are going to use later, that the tilde superscript refers to objects in the moving frame basis. In particular, for any operator $\mathcal O$ we define $\tilde {\mathcal O}=U^\dagger \mathcal O U$.  By examining \eref{eq:H_eff_rot_spin_half}, it is clear that the combination $\tilde{\mathcal H}_m=\tilde{\mathcal H}-\dot \phi \tilde{\mathcal A}_\phi$ plays the role of the Hamiltonian in the moving frame basis, and we refer to this operator as the moving frame Hamiltonian. Note that we can remove tilde signs here by doing the inverse unitary transformation to get $\mathcal H_m=\mathcal H-\dot\phi \mathcal A_\phi$, which is equivalent to projecting operators back to the original basis.  Let's try to understand a bit more about what this gauge potential does in our system by calculating its matrix elements. For instance,
\beq
\langle e | \mathcal A_\phi | g \rangle = \langle \downarrow | \tilde{\mathcal A}_\phi | \uparrow \rangle &=& \langle \downarrow | i \hbar U^\dagger (\partial_\phi U) | \uparrow \rangle
\\ &=& i \hbar \langle \downarrow | U^\dagger \left[ \partial_\phi ( U | \uparrow \rangle ) - U \cancelto{0}{\partial_\phi | \uparrow \rangle} \right]
\\ &=& i \hbar \langle e | \partial_\phi g \rangle = \frac{\hbar \sin \theta}{2} ~,
\label{eq:A_phi_eg}
\eeq
where we have used from the definition of $U$ that $U(\theta,\phi)|\uparrow \rangle = |g(\theta,\phi)\rangle$ and similarly $U(\theta,\phi)|\downarrow \rangle = |e(\theta,\phi)\rangle$. From the remaining matrix elements, it becomes clear that we can think of $\mathcal A$ as the derivative operator $\mathcal A_\phi = i \hbar \hat \partial_\phi$. As we will see later, this is a very general property defining the gauge potentials. It is straightforward to check by comparing the matrix elements that $\mathcal A_\phi=\hbar \sigma^z/2$, which is nothing but the angular momentum operator $S^z$. Indeed in this case adiabatic transformations of the Hamiltonian are simply rotations around $z$-axis generated by the angular momentum. Likewise it is easy to check that $\mathcal A_\theta=\hbar \sigma^y/2$.  Similarly, if instead of rotations we were to consider a particle in some potential which depends on $x_0-x$ and translate $x_0$, then we will see later that the (adiabatic) gauge potential with respect to $x_0$ is $\mathcal A_{x_0}=i\hbar \partial_{x_0}=-i\hbar \partial _x=p$, which is nothing but the momentum operator.

To see how the gauge potential connects to geometry, we note that its expectation value in the moving frame ground state $|\uparrow\rangle$ is the ground state Berry connection multiplied by Planck's constant, $A_\phi \equiv i \hbar \langle g |\partial_\phi g \rangle = -\hbar \sin^2(\theta/2)$ \cite{Berry1984_1}\footnote{Traditionally the Berry connection $A_\phi$ is defined without the factor of $\hbar$. However, it is the operator $\mathcal A_\phi=i\hbar \partial_\phi$ which has a well defined classical limit. As we are treating classical and quantum systems here on equal footing it is more natural to define $A_\phi$ as an expectation value of $\mathcal A_\phi$ and refer to it as the Berry connection.}. The Berry connection is related to both the ground state Berry (a.k.a. geometric) phase $\varphi_B$ and the Berry curvature $F_{\mu\nu} \equiv (\partial_\mu A_\nu - \partial_\nu A_\mu)/\hbar$ via
\be
\varphi_B = \frac{1}{\hbar} \oint \vect {A}_\lambda \cdot d \vect \lambda =\int F_{\mu \nu} d \lambda_\mu \wedge d \lambda_\nu ~,
\ee
where for the remainder of this review we use the convention that repeated indices are summed over, unless stated otherwise. More generally, one can think of the gauge potentials as connections defining a notion of parallel transport of wave functions in parameter space via the covariant derivative $D_\mu  = \partial_\mu + i \mathcal A_\mu / \hbar$ such that $D_\mu |\psi_n\rangle = 0$ for all energy eigenstates $|\psi_n\rangle$. This is the fundamental geometric definition that will later allow us to define curvature and distances via the covariance matrix of this connection.

\add{The form of the moving frame Hamiltonian [\eref{eq:H_eff_rot_spin_half}] immediately suggests another interesting application of the gauge potentials, namely counter-diabatic or transitionless driving~\cite{Demirplak2003_1,Demirplak2005_1, Berry2009_1}. Indeed the term responsible for transitions between energy levels in the instantaneous frame is $-\dot\phi \tilde{\mathcal A}_\phi$ because the Hamiltonian $\tilde H$ is diagonal by construction. Therefore if we add the term $\dot\phi \mathcal A_\phi$ to the lab frame Hamiltonian then the moving frame Hamiltonian will be just $\tilde H$ and there will be no transitions between energy levels. This observation defines the counter-diabatic Hamiltonian
\be
\mathcal H_{\rm CD}=\mathcal H(\phi)+ {\hbar \dot\phi \over 2}\sigma^z,
\label{h_cd_spin}
\ee
under which the system will remain in its instantaneous ground state for arbitrary smooth protocol $\phi(t)$. In this example, the second term has a simple interpretation. It is well known that in a rotating frame the Hamiltonian acquires an extra contribution equal to the product of the angular velocity and the angular momentum. The second term in \eref{h_cd_spin} simply counters this contribution, such that the rotating frame Hamiltonian is $\tilde {\mathcal H}=-h \sigma^z$
}

Next, let us use this simple example to discuss the connection of adiabatic gauge potentials to the dynamical response of the system to slow perturbations. A natural way to analyze this response is to go to the moving frame and then apply the framework of adiabatic perturbation theory \cite{Rigolin2008_1,DeGrandi2010_2}. The latter 
effectively treats the Galilean term $\tilde{\mathcal H}_1 = -\dot \phi \tilde{\mathcal A}_\phi$ in the moving frame Hamiltonian as a small perturbation around $\tilde{\mathcal H}_0= -h \sigma^z$. If we imagine gradually ramping up the velocity starting from $\dot \phi=0$ to some constant value then, up to order $\dot\phi^2$, the system will follow the ground state of the moving Hamiltonian $\mathcal H_m$. For constant velocity $\dot \phi$ this Hamiltonian is time independent and hence we can use the static perturbation theory to find the non-adiabatic corrections to transition probabilities and various observables in the original lab frame. For example, the transition amplitude to be in the instantaneous excited state $|e\rangle$ at first order in adiabatic perturbation theory is 
\be
a_e =\frac{\langle e | \mathcal H_1 | g \rangle}{E^0_{g} - E^0_{e}} = \dot \phi \frac{\langle \downarrow | \tilde{\mathcal A}_\phi | \uparrow \rangle}{2h} = \frac{\hbar \sin \theta}{4 h} \dot \phi ~.
\label{eq:a_e_pt_general}
\ee
If we think of the actual time-dependent wave function $|\psi(t)\rangle$ as the ground state of this weakly-perturbed Hamiltonian in the moving frame, then at lowest order in perturbation theory,
\be
|\psi(t)\rangle = a_g |g \rangle + a_e |e \rangle,
\ee
where as always the normalization condition gives $|a_g|\approx 1- |a_e|^2/2$ and the phase of $a_g$ is given by the sum of the dynamical and geometric phases. 

\begin{figure}
\includegraphics[width=0.3\columnwidth]{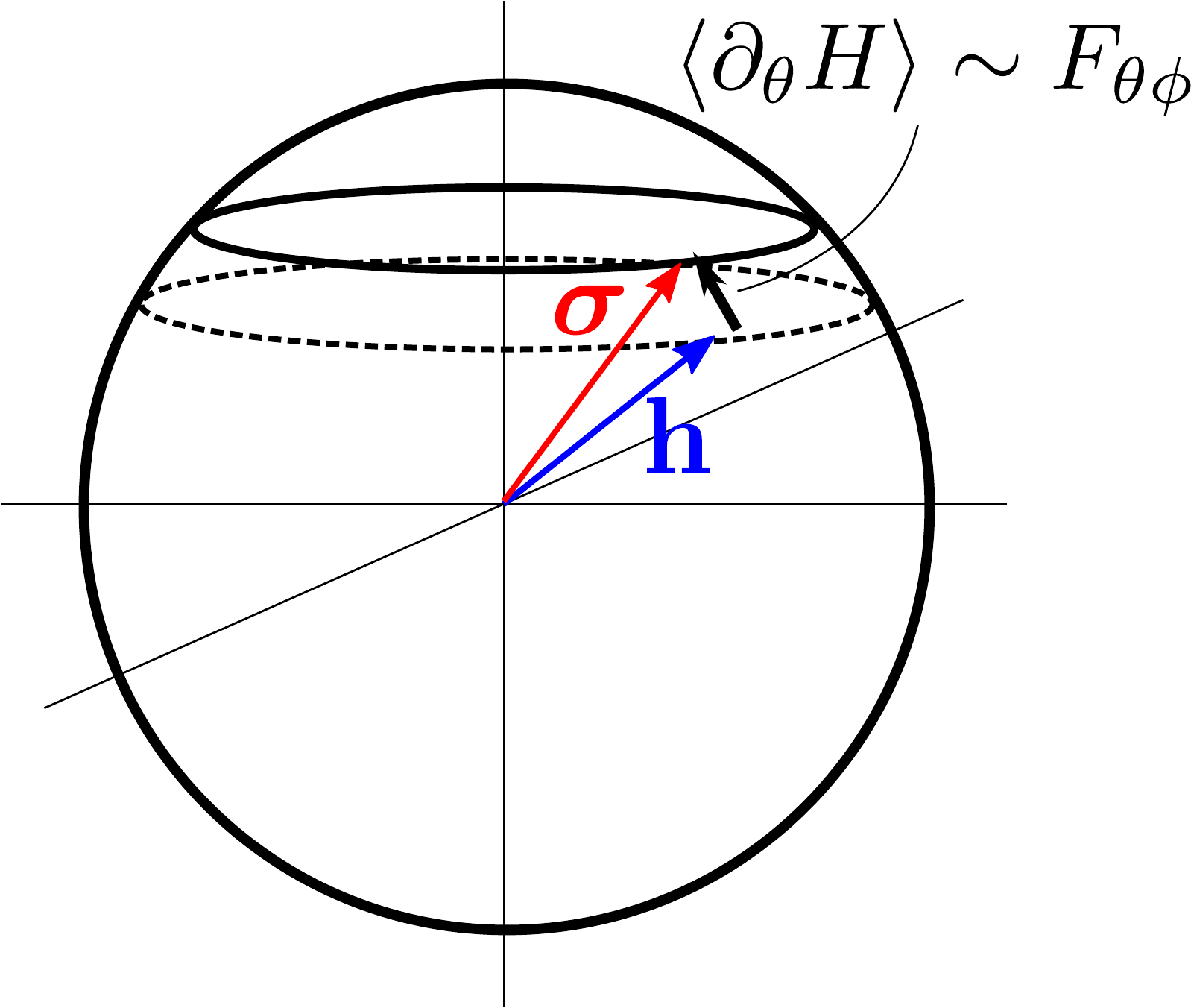}
\caption{Ramping the angle $\phi$ leads to a deflection in $\langle \partial_\theta \mathcal H \rangle$ which is proportional to the Berry curvature $F_{\theta \phi}$.}
\label{fig:berry_curv_spin}
\end{figure}

Let us use this perturbative result to calculate the expectation value of the operator $\mathcal M_\theta \equiv -\partial_\theta \mathcal H$. In thermodynamics the equilibrium expectation value of $\mathcal M_\theta$ is known as a generalized force with respect to $\theta$.  For example, if the Hamiltonian had a conventional form $\mathcal H=p_\theta^2 / 2m + V(\theta)$, then $\mathcal M_\theta = -\partial_\theta V$ and its expectation value is the average angular force (a.k.a. torque) acting on the particle. By analogy we extend the definition of the generalized force to non-equilibrium states and define $M_\theta(t)=\langle \psi(t) | \mathcal M_\theta| \psi(t) \rangle$ as a non-equilibrium generalized force. At leading order in the ramp rate $\dot \phi$, 
\be
M_\theta(t)\equiv\langle \psi(t) | \mathcal M_\theta | \psi(t) \rangle = \langle g | \mathcal M_\theta | g \rangle + a_e \langle g | \mathcal M_\theta | e \rangle + a_e^\ast \langle e | \mathcal M_\theta | g \rangle + O(\dot \phi^2) ~.
\ee
It is straightforward to check that the matrix elements of $\mathcal M_\theta$ are $\langle g | \mathcal M_\theta | g \rangle = \langle e | \mathcal M_\theta | e \rangle = 0$ and $\langle g | \mathcal M_\theta | e \rangle = \langle e | \mathcal M_\theta | g \rangle = -h$, so
\be
M_\theta(t) \approx 2 \hbar \left( \frac{\dot \phi \sin \theta}{4 h} \right) (-h) = -\dot \phi \frac{\hbar \sin \theta}{2} = \hbar F_{\theta \phi} \dot \phi  ~,
\label{eq:M_theta_single_qubit}
\ee
where in the last equation we have used the Berry curvature $\hbar F_{\theta \phi} = \partial_\theta A_\phi - \partial_\phi A_\theta = -\hbar \sin \theta / 2$ for the spin-1/2 ground state. We will see later that this identification of the Berry curvature times the ramp velocity as the leading non-adiabatic correction to the generalized force is a universal result. In turn this means that the Berry curvature underlies the Coriolis or Lorentz-type forces, which are well known from elementary physics, as illustrated in \fref{fig:berry_curv_spin}.

\begin{figure}
\includegraphics[width=0.3\columnwidth]{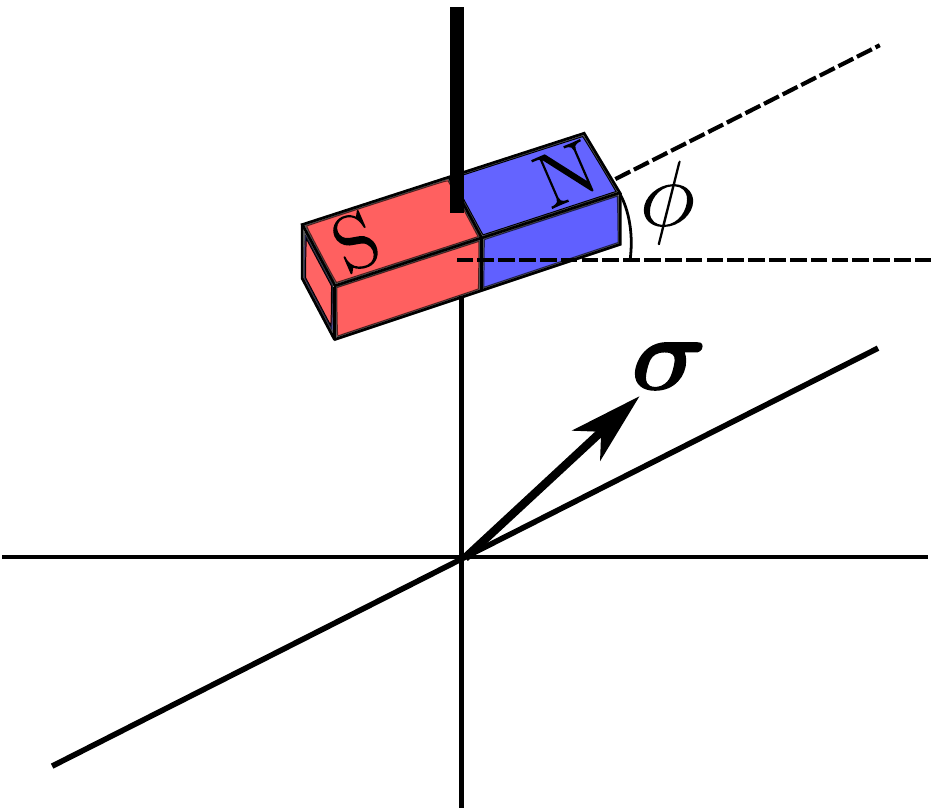}
\caption{Possible experimental realization of the rotating magnetic field. The rotation of the bar magnet is dressed by exciting the quantum spin and leads to mass renormalization (see text).}
\label{fig:mass_renorm_spin}
\end{figure}

In addition to the generalized force, we can look at other observables such as the energy. The leading correction to the energy is of order $\dot \phi^2$ and is given by
\begin{eqnarray}
\Delta E &=& \langle \psi | \mathcal H | \psi \rangle - \langle g | \mathcal H | g \rangle = (|a_g|^2-1)E_g + |a_e|^2 E_e = -|a_e|^2 E_g + |a_e|^2 E_e \nonumber
\\ &\approx& 2\hbar h \left( \dot \phi \frac{\sin \theta}{4h} \right)^2 = \hbar \dot \phi^2 \frac{\sin^2 \theta}{8 h} ~.
\end{eqnarray}
An interesting consequence arises if we ask the question ``where did this extra energy come from?'' For instance, consider a setup as illustrated in \fref{fig:mass_renorm_spin} in which the spin-1/2 is placed below a bar magnet that is rotating without friction around an axis inline with the spin. If the magnet has moment of inertia $I_0$ and the spin is in its ground state, then as we start to rotate the magnet, the spin will attempt to follow its rotation. By conservation of energy, the work $W$ done on the magnet in order to accelerate it to angular velocity $\dot \phi$ will be equal to the total energy change in the full magnet + spin system:
\be
W = \frac{1}{2} I_0 \dot \phi^2 + \Delta E_\mathrm{spin} (\dot \phi) = \frac{1}{2} \left( I_0 + \hbar^2 \frac{\sin^2 \theta}{4 h} \right) \dot \phi^2 ~.
\ee
Clearly the non-adiabatic excitations of the spin appear as an extra moment of inertia for the magnet, $\kappa = \hbar^2 \sin^2 \theta / 4h$, making it appear to have net moment of inertia $I_\mathrm{eff} = I_0 + \kappa$. The additional positive contribution $\kappa$ comes from the dressing of the magnet by the spin. Not surprisingly, this mass renormalization is stronger if the gap in the system gets smaller. So with this simple example, we have seen that doing perturbation theory in a slowly-moving frame yields important corrections to the dynamics, such as an effective force due to the Berry curvature and a renormalization of the moment of inertia of the macroscopic degree of freedom causing the parameter change.  

\begin{figure}
\includegraphics[width=0.1\columnwidth]{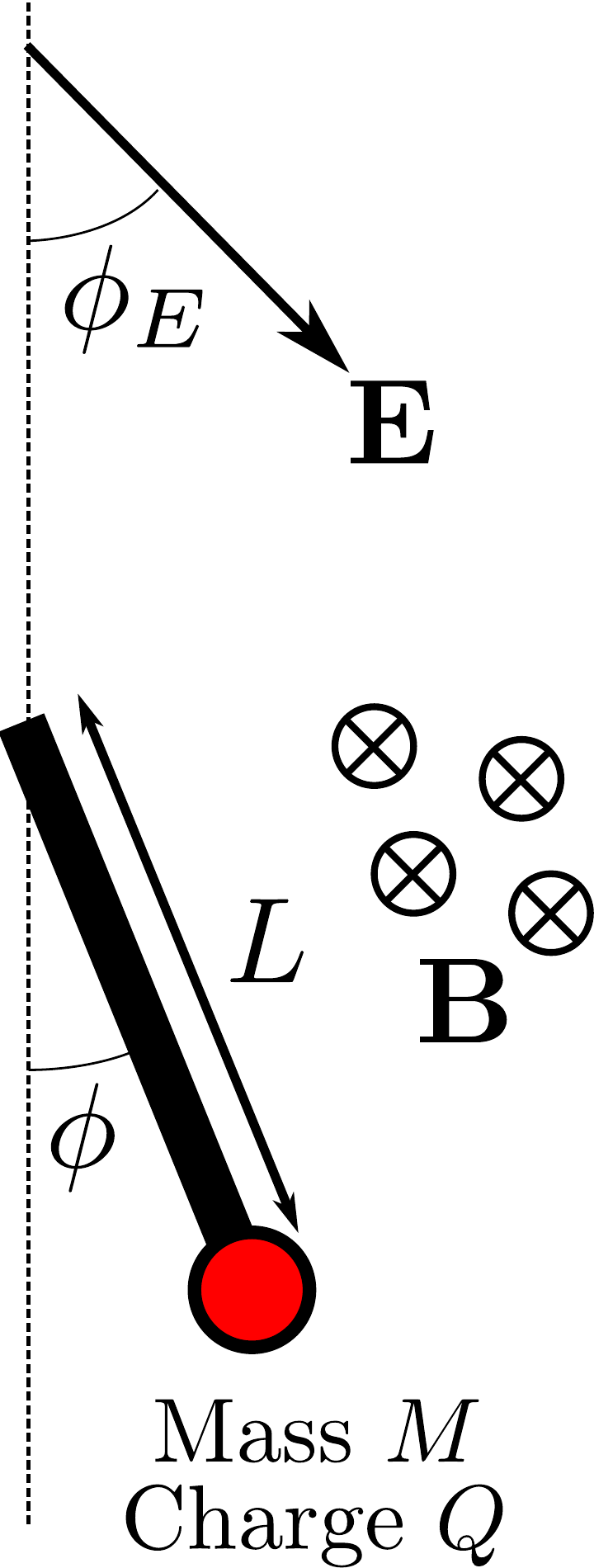}
\caption{A charged pendulum in crossed electric and magnetic fields.}
\label{fig:charged_pendulum}
\end{figure}

The second example we will consider in this introduction is a simple harmonic oscillator with an offset in both its position ($x_0$) and its momentum ($p_0$). This example is also relatively simple, but importantly has a well-defined classical limit. This system is described by the Hamiltonian:
\be
\mathcal H=\frac{(\hat p-p_0)^2}{2 m} + \frac{1}{2} m \omega^2 (\hat x-x_0)^2 ~.
\label{eq:H_shifted_sho}
\ee
We use hat-notation for position and momentum operators to distinguish them from the parameters $x_0$ and $p_0$. Translations in $x_0$ are fairly easy to generate by, for example, moving a harmonic trap or spring. It's not as obvious how one gets  a time-dependent $p_0$. One possibility is to consider a pendulum with a charged particle at the end of it in static electric and magnetic fields (see \fref{fig:charged_pendulum} and \rcite{Koch2007_1}). Then the angle of the electric field shifts the equilibrium position of the pendulum ($x_0$) and the magnetic field prefers a certain angular momentum ($p_0$).

\hrulefill

\exercise{\label{ex:euclid}Show that the setup in \fref{fig:charged_pendulum} gives the Hamiltonian in \eref{eq:H_shifted_sho} with $\hat x \to \hat \phi$, $\hat p \to \hat p_\phi$, $x_0=\phi_E Q E / (mg + Q E)$ and $p_0=L^2 Q B/c$ in the small angle limit $\phi_E,\phi \ll 1$.}

\exerciseshere

\hrulefill

As with the spin-1/2, we want to go to the moving frame in which we know how to diagonalize $\mathcal H$. We can do this with the unitary $U(x_0,p_0) = \mathrm{e}^{-i \hat p x_0 / \hbar} \mathrm{e}^{i \hat x p_0 / \hbar}$, in terms of which $\tilde {\mathcal H}=U^\dagger(x_0,p_0)\mathcal H(x_0,p_0)U(x_0,p_0)=\mathcal H(0,0)=\hbar \omega(\hat n+1/2) \equiv \mathcal H_0$. The moving Hamiltonian is thus 
\be
\mathcal H_m = \mathcal H_0 - \dot x_0 \tilde{\mathcal A}_{x_0} - \dot p_0 \tilde{\mathcal A}_{p_0} ~,
\ee
where 
\beq
\nn
\tilde{\mathcal A}_{x_0} &=& i \hbar U^\dagger \partial_{x_0} U  = i  \hbar \mathrm{e}^{-i \hat x p_0 /  \hbar} \mathrm{e}^{i \hat p x_0 /  \hbar} \partial_{x_0} \left( \mathrm{e}^{-i \hat p x_0 /  \hbar} \mathrm{e}^{i \hat x p_0 /  \hbar} \right) 
\\ \nn & = & i \mathrm{e}^{-i \hat x p_0 /  \hbar} (-i \hat p) \mathrm{e}^{i \hat x p_0 /  \hbar} = \hat p + p_0
\\\tilde{\mathcal A}_{p_0} &=& i  \hbar U^\dagger \partial_{p_0} U = -\hat x ~,
\label{eq:gauge_pot_sho}
\eeq
which stems from the fact that $\hat p$ ($-\hat x$) generates translations in position (momentum).

\begin{figure}
\includegraphics[width=0.3\columnwidth]{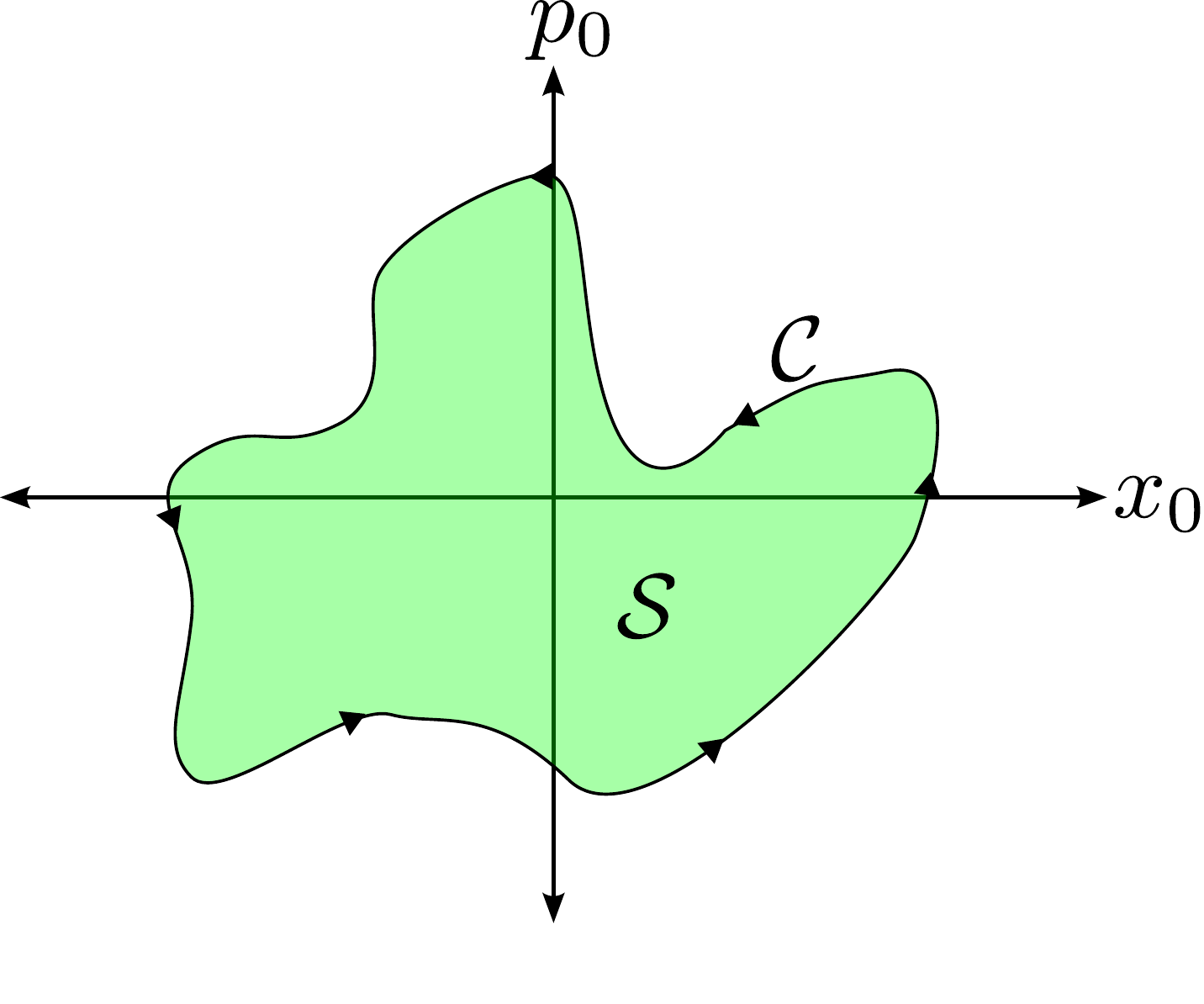}
\caption{The Berry phase of the harmonic oscillator is proportional to the area in phase space enclosed by the path $\left( x_0(t),p_0(t) \right)$.}
\label{fig:area_phase_space}
\end{figure}

For simplicity, let's consider this system for a fixed value of $p_0$. Then $\mathcal H_m = \mathcal H_0 - \dot x_0 (\hat p + p_0)$ and the amplitude to transition from the ground state $|0\rangle$ to the $n$-th eigenstate ($n\neq 0$) of the harmonic oscillator is, at first order in $\dot x_0$,
\be
a_n \approx \frac{\langle n | \left(-\dot x_0 (\hat p+p_0) \right) | 0 \rangle}{E_0 - E_n} ~.
\ee
The only non-zero matrix elements of the $\hat p$ operator connect the state $|0\rangle$ to the state $|1\rangle$, so $a_{n>1}=0$ and 
\be
a_1 \approx \dot x_0 \frac{\langle 1 | \hat p | 0 \rangle}{\omega} = {\dot x_0\over \omega} \bigl< 1 \bigr | \frac{i (\hat a^\dagger -\hat a)} {\ell \sqrt 2}  \bigl | 0 \bigr>= \frac{i \dot x_0}{\omega \ell \sqrt 2} ~,
\ee
where $\ell=\sqrt{ \hbar / m \omega}$ is the natural length scale of the oscillator, and $\hat a^\dagger$ and $\hat a$ are the standard creation and annihilation operators. Using the non-adiabatic corrections to the wave function we can easily find the leading non-equilibrium correction to the generalized force with respect to $p_0$, which should be proportional to the Berry curvature as suggested in \eref{eq:M_theta_single_qubit}:
\begin{multline}
\Delta F = \langle (-\partial_{p_0} \mathcal H) \rangle- \langle 0| (-\partial_{p_0} \mathcal H) |0 \rangle \approx  a_1^\ast \langle 1|(-\partial_{p_0} \mathcal H)|0\rangle+ a_1  \langle 0|(-\partial_{p_0} \mathcal H)|1\rangle\\
=-{2i \dot x_0\over \omega\ell \sqrt{2}}  \langle 1| \hat p|0\rangle=  \dot x_0 \cancelto{1}{\left(\frac{\hbar }{m \omega \ell^2}\right)}=\hbar F_{p_0 x_0} \dot x_0 
\end{multline}
For consistency one can compute Berry curvature directly from the Berry connection. Indeed , from \eref{eq:gauge_pot_sho} it is clear that $A_{x_0}=\langle 0| \mathcal A_{x_0}|0\rangle=p_0$ and $A_{p_0}=0$. Note the apparent asymmetry between the Berry connections is simply a gauge choice.\footnote{As with all gauge potentials, while the the expectation value of the gauge potential can depend on gauge choice, physical observables such as the generalized force are independent of it.}  Then $\hbar F_{p_0 x_0}=\partial_{p_0} A_{x_0}- \partial_{x_0} A_{p_0}=1$. If we consider an arbitrary closed path $\left( x_0(t),p_0(t) \right)$ as depicted in \fref{fig:area_phase_space}, the fact that the Berry curvature is $F_{p_0 x_0}= 1/\hbar$ means that we will get a Berry phase of $\varphi_B =  \hbar^{-1} \int_C A_\lambda \cdot d \lambda =\int_S dx_0 dp_0 F_{p_0 x_0} = \mathrm{Area}_S / \hbar$, i.e., the Berry phase is just area of the phase space trajectory enclosed by $(x_0, p_0)$ in units of $\hbar$. Similarly, the energy of excitations
\be
\Delta E \approx E_1 |a_1|^2 + E_0 (|a_0|^2 - 1) = (E_1 - E_0) |a_1|^2 = \frac{\dot x_0^2}{2} \left( \frac{1}{\omega \ell^2} \right) = \frac{m \dot x_0^2}{2}.
\ee
If the center-of-mass motion of the harmonic oscillator was generated by a trap, we know that by a conservation of energy argument analogous to that for the moment of inertia of the spin-1/2 particle, 
 the trap will feel heavier by an amount equal to the mass of the particle inside this trap. Here we see that this intuitive result comes from the (virtual) excitations of the particle created by the Galilean term.

\hrulefill

\exercise{Consider the harmonic oscillator Hamiltonian from the previous section but now initialized in an arbitrary energy eigenstate $|n\rangle$. Show that as $x_0$ is slowly ramped, the force $-\langle \partial_{p_0} \mathcal H \rangle$ and mass renormalization ($\kappa_{x_0}$) are the same as those in the ground state: $-\langle \partial_{p_0} \mathcal H \rangle = \dot x_0 \implies F_{p_0 x_0} = 1$ and $\kappa_{x_0} = m$.}

\exerciseshere
                   
\hrulefill

The harmonic oscillator can be also analyzed classically, where we just consider the same Hamiltonian with $x$ and $p$ as canonical phase space variables instead of operators. We can apply much of the same machinery to go to a moving frame, solve for corrections to the motion, etc. If the parameters $x_0$ and $p_0$ evolve along some generic path $\left( x_0(t),p_0(t) \right)$, then we can go to the moving frame by performing a canonical change of variables to $p'=p-p_0$ and $x'=x-x_0$. We anticipate that as in the quantum case, the effective Hamiltonian in the moving frame will be
\be
\mathcal H_m =\mathcal H_0-\mathcal A_{p_0}\dot p_0-\mathcal A_{x_0}\dot p_0= \mathcal H_0 + x' \dot p_0 - p' \dot x_0 ~,
\ee
where $\mathcal H_0=p'^2 / 2m + m \omega^2 x'^2 / 2$ and we used the classical limit for the gauge potentials introduced in \eref{eq:gauge_pot_sho}. As we explain in the next section, these gauge potentials in the classical language are simply generators of the canonical transformations to the moving frame, i.e., from $x,p$ to $x',p'$. The equations of motion in the moving frame are thus
\beq
\dot x' = \frac{\partial \mathcal H_m}{\partial p'} = \frac{p'}{m} - \dot x_0 
\label{eq:pprime_classical}
\\ \dot p' =  -\frac{\partial \mathcal H_m}{\partial x'} = - m \omega^2 x' - \dot p_0  
\label{eq:xprime_classical}
\eeq
Note that these equations can be directly obtained by first writing the lab-frame equations of motion and then shifting the phase space variables $x\to x'$, $p\to p'$. Let us again first consider the setup where $p_0$ remains constant and only $x_0$ slowly changes in time. Moreover we assume that we start in a stationary state $x(0)=x_0$ and $p(0)=p_0$. As we know from analytical mechanics, at leading order the adiabatic theorem tells us that the stationary orbit of a classical system with slowly changing parameters maps to another stationary orbit with same adiabatic invariants~\cite{Landau_Lifshitz_Mechanics}\footnote{We will show later that going to the moving frame first and then using conservation of adiabatic invariants is equivalent to finding leading non-adiabatic corrections in the lab frame.}. Since for the chosen initial condition the adiabatic invariant is zero, the adiabatic theorem simply states  that $\dot x'\approx 0$ and $\dot p'\approx 0$. Then \eref{eq:pprime_classical} implies that the particle moves at the same velocity as the potential, i.e., $p^\prime=m\dot x_0$, which is not surprising.  If instead $x_0=0$ and we only ramp $p_0$, then \eref{eq:xprime_classical} yields the slightly less obvious result that the particle is deflected to $x^\prime=-\dot p_0 / m \omega^2$. For the realization of the harmonic oscillator depicted in \fref{fig:charged_pendulum}, this just reflects the fact that a time-varying magnetic field generates an electric field by the Faraday effect which pushes the charged particle to one side. In our language this Faraday force is nothing but the effective Coriolis force due to the Berry curvature by analogy with the quantum case. For example, for a ramp of $p_0$, 
\be
\langle -\partial_{x_0} \mathcal H\rangle= m \omega^2 x'  \dot p_0=- \dot p_0 = \hbar F_{x_0 p_0} \dot p_0~,
\ee
where in the classical case angular brackets imply averaging over the adiabatically connected stationary distribution. Similarly, the additional energy for a ramp of $x_0$ is 
\be
\Delta E = \mathcal H(x=x_0,p=m\dot x_0) - \mathcal H(x=x_0,p=0) = \frac{1}{2} m \dot x_0^2 ~,
\ee
so as expected the mass of the trap generating the center-of-mass motion of the harmonic oscillator is dressed by an amount $\kappa_{x_0}=m$. \add{The harmonic oscillator can also be used to illustrate the ideas of counter-diabatic driving. However, we postpone the corresponding discussion until \sref{sec:shortcuts} because this setup naturally brings additional interesting subtleties related to utilizing the gauge freedom. }

The examples shown in this section illustrate how simple dynamical effects can be obtained using gauge potentials and adiabatic perturbation theory. We have seen that by going into a moving frame, one can derive both classically and quantum mechanically the leading corrections to generalized forces and the energy, from which the Coriolis force related to the Berry curvature and the mass renormalization emerge. \add{Using the spin example, we also illustrated the ideas of counter-diabatic driving allowing one to design  fast protocols in which transitions out of the ground state are suppressed.} Both of these systems are quite simple, and these results could have been obtained via a number of other methods. The power of this formalism comes from its generality. In what follows, we revisit the ideas in this section in their full generality, turning back to these two simple examples as useful illustrations throughout.

\section{Gauge potentials in classical and quantum Hamiltonian systems}
\label{sec:gauge_potentials}

\emph{{\bf Key concept:} Gauge potentials are generators of translations in parameter space. Adiabatic gauge potentials are a special subset of these which diagonalize the instantaneous Hamiltonian, attempting to leave its eigenbasis invariant as the parameter is changed. These adiabatic gauge potentials generate non-adiabatic corrections to the Hamiltonian in the moving frame.}

\subsection{Classical Hamiltonian systems}
\label{sec:gauge_pot_intro_classical}

Classical Hamiltonian systems are defined by specifying the Hamiltonian $\mathcal H$ in terms of a set of canonical variables $p_j, q_j$ satisfying the canonical relations
\be
\{ q_i, p_j\}=\delta_{ij},
\label{eq:classical_comm_rel}
\ee
where $\{\dots\}$ denotes the Poisson bracket:
\be
\{A(\vect p,\vect q), B(\vect p,\vect q)\}=\sum_j \left( {\partial A\over \partial q_j}{\partial B\over \partial p_j}-{\partial B\over \partial q_j}{\partial A\over \partial p_j}\right).
\label{poisson_bracket}
\ee
This choice of canonical variables is arbitrary, as long as they satisfy \eref{eq:classical_comm_rel}. There are therefore many transformations that preserve this Poisson bracket, such as the orthogonal transformation
\be
\vect q=R(\lambda) \vect q_0, \; \vect p=R(\lambda) \vect p_0 ~,
\label{ort_transf}
\ee
where $R$ is an orthogonal matrix ($R^T = R^{-1}$). A general class of transformations which preserve the Poisson brackets are known as canonical transformations~\cite{Landau_Lifshitz_Mechanics}.

In this work, we will mostly be interested in families of canonical transforms that depend continuously on some parameter(s) $\lambda$. It is easy to check that continuous canonical transformations can be generated by functions $\mathcal A_\lambda$ which we refer to as gauge potentials:
\beq
&&q_j(\lambda+\delta\lambda)=q_j(\lambda)-{\partial \mathcal A_\lambda(\lambda,\vect p,\vect q)\over \partial p_j} \delta\lambda ~\implies ~ \frac{\partial q_j}{\partial \lambda} = - \frac{\partial \mathcal A_\lambda}{\partial p_j} =  \{\mathcal A_\lambda,q_j \}\nonumber\\
&&p_j(\lambda+\delta\lambda)=p_j(\lambda)+{\partial \mathcal A_\lambda(\lambda,\vect p,\vect q)\over \partial q_j}\delta\lambda  ~\implies ~ \frac{\partial p_j}{\partial \lambda} = \frac{\partial \mathcal A_\lambda}{\partial q_j} =  \{\mathcal A_\lambda,p_j\},
\label{canonical_transformations}
\eeq
where $\lambda$ parameterizes the canonical transformation\footnote{We use the partial derivative notation for $dq/d\lambda$ and $dp/d\lambda$ because later we consider time evolution such that phase space variables will be functions of both $\lambda$ and $t$} and the gauge potential is an arbitrary function of $q$, $p$, and $\lambda$. We can then see that, up to terms of order $\delta\lambda^2$, the transformation above preserves the Poisson brackets:
\be
\{ q_i(\lambda+\delta\lambda), p_j(\lambda+\delta\lambda)\}=\delta_{ij}-\delta\lambda\left( {\partial^2 \mathcal A_\lambda\over \partial p_j \partial q_i}-{\partial^2 \mathcal A_\lambda\over \partial p_j\partial q_i}\right)+O(\delta\lambda^2)=\delta_{ij}+O(\delta\lambda^2).
\ee

Consider a simple example of such a continuous canonical transformation, $q_i(\vect X) = q_i(0) - X_i$, in which the position coordinate is shifted by $\vect X$ (we use the $\vect X$ notation instead of $\vect \lambda$ notation for the parameters here to highlight their meaning as the coordinate shift). Let's determine the components of the gauge potential $\mathcal A_{X_i}(\vect q, \vect p, \vect X)$ using \eref{canonical_transformations}. First, note that $p_i$ is independent of $\vect X$, meaning that $\partial \mathcal A_{X_i} / \partial q_j=0$. Meanwhile, the $X$-dependence of $q$ gives $\partial q_i / \partial X_j = -\delta_{ij} = -\partial \mathcal A_{X_j} / \partial p_i$. These equations are solved by $\mathcal A_{X_j} = p_j + C_j$, where $C_j$ are arbitrary constants of integration. The presence of these constants in this solution is the first example we will see of a gauge choice that gives these gauge potentials their name.  This is directly analogous to the gauge choice in electromagnetism; indeed one can show that the gauge potentials for canonical shifts of the momentum appear exactly as the electromagnetic vector potential [see \exref{ex:gauge_A_p}]. Gauge potentials generalize these ideas from electromagnetism to arbitrary parameters. 

For fixed canonical variables, Hamiltonian dynamics gives a particular canonical transformation parameterized by time
\be
\dot q_j=-\{ \mathcal H, q_j\}={\partial \mathcal H\over \partial p_j},\;
\dot p_j=-\{ \mathcal H, p_j\}=-{\partial \mathcal H\over \partial q_j}
\label{hamiltonian_equations}
\ee
Clearly these Hamiltonian equations are equivalent to \eref{canonical_transformations} with the convention $\mathcal A_t=-\mathcal H$. In the same way that Hamiltonians generate motion in time, we see that the gauge potentials $\mathcal A_\lambda$ are generators of motion in the parameter space. For instance, we saw that if $\lambda_i=X_i$ corresponds to shifts in position, it is generated by $\mathcal A_{X_i} = p_i$.

\hrulefill

\exercise{\label{ex:gauge_ort_2d} Show that the generator of  rotations around the z-axis,
\[
q_x(\theta)=\cos(\theta) q_{x0}-\sin(\theta) q_{y0},\; q_y(\theta)=\cos(\theta) q_{y0}+\sin(\theta) q_{x0},\;
\]
\[
p_x(\theta)=\cos(\theta) p_{x0}-\sin(\theta) p_{y0},\; p_y(\theta)=\cos(\theta) p_{y0}+\sin(\theta) p_{x0},\;
\]
is the angular momentum operator $\mathcal A_\theta=p_x q_y-p_y q_x$.}

\exercise{\label{ex:dilations} Another particularly important transformation is \emph{dilation}, which plays a central role in renormalization group physics \cite{Kadanoff1966_1,Wilson1975_1,Shankar1994_1} and the closely related gauge-gravity duality \cite{McGreevy2010_1,Ammon2015_1}. This transformation involves dilating real space by a factor $\lambda$, $q(\lambda)=\lambda q_0$, and shrinking momentum space by the same factor, $p(\lambda)=p_0/\lambda$. Show that dilations are canonical transformations, and find the gauge potential $\mathcal A_\lambda$ that generates them.\footnote{As we will see below, dilation transformations often result in rescaling the Hamiltonian, which in turn can be absorbed in rescaling (dilating) time.}}

\exerciseshere

\hrulefill

Canonical transformations are conventionally expressed through generating functions~\cite{Landau_Lifshitz_Mechanics}, which are usually defined as functions of one old variable ($q_0$ or $p_0$) and one new variable ($q(\lambda)$ or $p(\lambda)$). For example, one can use the generating function $G(q_0, q(\lambda), \lambda)$ such that
\[
p_0={\partial G\over \partial q_0},\; p(\lambda)=-{\partial G\over \partial q(\lambda)}.
\]
Differentiating the second equation with respect to $\lambda$ at constant $q(\lambda)$ we find
\be
{\partial p(\lambda)\over \partial \lambda}={\partial \mathcal A_\lambda\over \partial q(\lambda)}=-{\partial^2 G\over \partial q(\lambda)\partial \lambda}.
\ee
Clearly this equation can be satisfied if we choose
\be
\mathcal A_\lambda=-{\partial G\over \partial \lambda}\biggr|_{q_0, q(\lambda)}.
\label{eq:gauge_pot_gen_function}
\ee
Similarly one can check that the gauge potential can be expressed through derivatives of other generating functions expressed through $q_0, p(\lambda)$, $q(\lambda), p_0$ and $p_0, p(\lambda)$. For instance, defining 
\[
G_1(q_0, p(\lambda),\lambda)=G(q_0, q(\lambda), \lambda)+ q(\lambda) p(\lambda)
\] 
such that $q(\lambda)=\partial G_1/\partial p(\lambda)$ we can check that 
\be
\mathcal A_\lambda=-{\partial G_1\over \partial \lambda}\biggr|_{q_0, p(\lambda)}.
\label{eq:gauge_pot_gen_function1}
\ee

Let us illustrate these relations using the example of orthogonal transformations [\eref{ort_transf}]. Differentiating these equations with respect to $\lambda$ we find
\beq
&&{\partial \vect q\over \partial \lambda}={d R\over d\lambda} \vect q_0={d R\over d\lambda} R^T R\,\vect q_0={d R\over d\lambda} R^T \vect q=-{\partial \mathcal A_\lambda\over \partial \vect p}, \nonumber\\
&& {\partial \vect p\over \partial \lambda}={d R\over d\lambda} \vect p_0={d R\over d\lambda} R^T R\vect p_0={d R\over d\lambda} R^T \vect p={\partial \mathcal A_\lambda\over \partial \vect q}
\eeq
It is straightforward to check that the gauge potential for these orthogonal transformations can be chosen to be
\be
\mathcal A_\lambda=-\vect{p}^{\,T}  {d R\over d\lambda} R^T\,\vect q=-p_j {d R_{jk}\over d\lambda} R^T_{ki} q_i
\label{eq:gauge_pot_ort}
\ee
One can also check that the generating function for this orthogonal transformation is
\be
G_1(\vect q_0, \vect p(\lambda),\lambda)=\vect p^T R \vect q_0
\ee
such that 
\[
{\partial G_{1}\over \partial \vect q_0}=\vect p^T R=\vect p_0, \quad {\partial G_{1}\over \partial \vect p}=R \vect q_0=\vect q
\]
Then according to \eref{eq:gauge_pot_gen_function1}
\[
\mathcal A_\lambda=-{\partial G_1\over \partial \lambda}=-\vect p^T {dR\over d\lambda} \vect q_0=-\vect p^T {dR\over d\lambda}R^T \vect q,
\]
which is identical to \eref{eq:gauge_pot_ort}.

\hrulefill

\exercise{Using \eref{eq:gauge_pot_ort}, find the gauge potential $\mathcal A_\lambda$ corresponding to the orthogonal transformation [\eref{ort_transf}] given by the two-dimensional rotation around $z$-axis:
\[
R=\left(\begin{array}{cc} \cos \theta & -\sin \theta \\ \sin \theta & \cos \theta \end{array}\right)
\]
with parameter $\lambda = \theta$ such that
\[
\left(\begin{array}{c} q_x(\theta)  \\ q_y(\theta) \end{array}\right)=R(\theta) \left(\begin{array}{c} q_{x0}  \\ q_{y0} \end{array}\right)
\]
and similarly for $p_x$ and $p_y$. Show that you recover the result of \exref{ex:gauge_ort_2d}
}

\exercise{Generalize the previous exercise to three dimensional rotations. Namely use that in a three dimensional space they can be decomposed into a product of three elementary rotations parameterized by the Euler angles $\alpha,\beta,\gamma$~\cite{Landau_Lifshitz_Mechanics}:
\be
R=R_x(\alpha) R_y(\beta) R_z(\gamma),
\ee
where $R_x, R_y, R_z$ are the rotation matrices
\[
R_x(\alpha)=\left[
\begin{array}{ccc}
1 & 0 & 0\\
0 & \cos\alpha & -\sin\alpha\\
0 & \sin\alpha & \cos\alpha
\end{array}
\right],\,R_y(\beta)=\left[
\begin{array}{ccc}
\cos\beta & 0 & -\sin\beta\\
0 & 1 & 0\\
\sin\beta & 0 & \cos\beta
\end{array}
\right],\, R_z(\gamma)=\left[
\begin{array}{ccc}
\cos\gamma & -\sin\gamma & 0\\
\sin\gamma & \cos\gamma & 0\\
0 & 0 & 1
\end{array}
\right].
\]
Choosing the Euler angles as parameters according to \eref{eq:gauge_pot_ort} find the components of the gauge potentials $\mathcal A_\alpha,\, \mathcal A_\beta,\,\mathcal A_\gamma$ as the functions of the Euler angles.
Show that the gauge potentials corresponding to infinitesimal rotations around $x$, $y$ and $z$ axes, i.e., $\mathcal A_\alpha(\beta=0,\gamma=0)$, $\mathcal A_\beta(\alpha=0,\gamma=0)$ and $\mathcal A_\gamma(\alpha=0,\beta=0)$ are precisely the $x,\, y$ and $z$ components of the angular momentum.
}

\exerciseshere

\hrulefill

As we know well from electromagnetism, when we are dealing with waves it is often convenient to deal with complex canonical variables (wave amplitudes and conjugate momenta). Recalling that normal modes of waves are identical to harmonic oscillators, let us show how one can introduce this complex phase space variables for a single normal mode, parameterized by the parameter $k$ (which in a translationally invariant system would be the momentum). Once these variables are introduced we can use them in arbitrary systems, linear or otherwise. The Hamiltonian for each mode is
\be
\mathcal H_k={p_k^2\over 2m}+{m\omega_k^2\over 2} q_k^2.
\ee
Let us define new linear combinations~\cite{Polkovnikov2010_1}
\be
p_k=i\sqrt{m\omega_k\over 2} (a_k ^\ast-a_k),\; 
q_k=\sqrt{1\over 2m\omega_k}(a_k+a_k^\ast)
\ee
or equivalently
\be
a_k^\ast={1\over \sqrt{2}}\left(q_k\sqrt{m\omega_k}-{i\over \sqrt{m\omega_k}} p_k\right),\;
a_k={1\over \sqrt{2}}\left(q_k\sqrt{m\omega_k}+{i\over \sqrt{m\omega_k}} p_k\right).
\ee
We will refer to $a_k^\ast$ and $a_k$ as coherent state phase space variables, since the eigenstates of the corresponding quantum creation and annihilation operators are precisely coherent states. Next we compute the Poisson brackets of the complex wave amplitudes \be
\{a_k, a_k\}=\{a_k^\ast,a_k^\ast\}=0,\, \{a_k, a_k^\ast\}=-i.
\label{poisson_aastar}
\ee

To avoid dealing with the imaginary Poisson brackets it is convenient to introduce new coherent state Poisson brackets
\be
\{A, B\}_c=\sum_k \left( {\partial A\over \partial a_k}{\partial B\over \partial a_k^\ast}-{\partial B\over \partial a_k}{\partial A\over \partial a_k^\ast} \right) ~,
\ee
where as usual we treat $a$ and $a^\ast$ as independent variables. From this definition it is immediately clear that
\be
\{ a_k, a_q^\ast\}_c=\delta_{kq}.
\ee
Comparing this relation with \eref{poisson_aastar} we see that standard and coherent state Poisson brackets differ by the factor of $i$:
\be
\{\dots\}=-i \{\dots\}_c.
\ee
Infinitesimal canonical transformations preserving the coherent state Poisson brackets can be defined by the gauge potentials:
\be
i{\partial a_k\over \partial \lambda}=-{\partial \mathcal A_\lambda\over \partial a_k^\ast},\quad
i{\partial a^\ast_k\over \partial \lambda}={\partial \mathcal A_\lambda\over \partial a_k}.
\ee

We can write the Hamiltonian equations of motion for the new coherent variables.  For any function of time and phase space variables $A(q,p,t)$  (or equivalently $A(a,a^\ast,t)$) we have
\be
{dA\over dt}={\partial A\over \partial t}+{\partial A\over \partial q} \dot q +{\partial A\over \partial p} \dot p={\partial A\over \partial t}+\{ A, \mathcal H\}={\partial A\over \partial t}-i \{ A,\mathcal H\}_c.
\ee
Let us apply now this equation to coherent state variables $a_k$ and $a_k^\ast$. Using that they do not explicitly depend on time (such dependence would amount to going to a moving frame, which we will discuss later) we find
\be
i {d a_k\over dt}=\{a_k, \mathcal H\}_c={\partial \mathcal H\over \partial a_k^\ast},\;
i {d a_k^\ast\over dt}=\{a_k^\ast, \mathcal H\}_c=-{\partial \mathcal H\over \partial a_k}
\label{GP}
\ee 
These are also known as the Gross-Pitaevskii equations. Note that deriving these equations we did not assume any specific form of the Hamiltonian, so they equally apply to linear and non-linear Hamiltonians.

\hrulefill

\exercise{Check that any unitary transformation $\tilde a_k=U_{k,k'} a_k'$, where $U$ is a unitary matrix, preserves the coherent state Poisson bracket, i.e., $\{\tilde a_k,\tilde a_q^\ast\}_c=\delta_{k,q}$.}

\exercise{Verify that the Bogoliubov transformation \cite{Valatin1958_1,Bogoljubov1958_1}
\be
\gamma_k=\cosh(\theta_k) a_k+\sinh(\theta_k) a_{-k}^\ast,\, 
\gamma_k^\ast=\cosh(\theta_k) a_k^\ast+\sinh(\theta_k) a_{-k},
\label{bogoliubov}
\ee
with $\theta_k=\theta_{-k}$ also preserves the coherent state Poisson bracket, i.e.,
\be
\{\gamma_k,\gamma_{-k}\}_c=\{\gamma_k,\gamma_{-k}^\ast\}_c=0,\; \{\gamma_k,\gamma_k^\ast\}_c=\{\gamma_{-k},\gamma_{-k}^\ast\}_c=1.
\ee
Assume that $\theta_k$ are known functions of some parameter $\lambda$, e.g., the interaction strength. Find the gauge potential $\mathcal A_\lambda=\sum_k \mathcal A_{\lambda,\, k}$, which generates such transformations.}

\exerciseshere

\hrulefill

\subsection{Quantum Hamiltonian systems}
\label{sec:quantum_gauge_pot}

The analogues of canonical transformations in classical mechanics are unitary transformations in quantum mechanics. In classical systems these transformations reflect the freedom of choosing canonical variables while in quantum systems they reflect the freedom of choosing basis states.

The wave function\footnote{This discussion also directly extends to density matrices.} representing some state can be always expanded in some basis:
\be
|\psi\rangle=\sum_n \psi_n |n \rangle_0,
\ee
where $|n\rangle_0$ is some fixed, parameter independent basis. One can always make a unitary transformation to some other basis $|m(\lambda)\rangle=\sum_n U_{nm}(\lambda) |n\rangle_0$ or equivalently $|n\rangle_0=\sum_m U_{nm}(\lambda)^\ast |m(\lambda)\rangle$. Then $|\psi\rangle$ can be rewritten as
\be
|\psi\rangle=\sum_{mn} \psi_n U_{nm}^\ast |m(\lambda)\rangle=\sum_m \tilde \psi_m(\lambda) |m(\lambda)\rangle,
\ee
where $\tilde \psi_m(\lambda)=\langle m(\lambda)|\psi\rangle=\sum_n U_{nm}^\ast \psi_n$, which is equivalent to the vector notation
\[
\tilde \psi=U^\dagger(\lambda) \psi.
\]

We can introduce gauge potentials by analogy with the classical systems as generators of continuous unitary transformations, namely
\be
i \hbar \partial_\lambda |  \tilde \psi(\lambda) \rangle  = i \hbar \partial_\lambda \left(U^\dagger |\psi\rangle \right) = i \hbar \left(\partial_\lambda U^\dagger \right) \left( U |\tilde \psi\rangle \right) = -\tilde{\mathcal A}_\lambda |\tilde \psi \rangle ~,
\label{eq:A_lambda_def}
\ee
where we used the fact that $|\psi \rangle$ is independent of $\lambda$. We use the tilde notation in the gauge potentials to highlight that they act on the rotated wave function $|\tilde \psi\rangle$.
\be
\tilde{\mathcal A}_\lambda = -i \hbar \partial_\lambda U^\dagger U = i \hbar U^\dagger \partial_\lambda U = (\tilde{\mathcal A}_\lambda)^\dagger.
\label{eq:A_lambda_def_1}
\ee
The second equality follows from the fact that
\be
\partial_\lambda(U U^\dagger) = \partial_\lambda \mathds 1 = 0 ~\implies~ U \partial_\lambda U^\dagger =-\partial_\lambda U\, U^\dagger~.
\ee

As in the classical case, the gauge potential generates motion in parameter space. From \eref{eq:A_lambda_def_1} it follows that $\tilde{\mathcal  A}_\lambda$ is a Hermitian operator, which can be formally defined through its matrix elements:
\be
_0\langle n |\tilde{\mathcal A}_\lambda |m\rangle_0=i\hbar\, _0\langle n | U^\dagger \partial_\lambda U |m\rangle_0=i
\hbar \langle n(\lambda) |\partial_\lambda |m(\lambda)\rangle~,
\label{gauge_pot_matr}
\ee
so we can think of the gauge potential in the lab frame, $\mathcal A_\lambda = U \tilde{\mathcal A}_\lambda U^\dagger$, as just $i\hbar$ times the derivative operator $\partial_\lambda$:
\be
\mathcal A_\lambda=i\hbar\partial_\lambda,
\ee
which immediately follows from the fact that $\langle n(\lambda)| \mathcal A_\lambda |m(\lambda\rangle= {}_0\langle n|\tilde{\mathcal A}_\lambda| m\rangle_0$. 

\hrulefill

\exercise{\label{ex:gauge_translations}Verify that the gauge potential corresponding to the translations: $\tilde \psi(x)=\psi(\lambda+x)$ is the momentum operator. Similarly verify that the gauge potential for rotations is the angular momentum operator.}

\exercise{\label{ex:operator_derivative} It is often useful to think in terms of the action of $\tilde{\mathcal A}_\lambda$ on operators instead of wave functions. Consider the case where our basis-changing unitary takes a $\lambda$-dependent operator $\tilde{\mathcal O} (\lambda)$ to a $\lambda$-independent operator $\mathcal O = U \tilde{\mathcal O} U^\dagger = \mathrm{const}(\lambda)$.
\begin{itemize}
\item Show that $\tilde{\mathcal A}_\lambda = i \hbar U^\dagger \partial_\lambda U$ differentiates the operator $\tilde {\mathcal O}$: $[\tilde{\mathcal A}_\lambda, \tilde{\mathcal O}(\lambda)] = - i \hbar \partial_\lambda \tilde{\mathcal O}$. 
\item In the previous problem, shifting the position is equivalent to defining the new operator $X' = \lambda + X$. Show that $\tilde{\mathcal A}_\lambda = P$ satisfies the correct commutation relations, with $\tilde{\mathcal O} = X'$ and $\mathcal O = X$.
\item What is the momentum operator $P'$ in this new basis? Show that it also satisfies the appropriate commutation relations with $\tilde{\mathcal A}_\lambda=P$.
\end{itemize}}

\exercise{Consider the quantum version of the Bogoliubov transformations  discussed in the previous section [\eref{bogoliubov}]. Show that the quantum and classical gauge potentials coincide if we identify complex amplitudes $a_k$ and $a_k^\ast$ with the annihilation and creation operators respectively. Note that you can do this one of two ways:
\begin{itemize}
\item \emph{The easy way} --  Consider a unitary operator $U$ that rotates the $\theta$-dependent number operator $\gamma_k^\dagger \gamma_k$ to the original basis $a_k^\dagger a_k$. Show that $\tilde{\mathcal A}_\theta$ satisfies the appropriate commutation relation with $\gamma_k$, using the results of the previous exercise.
\item \emph{The hard way} --  Verify by direct computation on the $\theta$-dependent number eigenstates $|n_k,n_{-k}\rangle \propto \left( \gamma_k^\dagger \right)^{n_k} \left( \gamma_{-k}^\dagger \right)^{n_{-k}} |\Omega_\gamma\rangle$, where $|\Omega_\gamma\rangle$ is the $\gamma$ vacuum, that the matrix elements satisfy $\langle m_k, m_{-k} | \tilde{\mathcal A}_\theta | n_k, n_{-k} \rangle = i\hbar \langle m_k, m_{-k} | \partial_\theta | n_k, n_{-k} \rangle $.
\end{itemize}}

\exerciseshere
\hrulefill

\subsection{Adiabatic gauge potentials}

Up to now gauge potentials we introduced were generators of arbitrary continuous canonical (classical) or unitary (quantum) transformations. In these notes we will be particularly interested in a special class of such gauge potentials, which we call adiabatic. It is easier to introduce them using the language of quantum mechanics first and then extend their definition to classical systems.

Imagine that we are dealing with a family of Hamiltonians $\mathcal H(\lambda)$ parameterized by some continuous parameter $\lambda$. We will assume that the Hamiltonians are non-singular and differentiable. At each value of $\lambda$, these Hamiltonian are diagonalized by a set of eigenstates  $|m(\lambda)\rangle$, which we call the adiabatic basis.  When the parameters $\lambda$ are varied adiabatically (infinitely slowly), these states are related by the adiabatic theorem. For the time being, we assume no degeneracies ensuring that the basis states are unique up to a phase factor.\footnote{More generally one can define an adiabatic basis as a family of adiabatically connected eigenstates, i.e., eigenstates related to a particular initial basis by adiabatic (infinitesimally slow) evolution of the parameter $\lambda$. For example, if two levels cross they will exchange order energetically but the adiabatic connection will be non-singular.} These basis states $|m(\lambda)\rangle$ are related by a particular adiabatic unitary transformation and we call the associated gauge potentials \emph{adiabatic gauge potentials}. Such potentials satisfy several properties, which we are going to use later. 

First let us note that the diagonal elements of the adiabatic gauge potentials in the basis of $\mathcal H(\lambda)$ are, by definition, the Berry connections
\[
A_\lambda^{(n)}=\langle n(\lambda)|\mathcal A_\lambda|n(\lambda)\rangle=i\hbar \langle n(\lambda)|\partial_\lambda |n(\lambda)\rangle.
\] 
Recall that, up to a sign, the gauge potential with respect to time translations is the Hamiltonian itself, and its expectation value is negative the energy. Another important property of these gauge potentials can be found by differentiating the identity
\[
\langle m| \mathcal H(\lambda) |n\rangle = 0\; \mbox{for}\; n\neq m
\]
with respect to $\lambda$:
\beq
0  & = & \langle \partial_\lambda m | \mathcal H | n \rangle + \langle m | \partial_\lambda \mathcal H | n \rangle + \langle m | \mathcal H | \partial_\lambda n \rangle 
\\ \nn &=& E_n \langle \partial_\lambda m | n \rangle + E_m \langle m | \partial_\lambda n \rangle + \langle m | \partial_\lambda \mathcal H | n \rangle = (E_m - E_n) \underbrace{\langle m | \partial_\lambda n \rangle}_{-i/\hbar \langle m | \mathcal A_\lambda | n \rangle} + \langle m | \partial_\lambda \mathcal H | n \rangle 
\\ \implies ~~ \langle m | \mathcal A_\lambda | n \rangle &=& i \hbar\frac{\langle m | \partial_\lambda \mathcal H | n \rangle}{E_n - E_m} ~,
\label{eq:gauge_pot_matrix_elem}
\eeq
\add{where it is important to note that all quantities -- the eigenstates $|n\rangle$, the Hamiltonian $\mathcal H$ and the energies $E_n$ -- depend on $\lambda$}.
This relation can be also written in the matrix form~\cite{Jarzynski2013_1}
\be
i\hbar \partial_\lambda \mathcal H=[\mathcal A_\lambda, \mathcal H]-i \hbar {M}_\lambda,
\label{eq:gauge_pot_matrix_rel}
\ee
where
\be
M_\lambda=-\sum_n {\partial E_n(\lambda)\over \partial \lambda} |n(\lambda)\rangle\langle n(\lambda)|
\label{defM}
\ee
is an operator that is diagonal in the instantaneous energy eigenbasis, whose values are the generalized forces corresponding to different eigenstates of the Hamiltonian.

From \eref{eq:gauge_pot_matrix_rel} and the fact that $[\mathcal H,M_\lambda]=0$, we immediately see that the adiabatic gauge potentials satisfy the following equation:
\be
[\mathcal H, i\hbar \partial_\lambda \mathcal H-[\mathcal A_\lambda, \mathcal H]]=0.
\label{eq:gauge_pot_commutator}
\ee
This equation can be used to find the adiabatic gauge potentials directly without need of diagonalizing the Hamiltonian. Note that as the spectrum of the Hamiltonian is invariant under gauge transformations, $M_\lambda$ is also gauge invariant. One may use this to reverse the previous arguments and show that \eref{eq:gauge_pot_commutator} implies \eref{eq:gauge_pot_matrix_rel}.

Because energy eigenstates are not well-defined in classical systems, we cannot directly extend \eref{eq:gauge_pot_matrix_elem} to define the classical adiabatic gauge potential. However, we can instead use the matrix relation \eref{eq:gauge_pot_matrix_rel}, recalling that in the classical limit the commutator between two operators corresponds to the Poisson bracket between corresponding functions: $[\dots]\to i\hbar\{\dots \}$. Then classical adiabatic gauge potentials must satisfy~\cite{Jarzynski2013_1}
\be
-\partial_\lambda \mathcal H=M_\lambda -\{\mathcal A_\lambda,\mathcal H\},
\label{classical_adiabatic_gp_relation}
\ee
where $M_\lambda$ is the classical generalized force, which is formally defined as the average of $-\partial_\lambda \mathcal H$ over time. This can be seen by recalling that time-averaging is the classical analogue of the quantum average over stationary eigenstates. In the non-chaotic systems that we are focusing on, this time average is equivalent to the average over the stationary orbit containing phase space points $\vect p, \vect q$. Note that because $M_\lambda$ is an averaged object, it depends only on conserved quantities like energy. 

\eref{eq:gauge_pot_commutator} can likewise be immediately extended to classical systems, serving as a practical tool for finding adiabatic gauge potentials in classical systems
\be
\{\mathcal H, \partial_\lambda \mathcal H-\{\mathcal A_\lambda, \mathcal H\}\}=0.
\label{eq:gauge_pot_poisson_bracket}
\ee
This equation has a useful implication that the Poisson bracket of the ``new'' Hamiltonian $\mathcal H'\equiv \mathcal H(\lambda+\delta\lambda)$ written in ``new'' coordinates $q'=q(\lambda+\delta\lambda), \; p'=p(\lambda+\delta \lambda)$ with the ``old'' Hamiltonian $\mathcal H(\lambda)$ written in the ``old'' coordinates $q(\lambda), \; p(\lambda)$ vanishes:
\be
{\partial \mathcal H'\over \partial q'}{\partial \mathcal H\over \partial  p}-{\partial \mathcal H'\over \partial p'}{\partial \mathcal H\over \partial  q}=0.
\label{eq:gauge_pot_poisson_bracket1}
\ee
This relation implies that if one finds the canonical transformation which keeps the Hamiltonian effectively invariant as a function of $\lambda$, then the gauge potential generating this transformation is precisely the adiabatic gauge potential. 

To prove \eref{eq:gauge_pot_poisson_bracket1}, we start by noting that 
\be
\mathcal H'\equiv \mathcal H(q, p, \lambda+\delta\lambda)\approx \mathcal H(q,p,\lambda)+\partial_\lambda \mathcal H\delta \lambda=\mathcal H(q,p,\lambda)-M_\lambda \delta \lambda +\{\mathcal A_\lambda, \mathcal H\}\delta \lambda,
\ee
where the last equality follows from \eref{classical_adiabatic_gp_relation}. From the definition of the gauge potentials and the fact that the transformation from $q', p'$ to $q,p$ is inverse of that from $q,p$ to $q',p'$, one finds
\be
\mathcal H(q,p,\lambda)\approx \mathcal H(q',p',\lambda)-\{\mathcal A_\lambda, \mathcal H\}\delta \lambda.
\ee
Combining these two equations we see that 
\be
\mathcal H'\approx \mathcal H(q',p',\lambda)-M_\lambda\delta \lambda~.
\ee
Therefore to linear order in $\delta \lambda$,
\be
{\partial \mathcal H'\over \partial q'}{\partial \mathcal H\over \partial  p}-{\partial \mathcal H'\over \partial p'}{\partial \mathcal H\over \partial  q}=\{\mathcal H-M_\lambda\delta\lambda,\mathcal H\}=0,
\ee
where we used that $M_\lambda$ is a function of only the energy and other conserved quantities, and thus $\{M_\lambda,\mathcal H\}=0$.  

To illustrate this idea let us consider a simple example of the Hamiltonian
\be
\mathcal H={p_0^2\over 2m}+V(q_0-\lambda)
\ee
The canonical transformation $q(\lambda)=q_0-\lambda$ and $p(\lambda)=p_0$ clearly keeps the Hamiltonian effectively independent of $\lambda$:
\[
\mathcal H={p(\lambda)^2\over 2m}+V(q(\lambda))
\]
such that \eref{eq:gauge_pot_poisson_bracket1} is trivially satisfied. Thus in this case the corresponding gauge potential is the adiabatic gauge potential. If we take a slightly more complicated example
\be
\mathcal H={p_0^2\over 2m}+{\lambda^2 \over 2} q_0^2
\ee
and perform the canonical (dilation) transformation to new variables $q(\lambda)=q_0/\sqrt{\lambda}$, $p(\lambda)=p_0\sqrt{\lambda}$ then the resulting family of Hamiltonians reads 
\be
\mathcal H=\lambda\left({p(\lambda)^2\over 2m}+{1 \over 2} q(\lambda)^2\right).
\ee
These Hamiltonians are not identical at different values of $\lambda$, but they differ only by an overall scale factor. As a result \eref{eq:gauge_pot_poisson_bracket1} is still satisfied. Therefore the gauge potential corresponding to this particular canonical transformation is the adiabatic gauge potential. An additional example of an adiabatic gauge potential found in a similar way is discussed in \sref{sec:geom_classical_syst}.

\add{Since the main purpose of these lectures is to connect the geometric notion of gauge potentials to dynamical quantities that derive from the Hamiltonian, we will be almost exclusively concerned with the adiabatic gauge potential. Thus, for the remainder of the review, we will often use the term ``gauge potential'' to mean adiabatic gauge potential unless otherwise specified. In other words, unitary changes of the energy eigenbasis will play a special role throughout the remainder of these notes.}

\subsection{Hamiltonian dynamics in the moving frame: Galilean transformation}

Gauge potentials are closely integrated into Hamiltonian dynamics. We will see that they naturally appear not only in gauge theories like electromagnetism, but also in other problems where we attempt a time-dependent change of basis. We will come to issues of gauge invariance later, but for now we simply note that the equations of motion should be invariant under these gauge transformations. Indeed we can describe the same system using an arbitrary set of canonical variables in the classical language or an arbitrary basis in the quantum language.

\subsubsection{Classical systems}

Let us first consider the classical equations of motion of some system described by a Hamiltonian that possibly depends on time for the canonical variables $q_i(\lambda,t)$ and $p_i(\lambda,t)$, where as before the index $i$ runs over both the particles and spatial components of the coordinates and momenta. If $\lambda=\lambda_0$ is time independent, then we are dealing with normal Hamiltonian dynamics
\be
\left[{d q_i\over d t}\right]_{l}=\{q_i,\mathcal H\},\; \left[{d p_i\over d t}\right]_{ l}=\{p_i, \mathcal H\},
\label{hamilt_eom}
\ee
where the subindex $l$ implies that the derivative is taken in the lab frame at $\lambda=\lambda_0$. Now let us consider the moving frame, where $\lambda$ also depends on time, i.e., not only do the variables $q_i$ and $p_i$ evolve in time but also their very definition changes with time. Then using \eref{canonical_transformations} we find
\beq
&&\left[{d q_i\over d t}\right]_{ m}=\left[d q_i\over d t\right]_{ l}+\dot \lambda {\partial q_i\over \partial \lambda}=\{q_i,\mathcal H\}-\dot \lambda \{q_i,\mathcal A_\lambda\}=\{q_i,\mathcal H_m\}\nonumber\\
 &&\left[{d p_i\over d t}\right]_{m}=\left[d p_i\over d t\right]_{l}+\dot \lambda {\partial p_i\over \partial \lambda}=\{p_i,\mathcal H\}-\dot \lambda \{p_i,\mathcal A_\lambda\}=\{p_i,\mathcal H_m\},
 \label{eq:galclasseom}
\eeq
where the subindex $m$ at time derivative highlights that it is taken in the moving frame and we defined the effective moving frame Hamiltonian by generalizing the Galilean transformation
\be
\mathcal H_m=\mathcal H -\dot \lambda \mathcal A_\lambda~.
\label{galilean}
\ee
We thus see that the equations of motion in the moving frame preserve their Hamiltonian nature. If $\lambda$ stands for a shift of the $x$-coordinate of the reference frame, $q_x(\lambda)= q_x(\lambda_0)-\lambda$, then as we discussed earlier $\mathcal A_\lambda=p_x$ and the expression above reduces to the standard Galilean transformation. If $\lambda$ stands for the angle of the reference frame with respect to the lab frame, then $\mathcal A_\lambda$ is the angular momentum and we recover the Hamiltonian in the rotating frame. If $\lambda$ is the dilation parameter then
\be
\mathcal H_m=\mathcal H-{\dot\lambda\over \lambda} \sum_j q_j p_j ~,
\ee
using the form of the gauge potential derived in \exref{ex:dilations}.

It is instructive to re-derive the equations of motion in a moving frame using a slightly different, but equivalent approach. Consider the equations of motion in terms of the lab frame coordinates $q_0$ and $p_0$:
\be
{d q_0\over dt}=\{q_0, \mathcal H\},\; {d p_0\over dt}=\{p_0, \mathcal H\}
\label{hamilt_eom}
\ee
Now let us go to the moving frame, i.e., let us find the analogous equations of motion in terms of canonical variables $q(q_0,p_0,\lambda,t)$ and $p(q_0,p_0,\lambda,t)$. By the chain rule,
\be
{dq\over dt}={\partial q\over \partial t}+\dot \lambda {\partial q\over \partial \lambda}+\dot q_0 \frac{\partial q}{\partial q_0} + \dot p_0 \frac{\partial q}{\partial p_0}= {\partial q\over \partial t}+\dot \lambda \{A_\lambda,q\} - \{\mathcal H,q\} ~,
\ee
and similarly for $p$ (and for each component of multi-dimensional $\vect q$ and $\vect p$). For the majority of the cases we consider, the basis choice will not depend explicitly on time (only implicitly through $\lambda$), so that $\partial_t q = \partial_t p = 0$. Then we see that the equations of motion in the moving frame reduce to \eref{eq:galclasseom}.

\hrulefill

\exercise{\label{ex:gauge_A_p}Find the gauge potential $\mathcal A_p$ corresponding to the translations of the momentum $p=p_0+\partial_q g(q,\lambda)$, $q=q_0$, where $g(q,\lambda)$ is an arbitrary function of the coordinate $q$ and the parameter $\lambda$, which in turn can depend on time. Show that the moving frame Hamiltonian with the Galilean term amounts to the standard gauge transformation in electromagnetism, where the vector potential (which adds to the momentum) is transformed according to $\Lambda \to \Lambda +\nabla_q f$ and the scalar potential (which adds to the energy) transforms as  $V \to V-\partial_t f$. Find the relation between the gauge potential $\mathcal A_p$ and the function $f$. We use the notation $\Lambda$ for the electromagnetic vector potential to avoid confusion with the parameter-dependent gauge potential $\mathcal A$.}

\exerciseshere

\hrulefill

It is interesting to note that the Galilean transformation can be understood from an extended variational principle, where the equations of motion can be obtained by extremizing the action in the extended parameter space-time
\be
S=\int \left[ p_i\,dq_i - \mathcal H dt +\mathcal A_\lambda d\lambda\right]
\label{eq:gen_action}
\ee
with respect to all possible trajectories $p_i(\lambda,t)$, $q(\lambda,t)$ satisfying the initial conditions. The derivation is a straightforward generalization of the standard variational procedure found in most textbooks, cf. Ref.~\cite{Landau_Lifshitz_Mechanics}. Extremizing the action at constant time $t$ clearly gives back the canonical transformations [\eref{canonical_transformations}]. Extremizing this action at constant $\lambda$ with respect to time reproduces the Hamiltonian equations of motion in the lab frame. If we extremize the action along some space time trajectory $\lambda(t)$ such that $d\lambda=\dot \lambda dt$ we will clearly reproduce the Hamiltonian equations of motion with the Galilean term [\eref{galilean}]. 

While we are focusing in these notes on conventional canonical transformations,  it is easy to use this variational formalism to consider more general transformation where we treat time on equal footing with the other parameters. In particular, we can consider a transformation that maps $(t, \lambda)$ to new coordinates $(\tau, \mu)$.  When the parameters represent translations in the physical coordinates themselves, this class of transformations are space-time transformation that mix time and space degrees of freedom, such as Lorentz transformations in special relativity. For these extended class of transformations, both $\lambda$ and $t$ can be viewed as functions of $\mu$ and $\tau$. By noting that $d\lambda = {\partial \lambda \over \partial \mu}d\mu +  {\partial \lambda \over \partial \tau}d\tau$ and $dt = {\partial t \over \partial \mu}d\mu +  {\partial t \over \partial \tau}d\tau$, we can rewrite the action [\eref{eq:gen_action}] as
\be
S=\int \left[ p_i\,dq_i - d\tau \left( \mathcal H {\partial t\over \partial \tau}- \mathcal A_\lambda {\partial \lambda\over \partial\tau}\right)  +d\mu \left( \mathcal A_\lambda {\partial \lambda\over \partial \mu}-\mathcal H {\partial t\over \partial \mu}\right)\right]
\label{eq:gen_action1}
\ee
It is clear that in this generalized moving frame defined by $(\tau, \mu)$ the Hamiltonian and the gauge potential are given by
\[
 \mathcal H_\tau =\mathcal H {\partial t\over \partial \tau}- \mathcal A_\lambda {\partial \lambda\over \partial\tau},\quad
\mathcal A_\mu= \mathcal A_\lambda {\partial \lambda\over \partial \mu}-\mathcal H {\partial t\over \partial \mu}.
\quad 
\]
and the corresponding equations of motion are given by [compare to \eref{eq:galclasseom}]
\be
{\partial q_i\over \partial \tau}=\{q_i,\mathcal H_\tau\}-{d\mu\over d\tau} \{q_i, \mathcal A_\mu\},\quad
{\partial p_i\over \partial \tau}=\{p_i,\mathcal H_\tau\}-{d\mu\over d\tau} \{p_i, \mathcal A_\mu\}
\ee
For the choice, $t=\cosh(\theta) \tau-\sinh(\theta)\mu$ and $\mu=\cosh(\theta) \lambda-\sinh(\theta) \tau$, with $\theta$ being a constant, the expressions above reproduce Lorentz transformations.
Gauge transformations give us a lot of freedom to choose the moving frame that gives the simplest equations of motion by utilizing symmetries of the Hamiltonian; they allow us to map superficially different problems onto each other, \add{map time dependent to time independent problems, and so on}. 

So far our discussion of equations of motion focused on arbitrary gauge transformations. In these notes we, however, are particularly interested in adiabatic gauge transformations. Dynamics in such special adiabatic moving frames is ideally suited for developing adiabatic perturbation theory (discussed in detail in \sref{sec:emergent_newtonian_dynamics}). For now, let us show that the adiabatic gauge potentials are responsible for non-adiabatic corrections to the energy change of the system:
\be
E(t)=\int dq_i\, dp_i\, \mathcal H (q_i, p_i,\lambda) \rho(q_i, p_i, t),
\label{eq:E(t)}
\ee
where $ \rho(q_i, p_i, t)$ is the time dependent density matrix. There is a word of caution needed in understanding this integral as it is usually the subject of confusion. The Hamiltonian and the integration variables here go over all phase space and in this sense they are time independent, while the density matrix evolves according to the Hamiltonian dynamics.  
Thus, if we choose our integration variables $q$ and $p$ to denote lab-frame coordinates, we get
\begin{eqnarray*}
{d\mathcal H (q_i, p_i,\lambda)\over dt}&=&{\partial \mathcal H\over \partial \lambda}\dot \lambda=\dot \lambda \left(\{\mathcal A_\lambda,\mathcal H\}-M_\lambda\right) \\ 
\frac{d \rho(q_i, p_i, t)}{dt} &=& \{\rho,H\}
\end{eqnarray*}
where we have rewritten $\partial_\lambda \mathcal H$ using \eref{classical_adiabatic_gp_relation}. Putting this back into \eref{eq:E(t)} and employing the cyclicity of the integral
\[
\int dq dp\, A(q,p)\{B(q,p),C(q,p)\}=\int dq dp\, B(q,p)\{C(q,p),A(q,p)\}~
\]
for any $A$, $B$, and $C$, we find
\begin{eqnarray}
\nonumber
{dE\over dt}&=&\int dq_i dp_i\, \left[ \dot \lambda \left( \{\mathcal A_\lambda, \mathcal H \} - M_\lambda \right) \rho + \mathcal H \{\rho, \mathcal H\} \right]
\\ & = & -\dot \lambda \int dq_i dp_i\,\rho\, M_\lambda + \dot \lambda \int dq_i dp_i\, \mathcal A_\lambda \{\mathcal H,\rho \} + \int dq_i dp_i \rho \cancelto{0}{\{\mathcal H,\mathcal H\}} ~.
\label{eq:dE_dt}
\end{eqnarray}
The first term here gives the standard adiabatic work averaged over the distribution function, as $M_\lambda$ is the generalized force. The second term 
gives the non-adiabatic corrections. In particular, it vanishes to the leading order in $\dot\lambda$ if the initial density matrix is stationary, since then $\{\rho_0,\mathcal H\}=0$. So the leading non-adiabatic contribution in the last term in \eref{eq:dE_dt} due to the adiabatic gauge potential is of order $\dot\lambda^2$  or higher.

\subsubsection{Quantum systems}

Very similar analysis goes through for the quantum systems. Interestingly the derivations for quantum systems are even simpler than for classical ones. Consider the Schr\"odinger equation
\be
i \hbar d_t | \psi \rangle=\mathcal H | \psi \rangle
\ee
after the transformation to the moving frame: $|\psi \rangle=U(\lambda) |\tilde \psi \rangle$:
\be
i \hbar \dot\lambda (\partial_\lambda  U) |\tilde \psi \rangle+i \hbar U \partial_t |\tilde \psi\rangle= \mathcal H U |\tilde\psi\rangle
\ee
Multiplying both sides of this equation by $U^\dagger$ and moving the first term in the L.H.S. of this equation to the right we find
\be
i \hbar d_t |\tilde \psi \rangle=\left[U^\dagger \mathcal H U-\dot \lambda \tilde{\mathcal A}_\lambda\right] |\tilde \psi\rangle=\left[\tilde{ \mathcal H} -\dot \lambda \tilde{\mathcal A}_\lambda\right] |\tilde \psi\rangle=\tilde{\mathcal H}_m|\tilde \psi\rangle~.
\label{eq:schr_eq_moving_frame}
\ee
Here $\tilde{\mathcal H} = U^\dagger \mathcal H U$ is the original Hamiltonian written in the moving basis while $-\dot \lambda \tilde{\mathcal A}_\lambda$ is the Galilean term. The moving Hamiltonian $\mathcal H_m=\mathcal H-\dot\lambda \mathcal A_\lambda$ thus retains the same form as in classical systems.

In the adiabatic moving frame it is easy to get the analogue of \eref{eq:dE_dt} by differentiating energy with respect to time and using \eref{eq:schr_eq_moving_frame}:
\be
{dE\over dt}={d\over dt}\langle \tilde \psi| \tilde {\mathcal H}|\tilde \psi\rangle = \dot \lambda \langle \tilde \psi| \partial_\lambda\tilde {\mathcal H}|\tilde \psi\rangle-{i\over \hbar}\dot \lambda \langle \tilde \psi | [\tilde{\mathcal A}_\lambda, \tilde{\mathcal H}|\tilde \psi\rangle=-\dot \lambda M_\lambda--{i\over \hbar}\dot \lambda \langle  \psi | [\mathcal A_\lambda, \mathcal H| \psi\rangle.
\ee
Because the Hamiltonian $\tilde {\mathcal H}$ is diagonal in the instantaneous basis, the expectation value $\langle \tilde \psi| \partial_\lambda\tilde {\mathcal H}|\tilde \psi\rangle$ is nothing but negative the generalized force defined in \eref{defM}. We also note that expectation value of any operator is invariant under choice of basis so we can remove tilde sign in all final expressions. Let us point out that the second term vanishes at leading order in $\dot\lambda$ if the initial state is an energy eigenstate. Then up to higher orders in $\dot\lambda$, $|\psi\rangle$ remains the eigenstate of the instantaneous Hamiltonian and the expectation value of the commutator $[\mathcal A_\lambda,\mathcal H]$ vanishes. It is easy to see that these considerations apply to mixed states as well.

From \eref{eq:schr_eq_moving_frame} we see that one gets a generalized Galilean correction to the Hamiltonian which is proportional to the gauge potential $\mathcal A_\lambda$. Later in these lectures, we will see many ways in which this term manifests in the dynamics of closed systems. We will do so after discussing in more detail the geometric properties that are encoded in the $\mathcal A_\lambda$ operator.

\subsection{Counter-diabatic driving}
\label{sec:shortcuts}

\add{We will conclude this section by briefly discussing another interesting application of the adiabatic gauge potentials, which has been termed in literature in three different ways: counter-diabatic driving, shortcuts to adiabaticity and transitionless driving~\cite{Demirplak2003_1,Demirplak2005_1, Berry2009_1, Campo2012_1,Campo2013_1,Torrontegui2011_1,Jarzynski2013_1,Deffner2014_1,Acconcia2015_1,Karzig2015_1, SelsArxiv2016}. We will stick with the term counter-diabatic driving, which was introduced in the first paper by M. Demirplak and S. Rice~\cite{Demirplak2003_1,Demirplak2005_1}. We will not attempt to review this already extensive literature, instead focusing on two simple but illustrative examples. In the next section we will comment how these ideas can be extended to complex systems using approximate adiabatic gauge potentials. For the remainder of this section we will use the quantum language, having already explained the equivalence to classical systems.}

In the previous section we discussed that the Galilean term is the one causing transitions between energy levels of the original Hamiltonian $\mathcal H$. In order to eliminate these transitions\add{,} it suffices to simply add a counter-diabatic (or counter-Galilean) term $\dot \lambda \mathcal A_\lambda$ to the driving protocol such that the system evolves under the Hamiltonian
\be
\mathcal H(\lambda)+\dot \lambda \mathcal A_\lambda(\lambda).
\ee
\add{Then if we go to the moving frame of the original Hamiltonian $\mathcal H$,} the effective Hamiltonian will be
\be
\tilde {\mathcal H}_m=\tilde {\mathcal H}-\dot \lambda \tilde {\mathcal A}_\lambda+\dot\lambda \tilde {\mathcal A}_\lambda=\tilde {\mathcal H}.
\ee
By definition $\tilde {\mathcal H}$ is the diagonalized version of $\mathcal H$. Thus, if the system is initially prepared in a stationary distribution, say the ground state, it will always remain in such a stationary distribution no matter how fast the protocol is. \add{We now study this for a variety of examples to see how counter-diabatic driving can be realized in practice.}

\subsubsection{Particle in a moving box}

We will start from the simplest and most intuitive example of a particle of mass $m$ confined to some potential $V(q-X)$ which depends on the parameter $X$, say the minimum of this potential. \add{This setup generalizes our earlier example from \sref{sec:invitation} beyond a harmonic potential}. We want to move this potential from some initial point $X_0$ to final point $X_1$ without exciting the particle inside. From \exref{ex:gauge_translations} we know that the gauge potential corresponding to translations is the momentum operator: $\tilde{\mathcal A}_X=\mathcal A_X=\hat p$. Thus the counter-diabatic term should be $\dot X \hat p$ such that the desired time-dependent Hamiltonian is
\be
\mathcal H_{\rm CD}={\hat p^2\over 2m}+V(q-X(t))+\dot X(t) \hat p.
\label{eq:counter_galilean1}
\ee
\add{This Hamiltonian is hard to realize because one needs to couple directly to the momentum operator [cf. \exref{ex:euclid}]. This is not always easy, especially for neutral particles. However, one can bring this Hamiltonian to a more familiar form by making another gauge transformation corresponding to the momentum shift $\hat{p}\to \hat p'=\hat p+ m\dot X$ [see \exref{ex:gauge_A_p}]} resulting in a moving frame Hamiltonian
\be
\mathcal H'_{\rm CD}={\hat p'^2\over 2m}+V(q-X(t))-m \ddot X(t) q-{m \dot X^2(t)\over 2}.
\label{eq:counter_galilean2}
\ee
The last term here, which comes from completing the square\add{,} can be omitted as it does not depend on $q$ and $p$. The third term is nothing but an effective gravitational field, which compensates the acceleration of the system. In the classical language this term ensures that in the accelerating frame there are no additional forces acting on the particle due to the acceleration.

\add{It is important to emphasize that the Hamiltonians $\mathcal H'$ and $\mathcal H$ are related by the unitary transformation, which defines the gauge choice. Therefore, under evolution with $\mathcal H'_{\rm CD}$ the system {\em will not} follow the ground state of the desired Hamiltonian $\mathcal H$, but rather the ground state of the gauge equivalent Hamiltonian $\mathcal H^\prime$. This is intuitively clear: the ground state of $\mathcal H$ at any fixed $X$ has a zero momentum while the particle following the ground state of $\mathcal H'$ clearly has non-zero velocity and momentum in the lab frame. However, note that $\hat p$ and $\hat p'$ coincide whenever $\dot X=0$, which is equivalent to the statement that the gauge transformation between the Hamiltonians $\mathcal H$ and $\mathcal H'$ reduces to the identity.  Therefore the protocol in \eref{eq:counter_galilean2} results in the system following the ground state of $\mathcal H$ only at these special points of zero velocity $\dot X=0$. } Moreover if one chooses the protocol with both vanishing velocity and acceleration at the initial and final points ($\dot X=0,\; \ddot X=0$), then the counter-diabatic drive is smooth and does not require any jump of the potential at the beginning or end of the ramp.

\subsubsection{Particle in a time-dependent vector potential}

We can similarly consider a  Hamiltonian with time-dependent momentum shift, which is equivalent to a charged particle in a time-dependent electromagnetic field (cf. \fref{fig:mass_renorm_spin}):
\be
\mathcal H={(\hat p-P(t))^2\over 2m}+V(q).
\ee
As we discussed  earlier in \eref{eq:gauge_pot_sho}, the gauge potential corresponding to the momentum shift is $\mathcal A_P=-q$ therefore the counter-diabatic Hamiltonian in this case will be
\be
\mathcal H_{\rm CD}={(\hat p-P(t))^2\over 2m}+V(q)-\dot P(t) q.
\ee
This example is even easier to interpret than the previous one; the compensating term in the counter-diabatic protocol simply reduces to adding an electric field opposing the one created by the time-dependent vector potential. 

\subsubsection{Scale invariant driving}

Let us now consider  a slightly more complicated example of a classical one-dimensional non-relativistic particle in an external potential $V(q,\lambda)$ described by the Hamiltonian
\[
\mathcal H={p^2\over 2m}+V(q,\lambda).
\]
Assuming that the parameter $\lambda$ changes in time according to some protocol, our goal will be to find the corresponding adiabatic gauge potential $\mathcal A_\lambda$ and hence the counter-diabatic term $\dot\lambda \mathcal A_\lambda$. Let us note as earlier that in general chaotic systems, the adiabatic gauge potentials do not exist~\cite{Jarzynski1995_2}. So the general problem does not have a solution; instead one can either find approximate solutions for $\mathcal A_\lambda$ (see \sref{sec:approx_agp}) or find particular protocols -- such as translations -- where a solution exists. We will focus on a latter possibility here, finding a protocol for which $\mathcal A_\lambda$ takes a simple form. Although we consider a one dimensional system, the protocol we discuss can be straightforwardly generalized to higher-dimensional cases.

The adiabatic gauge potential should satisfy \eref{eq:gauge_pot_poisson_bracket}. A sufficient condition to satisfy this equation clearly is
\be
\partial_\lambda V-\{\mathcal A_\lambda,\mathcal H\}=\xi \mathcal H\quad \Leftrightarrow\;
\partial_\lambda V-{\partial \mathcal A_\lambda\over \partial q}{p\over m}-{\partial V\over \partial q}{\partial \mathcal A_\lambda\over \partial p}=\xi\left({p^2\over 2m}+V\right).
\label{eq:shortcuts1}
\ee
where $\xi\equiv \xi(\lambda)$ is an arbitrary function of $\lambda$. Let us seek an adiabatic gauge potential of the form $\mathcal A_\lambda=p f(q,\lambda)$, where $f$ is an arbitrary function. Then \eref{eq:shortcuts1} reduces to
\be
\partial_\lambda V+{\partial V\over \partial q} f-{p^2\over m} {\partial f\over \partial q}=\xi\left({p^2\over 2m}+V\right),
\ee
which is obviously satisfied if we require that
\be
{\partial f\over \partial q}=-{\xi\over 2},\quad \partial_\lambda V+{\partial V\over \partial q} f=\xi V.
\label{eq:shortcuts2}
\ee
The first equation implies that $f=-\xi q /2+\eta$, where $\eta(\lambda)$ is another arbitrary function of $\lambda$.  Substituting this to the second equation above gives
\be
 \partial_\lambda V-{\xi\over 2}q  {\partial V\over \partial q} +{\partial V\over \partial q}\eta=\xi V,
\label{eq:what_V_must_satisfy}
\ee
which clearly puts a constraint on the possible driving protocols which can be made to satisfy our ansatz for $\mathcal A_\lambda$. 

A particular form of the potential which satisfies this constraint realizes the so-called scale-invariant driving protocol~\cite{Deffner2014_1}:
\be
V(q, \lambda)={1\over \gamma^2(\lambda)}V_0\left({q-X(\lambda)\over \gamma(\lambda)}\right).
\ee
where $\gamma(\lambda)$ and $X(\lambda)$ correspond to squeezing and translations of the potential respectively. Plugging this ansatz into \eref{eq:what_V_must_satisfy} we find
\be
-{2\over \gamma^3}{d\gamma\over d\lambda} V_0-{q-X\over \gamma^4}{d\gamma\over d\lambda}V_0'-{1\over \gamma^3}{dX\over d\lambda} V_0' -{\xi\over 2\gamma^3} q V_0' +{\eta\over \gamma^3} V_0'={\xi\over \gamma^2} V_0, 
\ee
which can only hold for generic function $V_0$ if
\be
\xi(\lambda)=-{2\over \gamma}{d\gamma\over d\lambda}, \quad \eta(\lambda)={dX\over d\lambda}-{X\over \gamma}{d\gamma\over d\lambda}.
\label{eq:xi_eta}
\ee
Thus, for arbitrary protocols $X(\lambda)$ and $\gamma(\lambda)$, the adiabatic gauge potential is
\be
\mathcal A_\lambda=-{\xi(\lambda)\over 2} q p+\eta(\lambda)p 
\ee
\add{such that the counter-diabatic Hamiltonian} reads
\be
\mathcal H_{\rm CD}={p^2\over 2m}+{{1\over \gamma^2(\lambda)}V_0\left({q-X(\lambda)\over \gamma(\lambda)}\right)}-\dot\lambda p\left({\xi(\lambda)\over 2} q-\eta(\lambda)\right).
\ee
As in the previous example one can shift the momentum $p$ to
\[
\tilde p=p-m \dot\lambda \left({\xi(\lambda)\over 2} q-\eta(\lambda)\right)
\]
resulting in a gauge equivalent Hamiltonian
\be
\mathcal H_{\rm CD}'={\tilde p^2\over 2m}+{{1\over \gamma^2(\lambda)}V_0\left({q-X(\lambda)\over \gamma(\lambda)}\right)}-
{m \dot\lambda^2\over 2} \left({\xi(\lambda)\over 2} q-\eta(\lambda)\right)^2+{m\over 4} {d\over dt} (\dot \lambda \xi(\lambda)) q^2-m {d\over dt} (\dot \lambda \eta(\lambda)) q.
\ee
As before $\tilde p$ coincides with $p$ at the points of zero velocity $\dot\lambda=0$, so this is where adiabatic (transitionless) driving of the original Hamiltonian occurs. 

The path in parameter space $\gamma(\lambda)$ and $X(\lambda)$ can be arbitrary. In particular, choosing $\gamma(\lambda)=0$ and $X(\lambda)=\lambda$ we see from \eref{eq:xi_eta} that $\xi=0$ and $\eta=1$, such that we reproduce the previous example of a particle in a moving box. If we choose the squeezing protocol where $\gamma=\lambda$ and $X=0$, we instead find that $\xi=-2/\lambda$ and $\eta=0$, such that the counter-diabatic Hamiltonian  is
\be
\mathcal H_{\rm CD}'={\tilde p^2\over 2m}+{1\over \lambda^2}V_0(q/\lambda)-\frac{m}{\add{2}} {\ddot \lambda \over \lambda} q^2,
\label{eq:hamiltonian_shortcut_dilation}
\ee
showing that one only has to introduce an additional harmonic potential. Interestingly, if $V_0(q)$ is harmonic itself, then the counter-diabatic protocol simply affects the time dependence of the effective spring constant. So we see that for the scale-invariant case, one may obtain transitionless driving by adding a rather simple potential, even in the more complicated case where the path consists of a combination of translations and dilations.

\hrulefill

\exercise{Write down explicitly the Hamiltonian equations of motion corresponding to \eref{eq:hamiltonian_shortcut_dilation}. Check that under substitutions $dt=\lambda^2 d\tau$, $q'=q /\lambda$ and $p'=\tilde p \lambda$ the equations of motion become independent of $\lambda$ for any time dependence $\lambda(t)$:
\be
{dp'\over d\tau}=-\partial_{q'} V(q'),\quad {d q'\over d\tau}={p'\over m}.
\ee
Based on this, we may conclude that the evolution of the system is strictly adiabatic, which manifests in exact conservation of the adiabatic invariants, zero heat generation and complete reversibility for any cyclic function $\lambda(t)$.}

\exerciseshere

\hrulefill

Interestingly, the ansatz we chose for $\mathcal A_\lambda$ is far from unique. For instance, \eref{eq:shortcuts1} allows for another simple ansatz:
\be
\mathcal A_\lambda=p f(q,\lambda)+{\zeta\over 3m} p^3,
\label{eq:A_lam_p_cubed}
\ee
where $\zeta(\lambda)$ is another function of $\lambda$. Then instead of \eref{eq:shortcuts2} we get
\be
{\partial f\over \partial q}+\zeta {\partial V\over \partial q}=-{\xi\over 2},\quad \partial_\lambda V+{\partial V\over \partial q} f=\xi V,
\ee
such that $f(q,\lambda)=-\xi(\lambda) q/2+\eta(\lambda)-\zeta(\lambda) V(q)$ and hence
\be
 \partial_\lambda V-{\xi(\lambda)\over 2}q  {\partial V\over \partial q}+\eta(\lambda) {\partial V\over \partial q}-\zeta(\lambda) V {\partial V\over \partial q} =\xi(\lambda) V~.
\ee
\add{The functions $\xi(\lambda)$, $\eta(\lambda)$, and $\zeta(\lambda)$ can have arbitrary dependence on $\lambda$. If the potential $V(q,\lambda)$ satisfies the differential equation above, then counter-diabatic driving is again possible by adding a counter-diabatic term proportional to $\mathcal A_\lambda$ of the form in \eref{eq:A_lam_p_cubed}. A particular choice of the functions -- $\xi(\lambda)=0$, $\zeta(\lambda)=-1$, and $\eta(\lambda)=0$ -- yields the dispersionless Korteweg-de Vries (KdV) equation for the potential $V$~\cite{Okuyama2017_1}:
\be
 \partial_\lambda V+V{\partial V\over \partial q}=0,
\ee
which is intimately related to integrability of the model. The non-linear terms in this gauge potential yield a crucial difference between quantum and classical systems which shows up as a dispersion term in the KdV equation. Nevertheless, it was shown in Ref.~\cite{Okuyama2017_1} that both quantum and classical systems admit counter-diabatic driving. In general,
understanding which driving protocols correspond to tractable adiabatic gauge potentials even for such few-body problems is an active and vital topic in the field}.

\subsubsection{Klein-Gordon theory with a time-dependent mass}

Finally, let us apply the ideas of counter-diabatic driving to a somewhat less trivial example of an extended system. Specifically we will analyze a massive Klein-Gordon theory with time-dependent mass. It is intuitively expected that in a gapped system the adiabatic gauge potential should be localized within the length scale inversely proportional to the mass (the Compton wavelength). Let us demonstrate here that this is indeed the case.

We consider the system described by the following Hamiltonian:
\be
\mathcal H=\frac{1}{2} \int d^dx [\Pi^2(x) +(\nabla\Phi(x))^2+m^2(t) \Phi^2(x)],
\ee
where $d$ is the spatial dimensionality and $\Pi$ and $\Phi$ are canonically conjugate fields satisfying standard commutation relations: $[\Pi(x),\Phi(x')]=-i \delta(x-x')$ if the fields are quantum (we set $\hbar=1,$ in this section) and $\{\Pi(x),\Phi(x')\}=-\delta(x-x')$ if the fields are classical. Because the Hamiltonian is harmonic, the analysis of quantum and classical systems is identical. Since the above Hamiltonian is translationally invariant, it is convenient to work in momentum space, where the Hamiltonian reads
 \be
\mathcal H=\frac{1}{2} \int \frac{d^dk}{(2\pi)^d} [\Pi_k \Pi_{-k} +\omega_k^2(t) \Phi_k \Phi_{-k}],
\ee
with $\omega_k^2=k^2+m^2(t)$. The problem is now reduced to a set of independent parametric harmonic oscillators. This system was already considered in \exref{ex:dilations}. Using those results we find that the adiabatic gauge potential reads 
\be
\label{eq:A_hom_KG}
\mathcal{A}_{m^2}=- \int \frac{d^dk}{(2\pi)^d}  
\frac{\alpha_k}{2} [\Pi_k \Phi_{-k} + \Phi_k \Pi_{-k}] \quad {\rm with} \quad \alpha_k=\frac{1}{4}\frac{\partial_{m^2} \, \omega_k^2}{ \omega_k^2}
\ee
By transforming back to real space this becomes
\be
\mathcal{A}_{m^2}=-\frac{1}{8}   \iint d^dxd^dx' K(\mathbf{x-x'})   [\Pi_x \Phi_{x'} + \Phi_x \Pi_{x'}],
\ee
where the kernel $K(\mathbf{x}-\mathbf{x}^\prime)$ is the Green's function of the screened Laplace equation:
\be
K(\mathbf{x})=\int \frac{d^dk}{(2\pi)^d}  \frac{\mathrm{e}^{i\mathbf{k} \cdot \mathbf{x}}}{k^2+m^2} \quad \Longleftrightarrow \quad (-\nabla^2+m^2)K=\delta(x) 
\ee
The exact expression depends on the dimensionality $d$; for $d=3$ it becomes the Yukawa potential: 
\be
K(\mathbf{x})=\frac{\mathrm{e}^{-m |\mathbf{x}|}}{4\pi |\mathbf{x}|}.
\ee
Note that for any $d$ the kernel $K(x)$ decays exponentially at large separation $|\mathbf{x-x'}|$ as long as the mass is finite. The kernel is thus local on the scale of the Compton wavelength $1/m$ as we originally anticipated. In the limit of vanishing mass the adiabatic gauge potential becomes long-ranged (even divergent in the infrared below two dimensions), reflecting non-locality of adiabatic transformations in gapless (critical) systems. 

The counter-diabatic Hamiltonian associated with the time-dependent mass is thus:
\be
\mathcal{H}_{\rm CD}=\frac{1}{2} \int \frac{d^dk}{(2\pi)^d} \left[\Pi_k \Pi_{-k} +\omega_k^2(t) \Phi_k \Phi_{-k}-\zeta_k(\Pi_k \Phi_{-k} + \Phi_k \Pi_{-k})\right]; \quad \zeta_k={1\over 4\omega_k^2} {d m^2\over dt}. 
\ee
As in the single particle case, we can shift the momenta \add{ $\Pi_k\rightarrow \Pi_k +\zeta_k\Phi_k$} and find the gauge equivalent Hamiltonian, which does not contain cross terms
\be
\mathcal{H}'_{\rm CD}=\frac{1}{2} \int \frac{d^dk}{(2\pi)^d} \left[\Pi_k \Pi_{-k} +\Omega_k^2(t) \Phi_k \Phi_{-k} \right], \quad {\rm with} \quad \Omega_k^2=\omega_k^2-\zeta_k^2+\partial_t \zeta_k.
\ee
Since all the terms have a different k-dependence, the counter-diabatic Hamiltonian can not be realized by simply modulating some global coupling. However, if we do a small $k$ expansion, then this Hamiltonian approximately maps back to the Klein-Gordon theory with both a time-dependent mass and time-dependent speed of light: 
\be
\Omega^2_k(t)=m^2_{{\rm eff}}(t)+v^2_{{\rm eff}}(t)k^2+O(k^4).
\ee
The latter can be removed by an additional scale and gauge transformation. By a uniform scaling of all the fields, $\Pi_k \rightarrow \Pi_k \alpha_t$ and $\Phi_k \rightarrow \Phi_k/ \alpha_t$, the transformed Hamiltonian becomes
\be
\mathcal{H}''_{\rm CD}=\frac{\alpha^2(t)}{2} \int \frac{d^dk}{(2\pi)^d} \left[\Pi_k \Pi_{-k} +\frac{\Omega_k^2(t)}{\alpha^4(t)} \Phi_k \Phi_{-k}+\frac{\partial_t \alpha(t)}{2\alpha^3(t)}(\Pi_k \Phi_{-k} + \Phi_k \Pi_{-k})  \right].
\ee
Setting $\alpha^2(t)=v_{{\rm eff}}(t)$ makes the effective speed of light constant. A gauge transform $\Pi_k\rightarrow \Pi_k -(\dot{v}_{{\rm eff}}/4v_{{\rm eff}})\Phi_k$ finally results in a Hamiltonian which is of exactly the same form as the original one. Note that the prefactor in front of the Hamiltonian does not cause any transitions and simply tells us that proper time should be measured as $ds=v_{\rm eff} dt$.

For completeness, let us generalize the above results to general inhomogeneous mass distributions considering a more general Hamiltonian 
\be
\mathcal H=\frac{1}{2} \int d^dx \Pi_x^2 +\frac{1}{2} \iint d^dxd^dx' \Phi_x \, V_{x,x'}(\lambda(t)) \,\Phi_x'.
\ee
Since the Hamiltonian is still harmonic and real, the most general adiabatic gauge potential must be also harmonic and imaginary
\be
\mathcal{A}_\lambda=  {1\over 2} \iint d^dxd^dx'  (\Pi_x K_{x,x'}  \Phi_{x'}+ \Phi_{x'} K_{x,x'}\Pi_{x'}),
\ee
where $K_{x,x'}$ is the real symmetric kernel.
Substituting this ansatz into \eref{eq:gauge_pot_commutator} and explicitly evaluating all commutators one can show that the kernel $K_{x,x'}$ must satisfy the following equation:
\be
\int dx'' (K(x,x'') V(x'',x')+V(x,x'') K(x'',x'))={1\over 2} \partial_\lambda V(x,x')
\label{eq:gauge_pot_inhomogeneous_chain}
\ee
Although closed form solutions to this equation only exist for a handful of special examples, one can formally solve it by discretizing space and using eigenvectors $\{\left| n \right\rangle\}$ and eigenvalues $\{\epsilon_n \}$ of the discretized matrix $V$:
\be
\left\langle n \right| K \left| m \right\rangle= -\frac{1}{2}  \frac{ \left\langle n \right| \partial_\lambda V \left| m \right\rangle}{\epsilon_n+\epsilon_m}.
\ee
In the translationally invariant case $\partial_\lambda V$ is diagonal in momentum space and we immediately recover the homogeneous result.

\hrulefill

\exercise{\label{ex:gauge_pot_inhomogeneous chain} Perform the missing steps in the derivation of \eref{eq:gauge_pot_inhomogeneous_chain}. }

\exerciseshere

\hrulefill

\section{Approximate adiabatic gauge potentials in complex systems}
\label{sec:approx_agp}

\emph{{\bf Key concept:} Adiabatic gauge potentials can be computed from a minimization principle, which in turn can be used to develop variational methods for finding approximate gauge potentials. These variational gauge potentials are useful for creating approximate counter-diabatic protocols and finding approximate eigenstates of complex interacting Hamiltonians.}

We have seen from \eref{eq:gauge_pot_matrix_elem} that the adiabatic gauge potentials maybe ill-defined if the energy spectrum is dense and there are non-zero matrix elements of the generalized force operator, $-\partial_\lambda \mathcal H$, between nearby eigenstates. This is known as the problem of small denominators. If these divergences happen at isolated points such as phase transitions, then they can be easily dealt with as we will discuss later. But in generic chaotic systems, the situation is more subtle and requires careful regularization. Intuitively, this issue arises because the exact adiabatic gauge potential would allow one not only to adiabatically follow the ground state, but also excited states of the system. But the widely-accepted eigenstate thermalization hypothesis implies that those are essentially equivalent to random vectors in the Hilbert space, which are exponentially susceptible to tiny perturbations~\cite{DAlessioArxiv2015_1}. Following these states would require exponential fine tuning of $\mathcal A_\lambda$. More formally the eigenstate thermalization hypothesis \cite{Deutsch1991_1,Srednicki1994_1,Srednicki1996,Rigol2008_1,DAlessioArxiv2015_1} states that the off-diagonal matrix elements appearing in the numerator in \eref{eq:gauge_pot_matrix_elem} scale as $\exp[-S/2]$, where $S$ is the extensive thermodynamic entropy of the system, while the energy denominator for nearby states scales as $\exp[-S]$. Therefore the matrix elements of $\mathcal A_\lambda$ between nearby energy eigenstates are exponentially divergent with the system size, scaling as $\exp[S/2]$. However, even in generic chaotic systems, thermodynamic adiabaticity is a useful and well-defined limit. So our goal must be not finding an exact fine-tuned adiabatic gauge potential, but rather a good approximation which would allow one to eliminate or significantly reduce dissipation in the system and, in particular, to nearly adiabatically follow the ground state of the system. Note also that, while this argument is primarily quantum mechanical, it has also been shown~\cite{Jarzynski1995_2} that adiabatic gauge potentials are similarly divergent in classical chaotic systems without reference to quantum mechanics.

In this section, we will discuss methods for finding approximate adiabatic gauge potentials. We begin by reformulating \eref{eq:gauge_pot_commutator} for the gauge potential as a least action principle, which will prove useful for concrete calculations. We proceed to show how variational minimization of this action leads to a class of approximate gauge potentials which do not suffer suffer from issues of small denominators. We show how similar methods may be used to obtain perturbative approximations to the adiabatic gauge potentials. Finally, we illustrate the usefulness of these approximations by deriving variational gauge potentials for a variety of many-body systems: interacting chains of two or more spins, the quantum XY chain, and an impurity in a Fermi gas. In addition to deriving approximate gauge potentials, we show how they may be used to obtain approximate counter-diabatic protocols, ground states, and excited states for these complicated strongly-interacting systems.

\subsection{Adiabatic gauge potentials from the least action principle}

While Eqs.~(\ref{eq:gauge_pot_commutator}) and (\ref{eq:gauge_pot_poisson_bracket}) defining the adiabatic gauge potential cannot always be solved exactly, we will show how they may be used to derived a useful and tractable variational ansatz. To do so, let us begin by reformulating these equations as a minimum action principle, which serves as a seed for developing efficient approximate schemes. We will focus our derivation on the more general case of quantum systems, keeping in mind that in the classical limit one simply has to substitute the commutators with Poisson brackets and traces of operators with averages over classical phase space.

Define the Hermitian operator $G_\lambda(\mathcal X)$ as
\be
G_\lambda(\mathcal X) = \partial_\lambda \mathcal{H}+\frac{i}{\hbar} \left[ \mathcal X, \mathcal{H} \right],
\ee
where the argument $\mathcal X$ is itself a Hermitian operator. Then \eref{eq:gauge_pot_matrix_rel} determining the adiabatic gauge potential $\mathcal A_\lambda$ simply reads $G_\lambda(\mathcal A_\lambda)=-M_\lambda$. Instead of directly solving for $\mathcal A_\lambda$, we may reformulate it as a problem of minimization the operator distance between $G_\lambda(\mathcal X)$ and $-M_\lambda$ with respect to $\mathcal X$. Since $G_\lambda$ is linear in $\mathcal X$ it is natural to use the Frobenius norm for defining the distance, i.e., 
\be
D^2(\mathcal X)=\mathrm{Tr} \left[\left( \partial_\lambda \mathcal{H}+\frac{i}{\hbar} \left[ \mathcal X, \mathcal{H} \right] +M_\lambda \right)^2\right]= \mathrm{Tr} \left[ \left(G_\lambda +M_\lambda \right)^2 \right] = \mathrm{Tr} \left( G_\lambda^2 \right) + \mathrm{Tr} \left(M_\lambda^2 \right) + 2 \mathrm{Tr} \left( M_\lambda G_\lambda \right).
\ee 
Clearly the distance is indeed minimal (zero) when $G_\lambda=-M_\lambda$. Tracing in the energy eigenbasis and using cyclic properties of the trace,  one finds
\begin{eqnarray}
\mathrm{Tr} \left( M_\lambda G_\lambda \right) & =& \mathrm{Tr} \left( M_\lambda \partial_\lambda \mathcal H \right) + \frac{i}{\hbar} \mathrm{Tr} \left( M_\lambda \left[ \mathcal X, \mathcal H \right]\right)  =  - \mathrm{Tr} \left( M_\lambda^2 \right)-\frac{i}{\hbar}\cancelto{0}{\mathrm{Tr} \left( [M_\lambda, \mathcal H] \mathcal X \right)}
\\ D^2(\mathcal X) & =& \mathrm{Tr} \left[G^2_\lambda(\mathcal X)\right]- \mathrm{Tr} \left[M_\lambda^2 \right],
\label{eq:distance_gauge}
\end{eqnarray}
where we used that the operator $M_\lambda$ commutes with $\mathcal H$ and that $\mathrm{Tr} \left( M_\lambda \partial_\lambda \mathcal H \right)=-\mathrm{Tr}  \left( M_\lambda^2 \right)$, which becomes obvious if we explicitly write the trace in the eigenbasis of the Hamiltonian.  Since the generalized force term does not depend on $\mathcal X$, it does not affect the minimization. Hence, minimizing the distance is equivalent to minimizing the norm of $G_\lambda$. One can thus consider the norm of $G_\lambda$ as the action associated with the gauge potential:
\be
\label{eq:action_gauge}
S=\mathrm{Tr} \left[G^2_\lambda(\mathcal X)\right].
\ee
It can easily be minimized by expanding $\mathcal X$ in some operator basis. The distance is minimized whenever $\mathcal{A}_\lambda$ satisfies:
\be
\frac{\delta S}{\delta \mathcal X}\bigg|_{\mathcal X=\mathcal A_\lambda}=0 \quad \Rightarrow \quad \left[ \mathcal{H}, \partial_\lambda \mathcal{H}+\frac{i}{\hbar} \left[ \mathcal{A}_\lambda, \mathcal{H} \right] \right]=0~,
\label{eq:least_action_principle}
\ee
which, as anticipated, is just \eref{eq:gauge_pot_commutator}.

An even a simpler way to see that $\mathcal X=\mathcal A_\lambda$ minimizes the distance between $G_\lambda(\mathcal X)$ and $M_\lambda$ is by noting that the diagonal elements of $G_\lambda$ in the eigenbasis of $\mathcal H$ do not depend on $\mathcal X$: $\langle n|G_\lambda(\mathcal X) |n\rangle=\partial_\lambda E_n$. Therefore by minimizing the norm of $G_\lambda$ we are minimizing the sum off-diagonal elements of $G_\lambda$. For $\mathcal X=\mathcal A_\lambda$, $[G_\lambda,\mathcal H]=0$ by construction, so all off-diagonal elements of $G_\lambda$ are equal to zero and hence the distance $D(\mathcal X)$ reaches the absolute minimum. 

The action $\mathcal S$ itself has a simple physical interpretation \add{in terms of the } transition rate from a particular energy level to all other states under white noise modulation of $\lambda$ in the presence of the compensating term $\dot\lambda \mathcal X$. To see this\add{,} let us consider a Hamiltonian
\be
\mathcal H_X=\mathcal H(\lambda)+\dot\lambda \mathcal X,
\label{eq:cd_ham_X}
\ee
where $\lambda=\lambda_0+\epsilon(t)$ and $\epsilon(t)$ is an infinitesimal white noise $\overline{\epsilon(t)\epsilon(t')}=\kappa \delta(t-t')$. We can expand the Hamiltonian $\mathcal H_X$ in small modulation amplitude $\lambda(t)=\lambda_0+\epsilon(t)$:
\be
\mathcal H_X\approx \mathcal H(\lambda_0)+\epsilon\, \partial_\lambda \mathcal H(\lambda_0)+\dot\epsilon \mathcal X.
\ee
Next we will use the standard Fermi-Golden rule expression for the transition rate $\Gamma_n$ from the eigenstate $|n\rangle$ with energy $E_n$ to all other states~\cite{Clerk2010}:
\begin{multline} 
\Gamma_n=\int_{-\infty}^{\infty} d\omega\,\mathcal S_{\epsilon}(\omega) \sum_{m\neq n} \left|\langle n| \left(\partial_\lambda \mathcal H-i\omega\mathcal X \right)| m\rangle\right|^2 \delta (E_m-E_n-\omega)\\
=\int_{-\infty}^{\infty} d\omega\,\kappa \sum_{m\neq n} \left|\langle n| \left(\partial_\lambda \mathcal H+{i\over \hbar}[\mathcal X,\mathcal H]\right) | m\rangle\right|^2 \delta (E_m-E_n-\hbar\omega)=\kappa \langle n| G_\lambda^2(\mathcal X)|n\rangle-\kappa \langle n |\partial_\lambda \mathcal H|n\rangle^2. 
\label{eq:Gamma_n}
\end{multline}
Here $\mathcal S_{\epsilon}(\omega)=\kappa$ is the spectral density of the noise $\epsilon(t)$, which is frequency independent for white noise. Averaging $\Gamma_n$ over all eigenstates yields the average lifetime, which is clearly proportional to \eref{eq:distance_gauge}. \add{Minimizing the average transition rate is thus equivalent to minimizing the action in \eref{eq:action_gauge}.}

The lifetime of eigenstates especially in many-particle systems is not a physically observable quantity. A more physical measure of dissipation is the heating rate. But the latter is always zero if we consider states in the middle of the spectrum, which effectively correspond to infinite temperature. The easiest way to characterize dissipation that remains valid at infinite temperature is to instead look at the rate of change of the energy variance $\sigma_E^2$, which we term here the dissipation rate. If the system stays in the eigenstate, then there is no dissipation and the energy variance remains zero. Otherwise the dissipation rate is positive and the energy variance grows in time. In macroscopic ergodic systems satisfying the eigenstate thermalization hypothesis\add{,} the heating rate and the dissipation rate are proportional to each \add{other} because of the fluctuation-dissipation relation with the proportionality constant equal to half the inverse temperature of the eigenstate~\cite{DAlessioArxiv2015_1}. Within Fermi\add{'s} \add{g}olden \add{r}ule\add{,} the dissipation rate starting from the eigenstate $|n\rangle$ is obtained from \eref{eq:Gamma_n} \add{by} multiplying the transition probabilities by $\omega^2$:
\be
{d \sigma_E^2\over dt}=\int_{-\infty}^{\infty} d\omega\,\mathcal S_{\epsilon}(\omega) \sum_{m\neq n} \omega^2 \left|\langle n| \left(\partial_\lambda \mathcal H-i\omega\mathcal X\right)| m\rangle\right|^2 \delta (E_m-E_n-\omega)=\add{-\kappa \langle n| [G_\lambda(\mathcal X), \mathcal H]^2|n\rangle},
\label{eq:sigma_E_n}
\ee
Clearly the dissipation rate averaged over all eigenstates $|n\rangle$ defines the action given by the Frobenius norm of the commutator of $G_\lambda(\mathcal X)$ with the Hamiltonian. Thus the average dissipation rate for a system subject to external white noise in  $\lambda$ and in the presence of the compensating term from \eref{eq:cd_ham_X} is equal to the square error of \eref{eq:least_action_principle} when $\mathcal{X}$ is used as an approximate solution to \eref{eq:least_action_principle}.

\subsection{Variational gauge potentials}
\label{sec:variational_gauge_potentials}

The least action principle [\eref{eq:least_action_principle}] often serves as a useful way to calculate $\mathcal A_\lambda$, as we will see throughout the following sections. More importantly, it establishes a variational principle for the gauge potential based on the inequality $S(\mathcal X) \geq S(\mathcal A_\lambda)$. The  benefit of such a variational principle is that it allows us to construct approximate solutions to problems in which \eref{eq:gauge_pot_commutator} is hard to solve analytically. We will see throughout the following sections that this variational method is both useful and tractable, even for complicated interacting Hamiltonians.

Before moving on to a simple example, let us note that the trace norm in the action is similar to the infinite temperature norm as we are summing over all the eigenstates of $\mathcal H$ with equal weight. Very often we are interested only in the low energy manifold, as for example in trying to find the approximate counter-diabatic driving required to keep the system close to the ground state. If we are dealing with quantum or classical systems with unbounded spectra, the Frobenius norm of the operators may also be ill-defined, requiring some cutoff regularization. In such situations we may instead define the finite temperature action:
\begin{equation}
\mathcal{S}\left(\mathcal X,\beta \right)=\left< G_\lambda^2(\mathcal X)\right>-\left< G_\lambda(\mathcal X)\right>^2,
\label{eq:action_gauge1}
\end{equation}
where $\langle \dots \rangle$ stands for the averaging with respect to the thermal density matrix $\rho(\beta)={1\over Z}\exp[-\beta \mathcal H]$. The original Frobenius norm in \eref{eq:action_gauge} can be recovered  as the infinite temperature limit $(\beta\to 0)$ of \eref{eq:action_gauge1} up to an addition of $\langle G_\lambda(\mathcal X)\rangle^2=\langle M_\lambda\rangle^2$, which does not depend on $\mathcal X$. In the zero temperature limit $\beta\to\infty$\add{,} the action in \eref{eq:action_gauge1} reduces to the variance of $G_\lambda$ in the ground state. Clearly the exact gauge potential minimizes \eref{eq:action_gauge1} for any fixed temperature $\beta$. However, minima of the action restricted to particular variational manifolds will generally depend on $\beta$. In these notes we will restrict the discussion of complex systems to lattice models, where the infinite temperature norm can be used. Analysis of the finite temperature action will be left for future work.

\hrulefill
\exercise{\label{ex:gauge_A_p} Consider the following time-dependent Hamiltonian for a two-level system: $\mathcal{H}=\Delta \sigma^z +h(t) \sigma^x$. Find the gauge potential $\mathcal{A}_h$ corresponding to a change in the x-magnetic field $h$ by minimizing the action in \eref{eq:action_gauge} using the ansatz $\mathcal{A}_h= r(\lambda)\sigma_y$.  Justify why this ansatz should lead to the exact result. Use the result to construct the generalized force $M_h$ and check that it agrees with the correct answer.}

\exerciseshere

\hrulefill

We will now illustrate how the variational principle can be used for finding the adiabatic gauge potential in an interacting spin system. Let us consider a system of two coupled identical spin one-half particles in a uniform magnetic field with Ising interactions:
\be
\mathcal{H}=- h \cos\theta (\sigma^z_1+\sigma^z_2)-h \sin\theta (\sigma^x_1+\sigma^x_2)- J_z\sigma^z_1\sigma^z_{2},
\label{eq:hamiltonian_two_spins}
\ee
with $\hbar=1$ throughout this section. The Hamiltonian above corresponds to a real matrix in the $z$-basis, so the adiabatic gauge potential can be chosen to be imaginary, and hence contain an odd number of $\sigma^y$ operators.  As a first variational ansatz for $\mathcal A_\theta$, we will choose the non-interacting form 
\be
\mathcal X={\alpha\over 2} (\sigma^y_1+\sigma^y_2).
\ee
It is then straightforward to compute the function $G(\mathcal X)$:
\be
G(\mathcal X)=\partial_\theta \mathcal H+i[\mathcal X,\mathcal H]=h\sin\theta(1-\alpha) (\sigma^z_1+\sigma^z_2)-h\cos\theta(1-\alpha)(\sigma^x_1+\sigma^x_2)+\alpha J_z (\sigma^x_1\sigma^z_2+\sigma^z_1\sigma^x_2).
\ee
To evaluate the action, $S(\alpha)=\mathrm{Tr}\left[G^2(\mathcal X)\right]$, we note that all Pauli matrices are traceless, so only even powers of Pauli matrices contribute. Therefore
\be
\add{{S(\alpha)\over 4}=2 h^2 (1-\alpha)^2+2\alpha^2 J_z^2},
\ee
where the overall factor of four in the denominator comes from the trace of the identity matrix. Minimizing this action \add{with respect to $\alpha$,} we find the optimal value $\alpha^\ast$:
\be
\left. \frac{\partial S}{\partial \alpha} \right|_{\alpha^\ast} = 0 ~ \implies ~ \alpha^\ast={h^2\over h^2+J_z^2},
\ee
leading to the following variational gauge potential
\be
\mathcal A_\theta^\ast={1\over 2}{h^2\over h^2+ J_z^2}(\sigma^y_1+\sigma^y_2).
\label{eq:A_variational_two_spins}
\ee
We use the $\ast$-notation to highlight that this gauge potential represents the best variational solution but it is not generally exact because our variational manifold is restricted to single spin operators. Nevertheless this variational solution is very instructive as it clearly shows that the spin-spin coupling affects the magnitude of the gauge potential compared to the non-interacting case, suppressing it.

We can improve the variational ansatz by adding two-spin terms into the variational manifold
\be
\mathcal X={\alpha\over 2} (\sigma^y_1+\sigma^y_2)+{\beta\over 2} (\sigma^y_1\sigma^x_2+\sigma^x_1\sigma^y_2)+{\gamma\over 2}(\sigma^y_1\sigma^z_2+\sigma^z_1\sigma^y_2)
\ee
Evaluating the additional commutators of the two spin terms with the Hamiltonian, we  find
\begin{multline}
G(\mathcal X)=h\sin\theta(1-\alpha) (\sigma^z_1+\sigma^z_2)+\left(\gamma J_z-h\cos\theta(1-\alpha)\right)(\sigma^x_1+\sigma^x_2)+(\alpha J_z-\beta h \sin\theta+\gamma h\cos\theta) (\sigma^x_1\sigma^z_2+\sigma^z_1\sigma^x_2)\\+2\beta h \cos\theta\, \sigma^x_1\sigma^x_2-2\gamma h \sin\theta\, \sigma^z_1\sigma^z_2-2 h(\beta  \cos\theta-\gamma  \sin\theta) \sigma^y_1\sigma^y_2.
\label{eq:G_two_spins}
\end{multline}
From this expression one can easily compute the action by noting that all cross-terms in $G^2(\mathcal X)$ are traceless. Thus, 
\begin{multline}
\add{S(\alpha,\beta,\gamma)\over 8}=h^2 (1-\alpha)^2-2\gamma(1-\alpha) J_z h\cos\theta+\gamma^2 J_z^2+(\alpha J_z-\beta h \sin\theta+\gamma h\cos\theta)^2\\+4\beta^2 h^2 \cos^2\theta + 4\gamma^2 h^2 \sin^2\theta-4\beta\gamma h^2 \sin\theta\cos\theta.
\label{eq:action_two_spins}
\end{multline}
In general, the variational action is a quadratic function of the variational parameters so variational optimization simply reduces to solving a set of linear equations whose rank is equal to the number of variational parameters. Here it gives
\be
\begin{pmatrix}
 h^2+J_z^2 & -J_zh\sin\theta & 2J_z h\cos\theta \\
 -J_z h\sin\theta & h^2(3\cos^2\theta +1) & -3 h^2\cos\theta\sin\theta \\
 2J_z h\cos\theta & -3h^2\cos\theta\sin\theta & h^2(3\sin^2\theta+1)+J_z^2
\end{pmatrix}
\begin{pmatrix}
\alpha^\ast \\
\beta^\ast \\
\gamma^\ast
\end{pmatrix}= 
\begin{pmatrix}
h^2 \\
0 \\
J_z h\cos\theta
\end{pmatrix}
\label{eq:two_spin_secular_adiabatic}
\ee
This equation can be solved to obtain the variational parameters $\alpha^\ast$, $\beta^\ast$, $\gamma^\ast$. For example, for $\theta=0$ (field pointing along the $z$-direction), we get
\[
\mathcal A_\theta(\theta=0)={1\over 2}{h^2\over h^2-J_z^2} (\sigma^y_1+\sigma^y_2)-{1\over 2}{J_z h\over h^2-J_z^2} (\sigma^y_1\sigma^z_2+\sigma^z_1\sigma^y_2).
\]
Note that we left the asterisk off of $\mathcal A_\theta$ because our variational ansatz exhausted all possible terms odd in $\sigma^y$ and hence the variational solution must coincide with the exact result [see \exref{ex: G_two_spins}]. Not surprisingly the gauge potential becomes singular at $J_z=\pm h$ because at $\theta=0$ the symmetric eigenstates become degenerate (e.g., $(|\uparrow \downarrow\rangle+|\downarrow\uparrow\rangle)/\sqrt{2}$ becomes degenerate with $|\downarrow\downarrow\rangle$ for $J_z=h$) and this degeneracy is lifted by applying the $x$-magnetic field. Likewise for $\theta=\pi/2$ we find
\[
\mathcal A_\theta(\theta=\pi/2)={1\over 2} (\sigma^y_1+\sigma^y_2)+{1\over 2}{J_z \over h} (\sigma^y_1\sigma^x_2+\sigma^x_1\sigma^y_2).
\]
Now the gauge potential becomes singular at $h\to 0$ because of degeneracy between $|\downarrow\downarrow\rangle$ and $|\uparrow\uparrow\rangle$ states lifted by a small magnetic field in the $z$-direction.

\hrulefill
\exercise{\label{ex: G_two_spins} Verify \eref{eq:G_two_spins}. Compare the variational gauge potential obtained by the minimization of the action in \eref{eq:action_two_spins} with the exact result, which can  be obtained from \eref{eq:gauge_pot_matrix_elem}.}

\exercise{Compute the variational gauge potential 
$\mathcal A_{J_z}$ using single spin and two spin approximations. Compare with the exact result. }

\exerciseshere

\hrulefill

\subsection{Perturbative gauge potentials}
\label{sec:perturbative_agp}

In lieu of the variational approach, it is sometimes more convenient to start from a simple limit, such as a non-interacting model, where the gauge potential can be found exactly and use perturbative methods to correct it. The general idea of such a perturbative method is very similar to that used to find approximate integrals of motion in a weakly interacting systems~\cite{Moeckel2008,Kollar2011}. In that work, the interacting eigenstates $|\tilde{n}\rangle$ are found from the non-interacting eigenstates $|n\rangle$ via unitary transform $|\tilde{n}\rangle=\mathrm{e}^{-S}|n\rangle$, where $S=gS_{1}+\frac{1}{2}g^{2}S_{2}+\ldots$ is anti-Hermitian. Then one tries to perturbatively find $S$ and use it to dress the integrals of motion. In our language $S$ is directly related to the adiabatic gauge potential.

Instead of developing a fully general approach let us demonstrate how one can obtain the leading perturbative correction using the two-spin model in \eref{eq:hamiltonian_two_spins}. We will generalize this to an interacting spin chain in \sref{subsection:NonintegrableIsing}. Treating the coupling $J_z$ as a perturbation, let us find the leading perturbative correction to $\mathcal A_\theta$. Suppose that $U_0(\theta)$ is the unitary operator connecting unperturbed eigenstates corresponding to different angles $\theta$ with eigenstates corresponding to $\theta=0$ such that by definition
\[
\mathcal A_\theta^0=i (\partial_\theta U_0) U_0^\dagger.
\]
Next let us denote by 
\[
U_1(J_z,\theta)=\mathcal{P} \mathrm{e}^{-i\int_0^{J_z} dJ_z^\prime \mathcal A_{J_z^\prime}},
\]
where $\mathcal P$ stands for the path-ordered exponential, the unitary which connects the eigenstates of the Hamiltonian corresponding to fixed $\theta$ and different values of $J_z$. At leading order in perturbation theory, clearly
\[
U_1\approx 1-i J_z \mathcal A_{J_z}^0.
\]
The total unitary connecting the states at $J_z=0,\,\theta=0$ with arbitrary $J_z,\theta$ is then $U_1 U_0$, such that 
\be
\mathcal A_\theta=i (\partial_\theta U_1 ) U_1^\dagger+i U_1 (\partial_\theta U_0)U_0^\dagger  U_1^\dagger\approx \mathcal A_\theta^0+J_z\partial_\theta \mathcal A_{J_z}^0-i J_z \left[ \mathcal A_{J_z}^0, \mathcal A_\theta^0\right],
\label{eq:A_theta_pert_two_spins}
\ee
where the $0$ superscripts indicate that the gauge potential corresponds to the non-interacting limit $J_z=0$.

To find $\mathcal A_{J_z}^0$ exactly, we need to minimize the Frobenius norm of 
\[
G(\mathcal X)=\partial_{J_z}\mathcal H\bigl|_{J_z=0}+i[\mathcal X,\mathcal H_0]
\]
where $\mathcal H_0=-h\cos\theta(\sigma^z_1+\sigma^z_2)-h\sin\theta(\sigma^x_1+\sigma^x_2)$ is the non-interacting Hamiltonian. Because $\mathcal H_0$ is non-interacting and $\partial_{J_z}\mathcal H$ contains only two-spin terms, it is clear that the exact ansatz for $\mathcal X$ should also contain {\em only} two spin terms. Unlike in the previously discussed case of finite $J_z$ this is not an artifact of the two-site system but an exact statement, which applies to arbitrary system sizes. Choosing
\be
\mathcal X={\alpha\over 2} (\sigma^y_1\sigma^x_2+\sigma^x_1\sigma^y_2)+{\b\over 2} (\sigma^y_1\sigma^z_2+\sigma^z_1\sigma^y_2)
\label{eq:two_spin_ansatz}
\ee
we find
\be
G(\mathcal X)=-(2\beta h \sin\theta+1)\sigma^z_1\sigma^z_2+ 2\a h\cos\theta\, \sigma^x_1\sigma^x_2+2h (\beta \sin\theta-\alpha\cos\theta) \sigma^y_1\sigma^y_2+h(\b  \cos\theta-\a  \sin\theta) (\sigma^z_1\sigma^x_2+\sigma^x_1\sigma^z_2)
\ee
leading to the action
\be
{S(\alpha,\beta)\over 4}=1+ 4\beta h \sin\theta +8\beta^2 h^2 \sin^2\theta +8 \a^2 h^2\cos^2\theta-12\a\b h^2 \sin\theta\cos\theta+2\beta^2 h^2\cos^2\theta+2\a^2 h^2\sin^2\theta.
\ee
Minimizing this action yields the desired adiabatic gauge potential
\be
\mathcal A^0_{J_z}=-{3\cos\theta\,\sin^2\theta\over 8 h} (\sigma^y_1\sigma^x_2+\sigma^x_1\sigma^y_2)-{\sin\theta (3\cos^2\theta+1)\over 8h}(\sigma^y_1\sigma^z_2+\sigma^z_1\sigma^y_2).
\ee
Finally, plugging this expression into \eref{eq:A_theta_pert_two_spins}, we find
\be
\mathcal A_\theta\approx {1\over 2}(\sigma^y_1+\sigma^y_2)+{J_z\over h} {\sin\theta\over 4} (2-3\cos^2\theta) (\sigma^x_1\sigma^y_2+\sigma^y_1\sigma^x_2)+{J_z\over h} {\cos\theta\over 4} (1-3\cos^2\theta) (\sigma^z_1\sigma^y_2+\sigma^y_1\sigma^z_2)
\label{eq:A_theta_two_spin_pert}
\ee

\hrulefill

\exercise{\label{ex: pert_two_spins} Verify that \eref{eq:A_theta_two_spin_pert} agrees with the exact result, which can be obtained by solving the linear system in \eref{eq:two_spin_secular_adiabatic} perturbatively to linear order in $J_z$. }

\exerciseshere

\hrulefill

\subsection{Exact and variational gauge potentials for many-body systems}

Having introduced variational and perturbative methods for calculating approximate adiabatic gauge potentials, we will now show how they may be applied to understanding complex many-body systems. We will consider both interacting and  non-interacting systems, the latter of which can be analyzed analytically to highlight various important properties of gauge potentials. Specifically we will focus on the locality and convergence of the variational ansatz. We will discuss three characteristic examples: the non-integrable Ising chain, the quantum XY chain, and an impurity in a gas of free fermions. These examples illustrate some qualitative features of the adiabatic gauge potentials in both gapped and gapless regimes, as well as close to singularities like the Anderson orthogonality catastrophe for the impurity in a gas of fermions or near quantum critical points in the XY model. 

\subsubsection{Non-integrable Ising model}
\label{subsection:NonintegrableIsing}

Let us now extend the two-spin example from Secs.~\ref{sec:variational_gauge_potentials} and \ref{sec:perturbative_agp} to the Ising spin chain in a uniform magnetic field described by the Hamiltonian
\be
\mathcal{H}=-\sum_{j=1}^L h(\cos\theta\,\sigma_j^{z}+\sin\theta\,\sigma_{j}^{x})-J_z\sum_{j=1}^L\sigma_{j}^{z}\sigma_{j+1}^{z}
\label{eq:H_spin_with_ising}
\ee
with periodic boundary conditions. This Hamiltonian is already very complicated as it is known that except for $\theta=\pi n /2$ it is non-integrable, and thus has chaotic eigenstates satisfying the eigenstate thermalization hypothesis (cf. Ref.~\cite{Kim2013_1}). Therefore, in the thermodynamic limit the adiabatic gauge potential does not exist as a local operator that is analytic in the coupling constants, and thus it cannot be written in a closed form for large system sizes. However, due to the locality of the Hamiltonian, one can anticipate that except for some special points such as phase transitions, there should be an accurate local approximation to the adiabatic gauge potential. Therefore, we will extend the ansatz from the previous two-spin example and use it to find the variational adiabatic gauge potential:
\be
\mathcal X=\frac{1}{2}\sum_j \left( \alpha \sigma^y_j +\beta (\sigma^x_j\sigma^y_{j+1}+\sigma^y_j \sigma^x_{j+1}) +\gamma (\sigma^z_j \sigma^y_{j+1}+\sigma^y_j \sigma^z_{j+1}) \right).
\ee
As before only terms odd in $\sigma^y$ contribute to the variational ansatz because $\mathcal H$ is real. By further using the fact that the system is translationally invariant, we are only left with three undetermined parameters $\alpha,\beta,\gamma$. To find $G_\lambda(\mathcal X)$ we will need the following commutators:
\begin{eqnarray}
 i \sum_j\left[ \sigma^y_j, \mathcal{H} \right]&=&\sum_j 2h \left(\cos\theta \sigma^x_j -\sin\theta \sigma^z_j\right)+\sum_j 2J_z(\sigma^x_i\sigma^z_{i+1}+\sigma^z_i\sigma^x_{i+1}),\\
  i\sum_j\left[ \sigma^x_j\sigma^y_{j+1}+\sigma^y_j \sigma^x_{j+1}, \mathcal{H} \right]&=&-2h\sum_j \left(2\cos\theta\, (\sigma^y_i\sigma^y_{i+1}-\sigma^x_i\sigma^x_{i+1})+\sin\theta\, (\sigma^x_i\sigma^z_{i+1}+\sigma^z_i\sigma^x_{i+1}) \right) \nonumber \\
 & +& 2J_z\sum_j \left(\sigma^x_i\sigma^x_{i+1}\sigma^z_{i+2}+\sigma^z_i\sigma^x_{i+1}\sigma^x_{i+2}-\sigma^y_i\sigma^y_{i+1}\sigma^z_{i+2}
-\sigma^z_i\sigma^y_{i+1}\sigma^y_{i+2}\right),\\
  i\sum_j\left[ \sigma^z_j\sigma^y_{j+1}+\sigma^y_j \sigma^z_{j+1}, \mathcal{H} \right]&=&2h\sum_j \left(2\sin\theta\, (\sigma^y_i\sigma^y_{i+1}-\sigma^z_i\sigma^z_{i+1})+\cos\theta (\sigma^x_i\sigma^z_{i+1}+\sigma^z_i\sigma^x_{i+1}) \right) \nonumber \\
 & +& 4J_z\sum_j  \left(\sigma^x_i\sigma^z_{i+1}\sigma^x_{i+2}+\sigma^x_i\right).
\end{eqnarray}
From this $G_\lambda(\mathcal X)$ reads
\begin{multline}
G_\lambda(\mathcal X) = \sum_j \left( h\sin\theta (1-\alpha) \sigma^z_j+(2\gamma J_z-h\cos\theta (1-\alpha))\sigma^x_j  -2h(\beta\cos\theta-\gamma \sin\theta)\sigma^y_i\sigma^y_{i+1} \right)  \\
+\sum_j \left( (\gamma h\cos\theta-\beta h\sin\theta+\alpha J_z)(\sigma^x_i\sigma^z_{i+1}+\sigma^z_i\sigma^x_{i+1})+2\beta h\cos\theta\, \sigma^x_i\sigma^x_{i+1}-2\gamma h\sin\theta\, \sigma^z_i\sigma^z_{i+1}\right) \\
+J_z\sum_j \left(\beta (\sigma^x_i\sigma^x_{i+1}\sigma^z_{i+2}+\sigma^z_i\sigma^x_{i+1}\sigma^x_{i+2})-\beta(\sigma^y_i\sigma^y_{i+1}\sigma^z_{i+2}+\sigma^z_i\sigma^y_{i+1}\sigma^y_{i+2}) +2\gamma \sigma^x_i\sigma^z_{i+1}\sigma^x_{i+2} \right).
\end{multline} 
As before the tracelessness of the Pauli matrices reduces computing the action to summing squares of  coefficients in front of linearly independent operators:
\begin{multline}
\add{\frac{S(\alpha,\beta,\gamma)}{L\,2^L}}=h^2(1-\alpha)^2\sin^2\theta +(2\gamma J_z-h(1-\alpha)\cos\theta)^2+4h^2(\gamma \sin\theta-\beta\cos\theta)^2\\+2(h\gamma\cos\theta-h\beta\sin\theta+J_z\alpha)^2 
+ (2h\beta\cos\theta)^2+(2h\gamma\sin\theta)^2+4(J_z\beta)^2+(2J_z\gamma)^2,
\end{multline}
\add{where the factor $2^L$ in the denominator comes from the trace of the identity operator and an additional factor of $L$ comes from summing over $L$ identical contributions.}
Minimization leads to following linear equations, which only slightly differ from the two-spin system due to the presence of three-spin terms and periodic boundary conditions used in the full spin chain:
\begin{equation}
\begin{pmatrix}
 h^2+2J_z^2 & -2J_z h\sin\theta & 4J_z h\cos\theta \\
 -2 J_z h\sin\theta & 2 h^2(3 \cos^2\theta +1)+4J_z^2 & -6h^2\cos\theta\sin\theta \\
 4J_z h\cos\theta & -6h^2\cos\theta\sin\theta & 2h^2 (3\sin^2\theta+1)+8J_z^2)
\end{pmatrix}
\begin{pmatrix}
\alpha^\ast \\
\beta^\ast \\
\gamma^\ast
\end{pmatrix}= 
\begin{pmatrix}
h^2 \\
0 \\
2J_z h\cos\theta
\end{pmatrix}
\label{eq:spin_chain_secular_adiabatic}
\end{equation}
These equations can again be easily solved, giving the following variational adiabatic gauge potentials for two specific values of $\theta$:
\beq
\mathcal A_\theta^\ast(\theta=0)&=&{1\over 2}{h^4\over 7 J_z^4 + (h^2-J_z^2)^2}\sum_j \sigma^y_j+{1\over 2} {J_zh(2J_z^2-h^2)\over 7 J_z^4 + (h^2-J_z^2)^2}\sum_j (\sigma^y_j\sigma^z_{j+1}+\sigma^z_j\sigma^y_{j+1})\nonumber\\
\mathcal A_\theta^\ast(\theta=\pi/2)&=&{1\over 2}{h^2(2J_z^2+h^2)\over 3 J_z^4 + (h^2+J_z^2)^2}\sum_j \sigma^y_j+{1\over 2} {J_z h^3\over 3 J_z^4 + (h^2+J_z^2)^2}\sum_j (\sigma^y_j\sigma^x_{j+1}+\sigma^x_j\sigma^y_{j+1})
\eeq
This expression does not contain any singularities associated with the quantum phase transition for $h=\pm J_z$ and $\theta=0$ in the thermodynamic limit. This is due to the insufficiency of the local variational ansatz to capture singularities associated with long wavelength excitations. Nevertheless, away from the critical point we will show in the following sections that this ansatz allows accurate approximations for many objects of physical interest, including counter-diabatic driving protocols, many-body ground and excited state wave functions, the geometric tensor, and more. Let us also note that the variational ansatz also allows one to recover the results of perturbation theory in the limit of small coupling $J_z$ {[see \exref{ex:pert_thry}]. Indeed it is easy to see that at linear order in $J_z$ we can only generate one or two spin corrections to $\mathcal A_\theta$, at second order we can generate corrections involving up to three nearby terms, and so on. To find the perturbative result we note that $\beta$ and $\gamma$ should be linearly proportional to $J_z$ so we can simplify \eref{eq:spin_chain_secular_adiabatic} to
\begin{equation}
\begin{pmatrix}
 h^2+2J_z^2 & -2J_z h\sin\theta & 4J_z h\cos\theta \\
 -2 J_z h\sin\theta & 2 h^2(3 \cos^2\theta +1) & -6h^2\cos\theta\sin\theta \\
 4J_z h\cos\theta & -6h^2\cos\theta\sin\theta & 2h^2 (3\sin^2\theta+1))
\end{pmatrix}
\begin{pmatrix}
\alpha^\ast \\
\beta^\ast \\
\gamma^\ast
\end{pmatrix}\approx 
\begin{pmatrix}
h^2 \\
0 \\
2J_z h\cos\theta,
\end{pmatrix}
\label{eq:spin_chain_secular_adiabatic_approx}
\end{equation}
which in turn gives
\be
\alpha^\ast\approx 1, \quad \beta^\ast\approx \frac{J_z}{h} \frac{\sin\theta}{2} (2-3\cos^2\theta), \quad \gamma^\ast\approx \frac{J_z}{h} \frac{\cos\theta}{2} (1-3\cos^2\theta).
\label{eq:A_pert_spin_chain}
\ee
leading to
\be
\mathcal A_\theta\add{^\ast}\approx {1\over 2}\sum_j \sigma^y_j+{J_z\over h} {\sin\theta\over 4} (2-3\cos^2\theta) \sum_j (\sigma^x_j\sigma^y_{j+1}+\sigma^y_j\sigma^x_{j+1})+{J_z\over h} {\cos\theta\over 4} (1-3\cos^2\theta) \sum_j(\sigma^z_j\sigma^y_{j+1}+\sigma^y_j\sigma^z_{j+1}).
\label{eq:A_pert_spin_chain1}
\ee

\hrulefill

\exercise{Find the leading perturbative correction in $J_z$ to $\mathcal A_\theta$ for a full spin chain described by the Hamiltonian in \eref{eq:H_spin_with_ising}. Specifically argue that the two spin-ansatz generalizing \eref{eq:two_spin_ansatz} can be used to find  $\mathcal A_{J_z}^0$. Then use \eref{eq:A_theta_pert_two_spins} to find the perturbative expression for $\mathcal A_\theta$. Verify that the result agrees with that obtained from the variational approach in \eref{eq:A_pert_spin_chain1}. \label{ex:pert_thry}}

\exerciseshere

\hrulefill

\subsubsection{Quantum XY model}
\label{sec:gauge_pot_xy_chain}

Next let us analyze the quantum XY chain. This model is sufficiently simple that all calculations can be done analytically, yet it has a rich phase diagram serving as a prototype of quantum phase transitions and multi-criticality~\cite{Sachdev1999_1}. The model also provides important insight in the convergence properties of the variational procedure.

The quantum XY chain is described by the Hamiltonian
\be 
\mathcal H=-\sum_{j=1}^L \big[ J_x
\sigma_{j}^x \sigma_{j+1}^x + J_y
\sigma_{j}^y \sigma_{j+1}^y + h \sigma_{j}^z \big] ~,
\label{eq:H_XY_chain_orig}
\ee 
where $J_{x,y}$ are exchange couplings, $h$ is a transverse field, and we use periodic boundary conditions.
It is convenient to re-parameterize the model in terms of new couplings $J$
and $\gamma$ as
\be 
J_x=J \left( \frac{1+\gamma}{2} \right),\; J_y=J \left( \frac{1-\gamma}{2} \right)~,
\ee 
where $J$ is the energy scale of the exchange interaction and $\gamma$ is 
its anisotropy. We add an additional tuning parameter $\phi$, corresponding
to simultaneous rotation of all the spins about the $z$-axis by
angle $\phi/2$.
While rotating the angle $\phi$ has no effect on the 
spectrum of $\mathcal H$, it does modify the eigenstate wave functions.
To fix the overall energy scale, we set $J=1$. Since the Hamiltonian is invariant under the 
mapping $\gamma \to -\gamma$, $\phi \to \phi + \pi$, we also generally
restrict ourselves to $\gamma \geq 0$.

We now follow a standard set of tricks to solve such a Hamiltonian \cite{Sachdev1999_1}. Rewriting the spin Hamiltonian in terms of free fermions via a Jordan-Wigner transformation, $\sigma^z_j \sim 1-2c_j^\dagger c_j$ and $\sigma^+_j \sim \prod_{k<j} \sigma^z_k c_j$, $\mathcal H$ can be mapped to an effective non-interacting spin one-half model with
\be
\mathcal H=\sum_k \mathcal \psi^\dagger_k H_k \psi_k  \label{eq:hk_xxz};\quad\mathcal H_k=-\left(
\begin{array}{cc}
h-\cos(k) & \gamma \sin(k) \mathrm{e}^{-i\phi}\\ 
\gamma \sin(k)
\mathrm{e}^{i\phi} & -[h-\cos(k)]
\end{array}
\right),\nonumber 
\ee
where $\psi_k^\dagger=(c^\dagger_k,c_{-k})$ denotes the usual Nambu spinor. The details of this transformation can be found elsewhere\footnote{We note that the sign convention for $\phi$, i.e., which direction the spins are rotated around the $z$ axis, differs from Ref.~\cite{KolodrubetzPRB2013_1}. This, and other differences in gauge choice, are done to match the conventions of earlier spin-1/2 examples in this review. The choices have no effect on the final measurable quantities such as the metric tensor.}; for our purposes, it is important to note that it has allowed us to reduce the interacting spin model in \eref{eq:H_XY_chain_orig} to a non-interacting fermionic two-band model which is at half-filling (\fref{fig:xy_chain_bands}).

\begin{figure}
\includegraphics[width=0.6\columnwidth]{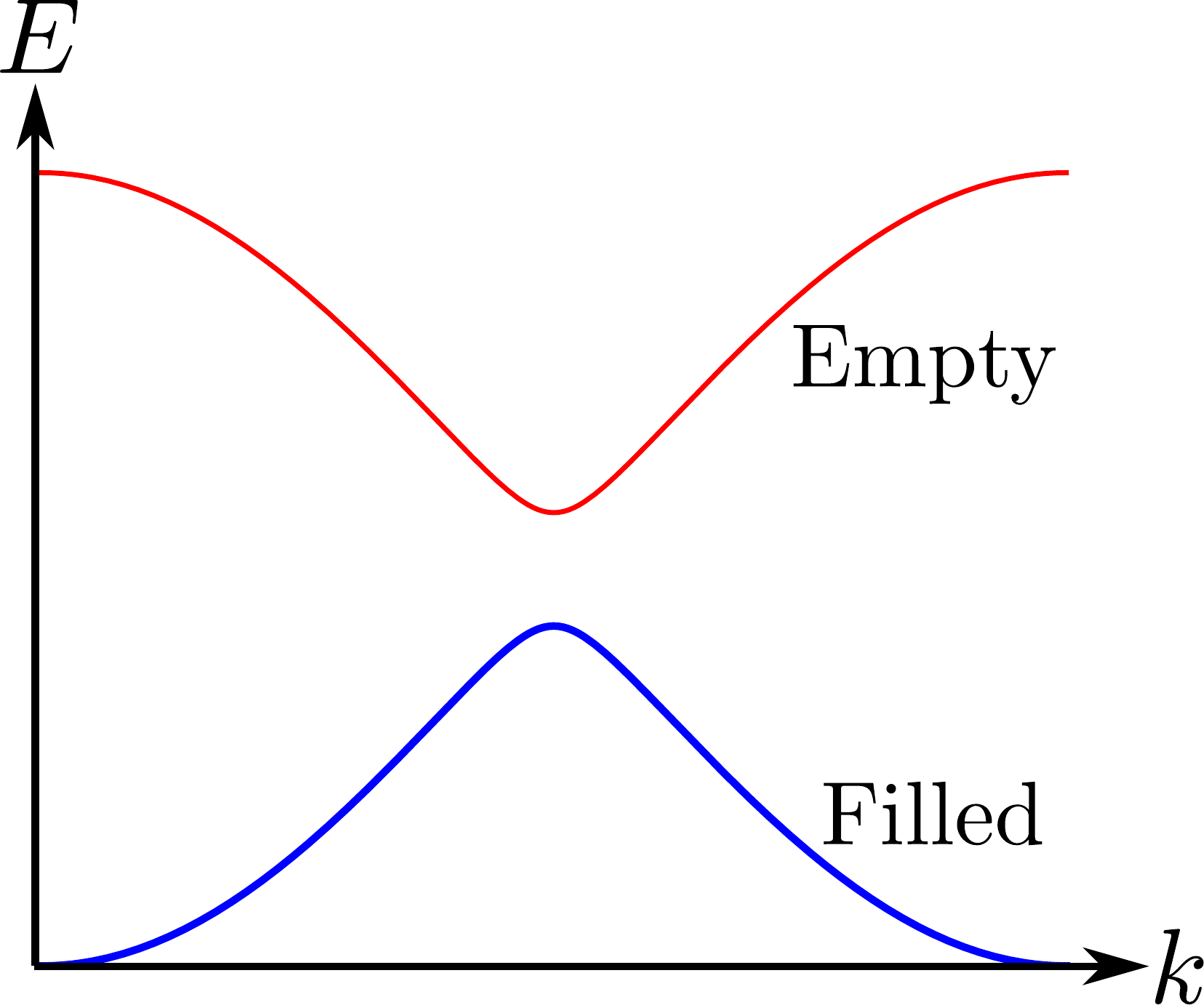}
\caption{\add{Illustration of the effective band structure of the XY chain that results after a Jordan-Wigner transformation. In the ground state, the  lower band is filled and the upper band is empty. Each mode $k$ can be excited by kicking the fermion into the upper band [see \eref{eq:hk_xxz}].}}
\label{fig:xy_chain_bands}
\end{figure}

Computing the exact gauge potential now simply amounts to finding the gauge potential for a set of uncoupled two-level systems described by $\mathcal{H}_k$. This can be done by any means described above, but since we want to compare the results to a variational approximation later, let us follow the minimal action approach and construct \eref{eq:action_gauge} for all three parameters $h$, $\gamma$, and $\phi$. By parameterizing the gauge potential 
\be
\mathcal{A}(k)=\frac{1}{2} \left(\alpha_x(k) \sigma_k^x+\alpha_y(k) \sigma^y_k+\alpha_z(k) \sigma_k^z \right),
\ee
the commutator of the gauge potential with the Hamiltonian becomes
\begin{eqnarray}
i\left[ \mathcal{A}(k), \mathcal{H}_k \right]=\left[ \alpha_y(h-\cos k)-\alpha_z\gamma \sin k \sin \phi  \right] \sigma^x_k + \left[\alpha_z \gamma \sin k\cos \phi- (h-\cos k)\alpha_x \right] \sigma_k^y \nonumber \\ +\gamma \sin k \left[\alpha_x \sin \phi -  \alpha_y  \cos \phi \right] \sigma^z_k.
\end{eqnarray}
Consequently the action for changing $h$ is given by:
\begin{eqnarray}
\frac{S_h}{2}=  \left[-1+\alpha_x  \gamma \sin k\sin \phi -  \alpha_y   \gamma \sin k\cos \phi \right]^2+\left[ \alpha_y(h-\cos k)-\alpha_z\gamma \sin k \sin \phi  \right]^2 \nonumber \\
+ \left[\alpha_z \gamma \sin k\cos \phi- (h-\cos k)\alpha_x \right]^2.
\end{eqnarray}
Minimizing the action yields the following particular solution:
\be 
\alpha_x(k)=\frac{\gamma \sin k \sin \phi}{(\cos k-h)^2+\gamma^2 \sin^2 k}, \quad \alpha_y(k)=-\frac{\gamma \sin k \cos \phi}{(\cos k -h)^2+\gamma^2 \sin^2 k}, \quad \alpha_z(k)=0,
\ee
such that the final gauge potential associated with the magnetic field $h$ reads
\be
\mathcal{A}_h= \frac{1}{2} \sum_k \frac{\gamma \sin k}{(\cos k -h)^2+\gamma^2 \sin^2 k}  \psi^\dagger_k \left( \sin \phi \sigma_k^x- \cos \phi \sigma^y_k \right) \psi_k.
\label{eq:gauge_pot_ising_exact}
\ee
The calculation for $\gamma$ and $\phi$ is completely analogous and results in
\begin{eqnarray}
\mathcal{A}_\gamma&=&  -\frac{1}{2} \sum_k \frac{\sin k(\cos k-h)}{(\cos k -h)^2+\gamma^2 \sin^2 k}  \psi^\dagger_k \left( \sin \phi \sigma_k^x- \cos \phi \sigma^y_k \right) \psi_k, \nonumber \\
\mathcal{A}_\phi&=& \frac{1}{2} \sum_k \frac{\gamma \sin k}{\cos k -h}  \psi^\dagger_k \left( \cos \phi \sigma_k^x+\sin \phi \sigma^y_k \right) \psi_k .
\end{eqnarray}

While the resulting gauge potentials are simple in the momentum-space fermionic representation, they are actually comprised of a series of long strings of spins in the original model. We have \add{found} the exact adiabatic gauge potential for the XY model, but we can only do this by virtue of its integrability. Thus, it is instructive to have a closer look at the real space representation of the gauge potentials, in particular in view of the previous discussion on variational gauge potentials in non-integrable spin chains, where we inevitably have to resort to approximate string expansions. We can now use the XY model to benchmark our results.

In order to more easily discuss the real space representation analytically, let us restrict the parameters to the transverse field Ising limit, $\gamma=1$ and $\phi=0$. First of all note that, by Fourier transform we have
\be
\psi_k^\dagger \sigma^y_k\psi_k=\frac{i}{L}\sum_{j,l} \sin(lk) (c^\dagger_jc^\dagger_{j+l}-c_{j+l}c_{j}).
\ee
Inverting the Jordan-Wigner transformation then results in the following expression for the fermions in terms of real spins
\be 
O_l=2i \sum_{j} (c^\dagger_jc^\dagger_{j+l}-c_{j+l}c_{j})=\sum_{j} \left( \sigma^x_j\sigma^z_{j+1}\ldots\sigma^z_{j+l-1}\sigma^y_{j+l}+\sigma^y_j\sigma^z_{j+1}\ldots\sigma^z_{j+l-1}\sigma^x_{j+l}\right),
\ee
where we introduced the symbol $O_l$ for an XY string of length $l$. Combining all of the above expressions, one arrives at the real-space representation of the gauge potential
\be
\mathcal{A}_h= \sum_l \alpha_l O_l \quad {\rm where} \quad \alpha_l=-\frac{1}{4 L}  \sum_k \frac{ \sin (k) \sin (lk)}{(\cos k -h)^2+ \sin^2 k}. 
\label{eq:TFI_string}
\ee
In the thermodynamic limit we can replace the sum with an integral, which yields
\be 
\alpha_l=-\frac{1}{4\pi}  \int  d k \frac{ \sin (k) \sin (lk)}{(\cos k -h)^2+ \sin^2 k} =-\frac{1}{8} \left\{ \begin{array}{l} h^{l-1} \quad {\rm for} \quad h^2<1 \\ h^{-l-1} \quad {\rm for} \quad h^2>1 \end{array} \right.
\label{eq:TFI_stringalpha}
\ee
This result was first obtained in~\cite{Campo2012_1}. Note that the coefficients $\alpha_l$ decay exponentially with the string length as long as the system is not critical. At the critical point, however, all strings have exactly the same weight. 

This treatment allows a variational expansion in terms of finite number of strings, i.e., we make the ansatz
\be
\mathcal{A}^\ast_h=\sum_{l=1}^M \alpha_l O_l.
\label{eq:A_h_ast_tfi}
\ee
By returning to k-space, this results in the fermionic representation
\be
\mathcal{A}^\ast_h=4\sum_k \left(\sum_{l=1}^M \alpha_l \sin (lk) \right) \psi_k^\dagger \sigma^y_k\psi_k.
\ee
Clearly if the sum goes up to $M=L$, we recover the \add{exact} result. But by truncating, we reduce the number of variational parameters to $M$. The action for the ansatz simply becomes
\be 
\frac{S_h^M}{2^L}=\sum_k \left[ \left( 1+8\sum_{l=1}^M \alpha_l \sin (lk) \sin (k) \right)^2+ \left(8\sum_{l=1}^M \alpha_l \sin (lk)  \right)^2 (h-\cos k)^2\right] .
\ee
After expanding the squares and summing over k-space, the expression takes the following simple form:
\be 
\frac{S_h^M}{2^{L+3}}=\alpha_1+4(h^2+1)\sum_{l=1}^M \alpha_l^2-4h\sum_{l=1}^M \alpha_l  (\alpha_{l+1}+\alpha_{l-1}).
\ee
Minimizing with respect to $\alpha_j$ yields the following set of equations
\be
-h(\alpha_{j-1}+\alpha_{j+1})+(1+h^2)\alpha_j=-\frac{1}{8} \delta_{j1}.
\ee
This problem is equivalent to a particle hopping on a chain with nearest neighbor hopping $h$, a constant potential $(1+h^2)$, and a source term at site $1$. It may be readily solved via discrete Fourier transform $\alpha_l=\sum_{k} \alpha_k \sin(lk)$, where $k=n\pi/(M+1)$ for $n$ from $1$ to $M$, giving
\be
\alpha_l=-\frac{1}{4 (M+1)}  \sum_k \frac{ \sin (k) \sin (lk)}{(\cos k -h)^2+ \sin^2 k} .
\label{eq:alpha_l_tfi_var}
\ee
Comparing this with the exact result, \eref{eq:TFI_string}, both expressions are identical if one replaces the
system size $L$ with the string length $M + 1$. So even though the sum might be infinite, the real space variational result is cut off by the length of the longest string used. Note that this implies that, when the system is not critical, the variational result for the smallest strings should quickly approach the exact result once the longest string exceeds the correlation length. This is consistent with our knowledge that longer strings are exponentially suppressed. However, close to the critical point all the coefficients are equally important and not taking into account strings longer than $M$ now strongly affects physical quantities. \add{Let us also emphasize that the variational ansatz coincides with the exact gauge potential for the system of size $M+1$, which differs from taking an exact result for the infinite chain and truncating it to the terms with the range up to $M$. In \sref{sec:tfi_variational_gs} we will analyze physical implications of this difference between the variational and the truncated exact solutions for state preparation.}

\subsubsection{Impurity in a Fermi gas}
\label{sec:impurity_fermi_gas}

For our final example, let us consider a gas of spinless fermions on a one-dimensional lattice subject to an external potential $V_j(\lambda)$:
\[
\mathcal H=-\sum_{j=-L/2}^{L/2}(c_{j}^{\dagger}c_{j+1}+h.c.)+ \sum_{j=-L/2}^{L/2} V_j(\lambda) c_{j}^{\dagger}c_{j},
\]
where $c_j^\dagger$ and $c_j$ are creation and annihilation operators and we assume open boundary conditions, though this is not particularly important for our discussion. The potential can, for instance, be an isolated impurity $V_j(\lambda)=\lambda \delta_{j0}$ with $\lambda$ representing the strength of the potential or a moving impurity, $V_j(\lambda)=V_0(\lambda-j)$, where $V_0$ is some potential that slowly varies in space \cite{SelsArxiv2016}.

To find the exact gauge potential, we will again minimize the action $S={\rm Tr}[G(\mathcal X)^2]$. Because the Hamiltonian $\mathcal H$ is real and non-interacting, the adiabatic gauge potential should be imaginary and non-interacting. So the exact ansatz minimizing $S$ should be
\be
\mathcal X=i \sum_{j,k} \alpha_{j,k}\left( c^\dagger_{k} c_j -h.c.\right),
\ee
where the variational parameters $\alpha_{j,k}=-\alpha_{k,j}$ are real and antisymmetric. While this ansatz is exact, it leads to analytically intractable system of coupled linear equations, which can only be solved numerically. Instead we will focus on the variational ansatz involving only local fermion hopping:
\begin{equation}
\mathcal X=i \sum_j \alpha_j\left( c^\dagger_{j+1} c_j -h.c.\right).
\label{eq:X_free_fermions}
\end{equation}

Explicitly evaluating the commutator of $\mathcal X$ with the Hamiltonian we find
\be
G(\mathcal X)=\sum_j \left(\partial_\lambda V_j - 2J \nabla \alpha_j\right) c_j^\dagger c_j+J\sum_j \nabla \alpha_j (c_{j+1}^\dagger c_{j-1}+c_{j-1}^\dagger c_{j+1})+\sum_j \nabla V_{j+1} \alpha_j (c_{j+1}^\dagger c_j+c_j^\dagger c_{j+1}),
\ee
where $\nabla \alpha_j=\alpha_j-\alpha_{j-1}$ \add{and $\nabla V_j=V_j-V_{j-1}$} stand for the lattice derivatives. As in the case of spins, the trace of any operator involving odd number of fermions on any site is equal to zero. There are only two even fermion combinations contributing to the action: ${\rm Tr}\left[ c_j^\dagger c_j\right]={\rm Tr}\left[(c_j^\dagger c_j)^2\right]=1$. Then, up to terms independent of $\mathcal X$, we have
\begin{eqnarray}
\nonumber
{\mathcal S(\{\alpha_j\})\over 2^{L-2}}&=&{\rm const}+\sum_j\left(\partial_\lambda V_j-2 J\nabla \alpha_j)\right)^2+2 J^2 (\nabla\alpha_j)^2+2 (\nabla V_{j+1})^2\alpha_j^2\\
&=&{\rm const} +\sum_j  \left(\partial_\lambda V_j \right)^2 +4J \alpha_j \partial_\lambda \left( \nabla V_{j+1} \right) + 6J^2 \left(\nabla\alpha_{j} \right)^2 +2\left( \nabla V_{j+1}\right)^2 \alpha_j^2.
\label{eq:action_free_fermions}
\end{eqnarray}
Minimizing the action with respect to $\alpha_j$ yields the following set of linear equations,
\begin{equation}
-3 J^2\Delta\alpha_j+\left( \nabla V_{j+1} \right)^2 \alpha_j= -J\partial_\lambda \left( \nabla V_{j+1} \right),
\label{eq:supl_vargauge_discrete}
\end{equation}
where $\Delta \alpha_j=\alpha_{j+1}-2\alpha_j +\alpha_{j-1}$ is the discrete Laplace operator.
This system can always be solved numerically by standard methods or analytically in some special cases. If the potential $V_j$ is smooth on the lattice scale, then discrete derivatives can be replaced by continuous ones such that the equation above becomes
\begin{equation}
-3J^2\partial_x^2 \alpha(x)+\left( \partial_x V(x,\lambda) \right)^2 \alpha(x)= -J\partial_\lambda\partial_x V(x,\lambda).
\label{eq:supl_vargauge}
\end{equation}

A particularly simple solution can be found for the linear potential: $V_j=\lambda j$. In this case $\nabla V_j=\lambda$ and one can solve \eref{eq:supl_vargauge_discrete} by a simple ansatz:
\be
\alpha_j=-{J\over \lambda^2}.
\ee
such that the variational gauge potential is the current operator:
\be
\mathcal A_\lambda^\ast=-i{J\over \lambda^2}\left( c^\dagger_{j+1} c_j -h.c.\right).
\label{eq:A_linear_pot}
\ee
This solution has to be modified near boundaries [see \exref{ex:linear_pot1}]. 

Another limit, where Eqs.~(\ref{eq:supl_vargauge_discrete}) and (\ref{eq:supl_vargauge}) can be solved explicitly corresponds to the weak potential $V_j\ll J$. In this case one can ignore the quadratic term in these equations such that, e.g., \eref{eq:supl_vargauge_discrete} reduces to
\be
-3 J^2\Delta\alpha_j= -J\partial_\lambda \left( \nabla V_{j+1} \right).
\label{eq:supl_vargauge_discrete_linearized}
\ee
The solution can be expressed through the single-particle Green's functions
\be
\alpha_j={1\over 3J}\sum_{j'} \Gamma_{jj'}\partial_\lambda \left( \nabla V_{j'+1} \right),
\label{eq:gauge_pot_greens_function}
\ee
where 
\be
\Gamma_{j,j'+1}-2\Gamma_{j,j'}+\Gamma_{j,j'-1}=\delta_{j,j'}.
\ee
Under Dirichlet boundary conditions for a system confined between $-L/2$ and $L/2$, the Green's function is a triangle (just like for the continuous Laplace equation) :
\be
\Gamma_{j,j'}=\frac{j j'}{L}+\frac{1}{2}|j'-j|-\frac{L}{4}.
\ee
For the impurity potential $V_j=\lambda \delta_{j0}$ we thus get
\be
\mathcal A_{\lambda,\, {\rm imp}}^\ast={i\over 6J}\sum_{j} \left(|1+j|-|j|-\frac{2j}{L}\right) (c_{j+1}^\dagger c_j-c_j^\dagger c_{j+1}).
\ee
While the overall amplitude of the Green's function grows with system size, the gauge potential does not and obtains a maximal strength of $1/6J$ around the impurity. Note however that, while the perturbation acts locally at site 0, the variational gauge potential is highly non-local. It is this non-locality of the gauge potential together with the presence of a Fermi-surface which is the cause of Anderson's famous ``orthogonality catastrophe,'' as will be discussed in detail in \sref{anderson_metric_tensor}. 

Finally, if the potential is a weak perturbation, we can directly use perturbation theory to compute the gauge potential. In this limit only the unperturbed energy and wave functions enter in the expression. Consequently for the above example we find
\be 
\mathcal{A}_\lambda= i \sum_{k\neq q} \frac{ \left\langle k | \partial_\lambda V_j | q\right\rangle }{\epsilon_k-\epsilon_q} c^\dagger_k c_q= i\sum_{j,l} \gamma_{j,l} \left( c^\dagger_{j+l} c_j- c^\dagger_j c_{j+l} \right)
\ee
where
\be 
\gamma_{j,l}=\frac{1}{2}\sum_{k\neq q} \frac{ \left\langle k | \partial_\lambda V_j | q\right\rangle }{\epsilon_k-\epsilon_q} \left\langle j+l |k\right\rangle \left\langle q |j \right\rangle,
\label{eq:anderson_pert}
\ee
$\epsilon_k=-2J\cos k$\add{,} and $\left| q\right\rangle$ denotes the single particle eigenstates of the bare hopping Hamiltonian. Note that\add{,} in contrast to our variational ansatz, the exact expression contains longer-range hopping terms. The length of the hopping is denoted by $l$. For Dirichlet boundary conditions $k=n \pi/ (L+1)$, with $n=1,2,\ldots,L$ and  $\left\langle k | x\right\rangle=\sqrt{2/L}\sin(k(x-L/2))$. Unfortunately the sum in \eref{eq:anderson_pert} can not simply be done analytically. By numerically performing the sum for $V_\lambda=\lambda \delta_{j0}$ one readily verifies that the gauge potential vanishes for all even hopping terms, i.e.\add{,} $\gamma_{j,2n}=0$ for all $n$. One also finds that the functional behavior of the nearest neighbor hopping term is identical to the variational result, albeit with different prefactor $\gamma_{j,1}=3/2\alpha_j$. The suppression can be understood from the counter-diabatic driving perspective; if one only has access to nearest neighbor terms, one can cancel excitations caused by inserting the potential. This however generates unwanted next-nearest neighbor terms. Those are again canceled out by next-next-nearest neighbor terms in the exact perturbative expression, but since they were not allowed in the variational solution, they will suppress the amplitude of the gauge potential.

\hrulefill

\exercise{Consider the infinite temperature action for the free fermion problem in \eref{eq:action_gauge1} with $\beta=0$. Show that this action is equivalent to \eref{eq:action_free_fermions}. Observe that in the infinite temperature action only physical connected terms show up, i.e., the contribution from the terms like $c_j^\dagger c_j c_k^\dagger c_k$ with $j\neq k$ automatically cancels. Show also that such terms are independent of $\mathcal X$ and thus the infinite temperature action is equivalent to \eref{eq:action_gauge}.}

\exercise{\label{ex:linear_pot}Show that \eref{eq:A_linear_pot} in fact gives the exact expression for $\mathcal A_\lambda$ for the linear potential apart from boundary terms. One can show this by adding the second nearest neighbor hopping to the variational ansatz in \eref{eq:X_free_fermions} and showing that all coefficients are identically equal to zero. From this its is easy to prove that all higher order hopping terms are zero as well.}

\exercise{\label{ex:linear_pot1}Find the effect of the boundaries on the variational gauge potential in \eref{eq:A_linear_pot} in the continuum limit. This can be done by solving \eref{eq:supl_vargauge} with open boundary conditions $\alpha(-L/2)=\alpha(L/2)=0$. Show that if $\lambda L\gg J$, your result reduces to \eref{eq:A_linear_pot} except near the boundaries.}

\exerciseshere

\hrulefill

\subsection{Variational gauge potentials for counter-diabatic driving and eigenstate targeting}
\label{sec:find_approx_eig}

The variational gauge potentials have a broad range of applications both experimental and theoretical. In these notes we will only touch upon a few such applications with the main goal to demonstrate the principle on relatively simple setups. In this section we will discuss their applications to designing approximate counter-diabatic driving protocols and finding approximate eigenstates of interacting systems. In the next section we will also briefly discuss applications for finding the approximate geometric tensor characterizing the ground state manifold. 

In \sref{sec:shortcuts} we showed that adiabatic gauge potentials can be used to design counter-diabatic (transitionless) driving protocols, which keep the system in exact  instantaneous eigenstate of the Hamiltonian for any time evolution. It is thus natural to expect that approximate gauge potentials can be used to suppress transitions between eigenstates and hence reduce dissipation in the system. Below we will show how this works in practice for the same set of examples we discussed above.

\subsubsection{Two coupled spins}

Consider the two-spin model from \eref{eq:hamiltonian_two_spins}. Using the variational gauge potential restricted to single-spin terms, we can construct a family of variational counter-diabatic protocols connecting eigenstates corresponding to different values of the angle $\theta$,
\be
\mathcal H_{\rm CD}^\ast=\mathcal H(\theta(t))+\dot\theta \mathcal A_\theta^\ast,
\label{eq:h_cd_two_spins}
\ee
where the gauge potential $\mathcal A_\theta^\ast$ is given by \eref{eq:A_variational_two_spins}. For simplicity we will consider a linear time dependence $\theta(t)=v t$, choose $h,J_z>0$, and focus on the ground state manifold. At $t=0$ the magnetic field is pointing along the $z$ axis and the ground state is given by $|\uparrow,\uparrow\rangle$. Propagating this state with the Hamiltonian in \eref{eq:h_cd_two_spins} we can get approximate eigenstates for any value of $\theta$. 

A particularly important limit of counter-diabatic driving corresponds to infinite velocity $v\to\infty$. In this case the second term in the Hamiltonian in \eref{eq:h_cd_two_spins} clearly dominates the time evolution and the Schr\"odinger equation can be rewritten as
\be
i\frac{d \psi(\theta)}{dt} =\dot\theta \mathcal A_\theta^\ast \psi(\theta)
\ee
Using the chain rule \add{$d\psi/dt=\dot \theta d\psi/d\theta$}, we can rewrite this equation as
\be
i {d \psi(\theta)\over d\theta}=\mathcal A_\theta^\ast \psi(\theta).
\label{eq:schr_eq_fast_two_spins}
\ee
Thus we see that the variational gauge potential serves as an approximate Hamiltonian generating adiabatic evolution in the coupling space. This is of course not surprising as the exact $\mathcal A_\theta$ is precisely the generator of adiabatic transformations. So the family of counter-diabatic Hamiltonians is nothing but interpolation between infinitesimally slow adiabatic evolution with the original Hamiltonian and the fast evolution with only the gauge potential acting on the system. 

For the initial state $|\psi_1(0)\rangle=|\uparrow,\uparrow\rangle$ and the variational gauge potential given by \eref{eq:A_variational_two_spins} we can solve \eref{eq:schr_eq_fast_two_spins} analytically to get
\be
\add{|}\psi_1^\ast(\theta)\add{\rangle}=\left(\cos\beta_\theta\, |\uparrow\rangle+\sin\beta_\theta\, |\downarrow\rangle\right)\otimes \left(\cos\beta_\theta\, |\uparrow\rangle+\sin\beta_\theta\, |\downarrow\rangle\right),
\ee
where 
\[
\beta_\theta={\theta\over 2} {h^2\over h^2+J_z^2}
\]
Likewise propagating the remaining two eigenstates in the symmetric sector $|\psi_2\rangle=(|\uparrow\downarrow\rangle+|\downarrow\uparrow\rangle)/\sqrt{2}$ and $|\psi_3\rangle=|\downarrow\downarrow\rangle$ according to \eref{eq:schr_eq_fast_two_spins} we obtain:
\be
|\psi_2^\ast(\theta)\rangle=\cos (2\beta_\theta) {|\uparrow\downarrow\rangle+|\downarrow\uparrow\rangle\over \sqrt{2}}-\sin (2\beta_\theta) {|\uparrow\uparrow\rangle+|\downarrow\downarrow\rangle\over \sqrt{2}},\quad
|\psi_3^\ast(\theta)\rangle=\left(\cos\beta_\theta\, |\downarrow\rangle-\sin\beta_\theta\, |\uparrow\rangle\right)\otimes \left(\cos\beta_\theta\, |\downarrow\rangle-\sin\beta_\theta\, |\uparrow\rangle\right).
\ee

Clearly the variational solution violates the periodicity of the ground state manifold with respect to $\theta$. For this reason it is not equivalent to the standard variational ansatz for $|\psi(\theta)\rangle$, which minimizes the energy. Nevertheless the counter-diabatic propagation has a clear advantage that it allows us to construct the whole manifold of eigenstates, which is beyond conventional variational methods. If we are interested only in ground state optimization, then one could use the zero-temperature norm for minimizing the action according to \eref{eq:action_gauge1}. This, however, yields a non-linear minimization problem because the ground state wave function itself depends on the variational parameters, and hence goes beyond the scope of these notes.

\subsubsection{Transverse field Ising model}
\label{sec:tfi_variational_gs}

The same story of course applies to the transverse field Ising model which we briefly studied as the $\gamma=1$ and $\phi=0$ limit of the XY model (see \eref{eq:H_XY_chain_orig} above). This example is analytically tractable, which makes it a nice system to study the convergence properties of these approximate counter-diabatic drives and eigenstate preparation protocols. It is again quite natural to try to prepare the ground state out of a fully polarized state, which in this model corresponds to the ground state of the system at infinite magnetic field,  denoted $\left| \psi (\infty)\right>$. One can generate the ground state at arbitrary $h$ out of this by simply applying the appropriate gauge field to it, i.e.,
\be 
\left| \psi (h_f)\right>= U(h_f,\infty)\left| \psi (\infty)\right> \quad {\rm where} \quad U (h_f,h_i)=\exp \left(-i \int_{h_i}^{h_f} d h \mathcal{A}_h \right)
\label{eq:ground_state_prep}
\ee
and $\mathcal{A}_h$ is given by \eref{eq:TFI_string}. Note that in general the exponential needs to be path-ordered but in this case the gauge potentials at different values of the magnetic field commute, \add{which follows \add{from the fact} that the variational gauge potential can be represented in momentum space as a sum of commuting $\sigma_k^y$ Pauli matrices [cf. \eref{eq:gauge_pot_ising_exact}] in the Ising $\phi=0$ limit}. By replacing $\mathcal A_\lambda$ by an approximation $\mathcal A_\lambda^\ast$ we arrive at an approximate ground state of our model $\left| \psi^\ast (h_f)\right>$. In order to establish how well we can approximate the ground state by this procedure using $\mathcal A_h^\ast$ from \sref{sec:gauge_pot_xy_chain} we look at two observables, namely the density of excitations on top of the ground state, \add{which is proportional to the logarithm of the fidelity of the wave function~\cite{DeGrandi2010_1},} and the fluctuations in the energy. Both are clearly positive quantities that vanish only when we recover the exact ground state. In Figs.~\ref{fig:TFI_dens_k} and \ref{fig:TFI_dens_M} we show results for preparing a state at $h_f=2$ and $h_f=0$. We first note that there is a stark contrast between the two; for example, the variational expansion converges exponentially for $h_f=2$ while it only converges algebraically for $h_f=0$. This is immediately linked to the presence of a critical point at $h=1$ (see \sref{sec:metric_xy_chain}), which results in slower convergence due to the Kibble-Zurek mechanism for protocols like $h_f=0$ that cross it \cite{Kibble1976_1,Zurek1985_1,Zurek2005_1,Polkovnikov2005_1}. However, as long as the Hamiltonian remains gapped, the expansion converges exponentially. Secondly we see that the variational expansion performs as well as a truncated form of the exact result, which is obtained by removing all strings of length greater than $M$ from the exact adiabatic gauge potential (dashed lines in Figs.~\ref{fig:TFI_dens_k} and \ref{fig:TFI_dens_M}). While one approximation has slightly more excitations, the other has a lower energy variance, as can be understood from \fref{fig:TFI_dens_k}. Abrupt truncation of the string expansion in real space results in oscillations in k-space which show up in the excitation probability. The variational result suffers much less from this but in turn has a slightly broader distributions at low energy. As a result, truncating the exact result causes fewer excitations overall than the variational result, but the excitations that exist cause larger energy fluctuations because they are higher momentum.

\begin{figure}

\includegraphics[width=0.7\linewidth]{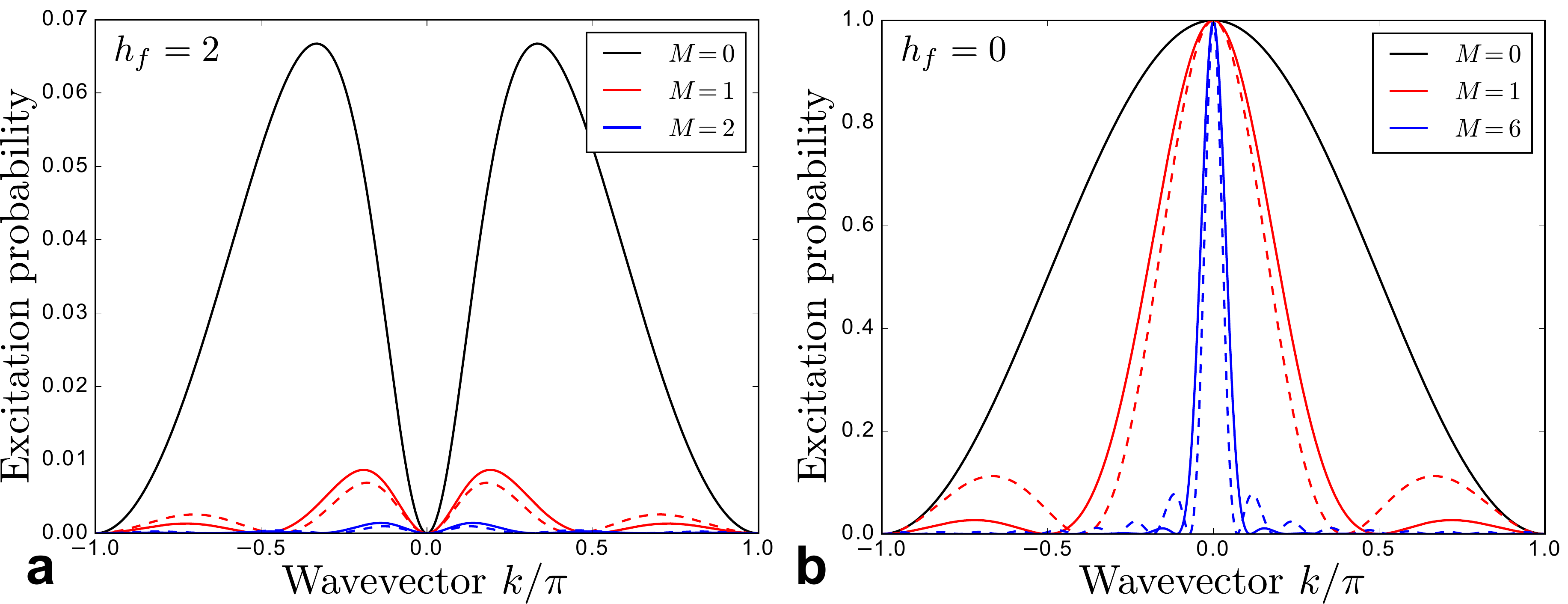}
\caption{\add{Momentum-resolved excitations above the ground state given by applying the variational state preparation method [\eref{eq:ground_state_prep}] starting from the fully polarized ground state at $h=\infty$ and integrating the variational gauge potential down to $h_f=2$ (a) or $h_f=0$ (b). The variational gauge potential is truncated at maximum string length $M$, where $M=0$ corresponds to $\mathcal A_h^\ast = 0$, i.e., no change of the initial wave function. Excitations are clearly much higher for $h_f=0$ because the system must cross a critical point at $h=1$. Solid lines correspond to the variational gauge potential from \eref{eq:alpha_l_tfi_var}, while dashed lines correspond to truncating the exact adiabatic gauge potential from \eref{eq:TFI_string}.}}
\label{fig:TFI_dens_k}

\end{figure}

\begin{figure}

\includegraphics[width=0.7\linewidth]{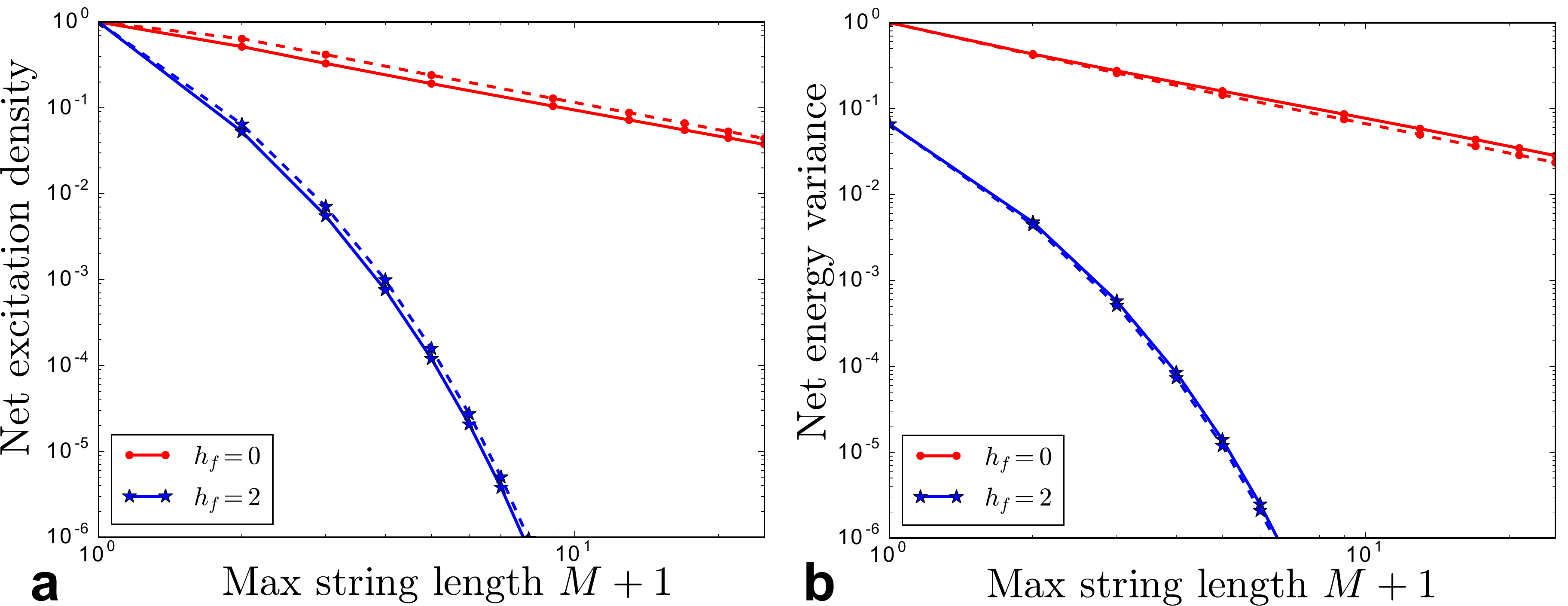}
\caption{\add{Dependence of variational preparation on string length $M$. (a) Excitations integrated over momenta as a function of $M$. Excitations die off exponentially in $M$ if one stays in the same phase ($h_f=2$) but become a power law upon crossing the phase transition ($h_f=0$). (b) Energy fluctuations of the variational ground state, showing the same behavior as the excitations in (a). The ground state preparation procedure is the same as in \fref{fig:TFI_dens_k}, where again solid lines use the variational and dashed lines use the truncated exact adiabatic gauge potential.}}
\label{fig:TFI_dens_M}

\end{figure}

\subsubsection{Non-integrable Ising model}

We can try exactly the same procedure to prepare ground states of the non-integrable Ising spin chain discussed in  \sref{subsection:NonintegrableIsing}. It is particularly interesting to prepare states at different angles $\theta$ out of the eigenstates at $\theta=0$, since the Hamiltonian is diagonal in $\sigma^z$ basis at that point and the eigenstates are trivial product states. The results of attempting to prepare arbitrary excited states in this strongly interacting model are shown in \fref{fig:TFI_dens2}, specifically looking at the energy fluctuations of a given variational eigenstate in the final Hamiltonian. A few trends are clear. First, as the number of spin terms included in $\mathcal A_\theta^\ast$ -- which we denote $M$ -- is increased, the average energy fluctuations of any given state decrease. This implies that the variational eigenstates are \add{confined to narrower and narrower microcanonical energy shells}. Similarly, the eigenstate-to-eigenstate fluctuations of the energy variance decrease with $M$ as well, which is consistent with the expectation from the eigenstate thermalization hypothesis. The latter is relevant \add{even though the} final Hamiltonian at $\theta=\pi/2$ is integrable because the protocol passes through non-integrable Hamiltonians for $\theta\in (0,\pi/2)$. Thus, it is impossible to target individual eigenstates with local counter-diabatic driving. At best one can hope to suppress energy fluctuations while populating all states within the relevant microcanonical shell, as confirmed by the data. Finally, we note that the energy fluctuations on average decrease as a function of the target energy density. This is perhaps not surprising, as the ground state will generally be much easier to target for this gapped protocol than eigenstates in the middle of the spectrum. Interestingly, there is no downturn at large energy densities, despite the fact that entropy density also decreases on that end of the spectrum and one might expect to easily prepare the maximally excited state. We leave further analysis of such state preparation in strongly interacting systems for future work. 

\begin{figure}
\includegraphics[width=0.6\linewidth]{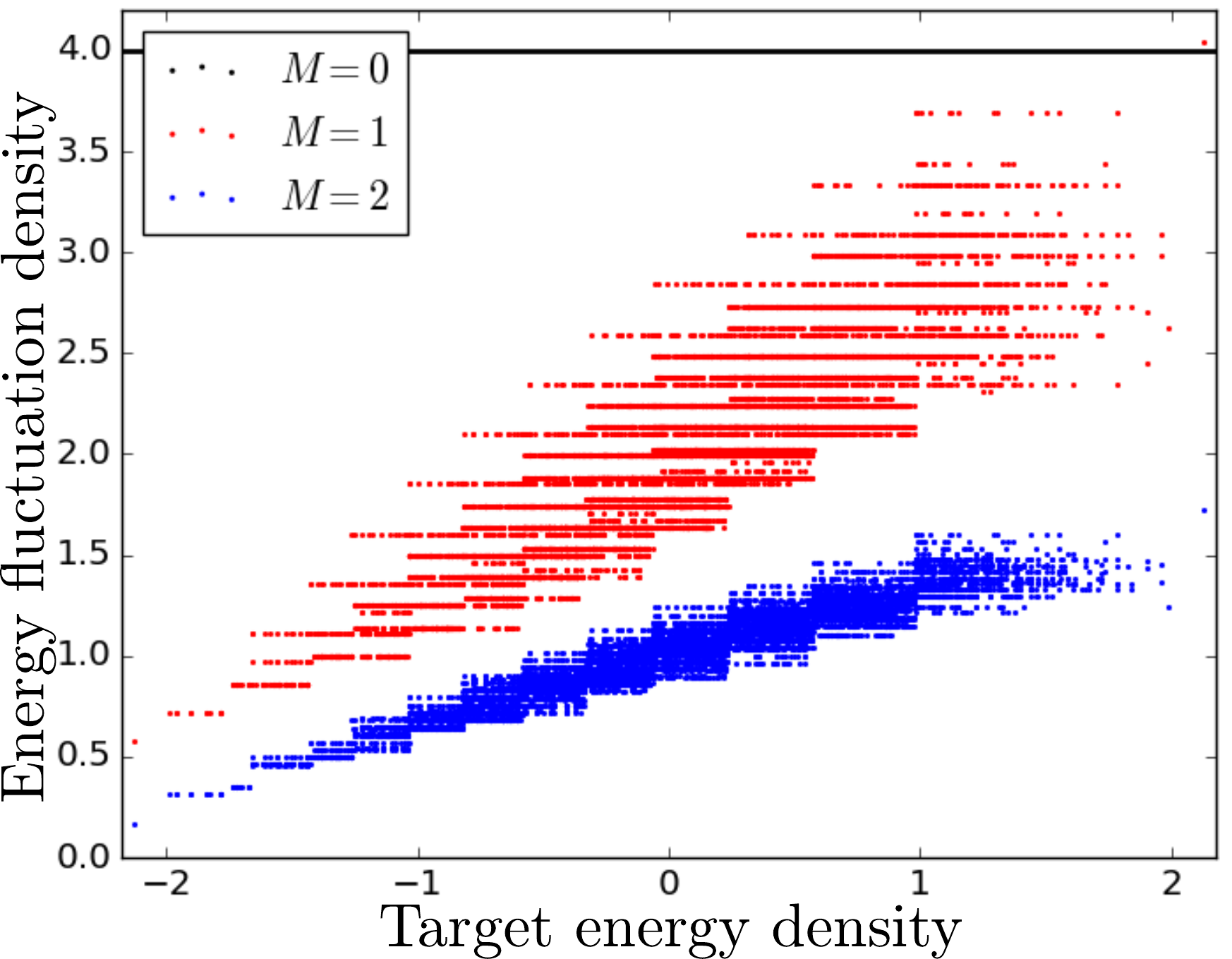}
\caption{\add{Energy fluctuations in the variational eigenstates of the Ising model at $\theta=\pi/2$ after evolving them out of $\sigma^z$ product states at $\theta=0$ according to \eref{eq:schr_eq_fast_two_spins} using the variational gauge potential $\mathcal A_\theta^\ast$ (see \sref{subsection:NonintegrableIsing}). Results are shown for chain of $14$ spins with $J=1$ and $h=2$. The red dots are the results for a single-spin variational gauge potential and the blue dots show the result for the two-spin ansatz. The black line is the energy variance for the initial eigenstates at $\theta=0$, which is independent of the state chosen within the $\sigma^z$ eigenbasis.}}
\label{fig:TFI_dens2}
\end{figure}

\section{Geometry of state space: Fubini-Study metric and Berry curvature}
\label{sec:geometric_tensor}

\emph{{\bf Key concept:} The geometry of the wave function in parameter space can be characterized by the geometric tensor, which is the covariance matrix of the gauge potentials. Its symmetric and antisymmetric parts define the Fubini-Study metric and the Berry curvature respectively. One can generalize these to classical systems using their representation as two-time correlation functions. }

\subsection{Geometry of the quantum ground state manifold}
\label{sec:geometry_gs_manifold}

Up till now, we have treated quantum and classical systems on an equal footing. In this section we will mostly focus on the geometric properties of the ground state manifold in quantum systems. For this reason, with the exception of \sref{sec:geom_classical_syst}, we restrict our discussion to the quantum case. We will return to classical systems later in the notes when we discuss non-adiabatic dynamical response.

\nocite{Provost1980_3}
The first notion of the quantum geometric tensor appeared in 1980 in \rcite{Provost1980_3}. Formally the geometric tensor is defined on any manifold of states smoothly varying with some parameter $\vect\lambda$: $|\psi_0(\vect\lambda)\rangle$.\footnote{From now on we will assume that the parameters can be multi-component.}  For now we will be primarily interested in the family of ground states of some Hamiltonian. We will assume that the ground state is either non-degenerate or, in the case of degeneracy, ground states are not connected by the matrix elements of generalized force operators $\mathcal{M}_\alpha\equiv \mathcal M_{\lambda_\alpha}=-\partial \mathcal H/\partial \lambda_\alpha$. The geometric tensor naturally appears when one defines the ``distance'' $ds$ between nearby states $|\psi_0(\vect\lambda)\rangle$ and $|\psi_0(\vect\lambda+d \vect \lambda)\rangle$:
\be
ds^2 \equiv 1-f^2=1-|\langle \psi_0(\vect\lambda)|\psi_0(\vect\lambda+d\vect\lambda)\rangle|^2,
\ee
where $f=|\langle \psi_0(\vect\lambda)|\psi_0(\vect\lambda+d\vect\lambda)\rangle|$ is the so called fidelity of the ground state. Note that $1-f^2$ is always positive.  Therefore, the Taylor expansion about $d\vect \lambda = 0$ does not contain any first order terms in $d\vect\lambda$, and starts with a quadratic term:
\be
ds^2= d\lambda_\alpha \chi_{\alpha\beta} d\lambda_\beta+O(|d\vect\lambda|^3),
\ee
where $\chi_{\alpha\beta}$ is as object known as the geometric tensor. To find this tensor explicitly let us note that $1-f^2$ is nothing but the probability to excite the system during a quantum quench where the parameter suddenly changes from $\vect\lambda$ to $\vect\lambda+d\vect \lambda$. \add{In other words,} $f^2$ is simply the probability to remain in the new ground state after this quench\add{, which is conserved under time evolution after the quench}. The amplitude of going to the excited state $|\psi_n\rangle$ is 
\be
a_n=\langle \psi_n(\vect\lambda+d\vect\lambda)|\psi_0(\vect\lambda)\rangle\approx d\lambda_\alpha \langle n|\overleftarrow \partial_\alpha |0\rangle = -d\lambda_\alpha \langle n|\partial_\alpha |0\rangle,
\ee
where the arrow over the derivative indicates that it acts on the left (derivatives without arrows implicitly act to the right). To shorten the notations, we introduce $\partial_\alpha\equiv \partial_{\lambda_\alpha}$ and $|n\rangle \equiv |\psi_n(\vect\lambda)\rangle$. Recall that [see \eref{gauge_pot_matr}]\footnote{As this section largely focuses on quantum systems, we set $\hbar=1$.}
\be
i\langle n |\partial_\alpha |m\rangle =\langle n|\mathcal A_\alpha |m\rangle.
\label{rel_AO}
\ee
Thus we see that the amplitude of going to the excited state at leading order in $d\vect\lambda$ is proportional to the matrix element of the gauge potential
\be
a_n= -d\lambda_\alpha \langle n|\partial_\alpha |0\rangle=i \langle n |\mathcal A_\alpha |0\rangle d \lambda_\alpha~.
\ee
Therefore the probability of transitioning to any excited state is given by taking a sum over $n\neq 0$:
\be
ds^2=\sum_{n\neq 0} |a_n^2|=\sum_{n\neq 0} d\lambda_\alpha d\lambda_\beta\langle 0|\mathcal A_\alpha |n\rangle \langle n| \mathcal A_\beta|0\rangle+O(|d\vect\lambda|^3)=d\lambda_\alpha d\lambda_\beta \langle 0|\mathcal A_\alpha \mathcal A_\beta|0\rangle_c+O(|d\vect\lambda|^3),
\ee
where the subscript $c$ implies that we are taking the connected correlation function (a.k.a. the covariance):
\be
\langle 0|\mathcal A_\alpha\mathcal A_\beta |0\rangle_c \equiv \langle 0|\mathcal A_\alpha\mathcal A_\beta |0\rangle-\langle 0|\mathcal A_\alpha |0\rangle \langle 0|\mathcal A_\beta |0\rangle.
\ee
This covariance precisely determines the geometric tensor introduced by Provost and Vallee \cite{Provost1980_3}:
\be
\chi_{\alpha\beta} \equiv\langle 0|\mathcal A_\alpha \mathcal A_\beta|0\rangle_c.
\label{geom_tens_def}
\ee
In terms of many-body wave functions the geometric tensor can be expressed through the overlap of derivatives:
\be
\chi_{\alpha\beta}=\langle 0|\overleftarrow \partial_\alpha \partial_\beta |0\rangle_c=\langle \partial_\alpha \psi_0|\partial_\beta\psi_0\rangle_c=
\langle \partial_\alpha \psi_0|\partial_\beta\psi_0\rangle-\langle \partial_\alpha \psi_0|\psi_0\rangle \langle \psi_0|\partial_\beta\psi_0\rangle.
\label{geom_tens_def1}
\ee
When $\hbar$ is not set to unity, the two definitions of the geometric tensor in Eqs.~(\ref{geom_tens_def}) and (\ref{geom_tens_def1}) differ by a factor of $\hbar^2$ as $\mathcal A_\alpha=i\hbar \partial_\alpha$. We will stick to \eref{geom_tens_def1} as the fundamental one because in this way it is always related to the distance between wave functions. The last term in this expression is necessary to enforce invariance of the geometric tensor under arbitrary global phase transformations of the wave function, $\psi_0(\vect\lambda)\to \exp[i\phi(\vect\lambda)]\psi_0(\vect\lambda)$, which should not affect the notion of the distance between different ground states.

\hrulefill

\exercise{Consider the global phase transformation $\psi_n(\vect\lambda)\to \exp[i\phi_n(\vect\lambda)]\psi_n(\vect\lambda)$ where $\phi_n(\vect \lambda)$ are smooth functions defined over the entire parameter manifold for each eigenstate $\psi_n$. What is the effect of this transformation on the gauge potential $\mathcal A$? Show that the ground state geometric tensor $\chi_{\alpha \beta}$ is invariant under this gauge transformation.}

\exerciseshere

\hrulefill

Note that in general the geometric tensor is not symmetric. Indeed because the operators $\mathcal A_\alpha$ are Hermitian one can show that $\chi$ is also Hermitian:
\be
\chi_{\alpha\beta}=\chi_{\beta\alpha}^\ast.
\ee
Only the symmetric part of $\chi_{\alpha\beta}$ determines the distance between the states; in the quadratic form
\[
ds^2=d \lambda_\alpha \chi_{\alpha\beta} d \lambda_\beta
\]
one can always symmetrize the indexes $\alpha$ and $\beta$ so that the antisymmetric part drops out. Nevertheless, both the symmetric and the anti-symmetric parts of the geometric tensor are very important. The symmetric part,
\be
g_{\alpha\beta}={\chi_{\alpha\beta}+\chi_{\beta\alpha}\over 2}={1\over 2}\langle 0 | ( \mathcal A_\alpha\mathcal A_\beta+\mathcal A_\beta \mathcal A_\alpha ) |0\rangle_c=\Re \langle 0 | \mathcal A_\alpha\mathcal A_\beta|0\rangle_c
\label{eq:Fubini-Study}
\ee
is called the Fubini-Study metric tensor.\footnote{Sometimes in the literature one understands the Fubini-Study metric to mean the complete metric in the projective Hilbert space where the number of parameters $\lambda_\alpha$ coincides with the dimension of the Hilbert space. Throughout this article, we mean by Fubini-Study the metric in parameter space defined by distances between wave functions, which can be thought of as the projection (or pullback) of the full Hilbert space metric onto the manifold of ground states.} The antisymmetric part of the geometric tensor defines the Berry curvature 
\be
F_{\alpha\beta}=i(\chi_{\alpha\beta}-\chi_{\beta\alpha})=-2\Im\chi_{\alpha\beta}=i\langle 0 |[\mathcal A_\alpha,\mathcal A_\beta]|0\rangle~,
\label{eq:Berry_curvature_geom_tens}
\ee
which we introduced earlier. The Berry curvature plays a crucial role in most known quantum geometric and topological phenomena.

Let us note that the Berry curvature can be expressed through the derivatives of the Berry connections:
\be
F_{\alpha\beta}=\partial_\alpha A_\beta-\partial_\beta A_\alpha,
\ee
where the Berry connection,
\be
A_\alpha=\langle 0 |\mathcal A_\alpha|0\rangle=i \langle 0|\partial_\alpha|0\rangle ~,
\ee
is just the ground state expectation of the gauge potential. One can easily check this through direct differentiation:
\be
\partial_\alpha A_\beta-\partial_\beta A_\alpha=i \langle 0 |\overleftarrow \partial_\alpha \partial_\beta|0\rangle-i \langle 0 |\overleftarrow \partial_\beta \partial_\alpha|0\rangle+i \langle 0 | \partial^2_{\alpha\beta}|0\rangle-i \langle 0 | \partial^2_{\beta\alpha}|0\rangle=i (\chi_{\alpha\beta}-\chi_{\beta\alpha})~.
\ee

It is well known that the Berry connection is directly related to the phase of the ground state wave function. Indeed if the wave function in position space, can be written as
\be
\psi_0=|\psi_0({\bf r},{\bm \lambda})| \exp[i\phi({\bm \lambda})]
\ee
we find that
\be
A_\alpha=-\int d{\bf r} |\psi_0|^2 \partial_\alpha \phi =- \partial_\alpha \phi.
\ee
Therefore the integral of $A_\alpha$ over a closed path $\mathcal C$ represents the total phase (Berry phase) accumulated by the wave function during the adiabatic evolution\cite{Pancharatnam1956_1,Berry1984_1}
\be
\varphi_B= \oint_{\mathcal C} \partial_\alpha \phi d\lambda_\alpha=-\oint_{\mathcal C} A_\alpha d\lambda_\alpha ~.
\ee
By Stokes' theorem, the same phase can be represented as the integral of the Berry curvature over the surface enclosed by the contour $\mathcal C$,
\be
\varphi_B=\int_S F_{\alpha\beta} d\lambda_\alpha\wedge d\lambda_\beta,
\ee
where the wedge product implies that the integral is directed. 

\begin{figure}
\includegraphics[width=0.5\columnwidth]{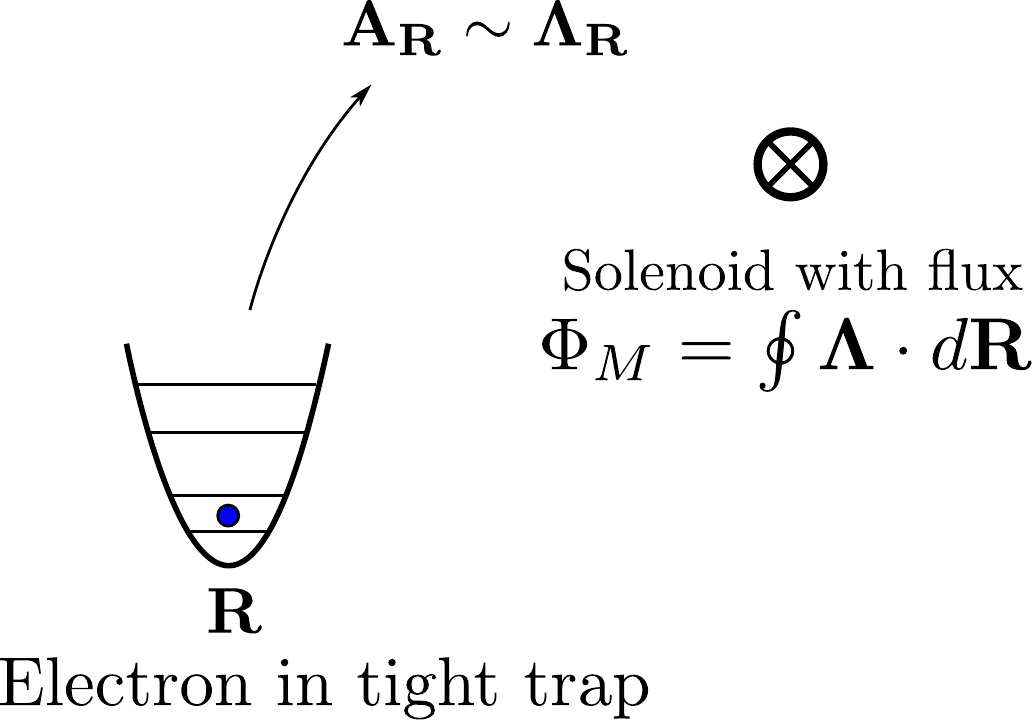}
\caption{Illustration of Aharonov-Bohm geometry being considered.}
\label{fig:aharonov_bohm}
\end{figure}

To get an intuition about the Berry curvature and the metric tensor let us consider two simple examples. First, following the original paper by Berry, let us consider the Aharonov-Bohm geometry (\fref{fig:aharonov_bohm}) \cite{Aharonov1959_1,Thouless1998_1}, namely a particle confined in a deep potential in the presence of a solenoid. The Hamiltonian for this system is
\be
\mathcal H={\left(\vect p-e \vect \Lambda(\vect r)\right)^2\over 2m}+ V(\vect r-\vect R),
\ee
where $\vect \Lambda$ is the electromagnetic vector potential (we use $\vect \Lambda$ to avoid the confusion with the Berry connection) and $V(\vect r-\vect R)$ is a confining potential near some point $\vect R$ outside the solenoid, where there is no magnetic field. For example, one can choose  $V(r)=m \omega^2 r^2 / 2$, which is simply the potential of a two-dimensional harmonic oscillator. Far from the solenoid, there is no magnetic field. Hence $\vect\nabla\times \vect \Lambda=0$, which implies that the vector potential can be written as a gradient of the magnetic potential $\Phi$ (cf. Ref.~\cite{Jackson1999_1}):
\[
\vect \Lambda=\vect \nabla \Phi\Rightarrow \Phi(\vect r,\vect R)=\int_{\vect R}^{\vect r} \vect \Lambda(\vect r^\prime) \cdot d\vect r^\prime ~.
\]
Because the vector potential is curl-free, the integral does not depend on the path, so the path can be a straight line. In principle, the lower limit of integration is arbitrary and does not have to be tied to $\vect R$. With this choice, however, it is guaranteed that whenever $\vect r$ is close to $\vect R$, the path does not cross the solenoid and thus does not break the curl-free requirement. One can easily check by explicitly plugging in the following functional form into the Schr\"odinger's equation that the vector potential can be locally eliminated by a gauge transformation:
\be
\psi_0({\bf r})=\tilde \psi_0({\bf r}-{\bf R}) \exp\left[i e \Phi({\bf r},{\bf R})\right].
\ee
Then the Hamiltonian for $\tilde \psi$ becomes independent of the vector potential and thus the wave function $\tilde\psi_0$, which is the ground state in the absence of the vector potential, can be chosen to be real. In this case the Berry connection with respect to the position of the trap $\vect R$ is, as we just discussed, the derivative of the phase with respect to $\vect{R}$:
\be
\vect A_{\vect R} =-e \partial_{\vect R} \Phi=e \partial_{\vect r} \Phi=e\vect \Lambda(\vect r).
\ee
More accurately one needs to average the vector potential over the wave function $\tilde \psi(\vect r-\vect R)$, but assuming that it is localized near $\vect R$, the averaging simply reduces to $\vect \Lambda(\vect R)$. Then the Berry phase for a cyclic path $\vect R(t)$ is $\varphi_B = \oint \vect A \cdot d \vect R = 2 \pi \Phi_M / \Phi_0$ if the path surrounds the solenoid, and zero if it does not, where $\Phi_0  =2\pi\hbar/e$ is the flux quantum of the electron. This follows directly from $\Phi_M=\oint \vect\Lambda(\vect R) \cdot d\vect R$ and the definition of the flux quantum with reinserted Planck's constant. 

From the above arguments we see that, up to fundamental constants, the Berry connection plays the role of the vector potential, hence the Berry phase assumes the role of the Aharonov-Bohm phase and the Berry curvature (curl of the Berry connection) plays the role of the magnetic field. We summarize this analogy in Table I. Note that the Berry curvature is more generally written as the arbitrary-dimensional curl, $F_{\alpha \beta} = \partial_\alpha A_\beta - \partial_\beta A_\alpha$, which is equivalent to representing the magnetic field as the off-diagonal components of the electromagnetic field-strength tensor. This analogy is very useful when we think about general parameter space and, as we will see later, this analogy is not coincidental. For example, we will see that like the magnetic field, the Berry curvature is the source of a Lorentz force.

\begin{table}
  \begin{center}
    \begin{tabular}{ c | c}
      {\bf Electromagnetism} & {\bf Quantum geometry} \\
      \hline
      \parbox[t]{7cm}{Vector potential\\$\vect A(\vect x)$} & 
      \parbox[t]{7cm}{Berry connection \\$\vect A (\vect \lambda)=i\langle \psi_0(\vect \lambda) | \nabla_\lambda \psi_0(\vect \lambda) \rangle $} \\
      \hline
      \parbox[t]{7cm}{Magnetic field/EM tensor\\$F_{ab}(\vect x)=\partial_{x_a} A_b (\vect x) - \partial_{x_b} A_a (\vect x) = \sum_c \epsilon_{abc} B_c$} &
      \parbox[t]{7cm}{Berry curvature\\$F_{\mu \nu}(\vect \lambda)=\partial_{\lambda_\mu} A_\nu (\vect \lambda) - \partial_{x_\nu} A_\mu (\vect \lambda)$} \\
      \hline
      \parbox[t]{7cm}{Aharonov-Bohm phase\\$\varphi_{AB}=\oint \vect A(\vect x) \cdot d \vect x$} & 
      \parbox[t]{7cm}{Berry phase \\$\varphi_{B}=\oint \vect A(\vect \lambda) \cdot d \vect \lambda$}
    \end{tabular}
    \caption{Comparison between electromagnetism and ground state (Berry) geometry in quantum mechanics. $\epsilon_{abc}$ is the Levi-Civita symbol.}
  \end{center}
\end{table}

Unlike the Berry phase/curvature, in this example the metric tensor does depend on details of the trapping potential $V(r)$. This is readily seen by considering that the particle is bound within some radius $\ell \ll R$ (for instance $\ell \sim 1/\sqrt{m \omega}$ for the harmonic oscillator). If we move the trap away from the solenoid by an amount $\Delta R$ such that $R\gg\Delta R \gg \ell$ the phase independent wave function $\tilde \psi_0(\vect r-\vect R)$ will rapidly change such that $\tilde \psi_0(\vect r-\vect R)$ and $\tilde \psi_0(\vect r-\vect R-\vect \Delta \vect R)$ will become almost orthogonal, while the phase $\phi$ will stay almost constant. So the metric tensor only slightly depends on the flux through the solenoid and instead strongly depends on how the ground state of $V$ changes with the position $\vect R$. For the isotropic harmonic oscillator, one easily sees that the metric tensor is isotropic: 
\be
g_{R_x R_x} = g_{R_y R_y} = \langle \mathcal A_{x} \mathcal A_x \rangle_c = \langle p_x^2 \rangle_c=\langle p_x^2\rangle = \frac{m \omega}{2}.
\ee
Going along a path, cyclic or not, one can define an invariant dimensionless length
\be
L_g = \oint ds = \int dt \sqrt{g_{\alpha \beta} \dot R_\alpha \dot R_\beta} = L \sqrt{\frac{m \omega}{2}} ~,
\ee
where $L = \int |\dot {\vect R}| dt$ is the length of the path in real space. Loosely speaking this length measures the number of orthogonal ground states traversed along the path. \add{A less hand-wavy interpretation of  $L_g$ is found in quantum information theory, where this distance may be related to the number of distinguishable states traversed for an optimal finite-strength measurement of the quantum system (cf.~Eq.~3 in \cite{Wootters1981}). We will show in \sref{sec:time_bounds} that the length $L_g$ also sets the minimum time required to move the particle around the solenoid without exciting it.}

Let us now analyze the geometry of another simple system, which we already encountered earlier: the quantum spin-1/2 in a magnetic field. As before, we choose parameters to be the angles $\theta$ and $\phi$ of the magnetic field. As a reminder, ground and excited states are [see \eref{eq:gs_es_qubit}]
\be
|g\rangle = \left( \begin{array}{c}  \cos (\theta / 2) \\ \mathrm{e}^{i \phi} \sin (\theta / 2)  \end{array} \right) ~,~|e\rangle = \left( \begin{array}{c}  \sin (\theta / 2) \\ -\mathrm{e}^{i \phi} \cos (\theta / 2)\end{array} \right) ~.
\ee
Direct evaluation of the geometric tensor for the ground state gives
\be
\chi_{\theta\theta}=\langle \partial_\theta g | \partial_\theta g \rangle - \langle \partial_\theta g | g \rangle \langle g | \partial_\theta g \rangle = {1\over 4},\; \chi_{\phi\phi}={1\over 4}\sin^2(\theta),\; \chi_{\theta\phi}={i\over 4}\sin(\theta) ~.
\label{eq:geom_tens_spin_half}
\ee
These expressions can also be obtained by calculating the covariance matrix of the gauge potentials,
\be
\mathcal A_\theta=i\partial_\theta=-{1\over 2}\tau_y,\; \mathcal A_\phi=i\partial_\phi=\frac{1}{2} \big(\sigma_z-1\big)=\frac{1}{2} \big(\tau_z \cos \theta + \tau_x \sin \theta -1\big) ~,
\label{eq:gauge_pot_spin_half}
\ee
which are generators of rotations in the $\theta$ and $\phi$ directions. Here the Pauli matrices $\tau$ are rotated to act in the basis of instantaneous eigenstates, e.g., $\langle e|\tau_x|g \rangle$=1. In this instantaneous basis the Hamiltonian is $\mathcal H=-h\tau_z$ (see \fref{fig:rot_inst_eig}). The equations above generalize to particles with arbitrary spin where instead of spin one-half operators like $\tau_y/2$, one uses the angular momentum operator $S_y$. 

\begin{figure}
\includegraphics[width=0.5\columnwidth]{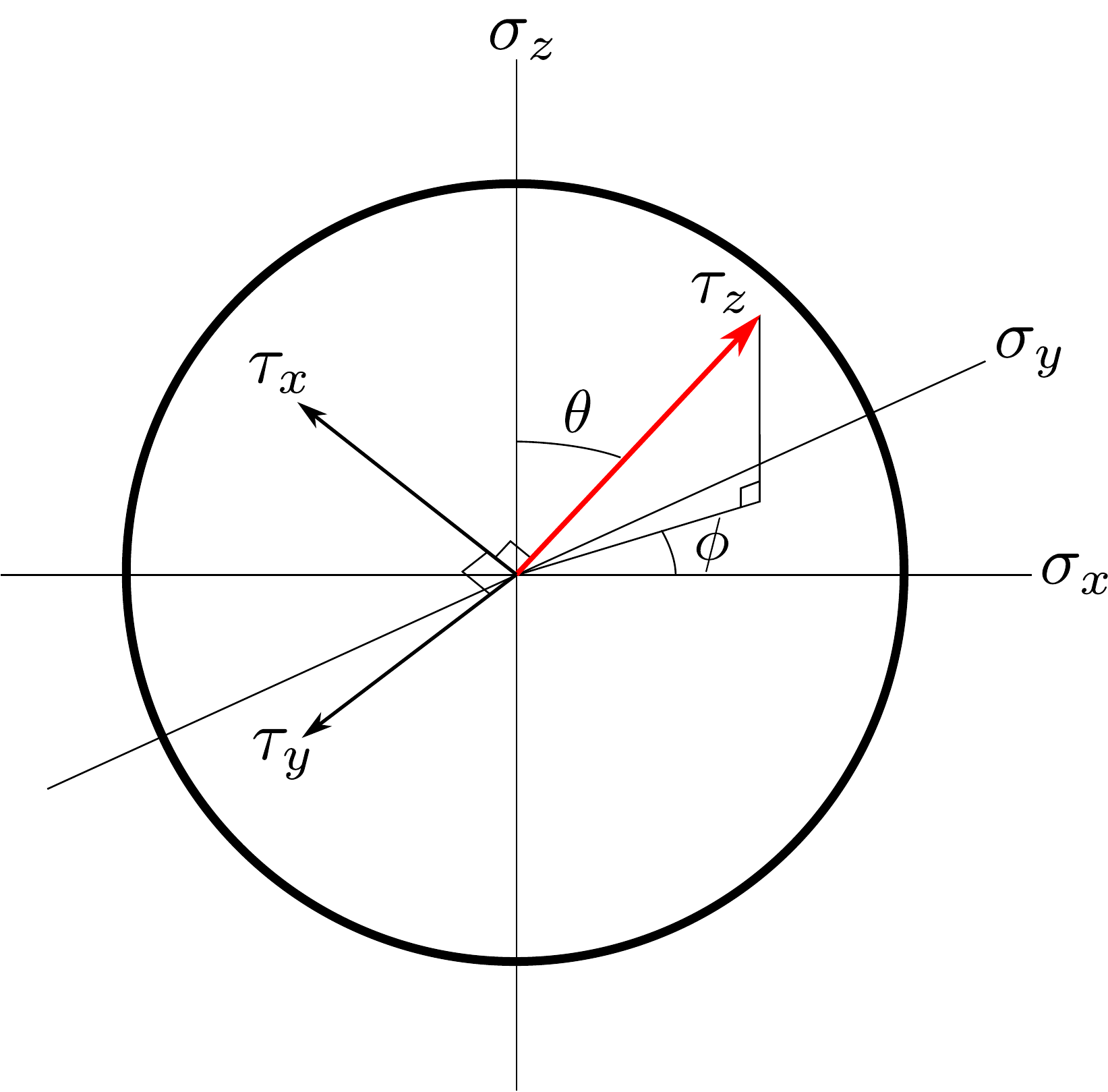}
\caption{Comparison of instantaneous eigenbasis ($\tau$) with the original one ($\sigma$). Rotations in the $\theta$ ($\phi$) direction correspond to rotations about the $-y'$ ($z$) axes, which are generated by $-\tau_y / 2$ and $\sigma_z/2$ respectively.}
\label{fig:rot_inst_eig}
\end{figure}

From the expression for the geometric tensor we see that the non-zero metric tensor components are
\be
g_{\theta\theta}={1\over 4},\; g_{\phi\phi}={1\over 4}\sin^2\theta,\; 
\label{eq:metric_components_spin}
\ee
and the Berry curvature is
\be
F_{\theta\phi}=\partial_\theta A_\phi-\partial_\phi A_\theta={1\over 2}\partial_\theta \cos(\theta)=-{1\over 2}\sin(\theta) = -F_{\phi \theta}~.
\ee
Note that the Fubini-Study metric for this model is equivalent to the metric of a sphere of radius $r=1/2$. It is interesting to note that for the excited state the metric tensor is the same while the Berry curvature has an opposite sign.

\hrulefill

\exercise{Calculate the covariance matrix of the spin-1/2 gauge potentials [\eref{eq:gauge_pot_spin_half}] and show that it gives the correct values for the geometric tensor.}

\exercise{\label{ex:geom_tens_sho} Find the geometric tensor of the shifted harmonic oscillator [\eref{eq:H_shifted_sho}] using the gauge potentials $\mathcal A_{x_0} = \hat p + p_0$ and $\mathcal A_{p_0} = -x$ [\eref{eq:gauge_pot_sho}].}

\exerciseshere

\hrulefill

\subsubsection{Relation to dissipative Kubo response}

The notion of distance between wave functions is very simple and intuitive but not directly measurable. However, we will now show that the geometric tensor can be related to a  standard Kubo susceptibility. Specifically it can be expressed through the unequal time correlation functions of physical operators in both real and imaginary times. We start by taking the geometric tensor in the so-called Kallen-Lehmann representation \cite{Kallen1952_1,Lehmann1954_1}, 
\be
\chi_{\alpha\beta}=\sum_{n\neq 0} \langle 0 |\mathcal A_\alpha|n\rangle\langle n|\mathcal A_\beta|0\rangle=\sum_{n\neq 0} {\langle 0|\partial_\alpha \mathcal H|n\rangle \langle n|\partial_\beta \mathcal H|0\rangle\over (E_n-E_0)^2}~,
\label{geom_tens_lehmann}
\ee
where the second equality follows from \eref{eq:gauge_pot_matrix_elem} and we assumed that the ground state is not degenerate. Let us use the following standard trick for connecting the Kallen-Lehmann representation of some observable to its unequal time correlation functions:
\be
{1\over (E_n-E_0)^2}=\int_{-\infty}^\infty d\omega {1\over \omega^2} \delta (E_n-E_0-\omega)=\int_{-\infty}^\infty \frac{d\omega}{\omega^2} \int_{-\infty}^{\infty} {dt\over 2\pi} \mathrm{e}^{-i (E_n-E_0-\omega)t}.
\ee
We can always add $\exp[-\epsilon |t|]$ to this integral to ensure convergence. Next we note that
\be
\langle 0| \mathrm{e}^{i E_0 t} \partial_\alpha \mathcal H \mathrm{e}^{-i E_n t} |n\rangle=
\langle 0|\partial_\alpha \mathcal H(t)|n\rangle
\ee
is the matrix element of the operator $\partial_\alpha \mathcal H$ in the Heisenberg representation. Plugging this into \eref{geom_tens_lehmann} we find
\be
\chi_{\alpha\beta}=\int_{-\infty}^\infty {d\omega\over 2\pi} {S_{\alpha\beta}(\omega)\over \omega^2}=\int_0^\infty {d\omega\over 2\pi} {S_{\alpha\beta}(\omega)+S_{\alpha\beta}(-\omega)\over \omega^2} ,
\label{eq:chi_alpha_beta_Sintegral}
\ee
where 
\be
S_{\alpha\beta}(\omega)=\int_{-\infty}^{\infty} dt \mathrm{e}^{i\omega t} \langle 0|
\partial_\alpha\mathcal H(t) \partial_\beta \mathcal H(0)|0\rangle_c.
\label{eq:S_alpha_beta}
\ee

This object $S_{\alpha\beta}(\omega)$ is just the Fourier transform of the observables' fluctuations. It is intricately related to standard Kubo linear response susceptibilities $\epsilon_{\alpha \beta}$ through the fluctuation-dissipation relation, which for the ground state reads~\cite{Jensen1991_1} (see also \aref{sec:kubo_metric_finite_temp}):
\be
S_{\alpha\beta}(\omega)=\left\{
\begin{array}{cc}
2\epsilon''_{\alpha\beta}(\omega) &\omega>0\\
0 & \omega<0.
\end{array}
\right.
\ee
In particular this relation implies that
\be
g_{\alpha\beta}=\int_0^\infty {d\omega\over 2\pi} {\epsilon''_{\alpha\beta}(\omega)+\epsilon''_{\beta\alpha}(\omega)\over \omega^2}.
\label{eq:metric_epsilon}
\ee
Thus the metric tensor can be directly measured from the symmetric part of $\epsilon''_{\alpha\beta}(\omega)$, which defines fluctuations (noise) and energy absorption~\cite{Jensen1991_1}. We note that a similar formula was derived independently in Ref.~\cite{HaukeArxiv2015_1}. For the special case of Bloch electrons, Neupert et al. have proposed using current fluctuations to measure the metric tensor as well \cite{Neupert2013_1}. As we will discuss in detail in \sref{sec:dyn_quantum_hall_effect}, the Berry curvature $F_{\beta\alpha}$ defines the Coriolis (or the Lorentz) force in parameter space. So it can be measured directly through the linear response of the generalized force $M_\beta$ to the ramp rate of the parameter $\lambda_\alpha$. 

\add{\subsubsection{Information theory and quantum speed limits}}
\label{sec:time_bounds}

Before moving on to global (topological) aspects of the geometric tensor, we would like to point out another important area where the metric tensor emerges naturally, namely quantum information theory. In that field, often going by the name quantum Fisher information or fidelity susceptibility, the metric tensor plays a fundamental role in information-theoretic distinguishability of states. One particularly important aspect of this is quantum parameter estimation, where it can be shown that the metric tensor sets a fundamental bound on the ability to determine an unknown parameter in the system \cite{Braunstein1994_1,Zanardi2008_1,Paris2009_1}. Therefore the ability to measure the metric tensor through fluctuations in quantum systems provides an important link between the theory of quantum information and practical experimental systems. \add{As we mentioned above, this allows us to interpret the distance between states defined by this metric in terms of distinguishable states traversed for an optimal finite-strength measurement of the quantum system.}

\add{Optimal measurements are intricately connected to optimal control, which we will now show provides a fundamental ``quantum speed limit'' on the ability of counter-diabatic driving or other similar protocols to drive the system between two states~\cite{margolus_98,Campbell2017,Funo2017,DeffnerArxiv2017}.}
Let us first make a general remark explaining what we mean by the speed limit. Consider a counter-diabatic protocol, which brings the system from initial state $|\psi_i\rangle$ to the target state $|\psi_f\rangle$ with $100\%$ fidelity in time $t^\ast$. It is clear that we can always scale the Hamiltonian by an arbitrary factor and reduce the time of the protocol by the same factor. So in order to define the maximum speed we have to fix the norm of the Hamiltonian or equivalently define the proper time. In order to do this, let us first rewrite the Schr\"odinger equation in dimensionless form:
\be
i\hbar {d\psi\over dt}=|| \mathcal H_{\rm CD}|| {\mathcal H_{\rm CD}\over ||\mathcal H_{\rm CD}||} \psi\;\Longleftrightarrow i{d\psi\over d\ell}={\mathcal H_{\rm CD}\over ||\mathcal H_{\rm CD}||} \psi,\quad \mbox{where}\; d\ell={||\mathcal H_{\rm CD}|| dt\over \hbar}.
\ee
Here $||\mathcal H_{\rm CD}||$ is the norm of the counter-diabatic Hamiltonian. As earlier we will stick to the Frobenius norm: $||\mathcal H_{\rm CD}||=\sqrt{{\rm Tr}[\mathcal H_{\rm CD}^2]}$. Note that the evolution with respect to $\ell$ happens with the unit norm Hamiltonian and thus is not affected by rescaling $\mathcal H_{\rm CD}$. Now the problem of finding the minimum length $\ell^\ast$:
\be
\ell^\ast={1\over \hbar}\int_{t_i}^{t_f} ||\mathcal H_{\rm CD}(t)|| dt
\ee
becomes non-trivial and defines the intrinsic speed limit. Note that $\ell^\ast$ is the intrinsic dimensionless length associated with the time evolution. We can only convert it into a physical time $\tau^\ast$ in the lab if we fix the norm of the Hamiltonian $\mathcal H_{\rm CD}$ to whatever value is appropriate for our laboratory setup. If we consider a setup in which the norm is restricted to $||\mathcal H_{\rm CD}||< \hbar\omega$, then the quantum speed limit is simply 
\be 
\tau^\ast=\frac{\ell^\ast}{ \omega}={1\over \hbar\omega}\int_{t_i}^{t_f} ||\mathcal H_{\rm CD}(t)|| dt.
\ee
Using the explicit form of the counter-diabatic Hamiltonian, $\mathcal H_{\rm CD}=\mathcal H+\dot\lambda \mathcal A_\lambda$, and changing the integration variables from $t$ to $\lambda$ in the definition of time, we see that
\be
\ell^{\ast}={1\over \hbar}\int_{\lambda_i}^{\lambda_f} d\lambda \left\| {\mathcal H\over \dot\lambda}+\mathcal A_\lambda\right\|.
\label{eq:time_bound_1}
\ee
The gauge potential can be always chosen to be orthogonal to the Hamiltonian in the sense that $||\mathcal H + a\mathcal A||^2=||\mathcal H||^2+a^2 ||\mathcal A||^2$, which follows from gauge invariance of $\mathcal A$ upon subtracting any terms commuting with (parallel to) $\mathcal H$. Now we trivially see that among all counter-diabatic protocols the \add{shortest evolution} is realized in the limit of infinite velocity $\dot\lambda\to\infty$, where the system is evolved with only the gauge potential $\mathcal A_\lambda$. This result is in fact very intuitive for some \add{of the} simple example\add{s} we analyzed earlier. \add{For instance,} consider a single spin in a magnetic field pointing initially along the $z$-axis, which we want to rotate in the $xz$-plane. The result we just discussed states that the fastest way to perform this rotation is to apply a magnetic field in the $y$-direction, which is the gauge potential in this case, and allow the spin to rotate by the desired angle. If the norm of the Hamiltonian is fixed (or bounded from above) then this evolution gives the fastest possible protocol to reach the desired ground state.

\add{The above expression for the quantum speed limit bears a strong resemblance to the Fubini-Study metric tensor \add{$g_{\lambda\lambda}$}.} This metric tensor defines the norm of the gauge potential $||\mathcal A||^2_0$. The only difference of this norm with the Frobenius norm is that instead of using the trace we are averaging $\mathcal A^2$ with respect to the ground state ($\langle 0|\mathcal A|0\rangle$ is zero for $\mathcal A$ orthogonal to $\mathcal H$). Alternatively one can define \add{$\mathcal A'=  \mathcal A \mathcal P_0+\mathcal P_0 \mathcal A$}, where $\mathcal P_0$ is the projector to the ground state manifold and note that \add{$2||\mathcal A||^2_0=||\mathcal A'||^2$}. Clearly $\mathcal A'$ has exactly the same effect on counter-diabatic driving as $\mathcal A$ if we are only interested in adiabatically following the ground state. Combining this discussion with \eref{eq:time_bound_1}, we see that the minimum time for the counter-diabatic protocol with restricted norm is
\be
\tau_{\rm min}=\frac{1}{\omega}\int_{\lambda_i}^{\lambda_f} \sqrt{g_{\lambda\lambda}} d\lambda.
\label{eq:time_bound_2}
\ee
This is nothing but the length of the segment connecting two points $\lambda_i$ and $\lambda_f$ in a curved manifold divided by norm of the maximum allowed Hamiltonian norm.

It is now straightforward to extend the analysis above to the multi-parameter space $\vect \lambda$. For any particular path $\vect \lambda(t)$ we already established that the minimum time is given by \eref{eq:time_bound_2}, where the integration is taken along this path. If we now minimize the time $\tau_{\rm min}$ we obtain that the shortest protocol corresponds to the counter-diabatic drive along the geodesic and the shortest time is set by the geodesic length:
\be
\tau_{\rm \min}=\frac{1}{\omega} {\rm min} \int \sqrt{g_{\alpha\beta}\, d\lambda_\alpha\, d\lambda_\beta}\add{.}
\ee
With this observation we will stop our brief detour into this very interesting and important topic, which goes beyond the scope of these notes and which contains many open questions. The main purpose of a rather brief discussion here was to highlight deep connections between problems of counter-diabatic driving, optimum state preparation and geometry of the ground state manifold \add{which we now discuss in much more detail.}

\subsection{Topology of the ground state manifold}

The geometric properties derived above give a local description of the wave functions living on the parameter manifold. From these local geometric properties, one can derive robust global properties of the manifold, i.e., its topology. In this section, we will discuss two types of topology that can be defined on the geometric tensor: the Chern number, which describes how the wave function wraps a closed parameter manifold via integrating the Berry curvature, and the Euler characteristic, which describes the topological shape of the Riemannian manifold encoded in the metric tensor.

As the Chern number has been extensively discussed in literature in many different contexts, we will mention it rather briefly and will concentrate more on the Euler characteristic, which has been discussed much less with respect to physical systems. We will also focus exclusively on two-dimensional manifolds, since the geometry and topology of higher-dimensional manifolds is much more complex and is often understood through various two-dimensional cuts. Please note that this section closely follows \rcite{KolodrubetzPRB2013_1}, and we refer interested readers there for more details.

\subsubsection{Basic definitions: Euler characteristic and Chern number}

The Euler characteristic of a (possibly open) manifold $\mathcal M$ is an integer equal to the integrated Gaussian curvature over the manifold with an additional boundary term:
\be
\xi(\mathcal M)=\frac{1}{2\pi} \left [\int_{\mathcal M} K dS+\oint_{\partial \mathcal M} k_g
dl \right].
\label{eq:gauss_bonnet}
\ee 
A standard notation for the Euler characteristic is $\chi$, but because we used this symbol for the geometric tensor, we will use $\xi$ instead. The two terms on the right side of \eref{eq:gauss_bonnet}
are the bulk and boundary contributions to the Euler characteristic
of the manifold.  We refer to the first term, 
\be
\xi_\mathrm{bulk}(\mathcal M) = \frac{1}{2\pi} \int_{\mathcal M} K dS ~,
\label{eq:chi_bulk}
\ee
and the second term,
\be
\xi_\mathrm{boundary}(\mathcal M) = \frac{1}{2\pi} \oint_{\partial \mathcal M} k_g dl ~,
\label{eq:chi_boundary}
\ee
as the bulk and boundary Euler integrals, respectively.  These terms, 
along with their constituents -- the Gaussian curvature ($K$), the geodesic curvature ($k_g$), the area
element ($dS$), and the line element ($dl$) -- are geometric invariants, meaning that
they remain unmodified under any change of variables.  More explicitly,
if the metric is written in first fundamental form as
\be
ds^2 = E d\lambda_1^2 + 2 F d\lambda_1 d\lambda_2 + G d\lambda_2^2 ~,
\ee
then these invariants are given by
\beq
\nn
K & = & \frac{1}{\sqrt g} \left[ \frac{\partial}{\partial \lambda_2} \left( 
\frac{\sqrt g \, \Gamma^2_{11}}{E} \right) - \frac{\partial}{\partial \lambda_1} \left(
\frac{\sqrt g \, \Gamma^2_{12}}{E} \right) \right] \\
\nn
k_g &=& \sqrt{g} G^{-3/2} \Gamma^1_{22} \\
\nn
dS &=& \sqrt g d\lambda_1 d\lambda_2 \\
dl &=& \sqrt G d\lambda_2 ~,
\label{eq:invariants}
\eeq
where $k_g$ and $dl$ are given for a curve of constant $\lambda_1$.  The metric 
determinant $g$ and Christoffel symbols $\Gamma^k_{ij}$ are 
\beq
g&=&EG-F^2 \\
\Gamma^{k}_{ij} &=& \frac{1}{2} g^{km} \left( \partial_j g_{im}
+ \partial_i g_{jm} - \partial_m g_{ij} \right) ~,
\eeq
where $g^{ij}$ is the inverse of the metric tensor $g_{ij}$.  As we see, the explicit expressions for the Euler characteristic are quite cumbersome, but they are known and unique functions of the metric tensor. A simple intuitive understanding of the Gaussian curvature of a two-parameter manifold comes from embedding the manifold in three dimensions. Then
\[
K={1\over R_1 R_2},
\]
where $R_1$ and $R_2$ are the principal radii of curvature, i.e., the minimal and the maximal radii of the circles touching the surface (see \fref{fig:principle_radii}). The geodesic curvature is the curvature of the boundary projected to the tangent plane, and is zero for a geodesic as the projection of the latter is locally a straight line. Thus, for example, the geodesic curvature of a great circle on a sphere is zero. For manifolds without boundaries like a torus or a sphere, the Euler characteristic simply counts the number of holes in the manifold. Thus for a sphere the Euler characteristic is $\xi=2$, for a torus $\xi=0$, and each additional hole gives an extra contribution of $-2$. 

\begin{figure}
\includegraphics[width=0.35\columnwidth]{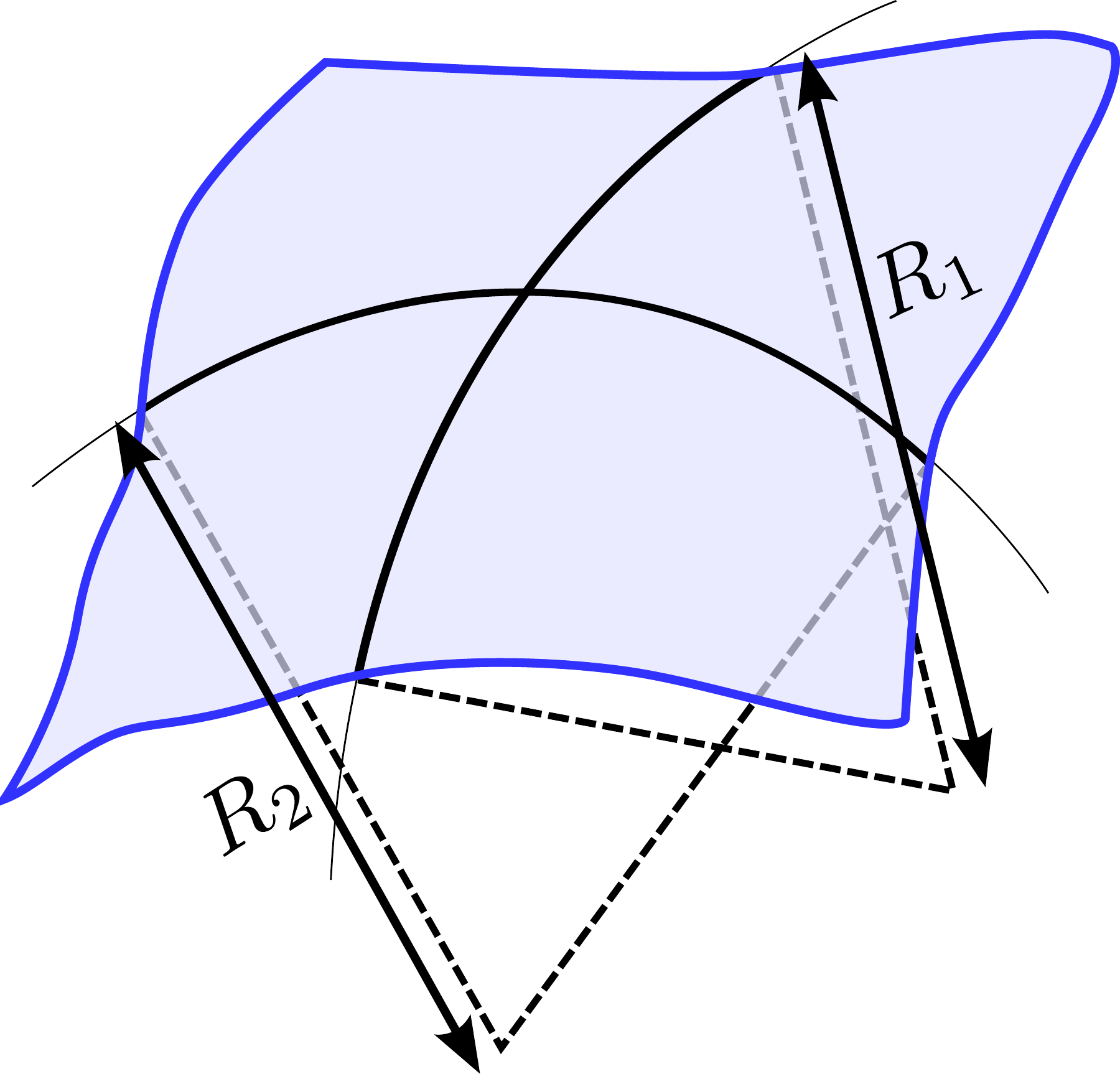}
\caption{Illustration of the principle radii $R_1$ and $R_2$ of a two dimensional manifold embedded in three dimensions.}
\label{fig:principle_radii}
\end{figure}

Another important topological invariant is the (first) Chern number, which is defined through the Berry curvature. To understand where it comes from, let us consider a closed manifold as shown in \fref{fig:spin_sphere} and choose an arbitrary closed contour on that sphere like the dashed line. Let us compute the Berry phase (flux) along this contour by two ways:
\be
\varphi_B^\mathrm{top}=\int_{S_\mathrm{top}} F_{\alpha\beta}\, d\lambda_\alpha \wedge d\lambda_\beta,\quad
\varphi_B^\mathrm{bottom}=-\int_{S_\mathrm{bottom}} F_{\alpha\beta}\, d\lambda_\alpha \wedge d\lambda_\beta,
\ee
where the minus sign in the second term appears because the top and bottom surfaces of the sphere bounded by the curve have opposite orientations with respect to this curve. Recall that $\varphi_B$ represents  the physical phase acquired by the wave function during the (adiabatic) motion in the parameter space. Since the wave function is unique the two phases should be identical up to an overall constant $2\pi n$. Thus we find that
\be
2\pi n=\varphi_B^\mathrm{top}-\varphi_B^\mathrm{bottom}=\oint_S F_{\alpha\beta}\, d\lambda_\alpha \wedge d\lambda_\beta
\ee
The integer $n$ is precisely the Chern number $C_1$ so we get
\be
C_1={1\over 2\pi}\oint_S F_{\alpha\beta}\, d\lambda_{\alpha} \wedge d\lambda_{\beta}~.
\ee

\begin{figure}
\includegraphics[width=0.4\columnwidth]{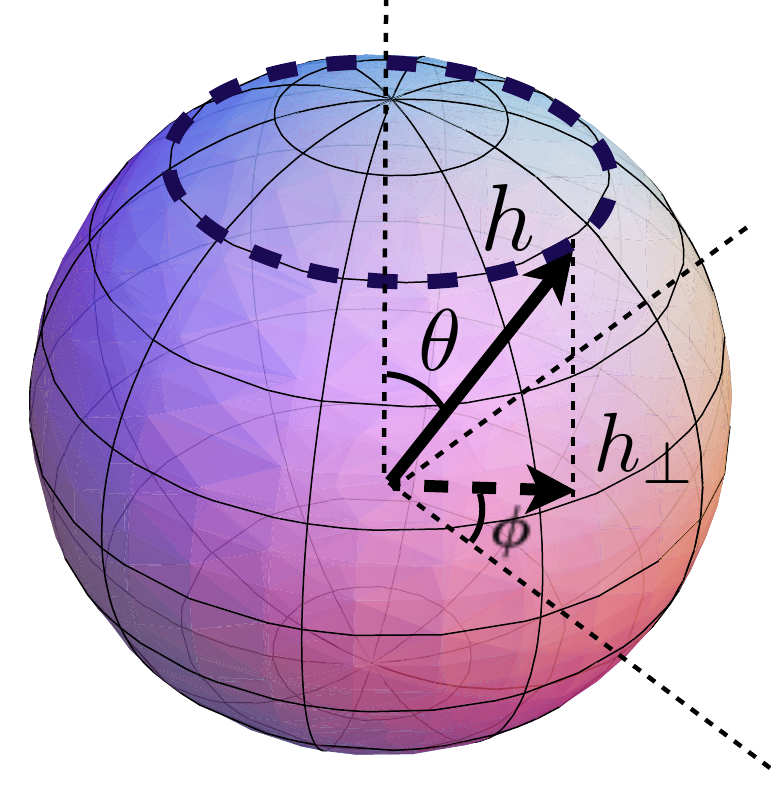}
\caption{Schematic representation of the spin in an external magnetic field, where the angles of the magnetic field $\theta$ and $\phi$ are the parameters. Figure reproduced with permission from Ref.~\citenum{Gritsev2012_1}.}
\label{fig:spin_sphere}
\end{figure}

Some intuition for the meaning of the Chern number can be obtained by returning to our electromagnetic analogy. We've seen that the Chern number is proportional to the Berry flux through a closed manifold $S$ in parameter space, which by Gauss's law for magnetism suggests that $C_1 \propto q_m$, the effective magnetic charge enclosed by the manifold. Indeed, it is known that if magnetic monopoles exist, they must be quantized \cite{Dirac1931_1}, which directly gives quantization of the Chern number. Berry showed that isolated degeneracies could act as sources of Berry curvature, and it is precisely the flux from these degeneracies that give rise to this topological invariant.

\subsubsection{Geometry and topology of a quantum spin-$1/2$}
\label{sec:topology_single_spin}

As our first example of these types of topology, let's pick up on the spin-1/2 in a magnetic field from the previous section. As before, the two-dimensional parameter space corresponds to the angles $(\theta, \phi)$ of the magnetic field with fixed magnitude, whose geometric tensor is given in \eref{eq:geom_tens_spin_half}. The diagonal real components of the geometric tensor $\chi_{\theta\theta}=g_{\theta\theta}=1/4$  and $\chi_{\phi\phi}=g_{\phi\phi}=1/4\sin^2(\theta)$ define a Riemannian metric which coincides with that of the sphere of the radius $1/2$ and constant Gaussian curvature $K=4$. The imaginary off-diagonal component of $\chi$ gives the Berry curvature: $F_{\phi \theta}={1/2}\sin(\theta)$. Thus we see that the Euler  invariant and Chern number are:
\beq
&&\xi={1\over 2\pi}\int K dS={1\over 2\pi} 4\int \sqrt{g\,}d\theta d\phi =2\\
&&C_1={1\over 2\pi} \int F_{\phi \theta}\, d\phi \wedge d\theta =1.
\eeq
The Euler characteristic implies that the metric topology of the spin-1/2 ground state in a rotating field is that of a sphere and the Chern number tells us that the wave function (i.e., the Bloch vector) ``wraps'' once if we adiabatically change the magnetic field over a full spherical angle. We can think of this Chern number as sourced by the degeneracy at magnetic field equal to zero, which our magnetic field sphere clearly encloses.

The example above can be generalized to an arbitrary spin $S$ in a magnetic field. The result is very simple: the spin-1/2 metric tensor is simply multiplied by $2S$:
\be
\chi_{\theta\theta}={S\over 2},\quad \chi_{\phi\phi}={S\over 2}\sin^2(\theta),\quad \chi_{\theta\phi}={iS\over 2}\sin(\theta).
\label{eq:geom_tens_spin_S}
\ee
The metric of the ground state manifold now coincides with that of the sphere of radius $\sqrt{S/2}$. The Euler characteristic, however, does not depend on the radius and thus we see that $\xi=2$ for any spin. Conversely, the Chern number is proportional to $S$: $C_1=2S$.

\hrulefill

\exercise{Prove \eref{eq:geom_tens_spin_S}. It may be useful to remember that $S_i$ is the generator of rotations about the $i$-axis for $i={x,y,z}$.}

\exercise{\label{ex:chern_insulator}The Chern number naturally appears in a band theory, where it is used to define various topological invariants and leads to numerous interesting physical effects such as topologically-quantized charge pumps \cite{Thouless1994_1}, the quantum Hall effect \cite{Thouless1982_1} and quantized spin-Hall effect in topological insulators \cite{Moore2007_1, Qi2010_1}. The Chern number for a (non-degenerate) band $\alpha$ is defined in a standard way: $C_1=\int_{BZ} dk_x dk_y F^\alpha_{k_x, k_y}$, where $F^\alpha_{k_x k_y} = \partial_{k_x} A^\alpha_{k_y} - \partial_{k_y} A^\alpha_{k_x}$ is the band Berry curvature and $A^\alpha_{k_j} = i \langle u_\alpha({\vect k}) | \partial_{k_j}  u_\alpha({\vect k}) \rangle$ is the band Berry connection. Here $|u_\alpha(\vect{k})\rangle$ are the Bloch wave functions corresponding to the band $\alpha$.

The simplest band model with a non-trivial topological structure  is two-dimensional with two atoms/orbitals per unit cell and complex hopping amplitudes such that the Hamiltonian reads
\be
\mathcal H=\sum_{k_x, k_y} \left(a^\dagger_{\bf k}, b^\dagger_{\bf k}\right)
\left(\begin{array}{cc}
h^z_{\bf k} & h^x_{\bf k}-i h^y_{\bf k}\\ 
 h^x_{\bf k}-i h^y_{\bf k} & -h^z_{\bf k}
\end{array}
\right)
\left(\begin{array}{c}
a_{\bf k}\\
b_{\bf k}
\end{array}
\right)+M \sum_{k_x, k_y} \left(a^\dagger_{\bf k}, b^\dagger_{\bf k}\right)
\left(\begin{array}{cc}
1 & 0\\ 
 0 & -1
\end{array}
\right)
\left(\begin{array}{c}
a_{\bf k}\\
b_{\bf k}
\end{array}
\right)
\ee
where $a^\dagger_{\bf k}\,, b^\dagger_{\bf k}, a_{\bf k}, b_{\bf k}$ are the momentum space fermion creation and annihilation operators corresponding to the two sublattices, $h^i_{\bf k}$ for $i=x,y,z$,  are smooth functions of $\bf k$ satisfying periodicity conditions $h^i_{{\bf k}+{\bf G}}=h^i_{\bf k}$, where $\bf G$ is the reciprocal lattice vector, and $M$ is the symmetry breaking field between two sublattices. For example for a Haldane model on a square lattice with a $\pi$-flux per plaquette and equal nearest neighbor and next nearest neighbor hopping $t$ we have~\cite{Neupert2011_1}:
\begin{eqnarray*}
&&h^z_{\bf k}=2t (\cos k_x-\cos k_y),\\
&&h^x_{\bf k}=t\left(\cos(\pi/4)+\cos(k_y-k_x-\pi/4)+\cos(k_y+\pi/4)+\cos(k_x-\pi/4)\right),\\
&&h^y_{\bf k}=t\left(-\sin(\pi/4)+\sin(k_y-k_x-\pi/4)+\sin(k_y+\pi/4)-\sin(k_x-\pi/4)\right).
\end{eqnarray*}

\begin{itemize}
\item Show that this problem can be mapped to the spin one half in an effective $\bf k$-dependent magnetic field of magnitude
\[
h_{\bf k}=\sqrt{(2t(\cos k_x-\cos k_y)+M)^2+4t^2 (1+\cos k_x \cos k_y)}
\]
and angles $\theta_{\bf k}$ and $\phi_{\bf k}$ defined according to
\[
\tan(\theta_{\bf k})={\sqrt{(h^x_{\bf k})^2+(h^y_{\bf k})^2}\over h^z_{\bf k}+M}={2t\sqrt{1+\cos k_x \cos k_y}\over 2t (\cos k_x-\cos k_y)+M},\; \tan \phi_{\bf k}={h^y_{\bf k}\over h^x_{\bf k}}.
\]
\item Identify the momenta corresponding to the north and south poles of the sphere. Argue that for $M=0$, region A labeled in \fref{fig:fbz_two_band_problem} maps to the top of the sphere and region B maps to the bottom of the sphere.
By arguing that the Chern number is invariant under parameterization from this mapping conclude that the Chern number of the lower band (corresponding to the ground state manifold) with respect to $k_x$ and $k_y$ is equal to one and the Chern number of the higher band (corresponding to the excited state manifold) is equal to negative one.

\begin{figure}[h]
\includegraphics[width=0.5\columnwidth]{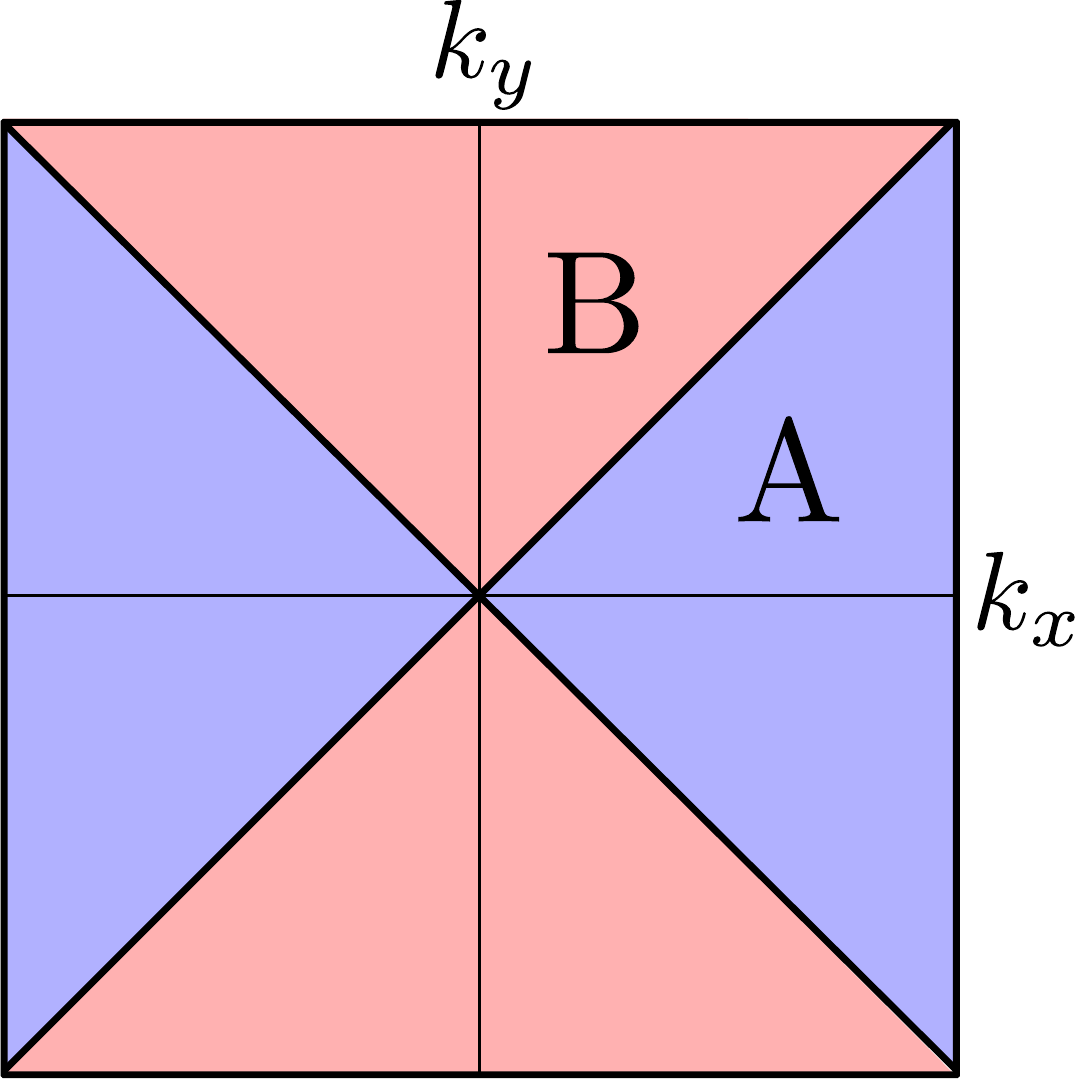}
\caption{First Brillouin zone illustrating the regions A and B that map to the two hemispheres.}
\label{fig:fbz_two_band_problem}
\end{figure}

\item Argue that both Chern numbers do not change with $M$ as long as $M<4t$ and that for $M>4t$ the band Chern numbers become zero.

\end{itemize}
 }

\exerciseshere

\hrulefill

We will not explore the Chern number further in this section, though we will return to it in the context of dynamical response in \sref{sec:dyn_quantum_hall_effect}. More recently, a few works have explored the ground state metric topology of assorted systems. We will now detail one such system, namely the metric of the quantum XY chain, closely following  \rcite{KolodrubetzPRB2013_1}.

\subsubsection{Geometry and topology of the quantum XY model}
\label{sec:metric_xy_chain}

Let us now analyze the geometric invariants for the quantum XY chain, which as we saw in \sref{sec:gauge_pot_xy_chain} is an integrable model on which many calculations may be done analytically. This model has a  rich phase diagram whose geometric properties we now explore. Its Hamiltonian is given by \eref{eq:H_XY_chain_orig}, :
\be 
\mathcal H=-\sum_{j=1}^L \big[ J_x 
\sigma_{j}^x \sigma_{j+1}^x + J_y
\sigma_{j}^y \sigma_{j+1}^y + h \sigma_{j}^z \big] ~,
\ee 
where we parameterize $J_{x,y}$ as
\be 
J_x=J \left( \frac{1+\gamma}{2} \right),\; J_y=J \left( \frac{1-\gamma}{2} \right)~.
\ee 
As before, we add tuning parameter $\phi$, corresponding
to simultaneous rotation of all the spins about the $z$-axis by
angle $\phi/2$.

\begin{figure}
\includegraphics[width=.6\linewidth]{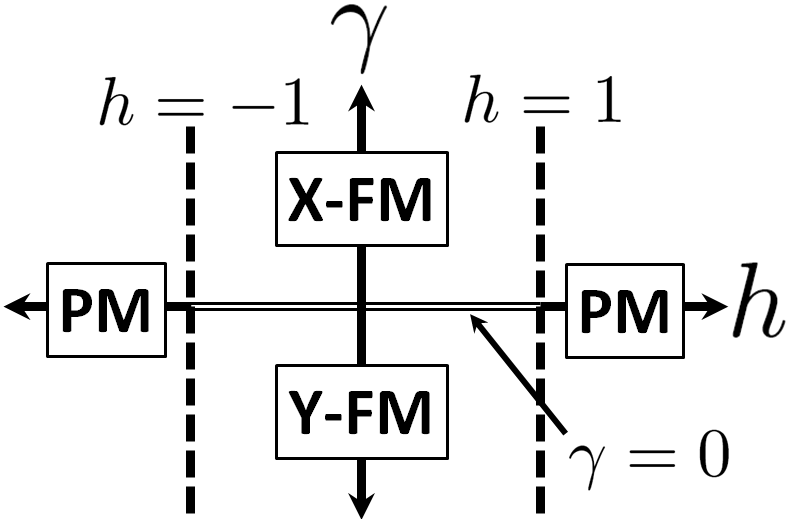}
\caption{Ground state phase diagram of the XY Hamiltonian [\eref{eq:H_XY_chain_orig}]
for $\phi=0$.  The rotation parameter $\phi$ modifies the Ising 
ferromagnetic directions, otherwise maintaining all features of the phase diagram.  As a function
of transverse field $h$ and anisotropy $\gamma$, the ground state undergoes
continuous Ising-like phase transitions between paramagnet and ferromagnet at $h=\pm 1$
and anisotropic transitions between ferromagnets aligned along 
X and Y directions (X/Y-FM) at $\gamma=0$.  Figure reproduced with permission from \rcite{KolodrubetzPRB2013_1}.}
\label{fig:phasediagram}
\end{figure}

The phase diagram for this model is shown in \fref{fig:phasediagram} (see Ref.~\cite{Damle1996_1} for details). There is a
phase transition between paramagnet and Ising ferromagnet at $|h|=1$ and
$\gamma \neq 0$. There is an additional critical line at the
isotropic point $\gamma=0$
for $|h|<1$. The two transitions meet at multi-critical points when $\gamma=0$
and $|h|=1$. Another notable line is $\gamma=1$, 
which corresponds to the transverse-field Ising (TFI) chain. 
Finally let us note that there are two other special lines $\gamma=0$ and $|h|>1$ 
where the ground state is fully polarized along the magnetic field and thus $h$-independent. 
These lines are characterized by vanishing susceptibilities including vanishing metric 
along the $h$-direction. One can show that such state is fully  protected by the rotational symmetry of the model and can be terminated only at the critical (gapless) point~\cite{KolodrubetzPRB2013_1}. The phase diagram is invariant under changes of the rotation
angle $\phi$. 

The exact and variational gauge potentials for this model were derived in \sref{sec:gauge_pot_xy_chain}. We note here that the exact adiabatic gauge potentials may be written as
\be
\mathcal A_\lambda = -\frac{1}{2} \sum_k \big( \partial_\lambda \theta_k \big) \tau_k^y,
\ee
where $\lambda=\{h,\gamma\}$ and $\tau_k^{x,y,z}$ are Pauli matrices that act in the 
instantaneous ground/excited state basis, i.e., the basis in which $H_k = -\sqrt{(h-\cos k)^2 + \gamma^2 \sin^2 k}~\tau_k^z$. Similarly, for the parameter $\phi$, 
\be
\mathcal A_\phi = \frac{1}{2} \sum_k \left[ \cos(\theta_k) \tau^z_k + \sin(\theta_k) \tau^x_k -1 \right].
\ee
These expressions make calculating the metric tensor simple. From
\be
g_{\mu \nu} = \frac{1}{2} \langle g|(\mathcal A_\mu \mathcal A_\nu +\mathcal A_\nu\mathcal A_\mu ) |g\rangle_c
\ee
we find
\be
g_{hh}={1\over 4}\sum_k
\left({\partial\theta_k\over\partial
  h}\right)^2,\;
 g_{\gamma \gamma}={1\over
  4}\sum_k \left({\partial\theta_k\over\partial
  \gamma}\right)^2,\; g_{h\gamma}={1\over
  4}\sum_k {\partial\theta_k\over\partial h}{\partial\theta_k\over
  \partial \gamma},\; g_{\phi\phi}={1\over 4}\sum_k
     \sin^2(\theta_k).\; 
\ee
The remaining two components of the metric tensor, $g_{h\phi}$ and $g_{\gamma \phi}$, are equal to zero.

The $g_{hh}$ component of the metric tensor, known as the fidelity susceptibility, has been computed analytically for finite size systems~\cite{Damski2013_1,Damski2015_1}. The remaining expressions can be analytically evaluated in the thermodynamic limit, where the summation becomes integration over momentum space. Calculating these integrals, one finds that
\beq 
&& \frac{g_{\phi\phi}}{\add L}= \frac{1}{8} \left\{
\begin{array}{cc}
\frac{|\gamma|}{|\gamma|+1}, &
|h|<1\\ \frac{\gamma^2}{1-\gamma^2}\left( \frac{|h|}
{\sqrt{h^2-1+\gamma^2}} - 1 \right) , &
|h|>1
\end{array}
\right.
\label{eq:g_phiphi}\nonumber\\
&&  \frac{g_{hh}}{\add L}= \frac{1}{16} \left\{\begin{array}{cc}
      \frac{1}{|\gamma| (1-h^2)}, & |h|<1\\ \frac{|h| \gamma^2}
{(h^2-1)(h^2-1+\gamma^2)^{3/2}}, & |h|>1
\end{array}\right. \label{eq:g_hh}\nonumber\\
&& \frac{g_{\gamma \gamma}}{\add L}= \frac{1}{16} \left\{\begin{array}{cc}
      \frac{1}{ |\gamma| (1+|\gamma|)^2}, & |h|<1\\ \left( \begin{array}{cc} 
	\frac{2}{(1-\gamma^2)^2} \Big[ \frac{|h|}{\sqrt{h^2-1+\gamma^2}} - 1 \Big] - \\
	 \frac{|h| \gamma^2}{(1-\gamma^2)(h^2-1+\gamma^2)^{3/2}}
\end{array} \right), & |h|>1
\end{array}\right. \label{eq:g_gg}\nonumber\\
&& \frac{g_{h \gamma}}{\add L}= \frac{1}{16} \left\{\begin{array}{cc} 0, &
|h|<1\\ \frac{-|h| \gamma}{h (h^2-1+\gamma^2)^{3/2}} , & |h|>1
\end{array}\right. \label{eq:g_hg}
\eeq
\add{Note that all components of the metric tensor are extensive, as expected, which mathematically comes from the replacement $\sum_k \to (L/2\pi)$.}

\add{While the remainder of this section will focus on exact results, let us briefly discuss the geometric tensor derived from  the variational gauge potentials found earlier [Eqs.~(\ref{eq:A_h_ast_tfi}) and (\ref{eq:alpha_l_tfi_var})]. In the language of operator strings, the geometric tensor for changing $h$ is given by
\be 
\frac{g_{h h}}{L}=\frac{1}{L} \left\langle \mathcal{A}_h \mathcal{A}_h \right\rangle=\sum_l (2\alpha_l)^2.
\ee
If we only go up to strings of length $M$ and use the exact coefficients in the thermodynamic limit [\eref{eq:TFI_string}], this results in
\be 
\frac{g_{hh}}{L}=\frac{1}{16} \left\{ \begin{array}{l} (1-h^2)^{-1} (1-h^{2M}) \quad {\rm for} \quad h^2<1 \\ h^{-2} (h^2-1) (1-h^{-2M}) \quad {\rm for} \quad h^2>1. \end{array} \right.
\ee
As expected, the metric tensor exponentially converges to the exact result with increasing string length $M$ as long as the system is not critical. At the critical point $h=1$, the approximate metric tensor becomes $M/16$ such that it only diverges linearly with $M$. Thus, critical properties only converge algebraically with $M$. Similar rounding of the phase transition is found due to finite size effects in Refs.~\citenum{Damski2013_1} and \citenum{Damski2015_1}.}

\begin{figure}
\includegraphics[width=.5\linewidth]{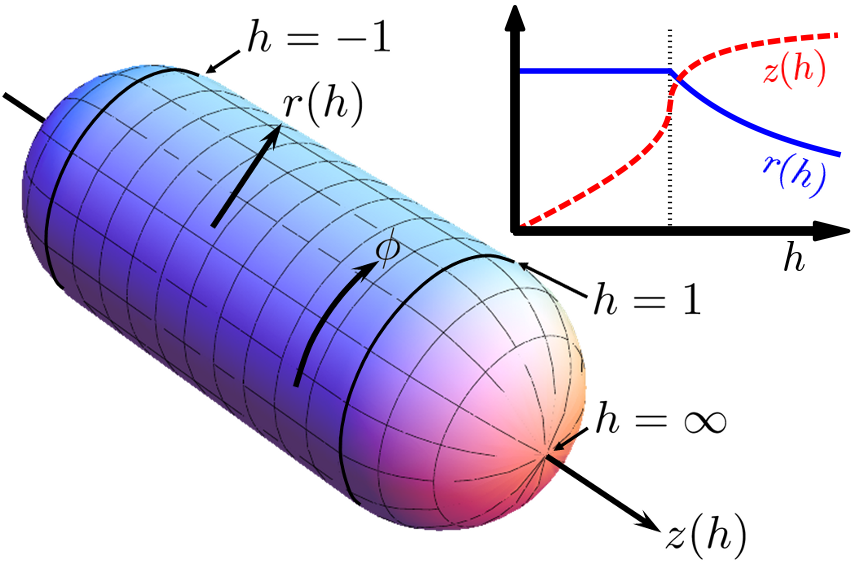}
\caption{Equivalent graphical representation of the phase diagram of
  the transverse field Ising model ($\gamma=1$) in the $h-\phi$ plane. 
  The ordered ferromagnetic phase maps to a cylinder of constant radius.
  The   disordered paramagnetic phases $h>1$ and $h<-1$ map to 
  the two hemispherical caps. The inset shows how the cylindrical
  coordinates $z$ and $r$ depend on the transverse field $h$. 
  Figure reproduced with permission from \rcite{KolodrubetzPRB2013_1}.}
\label{fig:ising_h_phi}
\end{figure}

Returning to the exact metric tensor, we can visualize the ground
state manifold by building an equivalent (i.e., isometric) surface
and plotting its shape. It is convenient to
focus on a two-dimensional manifold by fixing
one of the parameters. We then represent the two-dimensional 
manifold as an equivalent three-dimensional surface.
To start, let's fix the anisotropy parameter
$\gamma$ and consider the $h-\phi$ manifold. Since the metric tensor has cylindrical
symmetry, so does the equivalent surface. Parameterizing our shape
in cylindrical coordinates and requiring that
\be
dz^2+dr^2+r^2d\phi^2=g_{hh}dh^2+g_{\phi\phi} d\phi^2 ~,
\ee 
we see that
\be 
r(h)=\sqrt{g_{\phi\phi}},\;z(h)=\int_0^h dh_1
\sqrt{g_{hh}(h_1)-\left({dr(h_1)\over dh_1}\right)^2}.  
\ee 
Using \eref{eq:g_phiphi}, we explicitly find
the shape representing the XY chain. In the Ising limit ($\gamma=1$), we get
\be
r(h)={\add{\sqrt L} \over 4}~,~ z(h)={\add{\sqrt L} \arcsin(h) \over 4}~\mathrm{for}~|h|<1~;~~
r(h)={\add{\sqrt L} \over 4 |h|}~,~z(h)=\add{\sqrt{L}} \left( {\pi\over 8}{|h|\over h}+{\sqrt{h^2-1}\over  4 h} \right) ~\mathrm{for}~|h|>1.
\label{eq:cylinder_shape_TFI}
\ee
The phase diagram is thus represented by a
cylinder of radius $\add{\sqrt L}/4$ corresponding to the ferromagnetic phase
capped by the two hemispheres representing the
paramagnetic phase, as shown in \fref{fig:ising_h_phi}. It is easy to check
that the shape of each phase does not depend on the anisotropy
parameter $\gamma$, which simply changes the aspect ratio and radius of the
cylinder. Because of the relation $r(h)=\sqrt{g_{\phi\phi}}$, this 
radius vanishes as the anisotropy parameter $\gamma$ goes to zero.
By an elementary integration of the Gaussian curvature,
the phases have bulk Euler integral $0$ for the ferromagnetic cylinder
and $1$ for each paramagnetic hemisphere.  These numbers add up to $2$
as required, since the full phase diagram is topologically equivalent
to a sphere. From~\fref{fig:ising_h_phi},
it is also clear that the phase boundaries at $h=\pm 1$ are 
geodesics, meaning that the geodesic curvature (and thus 
the boundary contribution $\xi_\mathrm{boundary}$) is zero for a
contour along the phase boundary.  One can show that this boundary
integral protects the value of the bulk integral and vice versa.

In the Ising limit ($\gamma=1$), the 
shape shown in~\fref{fig:ising_h_phi}, can also be easily seen from directly computing the curvature $K$
using \eref{eq:invariants}.  Within the ferromagnetic phase, the 
curvature is zero -- no surprise, given that the metric is flat
by inspection.  The only shape with zero curvature and
cylindrical symmetry is a cylinder.  Similarly, within the 
paramagnet, the curvature is a constant $K=16/\sqrt{L}$, like that
of a sphere.  Therefore, to get cylindrical symmetry, the 
phase diagram is clearly seen to be a cylinder capped by two hemispheres.

We can also reconstruct an equivalent shape in the $\gamma-\phi$ plane. In
this case we expect to see a qualitative difference for $|h|>1$ and
$|h|<1$ because in the latter case there is an anisotropic phase
transition at the isotropic point $\gamma=0$, while in the former case
there is none. These two shapes are shown in
\fref{fig:anising_g_phi}. The anisotropic phase transition is manifest
in the conical singularity that develops at $\gamma=0$.\footnote{
We note a potential point of confusion, namely that a naive application of
\eref{eq:invariants} would seem to indicate that the curvature is a 
constant $K=4/\sqrt{L}$ in the ferromagnetic phase for $\gamma>0$, in which case the singularity at
$\gamma=0$ is not apparent.  However, a more careful derivation shows that
the curvature is indeed singular at $\gamma=0$: 
$K\sqrt{L}=4-8(1-\gamma)\frac{\partial^2}{\partial^2 \gamma} |\gamma|
=4-16\delta(\gamma)$, where $\delta(\gamma)$ is the Dirac delta function.}

The singularity at $\gamma=0$ yields a non-trivial bulk Euler integral
for the anisotropic phase transition. To see this, consider the bulk integral
\be 
\xi_\mathrm{bulk}(\epsilon)=\lim_{L\to\infty}\int_0^{2\pi} d\phi
\int_{\epsilon}^{\infty} d \gamma \, \sqrt{g(\gamma,\phi)}K(\gamma,\phi) ~.
\label{eq:euler_bulk}
\ee 
In the limit $\epsilon \to 0^+$,
this integral has a discontinuity as a function of $h$ at the phase transition, 
as seen in \fref{fig:anising_g_phi}. Thus, $\xi_\mathrm{bulk}
\equiv \xi_\mathrm{bulk}(\epsilon=0^+)$ can be used as a geometric
characteristic of the anisotropic phase transition. Direct calculation shows that  
$\xi_\mathrm{bulk}=1/\sqrt 2$ in the ferromagnetic phase and $\xi_\mathrm{bulk}=1$ in the 
paramagnetic phase. This non-integer geometric invariant is due to the existence of 
a conical singularity.

\begin{figure}
\includegraphics[width=0.7\linewidth]{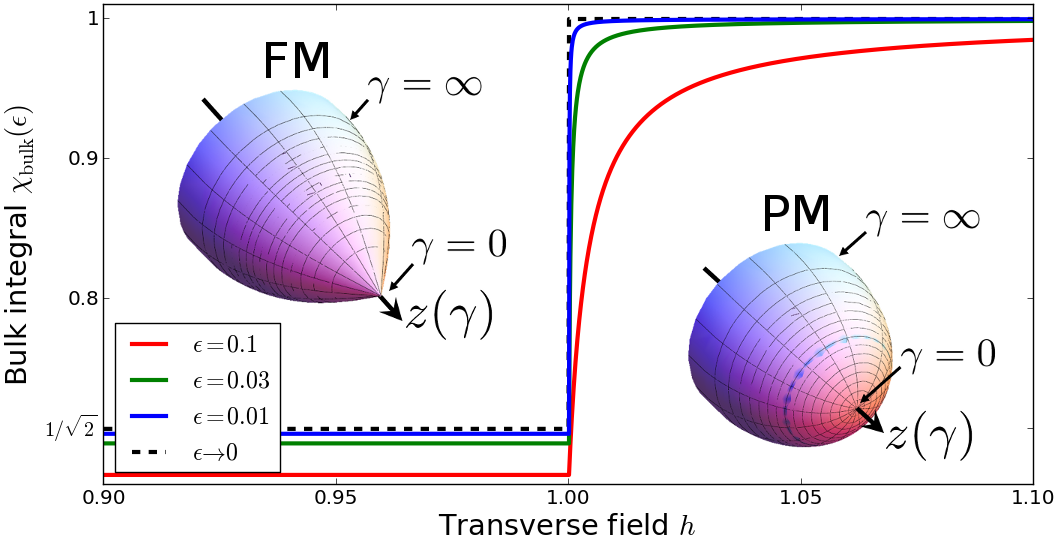}
\caption{(insets) Equivalent graphical representation of the phase diagram of
  the XY model in the $\gamma-\phi$ plane, where $\gamma\in[0,\infty)$ and
  $\phi\in[0,2\pi]$. The right inset shows the paramagnetic
  disordered phase and the left inset represents the
  ferromagnetic phase. It is clear that in the latter case
  there is a conical singularity developing at $\gamma=0$ which represents
  the anisotropic phase transition.
  The plots show bulk Euler integral $\xi_\mathrm{bulk}(\epsilon)$ as defined in \eref{eq:euler_bulk},
  demonstrating the jump in $\xi_\mathrm{bulk}$ at the phase transition
  between the paramagnet and ferromagnet in the limit $\epsilon \to 0^+$.
  Figure reproduced with permission from \rcite{KolodrubetzPRB2013_1}.}
\label{fig:anising_g_phi}
\end{figure}

A careful analysis shows that in both cases the bulk Euler characteristics are protected by the universality of the transition, i.e., if one adds extra terms to the Hamiltonian which do not qualitatively affect the phase diagram, then the bulk Euler characteristic does not change. The details of the proof are available in \rcite{KolodrubetzPRB2013_1}, but the basic idea is very simple. The sum of the bulk and the boundary Euler characteristics is protected by the geometry of the parameter manifold. As long as the boundary  of the manifold coincides with the phase boundary, all components of the metric tensor become universal~\cite{Venuti2007_1}. Therefore it is not surprising that the geodesic curvature also becomes universal and thus the boundary Euler characteristic is protected. As a result the bulk Euler characteristic is protected too. It is interesting that, unlike critical exponents, the bulk Euler characteristic truly characterizes the phase transition and does not depend on the parameterization. One can also analyze the Euler characteristic and the Gaussian curvature in the $h-\gamma$ plane~\cite{Zanardi2007_1, KolodrubetzPRB2013_1}. One finds additional non-integrable curvature singularities near the anisotropic phase transition and near the multi-critical point. This curvature singularity implies that the Euler characteristic of different phases becomes ill-defined and can no longer be used for their characterization.

These ideas may be readily extended to other phase transitions beyond the XY model. For the specific case of non-interacting Bloch bands, the Euler topology has already been explored in Refs.~\cite{Ma2013_1} and \cite{Yang2015_1}. More generally, the study of critical scaling of the quantum geometric tensor begin with pioneering work by Venuti and Zanardi \cite{Venuti2007_1, Zanardi2007_1}, which has spurred research on a wide variety of models \cite{Hamma2008_1,Gu2008_1,Garnerone2009_1,Albuquerque2010_1,Rezakhani2010_1,Gu2010_1,Mukherjee2011_1,Banchi2014_1,DeGrandi2011_1,
Kumar2012_1,Maity2015_1,Kumar2014_1}.
In addition, a great deal of work has been done developing related ideas for finite temperature systems, both quantum and classical, which should allow one to treat thermal phase transitions in a similar language \cite{Uhlmann1986_1,Ruppeiner1995_1,Crooks2007_1,Zanardi2007_2,Huang2014_1,Viyuela2014_1}.
In the next section we will describe one possible way to generalize these ideas to stationary mixed states such as the thermal Gibbs ensemble.

First, however, let us revisit the local information that the metric provides. One important property of the metric are its geodesics, i.e., paths $\lambda(s)$ that locally minimize the metric distance
\be
\ell[\lambda(s)]=\int_0^1  ds \sqrt{g_{\mu \nu} \dot \lambda^\mu \dot \lambda^\nu}
\label{eq:geodesic_length}
\ee
between two points $\lambda_i = \lambda(s=0)$ and $\lambda_f = \lambda(s=1)$, where $\dot \lambda \equiv d \lambda / ds$. 
As discussed in \sref{sec:time_bounds}, geodesics are related to fundamental bounds on the ability to control the system, such as quantum speed limits. Pictorially, geodesics can be visualized as shortest path between two points in the equivalent manifold [cf. \fref{fig:ising_h_phi} and \exref{ex:xy_geodesic}]. More generally, geodesics satisfy the geodesic equation
\be 
\delta \ell(\lambda)=0 \quad \Rightarrow \quad \ddot{\lambda}^\mu+\Gamma^\mu_{\alpha\beta} \dot{\lambda}^\alpha \dot{\lambda}^\beta=0,
\label{eq:geodesic_eqn}
\ee
where $\Gamma^\mu_{\alpha\beta}$  denote the Christoffel symbols. An important property of this geodesic length is that it is parameterization independent, i.e., $\ell[\lambda(s)] = \ell[\lambda(f(s))]$ for arbitrary monotonic function $f(s)$ satisfying $f(0)=0$ and $f(1)=1$.

Taking a page from general relativity, we want an affine parameterization of the curve. This is accomplished by minimizing the action~\cite{Misner1973}
\be
\mathcal S[\lambda(s)]=\int_0^1  ds\;g_{\mu \nu} \dot \lambda^\mu \dot \lambda^\nu,
\label{eq:geodesic_action}
\ee
whose Euler-Lagrange equations yield precisely the geodesic equation [\eref{eq:geodesic_eqn}]. Unlike the length $\ell$, the action $\mathcal S$ is only invariant to linear rescalings of $s$, and thus we can interpret $s$ physically as the scaled time $s=t/\tau_\mathrm{ramp}$ for a given protocol ramping from $\lambda_i$ to $\lambda_f$ in time $\tau_\mathrm{ramp}$. The geodesic action is clearly minimized by choosing $g_{\mu \nu} \dot \lambda_\mu \dot \lambda_\nu$ constant along the path, which helps simplify the geodesic equations. It also gives us important insight into the overall behavior of $\lambda(s)$; near points where the metric tensor becomes large, such as critical points, the speed $|\dot{\lambda}|$ goes down.

For a single parameter, the path is trivial and the geodesic equations are primarily useful for obtaining the parameterization $\lambda(s)$. Consider for example changing only magnetic field $h$ in the TFI chain. Conservation of $g_{h h} \dot{h}^2$ results in the following set of equations for the geodesic
\begin{eqnarray}
\frac{dh}{\sqrt{1-h^2}}=c \,ds \quad {\rm for} \quad h^2<1 \nonumber \\
\frac{ dh}{h\sqrt{h^2-1}}=c \, ds \quad {\rm for} \quad h^2>1, 
\end{eqnarray}
which yields the general solution
\be 
h(s)= \left\{ \begin{array}{l} \sin(c\,s+a) \quad {\rm for} \quad h^2<1 \\ \pm \left[\sin(c\,s+b)\right]^{-1} \quad {\rm for} \quad h^2>1. \end{array} \right.
\ee
Continuity of the curves near $h^2=1$ and boundary conditions can be used to fix all the constraints. For example, the geodesic that goes from $h=0$ to $h=\infty$ is
\be 
h(s)= \left\{ \begin{array}{l} \sin(\pi s) \quad {\rm for} \quad s<1/2 \\  \left[\sin(\pi-\pi s)\right]^{-1} \quad {\rm for} \quad s > 1/2 \end{array} \right.
\label{eq:geodesic_TFI}
\ee
resulting in the geodesic length 
\be
\ell= \pi \sqrt{L}/4,
\label{eq:geodesic_length_TFI}
\ee
\add{which also immediately follows from \eref{eq:cylinder_shape_TFI} and its equivalent graphical representation in \fref{fig:ising_h_phi}.} Note that, due to the extensivity of the metric tensor, geodesic lengths for all protocols $h_i \to h_f$ -- including those that do not cross any critical points -- nevertheless diverge as $\sqrt L$. This is expected for extensive systems and is quite similar to divergences that show up upon varying other control parameters, such as the local potential in the Anderson orthogonality catastrophe (see \sref{anderson_metric_tensor}). However, even though we are driving the system through a critical point, there is nevertheless a geodesic path that goes across it and in fact yields a finite length up to the overall $\sqrt L$ prefactor. This implies that the singularity of the metric tensor is sufficiently weak that $\sqrt{g}$ has an integrable singularity.  Moreover, close to any second order quantum critical point one can extract some universality in this geodesic. As discussed in Refs. \cite{Zanardi2007_1,Venuti2007_1,Rezakhani2010_1,DeGrandi2010_1}, in the vicinity of the critical point the metric scales as $g\sim |\lambda -\lambda_c|^{\nu d-2}$, where $d$  and $\nu$ are the spatial dimension and correlation length critical exponent respectively. Conservation of $g\dot{\lambda}^2$ immediately implies that geodesics behave as
\be 
\lambda(s)\approx \lambda_c+c (s-s_c)^{2/\nu d} \equiv \lambda_c + c(s-s_c)^\alpha.
\ee
Since $d\nu>0$, \emph{all} second order quantum critical points can be traversed by a geodesic, which furthermore has universal power law behavior in the vicinity of the critical point. Note that for the Ising transition in 1D, for which $d=\nu=1$, we indeed get quadratic scaling of the geodesic \add{[cf. \eref{eq:geodesic_TFI}]}.

\add{ In \sref{sec:time_bounds} we discussed that the minimum time required to follow the ground state is proportional to the geodesic length $\ell$. Let us now apply this result to the TFI model. Specifically let us consider a protocol where we initialize the Hamiltonian in the fully polarized state, i.e., in the ground state of the Hamiltonian
\be
\mathcal H_i=-h\sum_j \sigma^z_j.
\ee
We then apply a counter-diabatic driving protocol to ramp the system through the critical point to the ferromagnetic state, which is the ground state of the Hamiltonian
\be
\mathcal H_f=-J \sum_j \sigma^x_j \sigma^x_{j+1}
\ee
To derive a speed limit, we need to fix the norm of the driving field. The Frobenius norm of any extensive Hamiltonian scales as square root of the number of degrees of freedom. Therefore, we can bound the norm of the driving Hamiltonian by $||\mathcal H_{\rm CD}(t)|| \leq \omega \sqrt{L}$, where $\omega$ is some intensive energy scale. Then according to \eref{eq:time_bound_2}, the minimal time is
\be
\tau_{\rm min}={\hbar\over \omega}{\pi\over 4}.
\label{eq:tau_min_TFI}
\ee

This result is quite remarkable because it shows that the minimal time to ramp the TFI ground state across the phase transition is finite. Moreover, it exactly coincides with time to rotate independent non-interacting spins from the $x$ to the $z$ direction; the metric for $L$ non-interacting spins is $g_{\theta\theta}=L/4$ [\eref{eq:metric_components_spin}], so the geodesic length corresponding to a $\pi/2$ rotation is again given by \eref{eq:geodesic_length_TFI}, resulting in the same minimal time. This result might seem natural as the minimal time can not depend on how we prepare the system, so one might argue that it simply shows consistency of its geometric interpretation. But, importantly, the counter-diabatic driving protocol pushes the system through a path of highly-entangled ground states, rather than non-interacting product states. Furthermore, at all times counterdiabatic driving respects the $Z_2$ symmetry of the TFI model corresponding to flipping $\sigma^z_j\to -\sigma^z_j$ on each site. Therefore the ground state prepared by the counter-diabatic protocol is the true ground state of the TFI model, i.e., the macroscopic superposition (``Schr\"odinger cat'') of positive and negative magnetizations:
\[
|0_f\rangle={|\uparrow\uparrow\dots\uparrow\rangle+|\downarrow\downarrow\dots\downarrow\rangle\over \sqrt{2}}.
\]
This state is impossible to prepare without crossing a phase transition. So the fact that such a state can be prepared in the same time required to rotate a single spin is very counterintuitive. The subtlety is that the necessary counter-diabatic driving Hamiltonian to prepare this state becomes long range near the critical point. Finally, note that the minimal time to ramp the system to the critical point is just half of \eref{eq:tau_min_TFI}.

From this example, we see that geodesic length gives us an important physical constraint on the ability to prepare ground states with high fidelity in finite time. For relatively slow protocols, we will show in \sref{sec:meas_metric_tensor} that the conserved quantity $g\dot{\lambda}^2$ may be directly measured as excess energy fluctuations in the system. Therefore the metric tensor has important implications for both fast and slow dynamics and is a physical, measurable quantity.}

\hrulefill

\exercise{\label{ex:xy_geodesic} Calculate the length of the geodesic for the TFI model ($\gamma=1$) for a path starting from $(h_i,\phi_i=0)$  and ending at $(h_f,\phi_f)$. Without loss of generality, you may assume $0 < h_i < 1$, $1 < h_f$, and $0 < \phi_f < \pi$. You may find it useful to use the isometric shape illustrated in \fref{fig:ising_h_phi}, for which geodesics are known, by mapping the initial and final points to ones on the cylinder and sphere respectively.}

\exerciseshere

\hrulefill

\subsection{Geometric tensor of steady state density matrices}
\label{sec:geom_steady_state_dens_matr}

Having explored (global) topological properties of the ground state manifold, it is natural to ask how these ideas can be generalized to classical and/or finite temperature systems. To make the classical limit explicit we will reintroduce the Planck's constant $\hbar$ to all expressions. Previously we derived the geometric tensor for the ground state manifold, but clearly the arguments flow through trivially for arbitrary excited states $|\psi_m\rangle \equiv |m\rangle$:
\be
\chi_{\alpha\beta}^m={\langle m |  \mathcal A_\alpha \mathcal A_\beta | m \rangle -\langle m |  \mathcal A_\alpha | m\rangle \langle m | \mathcal A_\beta | m \rangle\over \hbar^2} =  \sum_{n\neq m} {\langle m|\partial_\alpha \mathcal H|n\rangle \langle n|\partial_\beta \mathcal H|m\rangle\over (E_n-E_m)^2} ~.
\label{eq:geom_tens_lehmann_exc_state}
\ee
While this may be relevant to microscopic or mesoscopic systems, it is generally very difficult to prepare excited energy eigenstates. We are therefore interested in exciting systems into some steady state density matrix. To ensure that it is stationary, we consider a density matrix of the form $\rho = \sum_n \rho_n |n\rangle \langle n|$. Then as before, we can define the geometric tensor as the covariance matrix of the gauge potentials: $\chi_{\alpha \beta} = \langle \mathcal A_\alpha \mathcal A_\beta \rangle_c/\hbar^2$. 

There remains a slightly subtle question that we must answer: what is the meaning of $\langle\cdots\rangle_c$ for a density matrix? Two natural solutions present themselves. The first option, $\langle A B \rangle_c = \langle A B \rangle - \langle A \rangle \langle B \rangle = \mathrm{Tr} \left[ \rho A B \right] - \mathrm{Tr} \left[ \rho A \right] \mathrm{Tr} \left[ \rho B \right]$ which we call the coherent connection, is the type of connected correlation function that appears in the theory of phase transitions, where it will often be singular near the transition. The second option, 
\be
\langle A B \rangle_c = \sum_n \rho_n \langle n | A B | n \rangle_c = \sum_n \rho_n \left( \langle n | A B | n \rangle - \langle n | A | n \rangle \langle n | B | n \rangle\right)
\label{eq:AB_c}
\ee 
at first seems much less natural, as it does not take the form of a simple operator expectation value. However, it will turn out that this second ``incoherent'' definition is the one which appears in the dynamical response in isolated systems and can be related to noise and dissipation (cf. \aref{sec:kubo_metric_finite_temp}). It is worth noting that the difference between the two ways of defining the connected correlation function is in their handling of the diagonal elements of the density matrix. Therefore, the anti-symmetric Berry curvature, which only depends on the off-diagonal part, does not care which definition we use. For the symmetric part of the correlation function, however, there is an important difference between the two options.\footnote{We note that if $A$ and $B$ are local operators describing some physical observables and $|n\rangle$ are eigenstates of an ergodic many-body Hamiltonian then the difference between the two definitions is small, vanishing in the thermodynamic limit. However, since we are not making any assumptions about taking the thermodynamic limit and generally are working with non-ergodic systems, we must be careful with choosing the right definition.} Let us also point out that the geometric tensor defined in this way does not correspond to a natural Bures distance between density matrices (see, e.g., Eq. (3) in Ref.~\cite{Zanardi2007_2}) so the relation of $\chi$ defined in this way with quantum information geometry becomes less clear. Nevertheless we will stick to this definition, as it naturally emerges in dynamical response, and leave discussion of the relationship between these two natural metrics for future work.

The natural extension of the geometric tensor to mixed stationary states is thus 
\be
\chi_{\alpha \beta} (\rho) = \sum_n \rho_n \chi_{\alpha \beta}^n ~.
\label{eq:chi_alpha_beta_rho}
\ee
As this is simply a sum over all eigenstates, it is trivial to write it as a response function by plugging in the expression for $\chi_{\alpha \beta}^n$ from Eqs.~(\ref{eq:chi_alpha_beta_Sintegral}) and (\ref{eq:S_alpha_beta}) with $|0\rangle \to |n \rangle$. For a finite temperature density matrix, it is similarly straightforward to see that this is connected to the dissipative part of linear response, which is shown in more detail in \aref{sec:kubo_metric_finite_temp}. 

In \exref{ex:geom_tens_sho}, we derived the metric tensor of the harmonic oscillator with respect to shifts in the position or momentum coordinate.  In the exercises below, we will see how this generalizes to finite temperature states. Let us analyze explicitly the metric of the harmonic oscillator with Hamiltonian $\mathcal H=p^2/2m + k x^2/2$ with respect to changing a slightly less trivial parameter: the spring constant $k$. If the mass is held fixed, then the generalized force with respect to changes of $k$ is $\partial_k \mathcal H = x^2/2$. Then the metric tensor for an arbitrary harmonic oscillator state $|n\rangle$ is 
\beq
\nonumber
g_{kk}^n &=& \sum_{m \neq n} \frac{\left| \left< m \right | (x^2 / 2) \left | n \right>\right|^2}{(E_n - E_m)^2} = \frac{\ell^4}{4}\sum_{m \neq n} \frac{\left| \left< m \right| (a+a^\dagger)^2 \left | n \right> \right|^2}{(E_n - E_m)^2} 
\\ &=& \frac{\ell^4}{16\hbar^2 \Omega^2} \left( \left| \left< n-2 \right | a^2 \left | n \right> \right|^2 + \left| \left< n+2 \right| (a^\dagger)^2\left | n \right> \right|^2 \right) = \frac{\ell^4}{8\hbar^2 \Omega^2} \left( n^2 + n + 1 \right) ~,
\label{eq:g_kk_n_sho}
\eeq
where $\Omega = \sqrt{k / m}$ and $\ell = \sqrt{\hbar/2m \Omega}$ are the natural frequency and length scales of the oscillator. Then for an arbitrary stationary state, the metric tensor is clearly 
\[
g_{kk}^\rho = {\ell^4 \over 8\hbar^2 \hbar^2\Omega^2} \left( \langle n^2 \rangle + \langle n \rangle + 1 \right)={1\over 32 \hbar^2 m^2\Omega^4} \left( \langle n^2 \rangle + \langle n \rangle + 1 \right). 
\]
For the Gibbs ensemble, $\rho_n = \mathrm{e}^{-\hbar\beta \Omega (n+1/2)}/Z$, one finds that $\langle n \rangle = 1/(\mathrm{e}^{\hbar\beta \Omega} - 1)$ and $\langle n^2 \rangle = (\mathrm{e}^{\hbar\beta \Omega} + 1) / (\mathrm{e}^{\hbar\beta \Omega}-1)^2$. In the high temperature or classical limit, $\beta \hbar \Omega \ll 1$, this reduces to 
\be
\hbar^2 g_{kk}^{T \gg \hbar \Omega} \to {\hbar^2\over 32 m^2\Omega^4}{2\over \hbar^2\beta^2\Omega^2} = \frac{(k_B T)^2 }{16 m^2\Omega^6}. ~
\label{eq:g_kk_quantum_hightemp}
\ee
We see that in the classical (high-temperature) limit it is the product $\hbar^2 g_{kk}=\langle \mathcal A_k^2\rangle_c$ which is well defined. 

We can also arrive to the result above by calculating the variance of the gauge potential. Note that the eigenstates of the harmonic oscillator are
\[
\psi_n(x)={1\over \sqrt{\ell}} \phi_n(x/\ell),
\]
where $\phi_n$ is the dimensionless eigenfunction of the oscillator expressed through the Hermite polynomials~\cite{Landau_Lifshitz_QM}. Differentiating this wave-function with respect to $k$ we find:
\be
\partial_k \psi_n(x)=-{1\over 2 \ell}{d \ell\over dk} \psi_n(x)-{x\over \ell}{d \ell\over dk}\partial_x \psi_n(x)=-{d \ell \over dk} {1+2 x\partial_x\over 2\ell}\psi_n(x).
\label{eq:derivative_k_harmonic_state}
\ee
Therefore 
\be
\mathcal A_k=i\hbar \partial_k={d \ell\over dk} {x \hat p+ \hat p x\over 2\ell}=-{1\over 4} {\ell\over k}\mathcal D,
\label{eq:gauge_pot_harmonic_spring}
\ee
where 
\be
\mathcal D={x \hat p+\hat p x\over 2\ell}
\label{eq:def_dilation_D}
\ee
is nothing but the quantum dilation operator [cf. \exref{ex:dilations}]. This is not surprising, as rescaling of the spring constant amounts to dilations. In the second quantized notation
\[
{x\hat p +\hat p x\over 2}={i\over 2} \left[(a^\dagger + a) (a^\dagger -a)+ (a^\dagger -a)(a^\dagger + a)\right]=i (a^\dagger a^\dagger-a a)
\]
Using this expression and substituting it into the definition of the metric tensor:
\[
\hbar^2 g_{kk}=\langle \mathcal A_k^2\rangle_c
\]
we can reproduce the expression for the metric tensor in \eref{eq:g_kk_n_sho}. In particular, in the classical limit, using the equipartition theorem we recover \eref{eq:g_kk_quantum_hightemp}:
\[
\langle \mathcal A_k^2\rangle_c={1\over 16 k^2} \langle x^2 p^2\rangle_c={1\over 16 k^2}{4m\over k }\left< {k x^2\over 2}\right>\left< {p^2\over 2m}\right>={m(k_bT)^2\over 16 k^3 }={(k_b T)^2\over 16 m\Omega^6}.
\]

\hrulefill

\exercise{Verify that the variance of  the gauge potential $\mathcal A_k$ in \eref{eq:gauge_pot_harmonic_spring} reproduces the metric tensor, \eref{eq:g_kk_quantum_hightemp}. }

\exercise{Repeat \exref{ex:geom_tens_sho} for the thermal state at temperature $T$. Check that at the zero temperature you reproduce the ground state geometric tensor. Find the asymptotic expression for the geometric tensor in the classical limit $T\gg \hbar\omega$.}

\exerciseshere

\hrulefill

\subsection{Geometric tensor in the classical limit}\label{sec:geom_classical_syst}

Having defined the thermal geometric tensor for a quantum system, we expect to be able to define a classical ($\hbar \to 0$) limit of the metric tensor that matches \eref{eq:g_kk_quantum_hightemp}. In the classical problem we have the stationary state $\rho(p,q) \propto \mathrm{e}^{-\beta \mathcal H(p,q)}$. Unfortunately, the definition of ``matrix elements'' of the operator $\partial_k \mathcal H$ is less clear, so we must resort to the dynamical definition of the geometric tensor given in Eqs.~(\ref{defM}), (\ref{eq:chi_alpha_beta_Sintegral}), (\ref{eq:S_alpha_beta}), and  (\ref{eq:chi_alpha_beta_rho}). The sum over eigenstates, $\sum_n \rho_n$, is replaced by an integral over phase space:
\be
S_{\alpha \beta}^{cl} (\omega)=\int_{-\infty}^{\infty} dt\, \mathrm{e}^{i\omega t} \int dp dq\, \rho(p,q) \left[ \partial_\alpha \mathcal H(p(t),q(t)) \partial_\beta \mathcal H(p,q)-M_\alpha(p,q) M_\beta(p,q) \right],
\label{eq:S_kk_rho}
\ee
where $p\equiv p(0)$, $q \equiv q(0)$, $M_\alpha(p,q)$ is the generalized force or the infinite time average of $-\partial_\alpha \mathcal H(p(t),q(t))$ starting from the initial conditions $q(0),p(0)$. This generalized force is nothing but the Born-Oppenheimer force emerging in the adiabatic approximation (see \sref{sec:emergent_newtonian_dynamics}). When doing this integral, one should think of integrating over $p$ and $q$ as integrating over initial conditions weighted by the probability $\rho(p,q)$. For instance, the ``Heisenberg'' operator $\partial_k \mathcal H (p(t),q(t)) = q(t)^2 / 2$ should be thought of as half the value of $q^2$ at time $t$ after starting at $t=0$ from the state $(p,q)$. Let us analyze the example from the previous section and find $g_{kk}$ for the harmonic oscillator in the thermal equilibrium. Time dependence of $q(t)$ for the oscillator is
\be
q(t) = q(0) \cos (\Omega t) + \frac{p(0)}{m \Omega} \sin (\Omega t)\equiv  q \cos (\Omega t) + \frac{p}{m \Omega} \sin (\Omega t) ~,
\ee
Therefore the generalized force
\be
M_k(p,q)=-\overline{q^2(t)\over 2}=-{q^2\over 4}-{p^2\over 4 m^2\Omega^2}=-{1\over 2m\Omega^2} \mathcal H(p,q)=-\frac{\mathcal H(p,q)}{2 k},
\label{eq:gen_force_cl_spring}
\ee
where the overline stands for time averaging. As expected the generalized force $M_k(p,q)$ only depends on conserved quantities, namely the Hamiltonian. Then the integrand appearing in the spectral function is given by 
\begin{multline}
\partial_k \mathcal H(p(t),q(t)) \partial_k \mathcal H(p,q)-M_k(q,p)^2 =\frac{1}{4} q^2 \left[ q \cos (\Omega t) + \frac{p}{m \Omega} \sin (\Omega t) \right]^2-{1\over 16}\left(q^2+{p^2\over m^2\Omega^2}\right)^2\\
 = \frac{1}{16} \left(q^4-{p^4\over m^4\Omega^4}\right) + {q^2\over 8}\left[\left( q^2 - \frac{p^2}{m^2 \Omega^2} \right) \cos (2 \Omega t) + \frac{q p}{m \Omega} \sin (2 \Omega t) \right].
\end{multline}
To calculate $S_{\alpha\beta}^{cl}(\omega)$ now according to \eref{eq:S_kk_rho} we have to average the expression above over the probability distribution and take the time integral.  Upon averaging over the equilibrium density matrix, the first, time-independent term vanishes because $\langle q^4\rangle=\langle p^4/(m^4\Omega^4)\rangle$. Similarly the last term averages to zero: $\langle q^3 p\rangle=0$. So the only non-zero contribution to the spectral function comes from the second term proportional to $\cos(2\Omega t)$. Because the integrals are all Gaussian, we may apply Wick's theorem to get
\[
\langle q^4\rangle=3 \langle q^2\rangle^2={12\over k^2}\left< {kq^2\over 2}^2\right>={12\over k^2} {(k_B T)^2\over 4}={3\over k^2} (k_B T)^2,\quad \langle q^2 p^2/m^2\Omega^2\rangle=\langle q^2\rangle^2=(k_B T)^2.
\]
Therefore
\be
S_{kk}^{cl} (\omega)={(k_B T)^2\over 4 k^2}\int_{-\infty}^{\infty} dt\, \mathrm{e}^{i\omega t} \cos(2\Omega t)={(k_B T)^2\over 4 k^2}\pi \left(\delta(\omega+2\Omega)+\delta(\omega-2\Omega)\right).
\ee
Then, via the dynamical definition of the geometric tensor [\eref{eq:chi_alpha_beta_Sintegral}],
\be
\hbar^2 g_{kk}=\int_0^\infty {d\omega\over 2\pi} {S_{kk}(\omega)+S_{kk}(-\omega)\over \omega^2}={(k_B T)^2\over 16 k^2 \Omega^2}={(k_B T)^2\over 16 m^2\Omega^6},
\ee
which indeed coincides with the classical limit of the quantum geometric tensor, \eref{eq:g_kk_quantum_hightemp}.

Let us now show how the same result can be reproduced using the language of the adiabatic gauge potentials. According to \eref{classical_adiabatic_gp_relation} the adiabatic gauge potential should satisfy
\be
-\partial_k \mathcal H(q,p)=M_k(q,p)-{\partial \mathcal A^{cl}_k\over \partial q}{\partial \mathcal H\over \partial p}+{\partial \mathcal A^{cl}_k\over \partial p}{\partial \mathcal H\over \partial q},
\ee
Using \eref{eq:gen_force_cl_spring}, the equation above reduces to
\be
{p^2\over 4 m k}-{q^2\over 4}=-{\partial \mathcal A^{cl}_k\over \partial q}{p\over m}+{\partial \mathcal A^{cl}_k\over \partial p}k q
\ee
It is easy to check that the desired adiabatic gauge potential is
\be
\mathcal A_k^{cl}=-{qp\over 4 k},
\label{eq:guage_pot_osc_cl}
\ee
which coincides with the earlier result \eref{eq:gauge_pot_harmonic_spring} in the classical limit and, as we already showed, reproduces the correct metric tensor. 

For simple cases like this, the gauge potentials can be found explicitly and can be much easier to work with   than correlation functions of the generalized forces. For more complicated situations such as the Duffing oscillator, one can imagine doing a similar construction numerically or iteratively and/or utilizing the correlation function of the generalized forces. In classical chaotic systems the gauge potentials and hence the geometric tensor will not necessarily converge~\cite{Jarzynski1995_2}. The issue comes from  a divergent low-frequency tail in the spectrum of generic observables due to the presence of diffusive modes. Physically these divergences are always cut off by either coupling to the bath or finite duration of the physical process. Introducing a consistent cutoff  for such systems is beyond the level of the present discussion and will be a subject of future research.

\subsection{Exact and variational geometric tensors for many-body systems}

We saw in \sref{sec:find_approx_eig} that one can target individual eigenstates by integrating the variational gauge potentials. Therefore, it is not surprising that one may also obtain a variational geometric tensor by differentiating these variational eigenstates. Here we show how to do for the non-integrable Ising chain and impurity in a Fermi gas explored earlier.

\subsubsection{Non-integrable Ising model}

Let us begin by examining the non-integrable Ising model discussion in \sref{subsection:NonintegrableIsing}. We already computed the single and two-spin variational gauge potentials, and now we will show how to use them  to construct the approximate geometric tensor and Berry curvature. Specifically, consider as before the Hamiltonian 
\be
\mathcal{H}(\theta,\phi)=-\sum_{j}h(\cos\theta\,\sigma_j^{z}+\sin\theta\cos\phi\,\sigma_{j}^{x}+\sin\theta\sin\phi\,\sigma_{j}^{y})-J_z\sum_{j}\sigma_{j}^{z}\sigma_{j+1}^{z}.
\label{eq:H_spin_with_ising_phi}
\ee
By rotating the system around the z-axis we can of course always make $\phi=0$, so the exact adiabatic gauge potential for $\phi$ is simply
\be 
\mathcal{A}_\phi=\frac{1}{2} \sum_j \sigma^z_j.
\ee
For $\phi=0$ the variational gauge potential with respect to $\theta$ was computed in \sref{subsection:NonintegrableIsing}, which crucially was possible without ever diagonalizing the Hamiltonian. To do this, we use the gauge potentials to prepare the ground state of the system at any value of $\theta,\phi$ out of a trivial state. For example, for $\phi=\theta=0$ (and positive $J_z$) the ground state is just a product state of the spins aligned with the magnetic field, which we denote $\left| \psi(0,0)\right\rangle$. Consequently we can write the approximate ground state at different angles as
\be 
\left| \psi^\ast(\theta,\phi) \right\rangle= \exp \left( -i \phi \mathcal{A}_\phi\right)\left| \psi^\ast(\theta,0) \right\rangle= \exp \left( -i \phi \mathcal{A}_\phi\right) \mathcal{P} \left[ \exp \left(-i \int_0^\theta  d \theta \mathcal{A}^\ast_\theta \right) \right] \left| \psi(0,0) \right\rangle,
\ee
where $\mathcal{A}^\ast_\theta$ is an approximate gauge potential for $\phi=0$. Note that in order to compute any geometric property we therefore also need to propagate $\mathcal{A}^\ast_\theta$ to finite $\phi$ by approximate rotation around the $z$-axis. Since the spins always undergo exactly the same rotations, it immediately follows that nothing explicitly depends on the angle $\phi$, only on $\theta$. Another direct consequence of this is that the Berry phase for $\theta$ vanishes and the metric tensor becomes diagonal. The system should however have a non-zero Berry curvature. 

We have used this procedure with the variational gauge potential from \sref{subsection:NonintegrableIsing} to find the variational geometric tensor. The results are depicted in \fref{fig:metricvariational} for a spin chain of length $L=18$. For comparison, we determine the exact value of the ground state geometric tensor (black dashed line) by truncating the Lehmann representation [\eref{geom_tens_lehmann}] to only incorporate the lowest 100 eigenstates and confirming convergence in the number of eigenstates included. As the number of spins $M$ in the variational ansatz is increased, the variational geometric tensor appears to converge towards the exact value. This is consistent with our expectations that the result should converge exponentially in $M$ because no critical points are crossed, as seen for the gauge potential itself in \fref{fig:TFI_dens_M}.

\begin{figure}
\includegraphics[width=\columnwidth]{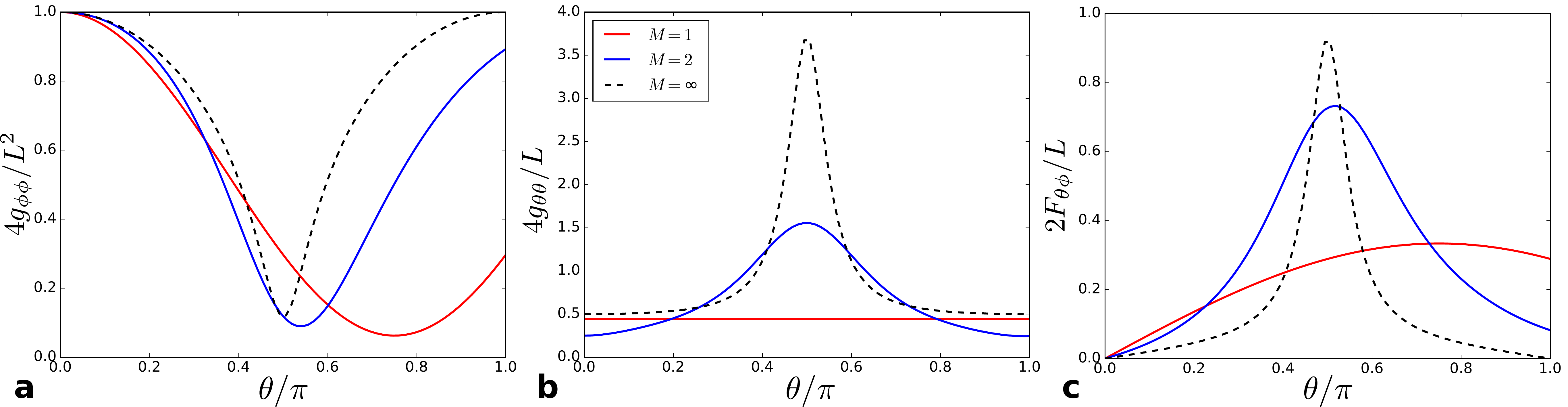}
\caption{\add{Ground state geometry for a non-integrable quantum Ising chain of length $L=18$ with $J=1$ and $h=2$. All panels show the single-spin variational result in red and the two-spin variational result in blue. For the dashed line we obtained the ground state and the first 100 excited states by Lanczos and computed the approximate metric tensor by simply truncating the sum over excited states to the first 100 states. Since the ground state is gapped, this is a good approximation of the exact result. Panel A shows the metric tensor $g_{\phi\phi}$, Panel B the metric $g_{\theta\theta}$ and Panel C the Berry connection $F_{\phi\theta}$.}}
\label{fig:metricvariational}
\end{figure}

\subsubsection{Impurity in a Fermi gas}
\label{anderson_metric_tensor}

Finally, consider the case of an impurity in a Fermi gas, which we saw in \sref{sec:impurity_fermi_gas} gives rise to a divergent metric tensor in the thermodynamic limit due to lack of an excitation gap. Another related consequence is the Anderson orthogonality catastrophe \cite{Anderson1967}, the phenomenon by which Anderson showed that a system would quickly become orthogonal to its initial ground state upon addition of a weak impurity. The emergent non-locality of the adiabatic gauge potential is intricately related to the orthogonality catastrophe.
We can see this more explicitly by evaluating the metric tensor, $g_{\lambda\lambda}$, for a filled Fermi sea $|\psi_0\rangle$ with Fermi momentum $k_{F}$:
\begin{eqnarray*}
g_{\lambda\lambda} & = & \langle\psi_{0}|\mathcal{A}_{\lambda}^{2}|\psi_{0}\rangle=-\frac{\hbar^{2}}{L^{2}}\sum_{k\neq k^{\prime}}\sum_{k^{\prime\prime}\neq k^{\prime\prime\prime}}\frac{\langle\psi_{0}|c_{k+}^{\dagger}c_{k^{\prime}+}c_{k^{\prime\prime}+}^{\dagger}c_{k^{\prime\prime\prime}+}|\psi_{0}\rangle}{(\cos k-\cos k^{\prime})(\cos k^{\prime\prime}-\cos k^{\prime\prime\prime})}\\
 & = & -\frac{\hbar^{2}}{L^{2}}\sum_{k\neq k^{\prime}}\sum_{k^{\prime\prime}\neq k^{\prime\prime\prime}}\frac{\Theta(k_{F}-k^{\prime\prime\prime})\Theta(k^{\prime\prime}-k_{F})\delta_{k^{\prime}k^{\prime\prime}}\delta_{kk^{\prime\prime\prime}}}{(\cos k-\cos k^{\prime})(\cos k^{\prime\prime}-\cos k^{\prime\prime\prime})}\\
 & = & (2\pi\hbar)^{2}\int_{0}^{k_{F}}dk\int_{k_{F}}^{\pi}dk^{\prime}\frac{1}{(\cos k-\cos k^{\prime})^{2}}.
\end{eqnarray*}
\add{The integrand above} diverges near $k=k^{\prime}$ and thus is dominated by terms
near the Fermi surface. Therefore, let us shift $k^{(\prime)} \to k^{(\prime)} - k_F$ and Taylor expand the denominator, using the momentum spacing $2\pi/L$ as a low momentum cutoff and $\Lambda\sim\pi$ as a high momentum cutoff. This gives
\begin{eqnarray*}
g_{\lambda\lambda} & \approx & \left(\frac{2\pi\hbar}{\sin k_{F}}\right)^{2}\int_{-\Lambda}^{-\pi/L}dk\int_{\pi/L}^{\Lambda}dk^{\prime}\frac{1}{(k-k^{\prime})^{2}}\\
 & \approx & \left(\frac{2\pi\hbar}{\sin k_{F}}\right)^{2}\ln\left(\frac{\Lambda L}{4\pi}\right).
\end{eqnarray*}
So we see that the metric tensor is also divergent in the thermodynamic
limit due to the long-range nature of $\mathcal{A}_{\lambda}$. The divergence is readily regulated by the finite size $L$ as well as the finite lattice spacing. Note that this \add{regularization} is quite similar to the methods we used to understand and regulate behavior of physical observables near critical points, since both the Fermi gas and a quantum critical system are gapless.~\footnote{Indeed, the impurity potential can be studied as a boundary perturbation of the conformal field theory given by linearizing the Fermi sea or even a non-Fermi Luttinger liquid about the Fermi surface. This is entirely analogous to the boundary critical theory of, for instance, the TFI chain.}
  
\section{Geometric tensor and non-adiabatic response}
\label{sec:measuring_geometric_tensor}

\emph{{\bf Key concept:} The geometric tensor appears naturally through response coefficients of the system to the rate of change of parameters $\dot\lambda_a$. The Berry curvature shows up as a Coriolis-type force while the metric tensor defines broadening of the energy distribution (energy variance). In the classical (high temperature) limit, the metric tensor also defines the leading non-adiabatic correction to the energy through renormalization of the mass.}

\subsection{Dynamical quantum Hall effect}
\label{sec:dyn_quantum_hall_effect}

We already noted in the first section that the gauge potentials appear in the Galilean term in the moving Hamiltonian:
\be
\tilde{\mathcal H}_{m}=U^\dagger \mathcal H U-\dot \lambda_\alpha \tilde{\mathcal A}_\alpha=\tilde {\mathcal H}-\dot \lambda_\alpha \tilde{\mathcal A}_\alpha.
\label{eq:H_general_moving_frame}
\ee
Then we introduced the geometric tensor $\chi$, which we found could be written as the covariance of the gauge potentials. In this section, we connect the dots between these observations by relating the geometric tensor to the dynamical response of physical observables.

We start by noting that the bare Hamiltonian in the moving frame $\tilde {\mathcal H}$ is diagonal and thus only produces shifts in the energies but does not couple them, so it is not responsible for the transitions between levels. Conversely the Galilean term generally has off-diagonal elements and thus causes transitions between levels. Near the adiabatic limit the Galilean term is small and thus can be treated as a perturbation. Because the gauge potentials are simultaneously responsible for the non-adiabatic response of the systems and for the geometry we just discussed in the previous section it is thus not very surprising that the response coefficients  can be related to the geometric tensor. The goal of this section is precisely to establish such connection.

Let us now consider the setup where the system is initially prepared at equilibrium (for concreteness in the ground state) at some initial value of the coupling $\vect \lambda_0\equiv \vect \lambda(t=0)$. Then the coupling starts changing in time. To avoid the need of worrying about initial transients, which can be done but makes the derivations more involved, we will assume that the rate of change of the coupling is a smooth function of time. Under this smooth transformation, at leading order in $|\dot\lambda|$ the system follows the ground state of the moving Hamiltonian $\mathcal H_{m}$. One can worry whether the adiabatic theorem applies to this Hamiltonian, which is still time-dependent; later we will give a more rigorous derivation of the result using the machinery of adiabatic perturbation theory (see Refs.~\cite{Rigolin2008_1, Gritsev2012_1} for more details). For now let us simply note, as we already did in the very first section, that the adiabatic approximation applied to the moving Hamiltonian encodes the leading non-adiabatic corrections beyond the standard adiabatic approximation, where the system follows the eigenstates of the instantaneous Hamiltonian $\tilde H$. Already at this level of approximation we can derive very important results such as emergence of the Coriolis force and the mass renormalization.

Applying first order perturbation theory to the moving frame Hamiltonian $\mathcal H_m$, the amplitude to transition to the excited state $|n\rangle$ of the bare Hamiltonian $\tilde{\mathcal H}$ due the Galilean term is given by
\be
a_n=\dot\lambda_\alpha {\langle n| \mathcal A_\alpha|0\rangle\over E_n-E_0}
\label{eq:ampl_n_apt}
\ee
One can alternatively understand this result as coming from the instantaneous measurement process viewed as a sudden quench, where the rate $\dot\lambda$ is quenched to zero. It is convenient to represent observables as generalized force operators conjugate to some other coupling $\lambda_\beta$:
\be
\mathcal M_\beta=-\partial_\beta\mathcal H.
\ee
The matrix elements of these objects already appeared in the definition of the geometric tensor so it is convenient to continue dealing with them. Generalized forces defined as expectation values of the generalized force operators, appear quite naturally in many problems. For example, the magnetization is a generalized force conjugate to the magnetic field, current is a generalized force conjugate to the vector potential, nearest neighbor correlation function can be viewed as a generalized force conjugate to the nearest neighbor hopping or interaction, etc. Indeed any observable $\mathcal O$ can be represented as some generalized force operator by adding a source term $-\lambda \mathcal O$ to the Hamiltonian. Taking the expectation value of $\mathcal M_\beta$ and using \eref{eq:gauge_pot_matrix_elem} for the matrix elements of the gauge potential, we find that
\begin{eqnarray}
\nonumber
M_\beta\equiv \langle \psi|\mathcal M_\beta |\psi\rangle & = & M_\beta^{(0)}- \sum_{n\neq 0} (a_n^\ast \langle n|\partial_\beta \mathcal H|0\rangle + a_n \langle 0 |\partial_\beta \mathcal H|n\rangle)
\\ &\approx&M_\beta^{(0)}+i\hbar  \dot \lambda_\alpha \sum_{n\neq 0} {\langle 0|\partial_\beta \mathcal H |n\rangle \langle n |\partial_\alpha \mathcal H|0\rangle-\langle 0|\partial_\alpha \mathcal H |n\rangle \langle n |\partial_\beta \mathcal H |0\rangle\over (E_n-E_0)^2}
\nonumber
\\&=&M_\beta^{(0)}+\hbar F_{\beta\alpha}\dot\lambda_\alpha, 
\label{dynamical_qhe}
\end{eqnarray}
where $M_\beta^{(0)}$ is the generalized force evaluated in the instantaneous ground state (i.e., in the adiabatic limit). This relation shows that the leading non-adiabatic (Kubo) correction to the generalized force comes from the product of the Berry curvature and the rate of change of the parameter $\vect\lambda$. Using our previous intuition that Berry curvature behaves as a magnetic field in parameter space, we see that this Kubo correction is the Lorentz (or the Coriolis) force in parameter space~\cite{Berry1989_1}. Because the integral of the Berry curvature over a closed parameter manifold is a quantized first Chern number, this effective Lorentz force leads to a quantized response, which one can term the dynamical quantum Hall effect~\cite{Gritsev2012_1}.

Related methods of understanding leading corrections to adiabaticity have been around since the early days of the quantum adiabatic theorem 
\cite{Born1928_1,Kato1950_1,Teufel2003_1}. In particular, similar notions in the language of Kubo response have been used to understand the response of quantum Hall systems \cite{Laughlin1981_1,Thouless1982_1,Simon1983_1,Niu1984_1,Avron1985_1,Avron1988_1,Avron1995_1,Read2011_1}, in semi-classical calculations of the anomalous Hall effect \cite{Karplus1954_1,Sundaram1999_1,Haldane2004_1} and in deriving ``molecular Aharonov-Bohm'' corrections to Born-Oppenheimer descriptions of molecules \cite{Mead1980_1,Moody1986_1,Zygelman1987_1,Berry1989_1}. While the results are similar to those described above, we emphasize here the generality of the results given by adiabatic perturbation theory, and in particular their applicability to parameters that are in no way connected to traditional Hall conductance.

\subsubsection{Quantum Hall effect}

Let us first illustrate that this relation does indeed reproduce the standard integer quantum Hall effect (QHE). We will make only two generic assumptions: (i) the ground state of the system is not degenerate (although degeneracies can lead interesting phenomena like the fractional QHE) and (ii) the Hamiltonian of the system can be represented in the form
\be
\mathcal H=\sum_{j=1}^N {\left(\vect p_j-e \vect \Lambda_j\right)^2\over 2m_j} + V(\vect r_1, \vect r_2,\dots \vect r_N),
\ee
where $V$ is an arbitrary momentum independent potential energy which can include both interactions between particles and an external potential. As before, we use the $\vect \Lambda_j\equiv\vect \Lambda(\vect r_j)$ notation for the vector potential to avoid confusion with the gauge potential. Let us assume that the vector potential consists of some static part (not necessarily uniform) representing a static magnetic field and an extra dynamic part representing the electric field in the system, where throughout this section we work in the Coulomb gauge, $\vect {\mathcal E}=\partial_t \vect \Lambda$. We will choose the components of the time-dependent vector potential as our parameters, i.e.,
\be
\lambda_x=\Lambda_x,\; \lambda_y=\Lambda_y. 
\ee
The generalized force with respect to $\lambda_y$ is
\be
 \mathcal M_y=-\partial_{\lambda_y}\mathcal H=\sum_j {e\over m_j} \left(p_j^{(y)}-e \Lambda_j^{(y)}\right)= \mathcal J_y,
\ee
which is the current operator along the $y$-direction. In the absence of the electric field there is no average current,  $\langle 0|\mathcal J_y|0\rangle=0$ so the dynamical Hall relation reads
\be
J_y=\hbar F_{\lambda_y \lambda_x} \dot{\lambda}_x = \hbar F_{\lambda_y \lambda_x} \mathcal E_x,
\ee
To find the Hall conductivity we note that the total current $J$ is related to the two-dimensional current density $j$ via
\be
J_y=L_x L_y j_y,
\ee
where $L_x$ and $L_y$ are the dimensions of the sample. Therefore the Hall conductivity $\sigma_{xy}=j_y/\mathcal E_x$ is related to the Berry curvature via
\be
\sigma_{xy}={\hbar F_{\lambda_x\lambda_y}\over L_x L_y}.
\ee

If we now focus on bulk response by considering a system with periodic boundary conditions (eliminating the edges), the parameter $\lambda_x$ can be gauged away once it reaches $\lambda_x^0 = 2 \pi \hbar / e L_x$, and similarly for $\lambda_y$. This corresponds to threading a flux quantum through the torus \cite{Laughlin1981_1}. Since the ground state returns to itself upon insertion of a flux quantum along either direction, this defines a closed manifold in $\lambda$ space on which we can define a Chern number. Furthermore, as $\lambda_{x,y}^0$ are very small and generally immeasurable in the thermodynamic limit, we can average over them to get the averaged conductance
\beq
\sigma_{xy}&\approx&{\hbar \overline{F_{\lambda_x\lambda_y}} \over L_x L_y} = \frac{\hbar}{L_x L_y} \frac{\int_0^{\lambda_x^0} d \lambda_x  \int_0^{\lambda_y^0} d \lambda_y F_{\lambda_x\lambda_y}}{\lambda_x^0 \lambda_y^0}
\\ &=& \frac{\hbar}{L_x L_y} \frac{2 \pi C_1}{\left( 2 \pi \hbar / e L_x \right) \left( 2 \pi \hbar / e L_y \right)}
\\ &=& C_1 \frac{e^2}{h} ~.
\eeq
Thus the quantization of the conductance in the quantum Hall effect can be thought of as the topological response to insertion of flux quanta along the two directions in the system~\cite{Niu1984_1}.

\hrulefill

\exercise{Show that for a system of free fermions in the thermodynamic limit with a gap between filled and unfilled bands, the many-body Berry curvature (and its Chern number) with respect to gauge potentials reduce to the sum of band Chern numbers defined in \exref{ex:chern_insulator}:
\be
F_{\lambda_x \lambda_y} = \frac{1}{\lambda_x^0 \lambda_y^0} \sum_\alpha \int_{FBZ} F^{\alpha}_{k_x k_y} d k_x d k_y ~,
\ee
where the integral is over the first Brillouin zone and the sum is over filled bands $\alpha$.}

\exerciseshere

\hrulefill

\subsubsection{Quantum spin-$1/2$}

The second example we discuss is our old friend, the spin-1/2 in a time-dependent magnetic field. Because this is a purely quantum system we will again set $\hbar=1$. Suppose that the spin is prepared in the ground state along a magnetic field whose angle then starts to change with time along, e.g., the $\theta$-direction. The generalized force along the orthogonal $\phi$-direction is just the $\phi$-component of the magnetization. In the adiabatic limit it is clearly zero since in this case the magnetization simply follows the magnetic field. The leading non-adiabatic correction is then given by the Berry curvature:
\be
M_\phi=\langle \mathcal M_{\phi}\rangle\approx F_{\phi\theta} \dot\theta~,
\ee
where $F_{\phi \theta} = \sin \theta / 2$ (see \sref{sec:invitation}). Similarly, if we again ramp the magnetic field in the $x-z$ plane ($\phi=0$), but now with a time dependent $x$-component and a time independent $z$-component, we have
\be
M_y=F_{yx} \dot h_x.
\label{eq:M_y_exp_spin_half}
\ee
Then by a standard transformation from spherical to Cartesian coordinates, we find 
\be
F_{yx} = \frac{F_{\phi \theta}}{h^2 \tan \theta} = \frac{\cos \theta}{2 h^2}~.
\ee
In \fref{fig:spin_qhe} we show numerically computed dependence of the transverse $y$-magnetization on the rate of change of the magnetic field $v$ for a particular protocol
\be
\mathcal H=-\sigma_z-h_x(t) \sigma_x,
\ee
where $h_x(t)=0.5+v t$. The transverse magnetization is computed at time $t=0$ and the initial condition corresponds to the ground state at large negative time $t=-100/v$.
\begin{figure}[ht]
\includegraphics[width=0.6\columnwidth]{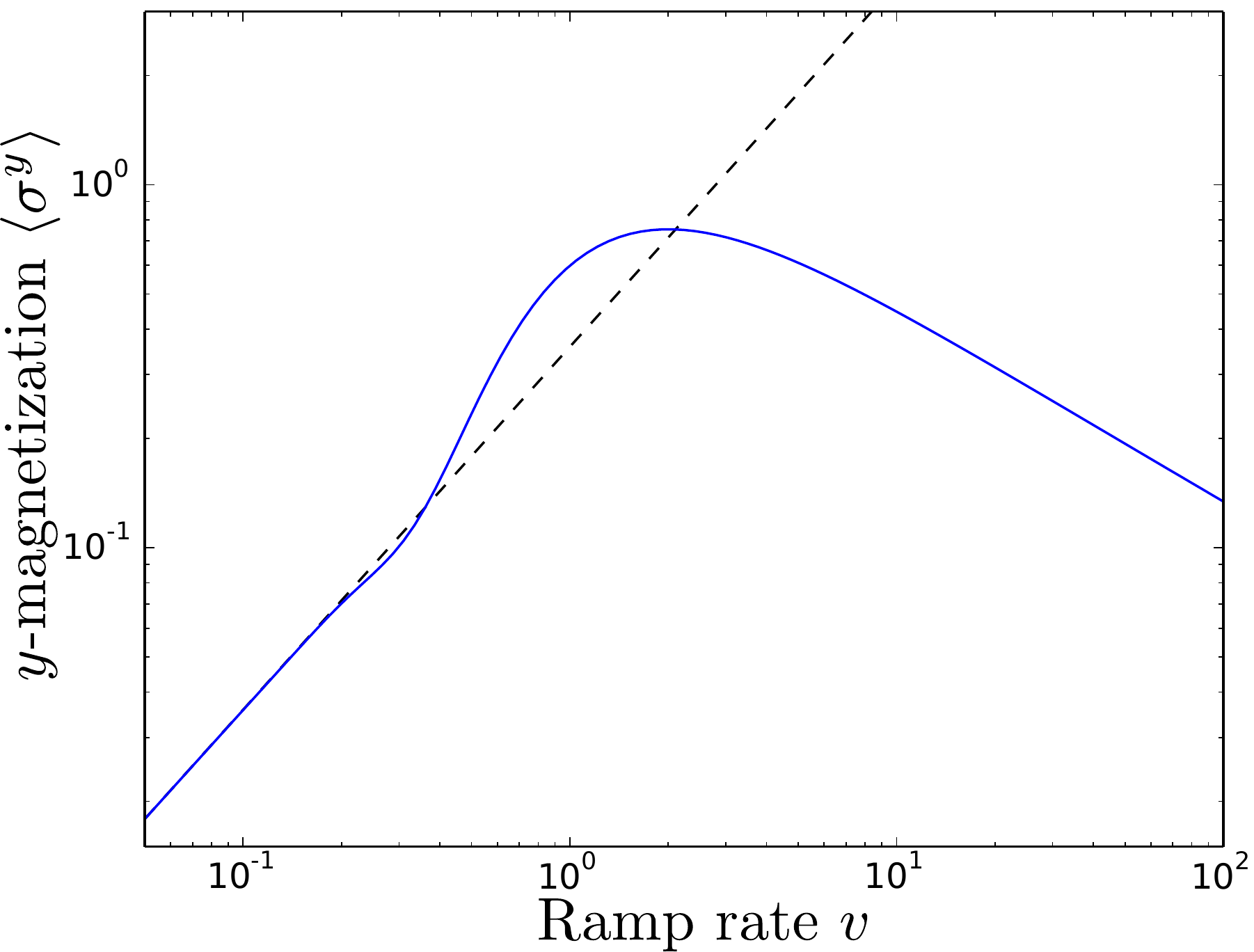}
\caption{Dependence of the transverse magnetization on the rate of change of the magnetic field along the $x$-direction (see text for details). The dashed line shows the expected low-velocity asymptote from the dynamical Hall effect, \eref{eq:M_y_exp_spin_half}. Adapted with permission from Ref.~\cite{Gritsev2012_1}}
\label{fig:spin_qhe}
\end{figure}
As is evident from the figure, at slow rates the dependence of the transverse magnetization on the rate is linear and the slope is exactly given by the Berry curvature.

Integrating the measured Berry curvature over the angles of the field, one can measure the Chern number, which we found to be $C_1=1$ for this example in \sref{sec:topology_single_spin}. Interesting, even within such a simple system, one can already observe a topological transition where the Chern number changes from $1$ to $0$. For this we can consider a slight modification into the Hamiltonian by adding a constant static magnetic field along the $z$ direction.
\be
\mathcal H=-{1\over 2}\left[ h_0\sigma_z+h_1\cos(\theta) \sigma^z+h_1\sin(\theta)\cos(\phi)\sigma^x +
h_1\sin(\theta)\sin(\phi)\sigma^y \right].
\ee
Then as one changes the magnetic field $\vect{h_1}$ along the sphere of constant radius at fixed $h_0$ we can have two different scenarios. First, $h_0<h_1$ still corresponds to the total magnetic field encircling the origin $h=0$ and thus produces a Chern number equal to one. The second scenario is realized when $h_0>h_1$. Then the total magnetic field does not enclose the origin and the Chern number is zero. The easiest way to see this is to take the limit $h_1\to 0$ and recall that the Chern number cannot change unless the surface crosses a gapless crossing point. This phase transition was recently observed in experiments on superconducting qubits \cite{Schroer2014_1}. Recall that the Chern number tells the magnetic monopole charge enclosed by our surface in parameter space. For the spin-1/2 we saw that the only monopole resided at ${\vect h}=0$ and carries charge 1. Therefore, one can interpret this topological transition as simply a shift of the surface in parameter space such that, for large $h_0$, it does not enclose the degeneracy at the origin. Interestingly this phase transition maps exactly to the phase transition in the Haldane model discussed in \exref{ex:chern_insulator} if one identifies angles of the magnetic field with the Bloch momenta. In this mapping the offset magnetic field $h_0$ plays the same role as the sublattice symmetry breaking parameter $M$ in the band model.

\subsubsection{Disordered quantum spin chain}

We have used a simple single-particle problem to illustrate the topological response of spins to a magnetic field. The situation becomes much more interesting if we consider interacting systems. In particular, following \rcite{Gritsev2012_1} we quote the numerical results for the Chern number computed through the non-adiabatic response for a disordered spin chain:
\be
\mathcal H=-\vect h \cdot \sum_{j=1}^L \zeta_j \vect\sigma_j-J\sum_{j=1}^{L-1}\eta_j \vect \sigma_j \cdot \vect \sigma_{j+1},
\ee
where $\zeta_j$ and $\eta_j$ are drawn from a uniform distribution in the interval $[0.75,1.25]$. We fix $|\vect h|=1$ and look into the Berry curvature associated with angles of the magnetic field $\theta$ and $\phi$ as a function of J (see \fref{fig:disordered_chain}). Because of the $SU(2)$ invariance of the system, as for a single spin the Chern number and the Berry curvature are simply different by a factor of 2.
\begin{figure}[ht]
\includegraphics[width=0.6\columnwidth]{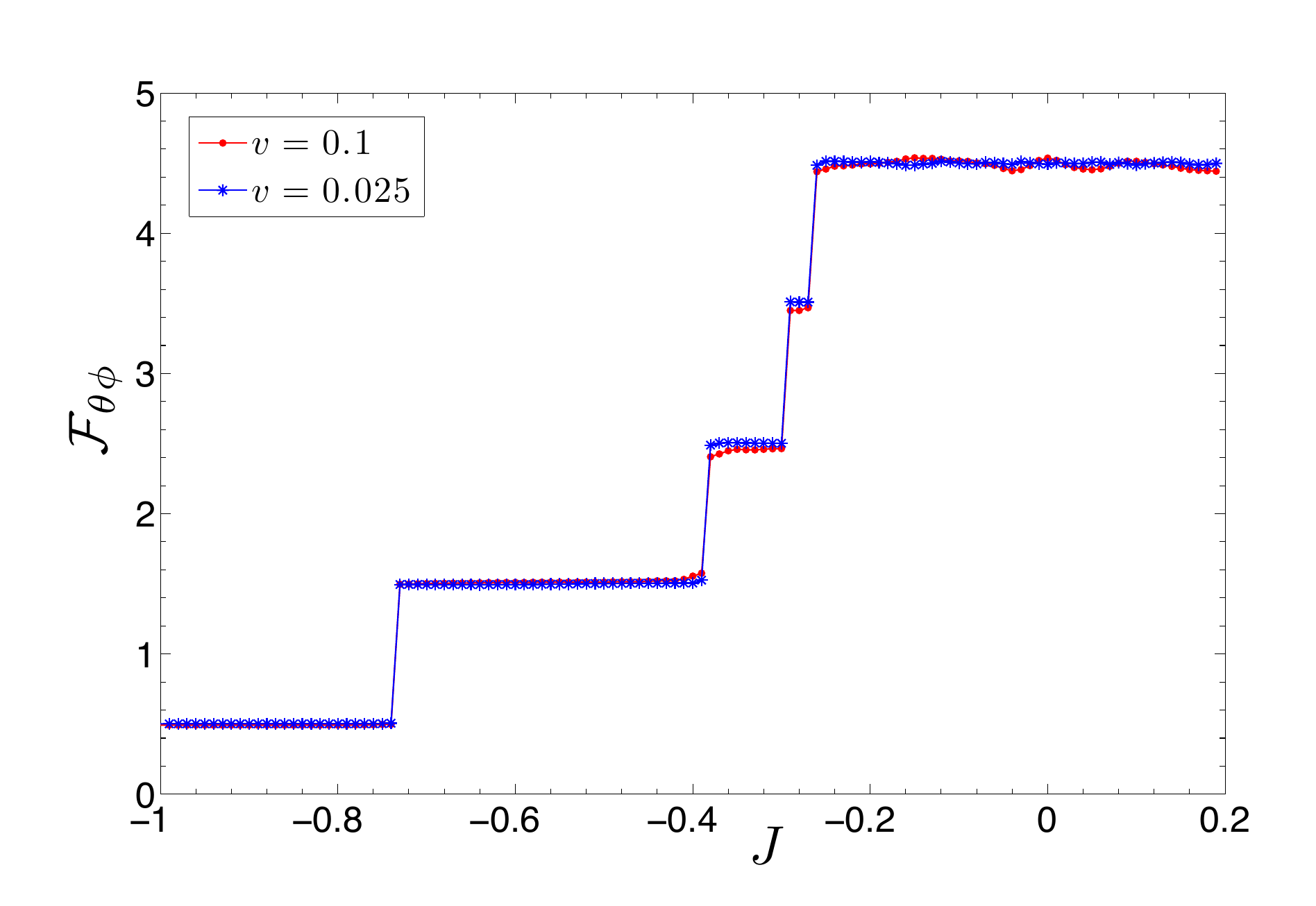}
\caption{Berry curvature at the equatorial plain $\theta=\pi/2$ for a disordered spin chain as a function of the coupling J for 9 spins. At large negative $J$ the system minimizes the total spin to $S=1/2$. The Berry curvature is also $1/2$, corresponding to Chern number equal to one. At small $J$ the system becomes polarized and the Chern number is $9$.  Figure reproduced with permission from \rcite{Gritsev2012_1}.}
\label{fig:disordered_chain}
\end{figure}
At large negative $J$, the system minimizes the total spin to $S=1/2$ and the Berry curvature is also $1/2$, corresponding to Chern number equal to one. At small $J$, the system becomes polarized and the Chern number is $L$ for a chain of length $L$. In between the Berry curvature and thus the Chern number changes in steps. If one breaks the $SU(2)$ invariance by considering, e.g., anisotropic interactions, the quantization of the Berry curvature disappears while the Chern number remains quantized. The minimal model for observing this is a two-spin system, which was recently realized experimentally also using superconducting qubits \cite{Roushan2014_1}. More recently, this has been extended experimentally to four-spin systems\cite{LuoArxiv2016_1}, demonstrating the applicability of these methods towards larger many-body systems where no other techniques may be used.

While a detailed discussion is beyond the scope of these notes, it is important to mention that these ideas are not only being extended to many-body systems, but also to more complicated systems with direct experimental relevance. One important avenue is understanding responses of open quantum systems, as most quantum systems have some non-negligible coupling to their environment. Non-adiabatic corrections to the dynamics of these systems are significantly more complicated, but they show some interesting connections to both the Berry curvature and the metric tensor \cite{Avron2012_1,Avron2012_2,Avron2011_1,Xu2014_2,Albert2015_1}. Another important direction is to understand the response of systems with degeneracies, as pioneered by Wilczek and Zee \cite{Wilczek1984_1}. Adiabatic perturbation theory in the presence of degeneracies has since been developed in a series of papers by Rigolin and Ortiz \cite{Rigolin2010_1,Rigolin2012_1,Rigolin2014_1}. Recently one of us has suggested to apply this to measure the second Chern number \cite{Avron1989_1,Avron1988_1,KolodrubetzArxiv2016_1}, a novel topological invariant that gives information about the non-Abelian Berry phase, a fundamental ingredient in active areas such as topological quantum computation. This non-Abelian topological invariant has subsequently been measured in a four-level system using hyperfine levels of ultracold atoms~\cite{SugawaArxiv2016}.

\hrulefill

\exercise{
Using two superconducting qubits, in \rcite{Roushan2014_1}, the authors are able to create Hamiltonian of the form
\be
\mathcal H = -B_r \hat{n}(\theta,\phi) \cdot  \big(\vect \sigma_1 + \vect \sigma_2\big)+B_0 \sigma^z_1 + g \big(\sigma^x_1 \sigma^x_2 + \sigma^x_1 \sigma^x_2 \big) ~,
\label{eq:H_two_qubits}
\ee
where $\hat n$ is a unit vector. For fixed magnitudes $B_r$, $B_0$, and $g$, they consider the Chern number with respect to the angles $\theta$ and $\phi$ as they encompass a sphere in parameter space. Here you will derive the theory behind some of the experimental results in the paper.
\begin{itemize}
\item Assume the system begins in its ground state at $\theta=0$, after which the angle $\theta$ is ramped slowly with time at fixed $\phi=0$. Use the dynamical quantum Hall effect to find an expression for the Berry curvature $F_{\theta \phi}$. Assuming the experimentalists are able to measure $\langle \vect \sigma \rangle$ for each qubit separately, what should they measure to find $F_{\theta \phi}$?
\item The Chern number is given by $C_1 = (2\pi)^{-1} \int d \theta d\phi F_{\theta \phi}$. For the given Hamiltonian, why is it sufficient to measure at $\phi=0$ instead of integrating over $\phi$?

\begin{figure}
\includegraphics[width=0.5\columnwidth]{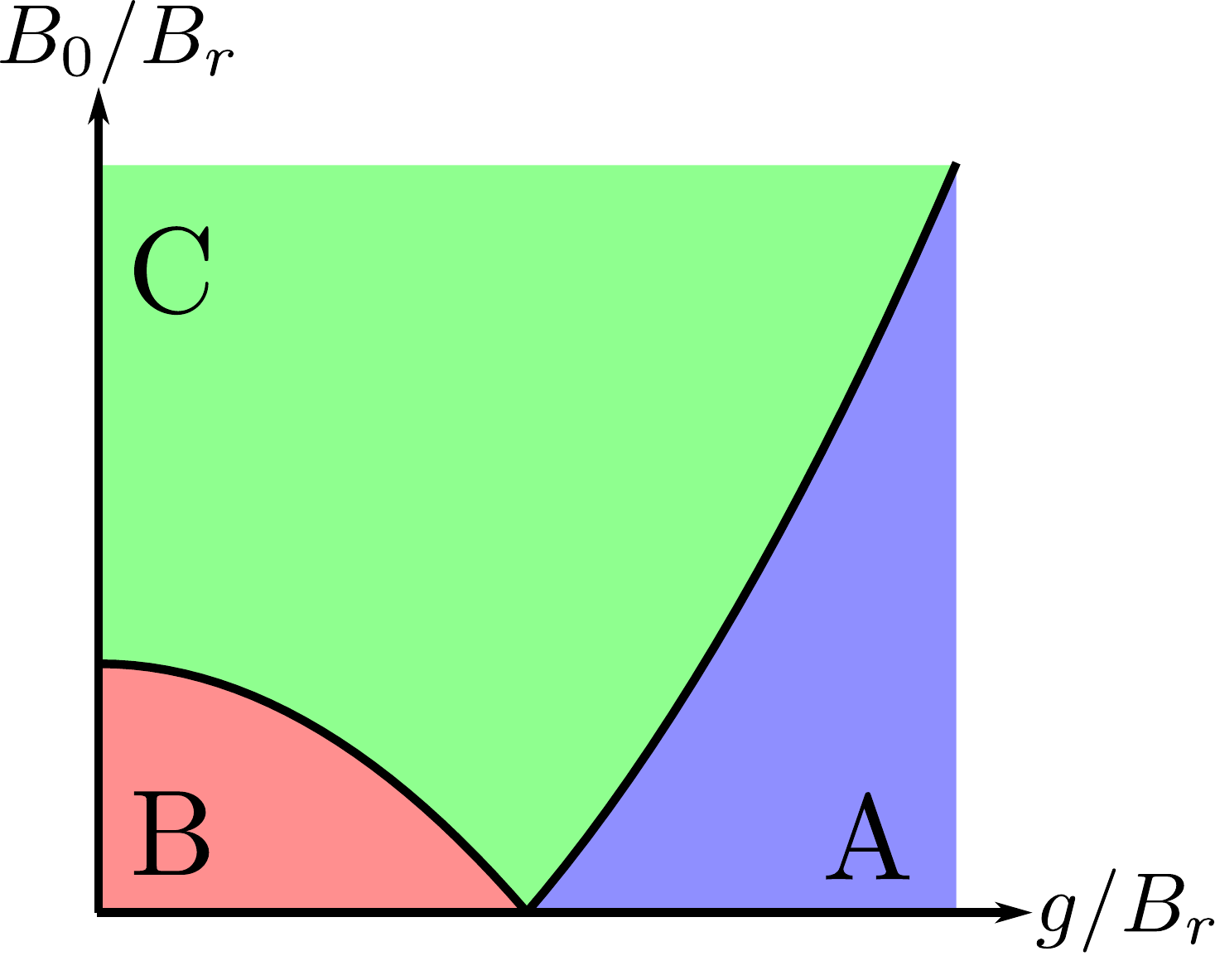}
\caption{Topological phase diagram of the two-qubit model in \eref{eq:H_two_qubits}.}
\label{fig:phase_diag_two_spin}
\end{figure}

\item The topological phase diagram of this model is depicted in \fref{fig:phase_diag_two_spin}. Let's begin by imagining that there are no interactions between the qubits ($g=0$). Using the solution of the single qubit above, what are the values of the Chern number in regions A and B?
\item Now turn off the ``pinning field'' $B_0=0$ and turn on very strong interaction, $g \gg B_r$. Argue that in this limit, deep in region C, the Chern number vanishes. Note that we have now found the Chern number in various limits of the phase diagram. Away from these limits, the math is much less trivial. Nevertheless, the Chern number remains perfectly quantized until a topological transition is reached, in which the gap above the ground state closes.
\item \emph{Bonus}: Find an analytical solution to the phase transition lines in \fref{fig:phase_diag_two_spin}. Hint: degeneracies are generally protected by a symmetry, so look along lines of high symmetry.
\end{itemize}}

\exercise{
\begin{figure}
\includegraphics[width=0.2\columnwidth]{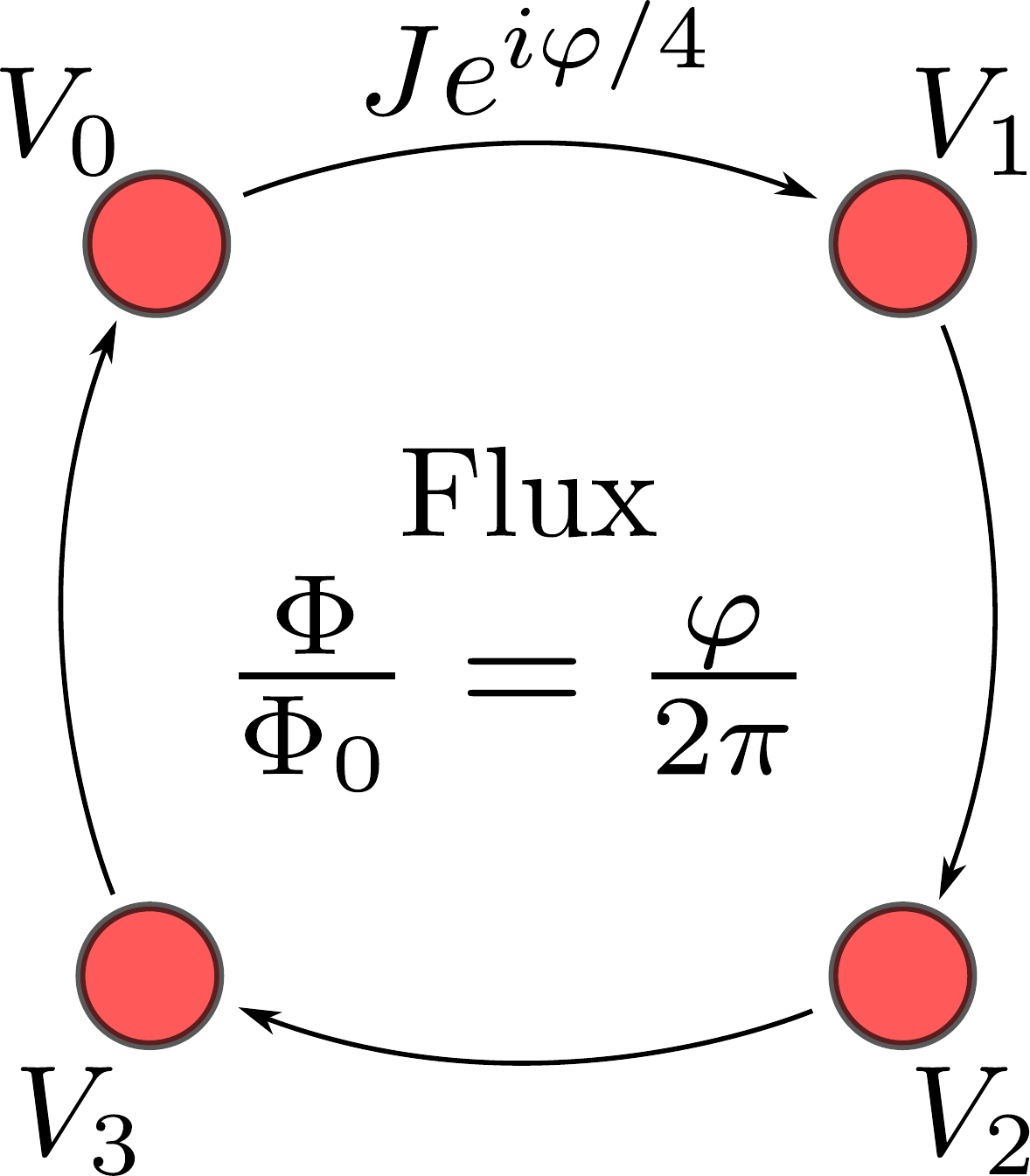}
\caption{Illustration of a single plaquette with flux, similar to what is realized in \rcite{Aidelsburger2013_1}.}
\label{fig:chern_plaquette}
\end{figure}

Let's consider another unusual situation where topology emerges. A quintessential model of topology in condensed matter systems is the Harper-Hofstadter model, where a magnetic flux is placed through each plaquette of a square lattice to create a lattice realization of the quantum Hall effect \cite{Hofstadter1976_1,Harper1955_1}. This has become particularly relevant recently as a route to realizing topological physics in systems of ultracold atoms \cite{Miyake2013_1,Aidelsburger2013_1,Aidelsburger2015_1,Kennedy2015_1}. Here we will show that topology manifests at the level of a single plaquette in such a model (\fref{fig:chern_plaquette}). Consider a single particle living on such a plaquette with flux $\Phi=\varphi \Phi_0 / 2\pi$ through it, such that the particle picks up phase $\varphi/4$ ($-\varphi/4$) each time it hops clockwise (counter-clockwise). Furthermore, put potentials $V_j$ on the four sites, setting $V_0=0$ without loss of generality. We will see that there is a non-zero Chern number with respect to the effective three-dimensional manifold defined by $x=\varphi-\pi$, $y=V_1-V_3$, and $z=V_2$. 
\begin{itemize}
\item In order to get a non-zero Chern number, we must first identify singularities that can act as sources of Berry curvature. Begin with $V_j=0$ for all $j$. Show that a degeneracy appears at $\varphi=\pi$. Then show that this degeneracy extends into a line of degeneracies for $V_1=V_3$.
\item Let's pick one point on this line of degeneracies by fixing $V_1=-V_3$. Then show that setting any of the above perturbations $x$, $y$, or $z$ to a small non-zero value breaks the degeneracy. Therefore, there exists an isolated degeneracy (a Berry monopole) at $x=y=z=0$.
\item Finally, consider a small sphere of radius $r$ in this parameter space, i.e., let $x=r \sin \theta \cos \phi$, $y=r \sin \theta \sin \phi$, $z=r\cos\theta$. Argue that the first Chern number with respect to the angles $\theta$ and $\phi$ is non-zero for small but non-zero $r$. Show numerically that its value can be measured using the dynamical Hall effect.
\end{itemize}
}

\exerciseshere

\hrulefill

\subsection{Metric tensor as a dynamical response}
\label{sec:meas_metric_tensor}

Originally Provost and Vallee thought that the metric tensor was a nice but unmeasurable mathematical object. On the other hand, it was very soon understood that the Berry curvature, i.e., its imaginary part, is responsible for many different physical phenomena such as the Aharonov-Bohm effect and the quantum Hall effect. In \sref{sec:geometry_gs_manifold} we already discussed that the ground state metric tensor can be expressed through the measurable imaginary part of the Kubo susceptibility [see \eref{eq:metric_epsilon}]. In \aref{sec:kubo_metric_finite_temp} we extended this relation to finite temperature density matrices. Here let us show that the metric tensor like the Berry curvature has a direct physical meaning as a non-adiabatic response coefficient.

Let us again use the result of the adiabatic perturbation theory for transition amplitudes [\eref{eq:ampl_n_apt}] and compute the energy variance due to the ramp rate:
\begin{multline}
\label{eq:deltaE2}
\Delta E^2=\langle \mathcal H^2\rangle-\langle \mathcal H\rangle^2=\sum_n |a_n|^2 E_n^2-\left(\sum_n E_n |a_n|^2\right)^2\\
= \sum_{\alpha\beta} \dot \lambda_\alpha \dot\lambda_\beta \left[\sum_{n\neq 0} (E_n-E_0)^2 {\langle 0|\partial_\alpha \mathcal H|n\rangle \langle n|\partial_\beta \mathcal H|0\rangle\over (E_n-E_0)^4}\right]+O(|\dot{\vect \lambda}|^3)=\hbar^2\sum_{\alpha\beta} \dot\lambda_\alpha g_{\alpha\beta} \dot\lambda_\beta+O(|\dot{\vect \lambda}|^3).
\end{multline}
So the metric tensor defines the leading non-adiabatic correction to the energy variance, which, by energy conservation, is equal to the variance of work done on the system  during the ramp $\delta w^2$. It is easy to see that this result is not tied to the ground state and applies to any other initial eigenstate. For mixed states with non-zero initial fluctuations metric tensor describes the increase in energy fluctuations due to the ramp, i.e.,
\be
\Delta E^2=\Delta E_{\rm ad}^2+\hbar^2\sum_{\alpha\beta} \dot\lambda_\alpha g_{\alpha\beta} \dot\lambda_\beta+O(|\dot{\vect \lambda}|^3),
\ee 
where $\Delta E_{\rm ad}$ is the width of the energy distribution for the adiabatic ramp with $|\dot{\vect \lambda}|^2\to 0$. This is closely connected with the quantum speed limit \cite{Fleming1973_1,Anandan1990_1,Vaidman1992_1}\add{, which we explored in more detail in \sref{sec:time_bounds}}.

In the high-temperature limit the average work and work fluctuations are not independent. The satisfy Einstein's relations, which are in turn derived from fluctuation theorems (see Refs.~\cite{Bunin2011_1, DAlessioArxiv2015_1}): $\delta w^2\approx 2 k_B T w$. Therefore in this classical or high-temperature case the metric tensor gives the leading non-adiabatic contribution to the energy:
\be
E\approx E_{ad}+ {{\hbar^2\over 2 k_B T}} \sum_{\alpha\beta} \dot\lambda_\alpha g_{\alpha\beta} \dot\lambda_\beta+O(|\dot{\vect \lambda}|^3)
\label{eq:heating_high_temp}
\ee
This non-adiabatic contribution to the energy is clearly proportional to the square of the velocity $\dot{\vect \lambda}$ and thus describes a correction to the kinetic energy associated with the parameter $\vect \lambda$. Therefore $\hbar^2 g_{\alpha\beta}/ (k_B T)$ plays the role of the mass renormalization of this parameter. We will derive this result more carefully in the next section. 

\hrulefill

\exercise{Using the leading order adiabatic perturbation theory as in \eref{eq:deltaE2}, prove \eref{eq:heating_high_temp} in the high-temperature limit assuming that the adiabatic (equilibrium) density matrix of the system is described by the Gibbs distribution: $\rho_n\approx 1/Z \exp[-E_n/(k_B T)]$. Hint: you can use the relation
\[
{\rho_n-\rho_m\over E_m-E_n}\approx {1\over k_B T} \rho_n,\quad \mbox{if}\quad k_B T\gg |E_n-E_m|.
\]}

\exerciseshere

\hrulefill

In passing we note that one can relate the metric tensor to the probability of doing zero work during an infinitesimal double quench \cite{KolodrubetzPRB2013_1}, which is connected to the well-known Loschmidt echo \cite{Silva2008_1}. These energy/work distributions are in principle measurable for a wide variety of systems, and in particular there has been a recent upswelling of progress in the field spurred by non-equilibrium fluctuation relations that also make reference to the work distribution \cite{Jarzynski1997_1}.

\section{Non-adiabatic response and emergent Newtonian dynamics}
\label{sec:emergent_newtonian_dynamics}

\emph{{\bf Key concept:} For slow macroscopic degrees of freedom $\vect \lambda$ coupled to fast degrees of freedom, Newtonian equations of motion emerge from leading non-adiabatic corrections to the Born-Oppenheimer approximation. In the classical or high-temperature limit, the emergent mass tensor is proportional to the metric tensor. In the quantum low-temperature limit, it is described by a related susceptibility expressed through the gauge potentials.}

\subsection{Adiabatic perturbation theory}

In the previous section, we argued that the leading non-adiabatic correction in $\dot{\vect\lambda}$ to the wave function of the system can be found from an assumption that the system follows the instantaneous ground state of the moving Hamiltonian $\tilde {\mathcal H}_m=\mathcal{\tilde H}-\dot \lambda_\alpha\tilde {\mathcal A}_\alpha$. From this we saw how leading non-adiabatic corrections to generalized forces and the energy broadening  connect the geometric tensor to response coefficients. In this section, we extend the previous analysis to a more general class of systems, which are not necessarily in a ground state and which might have gapless excitations. This chapter closely follows \rcite{DAlessio2014_1}. Our starting point will be the von Neumann equation for time evolution of the density matrix in the moving frame
\[
i\frac{d{\rho}}{dt}=\left[\mathcal H-\dot{\lambda}_{\alpha}\mathcal{A}
_{\alpha},\rho\right],
\]
where we remember that $\mathcal H$ is the diagonal Hamiltonian in the instantaneous basis. We again temporarily in this section set $\hbar=1$ in all intermediate formulas to simplify notations. Also for simplicity we drop the tilde signs in this section over the Hamiltonian, gauge potentials and other observables. As expressions for all resulting expectation values are gauge invariant (independent of the choice of frame), the tilde signs in the final expressions can be dropped anyway. As before, we will use standard perturbation theory (Kubo formalism), where the Galilean term plays the role of the perturbation, but now considering the full time dependent Hamiltonian. We will go to the interaction picture (i.e., the Heisenberg representation with respect to $\mathcal H$) via the time-dependent (diagonal) unitary $V(t) = \mathrm{e}^{-i \int^t \mathcal H (t') dt'}$:
\be
\rho=\mathrm{e}^{-i\int^t \mathcal H(t') dt'} \rho_H \mathrm{e}^{i\int^t \mathcal H(t') dt'}=V \rho_H V^\dagger,\;\mathcal A_\alpha=V A_{H,\alpha} V^\dagger~.
\label{eq:moving_frame}
\ee
Note that because the Hamiltonian $\mathcal H$ is diagonal by construction, and thus commutes with itself at different times, it remains unchanged in the interaction picture: $\mathcal H_H=\mathcal H$. For the same reason we do not need to worry about time-ordering in \eref{eq:moving_frame}, which illustrates the difference between the ``interaction'' picture above and the Heisenberg representation: in the latter one has to use the full time-dependent Hamiltonian (not its instantaneous diagonal part) and thus the time-ordered integral. 

In this interaction picture, the von Neumann equation becomes 
\begin{equation}
i\,\frac{d\rho_{H}}{dt}=-\dot{\lambda}_{\alpha}\left[\mathcal{A}_{H
,\alpha}(t),\rho_{H}(t)\right],\label{eq:rho}
\end{equation}
which is equivalent to the integral equation
\be
\rho_{H}(t)=\rho_{H}(0)+i\int_{0}^{t}dt'\dot{\lambda}_{\alpha}
(t')\left[\mathcal{A}_{H,\alpha}(t'),\rho_{H}(t')\right]
\label{int_equation}
\ee
Next we are going to utilize the standard linear response Kubo formalism to perturbatively solve this integral equation~\cite{Mahan2000_1}. As the Hamiltonian $\mathcal H$ is generally time-dependent, its spectrum explicitly depends on $\vect \lambda(t)$. However, this dependence is trivial because it only amounts to using phase factors $\phi_n=\int_0^t E_n(t') dt'$ instead of $\phi_n=E_n t$. To simplify derivations we will assume a $\vect\lambda$-independent spectrum for the remainder of this section and only comment in the end how one should modify the final expressions if this is not the case (see Ref.~\cite{DAlessio2014_1} for a more detailed derivation).

We assume that the system is initially prepared in a stationary state of the Hamiltonian $\mathcal H(\vect\lambda(0))$, after which one slowly turns on the ramping protocol. In the leading order in perturbation theory we can substitute the stationary density matrix into the R.H.S. of the integral equation~[\eref{int_equation}]:
\be
\rho_{H}(t)=\rho_0+i\int_{0}^{t}dt'\dot{\lambda}_{\alpha}
(t')\left[\mathcal{A}_{H,\alpha}(t'),\rho_0\right] + O(\dot \lambda^2),
\ee
where we used that $\rho_{H,0}=\rho_{0}$ because $\rho_0$ is stationary and hence commutes with $\mathcal H$.  From this we can find the linear response correction to the generalized forces:
\begin{equation}
\langle \mathcal M_\alpha(t)\rangle \approx M^{(0)}_\alpha+i\int_0^t dt'\,\dot\lambda_\beta(t') 
\langle [\mathcal M_{H,\alpha}(t),\mathcal A_{H,\beta}(t')]\rangle_0,
\label{eq:M_nu_orig}
\end{equation}
where $M^{(0)}_\alpha\equiv\langle \mathcal 
M_\alpha\rangle_0$ is the instantaneous generalized force. Evaluating the expectation value of the commutator in the co-moving basis and using \eref{eq:gauge_pot_matrix_elem} for the matrix elements of the gauge potential, we find
\begin{eqnarray}
\nonumber 
\langle [\mathcal M_{H,\alpha}(t),\mathcal A_{H,\beta}(t')]\rangle_0 &=& 
\sum_n \rho_n^0  \langle n | \mathcal M_{H,\alpha}(t) \mathcal A_{H,\beta}(t') | n \rangle - h.c.
\\ \nonumber
&=&\sum_{m \neq n} \rho_n^0  \langle n | \mathrm{e}^{i \mathcal H t} \mathcal M_{\alpha} (t) \mathrm{e}^{-i \mathcal H t} | m \rangle \langle m | \mathrm{e}^{i \mathcal H t'} \mathcal A_{\beta} (t') \mathrm{e}^{-i \mathcal H t'} | n \rangle - h.c.
\\ \nonumber
&=&\sum_{m \neq n} \rho_n^0  \mathrm{e}^{i(E_n-E_m)(t-t')} \langle n | \mathcal M_{\alpha}(t) | m \rangle \left( \frac{\langle m | \mathcal M_{\beta}(t') | n \rangle}{i (E_n-E_m)} \right) - h.c.
\\ &=&i \sum_{m \neq n} \frac{\rho_n^0-\rho_m^0}{E_m-E_n} \mathrm{e}^{i(E_n-E_m)(t-t')} \langle n | \mathcal M_{\alpha}(t) | m \rangle \langle m | \mathcal M_{\beta}(t') | n \rangle.
\end{eqnarray}
The time-dependence on the observables is a reminder that we are working in the instantaneous frame, which changes in time together with $\vect \lambda$. Substituting this expression back into \eref{eq:M_nu_orig} and switching to integration variable $t''=t-t'$, we find a general expression for the microscopic force:
\be
\langle \mathcal M_\alpha(t)\rangle= M_\alpha^{(0)}-\int_0^{t} dt''\dot\lambda_\beta(t-t'')\,\sum_{n\ne m} 
\frac{\rho_n^0-\rho_m^0}{E_m-E_n} \mathrm{e}^{i (E_m - E_n) t''} \langle m|\mathcal M_\alpha(t)|n\rangle \langle n |\mathcal M_\beta(t-t'')|m \rangle + O(\dot {\vect \lambda}^2)~.
\label{eq:gen_force_int}
\ee

This expression will generally hold for arbitrary systems as long as $\dot{|\vect \lambda|}$ is sufficiently small that the ${\dot {\vect \lambda}}^2$ term can be neglected. We can simplify this expression further
by using time scale separation. Recall that by assumption $\vect\lambda$ represents slow variables in the system. Mathematically this statement implies that the non-equal time correlation function of the generalized forces $\langle \mathcal M_{H,\alpha}(t'')\mathcal M_{H,\beta}(0)\rangle_{0, c}$ decays much faster than the characteristic time scale of  changing $\vect \lambda(t)$. Because we are interested in long time dynamics of the system, we expect therefore that the system will forget its long-time history: $\langle \mathcal M_{H,\alpha}(t)\mathcal M_{H,\beta}(t-t'')\rangle_{0, c}\to 0$ as $t''\gg \tau$, where $\tau$ is the characteristic relaxation time scale of fast degrees of freedom. Thus unless we are interested in short time transient dynamics $t\lesssim\tau$, we can extend the upper integration limit in \eref{eq:gen_force_int} to $\infty$. It is then natural to expand $\dot\lambda_{\mu}(t-t'')$ into a Taylor series near $t''=0$: $\dot\lambda_{\beta}(t-t'')\approx \dot\lambda_\beta(t)-t''\ddot\lambda_\beta(t)+\dots$. As we will see shortly, it is important to keep the first two terms in this expansion and all other terms, in most cases, describe unessential subleading corrections.\footnote{An important exception is a motion of a charged object in a vacuum, where the friction force is proportional to the third derivative of the coordinate. In this case one has to keep the next term to this expansion.} Similarly we can approximate $\mathcal M_\beta(t-t'') \approx \mathcal M_\beta(t)$, as the next order correction $\partial_{\alpha} \mathcal M_\beta(t)\dot\lambda_\alpha(t)$ will result in quadratic correction in $\dot{\vect \lambda}$ in \eref{eq:gen_force_int}. Then by grouping terms, we find
\be
\langle \mathcal 
M_\alpha(t)\rangle=M_\alpha^{(0)}-\dot\lambda_\beta(\eta_{\alpha\beta}-F_{\alpha\beta})
-\ddot \lambda_\beta(\kappa_{\alpha\beta}+F'_{\alpha\beta}) + O(\dddot {\vect \lambda}, \dot{\vect\lambda}^2)~,
\label{off_diag}
\ee
where we split the coefficients in front of $\dot\lambda_\beta$ and $\ddot\lambda_\beta$ into symmetric ($\eta_{\alpha\beta}$ and $\kappa_{\alpha\beta}$) and anti-symmetric ($F_{\alpha\beta}$ and $F'_{\alpha\beta}$) components. For instance,
\beq
&&\eta_{\alpha \beta}  = {1\over 2}\int_0^{\infty} dt''\,\sum_{n\ne m} 
\frac{\rho_n^0-\rho_m^0}{E_m-E_n} \mathrm{e}^{i (E_m - E_n) t''} \left[\langle m|\mathcal M_\alpha |n\rangle \langle n |\mathcal M_\beta |m \rangle +\alpha\leftrightarrow\beta\right],\nonumber\\
&&F_{\alpha \beta}=-{1\over 2} \int_0^{\infty} dt''\,\sum_{n\ne m} 
\frac{\rho_n^0-\rho_m^0}{E_m-E_n} \mathrm{e}^{i (E_m - E_n) t''}\left[ \langle m|\mathcal M_\alpha |n\rangle \langle n |\mathcal M_\beta |m\rangle-\alpha \leftrightarrow\beta\right],
\label{eq:first_order_dissip_curv_apt}
\eeq
where all matrix elements (and in general energies) as well as the eigenstates correspond to the instantaneous parameter value $\vect\lambda(t)$. 

It is now straightforward to evaluate the remaining integrals over $t''$. As usual one can regularize them by inserting small decaying exponential $\exp[-\delta t'']$ with infinitesimal positive $\delta$.  For instance, in \eref{eq:first_order_dissip_curv_apt}, one uses
\be
\int_0^\infty \exp[i (E_m-E_n) t''-\delta t''] dt''={1\over \delta -i (E_m-E_n)} \stackrel{\delta\to 0}{\xrightarrow{\hspace*{0.7cm}}} i P\left({1\over E_m-E_n}\right)+\pi\delta(E_n-E_m),
\ee
where $P$ stands for the principal value. Note that the first term is antisymmetric under the permutation of indexes $n$ and $m$, while the second is symmetric. Because, as is evident from \eref{off_diag}, the permutation of $n$ and $m$ is equivalent to the permutation of $\alpha$ and $\beta$, we see that the principal value determines the antisymmetric coefficient $F_{\alpha\beta}$ and the second, symmetric term determines $\eta_{\alpha\beta}$. Therefore 
\beq
\nn F_{\alpha\beta}(\lambda)&=&-i\sum_{n\neq m} \frac{\rho_n^0-\rho_m^0}{(E_m-E_n)^2} \langle m|\mathcal M_\alpha|n\rangle\langle 
n|\mathcal M_\beta|m\rangle\nonumber
\\ &=&i\sum_{n\neq m} \rho_n^0
{\langle n|\mathcal M_\alpha|m\rangle\langle m|\mathcal M_\beta |n\rangle-\langle n|\mathcal M_\beta|m\rangle\langle m|\mathcal M_\alpha|n\rangle\over (E_n-E_m)^2}~,
\eeq
where all energies and matrix elements are evaluated at $\lambda$. If we compare this expression with \eref{geom_tens_lehmann} and use that $F_{\alpha\beta}=i(\chi_{\alpha\beta}-\chi_{\beta\alpha})$ [cf. \eref{eq:Berry_curvature_geom_tens}], we will recognize $F_{\alpha\beta}$ is just the average of the Berry curvature over the adiabatic density matrix $\rho^{0}$.\

Similarly, for a thermal density matrix $\rho^{0}$ the symmetric part of the response coefficient is given by
\[
\eta_{\alpha\beta}={\pi\over k_B T}\sum_{n\neq m}\,\,\rho^0_{m}\langle m|\mathcal 
M_{\alpha}|n\rangle\langle n|\mathcal 
M_{\beta}|m\rangle\delta(E_{n}-E_{m}),
\]
where we used that for a thermal ensemble with $\rho_n^0 \propto \mathrm{e}^{-E_n/k_B T}$,
\be
\frac{\rho_n^0-\rho_m^0}{E_m-E_n}\to {1\over k_B T}\rho_n^0 
\ee
when $E_m\to E_n$. As we will see shortly, $\eta_{\alpha\beta}$ represents the friction force on the system. It is non-zero only if the system has gapless excitations. Therefore, at zero temperature or for a system with a discrete energy spectrum, the friction coefficient is always zero unless the system is gapless or quantum critical. Thus, for time being we set $\eta_{\alpha\beta}\to 0$.

In a similar spirit one can derive the other two coefficients. Let us use that
\begin{multline}
-\int_0^\infty t'\exp[i (E_m-E_n) t'-\delta t'] dt'=\partial_\delta \int_0^\infty \exp[i (E_m-E_n) t'-\delta t'] dt'\\=
-{1\over (\delta -i (E_n-E_m))^2}
\stackrel{\delta\to 0}{\xrightarrow{\hspace*{0.7cm}}}
{1\over (E_n-E_m)^2}-i\pi\delta'(E_n-E_m),
\end{multline}
Plugging this result into \eref{eq:gen_force_int} we see that now the off-shell term is symmetric, while the on-shell term is antisymmetric. The first (off-shell) term defines the coefficient $\kappa_{\alpha\beta}$, which as we will see shortly determines the mass renormalization
\be
\kappa_{\alpha\beta}=\sum_{n\neq 
m}\frac{\rho^0_{n}-\rho^0_{m}}{\left(E_{m}-E_{n}\right)^{3}}\,
\langle m|\mathcal M_{\alpha}|n\rangle\langle n|\mathcal 
M_{\beta}|m\rangle =\sum_{n\neq m}\frac{\rho^0_{n}-\rho^0_{m}}{E_{m}-E_{n}}\,\langle m|\mathcal A_\alpha|n\rangle\langle 
n|\mathcal A_\beta|m\rangle
\label{main_kappa}
\ee
At low temperatures $k_B T\to 0$ and hence $\rho_{n}^0\to \delta_{n0}$ this expression reduces to 
\be
\kappa_{\alpha\beta}\approx \hbar \sum_{m\ne0}\frac{\langle0|\mathcal 
M_{\alpha}|m\rangle\langle m|\mathcal 
M_{\beta}|0\rangle+\nu\leftrightarrow\mu}{\left(E_{m}-E_{0}
\right)^{3}}
\label{kappa_low},
\ee
while at high temperatures (or near the classical limit) we find
\be
\kappa_{\alpha\beta}\approx\frac{1}{2 k_B T}\sum_{n}\rho_{n}^0\,\left(\langle 
n|\mathcal A_\alpha \mathcal A_\beta|n\rangle_c+\alpha\leftrightarrow\beta\right)={\hbar^2\over k_B T}\, 
g_{\alpha\beta}
\label{kappa_high}
\ee
where $g_{\alpha\beta}$ is the Fubini-Study metric tensor for the finite temperature ensemble. We reintroduced the factor of $\hbar$ into the expression for the mass tensor to highlight that it has a well defined classical limit. It is straightforward to see that at any temperature the mass tensor $\kappa_{\alpha\beta}$ can be written as the integral of the connected imaginary time correlation function of the gauge potentials $\mathcal A_\alpha$ and $\mathcal A_\beta$:
\be
\kappa_{\alpha\beta}=\frac{1}{2\hbar}\int_0^{\hbar/k_B T} d\tau \langle \mathcal A_{H,\alpha}(-i\tau) 
\mathcal A_{H,\beta}(0)+\alpha\leftrightarrow\beta\rangle_{0,c}~,
\label{eq:mass_corr_function}
\ee
where 
\[
\mathcal A_{H,\alpha}(-i\tau)=\exp[\tau \mathcal H/\hbar]\mathcal A_\alpha \exp[-\tau \mathcal H/\hbar]
\]
is the imaginary time Heisenberg representation of the operator $\mathcal A_\alpha$. Then \eref{eq:mass_corr_function} immediately follows from \eref{main_kappa} if we use the identity
\begin{equation}
{1\over \hbar}\int_0^{\hbar/k_B T}\,d\tau\,\,\rho^0_n\,\,\mathrm{e}^{-(E_m-E_n)\tau/\hbar}=\frac{
\rho^0_n-\rho^0_m}{E_m-E_n}
\label{im_int}
\end{equation}

While we did not explain yet why the tensor $\kappa_{\alpha\beta}$ is related to mass, let us point out that its high temperature asymptotic is perfectly consistent with the equipartition theorem if $\alpha$ and $\beta$ describe macroscopic coordinates, say the position of the center of mass: $\alpha,\beta \in \{x,y,z\}$. In this case as we discussed earlier the gauge potential reduces to the total momentum operator of fast degrees of freedom $\mathcal A_\alpha=P_\alpha$ and thus, e.g., the $xx$ component of the mass according to \eref{kappa_high} satisfies
\be
\kappa_{xx}={1\over k_B T}\langle  P_x^2\rangle_c\quad \Leftrightarrow\quad {\langle P_x^2\rangle_c\over 2\kappa_{xx}}={k_B T\over 2},
\ee
which is indeed the famous equipartition theorem of the statistical physics. As a corollary to our derivation, we note that~\eref{eq:mass_corr_function} generalizes the equipartition theorem to quantum systems and applies at any temperatures. Perhaps a less trivial statement is that \eref{kappa_high} applies to all types of motion. For example, 
in a scale-invariant Hamiltonian where the slow parameter corresponds to dilations,
the additional mass in the classical limit is given by the product of the variance of the dilation operator and the inverse temperature, and similarly for whatever parameters the problem presents.

Finally, the antisymmetric tensor $F'$ is given by
\be
F'_{\alpha\beta}=-i \pi \sum_{n\neq 
m}\,\frac{\rho_n^0-\rho_m^0}{E_m-E_n} \langle 
n|\mathcal{M}_{\alpha}|m\rangle \langle 
m|\mathcal{M}_{\beta}|n\rangle\,\delta'(E_n-E_m).
\ee
Similar to $\eta_{\alpha\beta}$, this tensor is an on-shell contribution responsible for dissipation, but usually it is subleading to $\eta$. Like $F$, this tensor is always zero if the instantaneous Hamiltonian respects time-reversal symmetry. 

\subsection{Born-Oppenheimer approximation}

Up to now we were considering the parameter $\vect \lambda$ as an external slow field. This is usually justified when the back action of fast degrees of freedom is negligible. However, there are many instances where such back action can not be neglected despite the time scale separation. For example in atomic and molecular systems, as well as more complex materials, the motion of nuclei is much slower than the motion of electrons due to large mass difference, but the forces exerted by electrons on nuclei cannot be neglected. In systems with emergent macroscopic collective degrees of freedom like order parameters, dynamics of the latter can be much slower than that of microscopic degrees of freedom but yet it is entirely determined by interactions with these degrees of freedom (e.g., slow magnetization waves often originate from fast motion of electrons with different spin). In thermodynamic  heat engines, fast degrees of freedom such as atoms exert a macroscopic force on a macroscopic object such as a piston, causing motion of the piston due to its energy exchange with fast atoms. In all such situations it is natural to assume that fast degrees of freedom nearly adiabatically follow equilibrium corresponding to the instantaneous positions of the slow degrees of freedom. These ideas were first developed by Born and Oppenheimer in the context of atoms in 1927 and are known now as the Born-Oppenheimer approximation. Let us briefly discuss this approximation since it is the starting point of our further analysis. For simplicity we will assume that the slow degrees of freedom are classical, which is often justified since we are assuming they are macroscopic, while we will treat quantum degrees of freedom fully quantum mechanically.

Let us assume quite generally that the total Hamiltonian describing the degree of freedom $\vect\lambda$ and the rest of the system is 
\begin{equation}
\mathcal H_{tot}(\vect\lambda)=\mathcal H_{0}(\vect{\lambda})+\mathcal H(\vect{\lambda}),\label{eq:general}
\end{equation}
where $\mathcal H_{0}(\vect{\lambda})$ is the Hamiltonian describing the bare motion
of $ \vect{\lambda}$. The choice of  splitting $\mathcal H_{tot}$ between $\mathcal H_0$ and $\mathcal H$ is somewhat arbitrary and we can well choose $\mathcal H_0=0$ so that $\mathcal H_{tot}=\mathcal H$. However, for an intuitive interpretation of 
the results, it is convenient to assume that $\mathcal H_{0}(\vect{\lambda})$ represents a 
massive degree of freedom in some external potential $V(\vect{\lambda})$: 
\[
\mathcal H_{0}(\vect{\lambda})={1\over 2} p_\alpha m_{\alpha\beta}^{-1} p_\beta+V(\vect{\lambda}),
\]
where $m_{\alpha\beta}^{-1}$ is the inverse mass tensor. In the infinite mass limit ($||m_{\alpha\beta}||\rightarrow\infty$), $\vect{\lambda}$ 
represents an external (control) parameter whose dynamics is specified a priori. 
When $||m_{\alpha\beta}||$ is finite, $\vect{\lambda}$ is a dynamical variable and its dynamics 
needs to be determined self-consistently. The whole system can be described by coupled Hamiltonian equations of motion
\be
m_{\alpha\beta} \frac{d\lambda_\alpha}{dt}=p_\beta,\quad \frac{dp_\alpha}{dt}=-\frac{\partial 
V}{\partial\lambda_\alpha}+{\rm Tr} [\rho(t) \mathcal M_\alpha(\vect \lambda(t))], \quad\, i{d\rho(t)\over dt}=[\mathcal H(\vect \lambda(t)),\rho(t)].
\label{Hamilton}
\ee
Technically one can derive this equation from the path integral representation of the full quantum-mechanical evolution by taking the saddle point with respect to the classical field $\vect \lambda$ and treating other microscopic degrees of freedom fully quantum-mechanically. Alternatively, as was originally suggested by Born and Oppenheimer, one can assume that the quantum density matrix describing the full system factorizes into the product of density matrices for the slow degree $\vect\lambda$ and other degrees of freedom and then taking the classical limit for $\vect\lambda$. 

The key assumption of the Born-Oppenheimer approximation is that $\vect\lambda$ is slow, such that one can substitute the full density matrix $\rho(t)$ by its adiabatic limit $\rho_0$ and use this $\rho_0$ in the second equation in \eref{Hamilton}. Under these conditions the dynamics of $\vect \lambda$ is described by a motion in the modified potential:
\be
m_{\alpha\beta} \frac{d\lambda_\alpha}{dt}=p_\beta,\quad \frac{dp_\alpha}{dt}=-\frac{\partial 
V}{\partial\lambda_\alpha}+M_\alpha^{(0)}(\vect\lambda).
\ee
Due to the Feynman-Hellman theorem, in equilibrium
\[
M_\alpha^{(0)}(\vect \lambda)=-{\rm Tr}[\rho_0 \partial_\alpha\mathcal H]=-\partial_\alpha {\rm Tr}[\rho_0\,\mathcal H] ~\implies~\dot p_\alpha = -\partial_\alpha \big( V+\rm{Tr}[\rho_0 H] \big).
\]
So the slow degree of freedom effectively moves in the renormalized Born-Oppenheimer potential 
\be
V'(\vect\lambda)=V(\vect\lambda)+ {\rm Tr}[\rho_0\mathcal H(\vect\lambda)].
\label{eq:Born_Oppenheimer_potential}
\ee

\subsection{Emergent Newtonian dynamics}

While the Born-Oppenheimer approximation is very powerful for many systems, it completely misses non-adiabatic corrections to the density matrix. We already alluded to the fact that these corrections give rise to the Lorentz force, friction, mass renormalization and other effects, which we will briefly discuss below. To take these corrections into account and thus to go beyond the Born-Oppenheimer approximation, we simply need to combine the equations of motion [\eref{Hamilton}] and the non-adiabatic expansion of the generalized force [\eref{off_diag}]
\be
{d\over dt}\left[(m_{\alpha\beta}+\kappa_{\alpha\beta}+F'_{\alpha\beta})\dot \lambda_\beta\right]+(\eta_{\alpha\beta} 
-\hbar F_{\alpha\beta})\dot\lambda_\beta=-\frac{\partial V}{\partial\lambda_\alpha}+M_\alpha^{(0)}
\label{Newton}
\ee
up to terms of order $\dot \lambda^2$. The symmetric tensor in the first term in this equation represents the renormalized mass. Thus $\kappa_{\alpha\beta}$ indeed gives mass renormalization. The term $\eta_{\alpha\beta}\dot\lambda_\mu$ is clearly the dissipative force. The Berry curvature defines an analogue of the Coriolis or the Lorentz force and the other antisymmetric on-shell contribution encoded in $F'$ is effectively an antisymmetric friction term. In these notes we are focusing on quantum systems with discrete spectra. Therefore there are no on-shell contributions and hence we set $\eta$ and $F'$ to zero for the remainder of these notes. We also point out that within the accuracy of our expansion one can equally write the renormalized mass term as $\kappa_{\alpha\beta} \ddot\lambda_\beta$ or, as we did, as $d_t [\kappa_{\alpha\beta}\dot\lambda_\beta]$. Indeed it is easy to see that the difference between these two terms $d_\gamma \kappa_{\alpha\beta}\dot\lambda_\beta\dot\lambda_\gamma\sim O(|\dot{\vect \lambda}^2|)$. However, a more careful analysis shows that the mass renormalization terms gives a conservative contribution to the energy of the system, i.e., is given by the full derivative of the renormalized Hamiltonian and therefore writing it as in \eref{Newton} is more accurate~\cite{DAlessio2014_1}.

In the absence of dissipative contributions it is easy to check that the equations of motion [\eref{Newton}] come from the Lagrangian: 
\be
\mathcal{L}=\frac{1}{2}\,\dot\lambda_\alpha\,(m+\kappa)_{\alpha\beta}\,\dot\lambda_\beta + 
\dot\lambda_\beta\,A_{\beta}(\vect\lambda)-V'(\vect\lambda),
\label{lagrangian}
\ee
where 
\[
A_{\beta}(\vect\lambda)=Tr[\rho_0 \mathcal A_\beta] 
\]
is the equilibrium Berry connection and $V'$ is the Born-Oppenheimer potential in \eref{eq:Born_Oppenheimer_potential}. In the zero temperature case, the Berry connection reduces to the ground state Berry connection and the Born-Oppenheimer potential reduces to the sum of the bare potential and the instantaneous ground state energy of the system at given $\vect\lambda$. From the Lagrangian, \eref{lagrangian}, we can define the canonical momenta conjugate to the coordinates $\lambda_\nu$:
\be
p_\alpha\equiv\frac{\partial\mathcal{L}}{\partial 
\dot{\lambda}_\alpha}=(m_{\alpha\beta}+\kappa_{\alpha\beta})\dot{\lambda}_\beta+A_\alpha(\vect\lambda)
\ee
and the emergent Hamiltonian:
\be
\mathcal{H}_{\lambda}\equiv\dot\lambda_\alpha\,p_\alpha-\mathcal{L}=\frac{1}{2} (p_\alpha-A_\alpha) 
(m+\kappa)_{\alpha\beta}^{-1} (p_\beta-A_\beta)+V'(\vect\lambda). 
\label{eq:nonadiabatic_hamiltonian}
\ee
Clearly the equilibrium Berry connection term plays the role of the vector potential. Thus we see that the formalism of effective Hamiltonian dynamics for arbitrary macroscopic degrees of freedom is actually emergent.  Without the mass renormalization this (minimal coupling) Hamiltonian was derived earlier~\cite{Berry1989_1}. Away from the ground state the dissipative tensors ($\eta$ and $F'$) are, in  general, non-zero and it is not possible to  reformulate \eref{Newton} via the Hamiltonian or Lagrangian formalism. 

\hrulefill

\exercise{Verify explicitly that the Lagrangian and the Hamiltonian equations of motion given by the Lagrangian [\eref{lagrangian}] and the Hamiltonian [\eref{eq:nonadiabatic_hamiltonian}] are equivalent to Newtonian equations of motion [\eref{Newton}], assuming that there are no dissipative contributions, i.e., $\eta=F'=0$.}

\exerciseshere

\hrulefill

\subsection{Beyond Newtonian dynamics: the snap modulus}
\label{sec:snap}

Within the developed formalism we can continue the non-adiabatic expansion for the generalized force, \eref{eq:gen_force_int}. To simplify the analysis in this section let us assume that the parameter $\lambda$ is single component and thus all antisymmetric contributions vanish. As was mentioned in an earlier footnote, in this way one can recover the third derivative friction term, which describes dissipation due to radiation in Lorentz-invariant systems. Since we are focusing here on non-dissipative systems with a discrete spectrum this term will be zero. So the next non-zero term will appear if we go to the fourth derivative in $\lambda$. Such high-derivative term might look totally irrelevant given our assumption of time scale separation. But it has very important implications defining the leading correction to the Newtonian dynamics and thus showing the regime of its validity. Furthermore, as we will demonstrate later [see discussion below  \eref{eq:snap_modulus}] this term is closely related to the Unruh effect for accelerated photons confined to the cavity and has interesting observable physical consequences for the dynamics of the cavity.

It is straightforward to see that continuing the Taylor expansion in $t''$ in \eref{eq:gen_force_int} we find up to the fourth order (and in the absence of dissipative odd derivative terms) 
\be
\langle \mathcal M_\alpha\rangle\approx  M_\alpha^{(0)}-\kappa\ddot\lambda+\zeta {d^4\lambda\over dt^4},
\ee
where
\be
\zeta = \sum_{n \neq m} \frac{\rho_n - \rho_m}{(E_m - E_n)^3} |\langle m | \mathcal A_\lambda|n \rangle|^2= 
\sum_{n \neq m} \frac{\rho_n - \rho_m}{(E_m - E_n)^5} |\langle m | \mathcal M_\lambda|n \rangle|^2.
\label{eq:snap_modulus}
\ee
Following the definition of the fourth derivative of the position as snap~\cite{Visser2004_1} we term the coefficient $\zeta$ as the {\em snap modulus}.

\hrulefill

\exercise{Derive \eref{eq:snap_modulus}.}

\exercise{Derive the microscopic expression for the dissipative contribution entering the generalized force, which is proportional to $\dddot\lambda$. Show that if the temperature is positive it can only lead to dissipation of the bare energy of $\lambda$, i.e., due to this term $d\mathcal H_\lambda/dt \leq 0$, where $\mathcal H_\lambda$ is given by \eref{eq:nonadiabatic_hamiltonian}.}

\exerciseshere

\hrulefill

Substituting this generalized force into the equations of motion, \eref{Hamilton}, we get
\begin{equation}
M \ddot \lambda  =  -\partial_\lambda V' - \kappa \ddot \lambda + \zeta \lambda^{(4)}~.
\end{equation}
Multiplying by the velocity $\dot \lambda$ and rearranging, this becomes
\begin{eqnarray*}
0 & = & \dot \lambda \partial_\lambda V' + (M+\kappa)\ddot \lambda \dot \lambda - \zeta \lambda^{(4)} \dot \lambda \\ 
& \approx & \frac{d}{dt} \left( V' + \frac{M+\kappa}{2}\dot \lambda^2 - \frac{\zeta}{2} \left( 2 \dot \lambda \dddot \lambda - \ddot \lambda^2 \right) \right)~,
\end{eqnarray*}
up to terms of order $\dot \lambda^3$. Equivalently there is an emergent energy conservation law with 
\be
\mathcal E_\lambda=K+V'(\lambda)={\rm const},
\ee
where the kinetic energy in the presence of the snap modulus reads
\be
K={M+\kappa\over 2}\dot\lambda^2+{\zeta\over 2} \left(\ddot\lambda^2-2\dot\lambda\ddot\lambda\right)
\label{kin_energy}
\ee
Completing the square and ignoring the higher order term $\dddot \lambda^2$ one can approximately rewrite the kinetic energy as
\be
K\approx {M+\kappa\over 2}\left(\dot\lambda-{\zeta\over \kappa+M}\dddot\lambda\right)^2+{\zeta\over 2}\ddot\lambda^2
\ee
so that the third derivative term plays a role similar to the gauge potential. In this derivation we assumed for simplicity that both $\kappa$ and $\zeta$ are independent of $\lambda$. One can check that if this is not the case the correct equations of motion follow from the conservation of the energy $\mathcal E_\lambda$. This energy function does not represent a Hamiltonian any longer since it explicitly depends on higher order derivatives. Nevertheless one can define the Lagrangian and get the equations of motion from extremizing the action.

\section{Examples of emergent Newtonian dynamics}
\label{sec:newtonian_dynamics_examples}

We will now illustrate the emergent Newtonian dynamics formalized in the previous section with a few simple examples. First, we consider a particle in a box whose walls are allowed to move. We find that the excitations of the particle caused by motion of the walls dress the mass of the walls. The value of mass dressing depends on the nature of the wall motion (translational, dilational, etc.). Then we extend this concept to a massless particle, where now the mass dressing is found to depend on the energy of particle in the box. We then proceed to a simple many-body example, the classical version of a central spin problem, where excitations of the spin bath dress the central spin's moment of inertia. Finally, we show that emergent Newtonian dynamics can occur entirely internally by showing how the BCS gap of a superconductor can be treated as a semi-classical degree of freedom whose dynamics is modified by excitations of the superconducting quasiparticles. Throughout this section, we will explicitly insert all factors of $\hbar$ to better see when these effects could be observed in realistic systems.

\subsection{Particle in a moving box}

Let us begin by considering a massless spring connected to a wall, as illustrated in the left panel of \fref{fig:piston}. We imagine that a quantum particle of mass $m$ is initially prepared in the ground state of the  confining potential.
As in the previous example we will compute how the mass of a classical object 
(the wall) coupled to a quantum environment (the particle in the well) is 
renormalized, which in practice could be measured by, for example, a change in the oscillation frequency of the spring.

\begin{figure}[ht]
{\bf \Large a}\includegraphics[width=0.45\columnwidth]{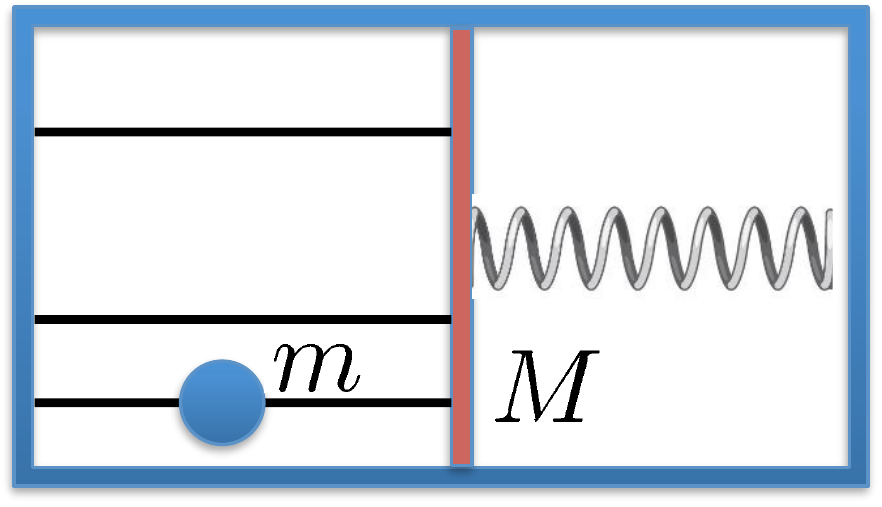}\hskip 0.5cm
{\bf \Large b}\includegraphics[width=0.45\columnwidth]{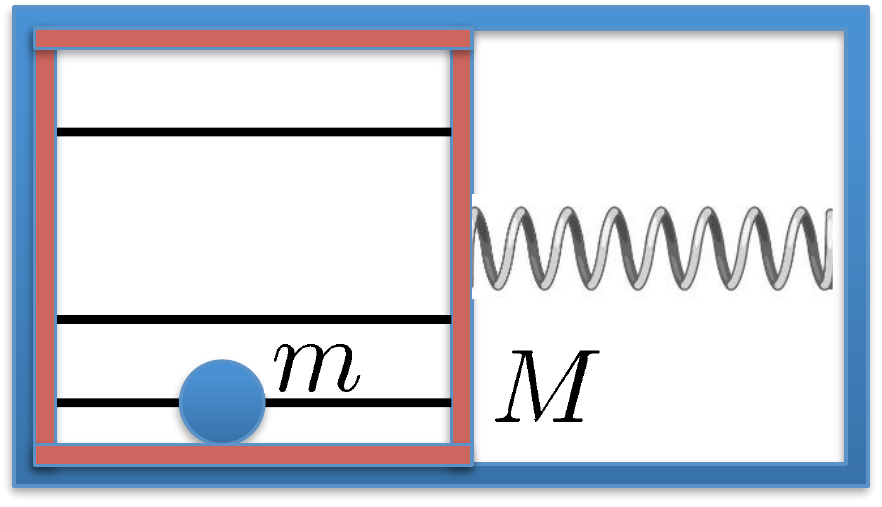}
\caption{(Color on-line) Schematic of a quantum piston. 
a) The spring is connected to a wall of the potential in which a quantum 
particle of mass $m$ is initially confined into the ground state.
b) As in (a) but now the spring is connected to the whole potential well which 
moves rigidly.
The horizontal black lines represent the low energy wave functions of the 
quantum particle in the confining potential.}
\label{fig:piston}
\end{figure}

According to \eref{kappa_low} the mass renormalization is given by
\be
\kappa_{RR}=2 \hbar^2 \sum_{n\neq 0} \frac{|\langle n| 
\mathcal{M}_\lambda|0\rangle|^2}{(E_n-E_0)^3},
\label{eq:kappa_R}
\ee
where $\lambda=X_R$ is the position of the right potential wall.
We approximate the confining potential as a very deep square well potential:
\be
\mathcal H=\frac{p^2}{2 m} + V \big( \Theta(X_L-x) + \Theta(x-X_R) \big)~.
\ee
Then $\mathcal{M}_\lambda\equiv-\partial_\lambda \mathcal H=V\delta(x-X_R)$ and we find
\be
\kappa_{RR}=2 \hbar^2 \sum_{n\neq 0} \frac{V^2 |\psi_0(X_R)|^2 |\psi_n(X_R)|^2}{(E_n-E_0)^3}.
\ee
Using the well known result for a deep but finite square well potential
\[
|\psi_n(X_R)|=\sqrt{\frac{2}{L}}\,\,\sqrt{\frac{E_n}{V}},
\]
where the factor of $\sqrt{2/L}$ comes from the normalization of the wave-function in a square potential of length $L$, we obtain
\be
\kappa_{RR}=2 \hbar^2 \left(\frac{2}{L}\right)^2 \sum_{n\neq 0} \frac{E_0 E_n}{(E_n-E_0)^3}.
\label{KR}
\ee
Substituting 
\[
E_n = \frac{\hbar^2 k_n^2}{2m},\quad k_n=\frac{n+1}{L}\,\pi,\quad\forall n\ge0
\]
we arrive at
\be
\kappa_{RR}=m\,\frac{16}{\pi^2}\sum_{n\geq 1} \frac{(n+1)^2}{[(n+1)^2-1]^3} 
=m\,\frac{2\pi^2-3}{6\pi^2}\approx 0.28 m.
\label{eq:kappa_RR_gs}
\ee
The result is identical if we connect the piston to the left wall, i.e., $\kappa_{LL}=\kappa_{RR}$.

\hrulefill

\exercise{\label{ex:dilation_mass_gauge}Derive the result \eref{eq:kappa_RR_gs} using the gauge potential. In particular, repeat steps similar to the ones for the harmonic oscillator leading to \eref{eq:gauge_pot_harmonic_spring} to find the gauge potential corresponding to moving $X_R$. Then using this gauge potential and \eref{kappa_low} compute the mass correction of the piston.}

\exerciseshere
\hrulefill

\subsubsection{Translations}

Now let us consider a slightly different setup where the spring connects to the 
whole box  (see \fref{fig:piston}b and \fref{fig:asymmetric_mass}) so that $\lambda=X_+$ now indicates the 
center of mass of the well. From Galilean invariance we expect $\kappa=m$. In fact, since now both 
potentials are moving, our expression gives
\[
\mathcal{M}_+=-\partial_{+} \mathcal H=V (\delta(x-X_R)-\delta(x-X_L)),
\]
where $X_L$ and $X_R$ are the left and right positions of the walls. Thus using 
\eref{eq:kappa_R} we obtain
\be
\kappa_{++}=2 \hbar^2 \sum_{n\neq 0} \frac{V^2(\psi_0(X_L)\psi_n(X_L)-\psi_0(X_R) \psi_n(X_R))^2}{(E_n-E_0)^3}.
\label{eq:kappa+}
\ee
Since in a symmetric potential well $\psi_n(X_R)=(-1)^n\psi_n(X_L)$, only the odd terms contribute in the equation above.
Following the same line of reasoning as before we arrive at (note the extra factor of $4$ with respect to \eref{KR})
\be
\kappa_{++}=2 \hbar^2 \left(\frac{2}{L}\right)^2 4 \sum_{n=odd} \frac{E_0 E_n}{(E_n-E_0)^3}  \\
=m \frac{64}{\pi^2} \sum_{n=odd} \frac{(n+1)^2}{[(n+1)^2-1]^3}=m.
\label{eq:kappa_++_potential}
\ee
So indeed we recover the expected result. This simple calculation illustrates  that we can understand the notion of the mass as a result of virtual  excitations created due to the acceleration of the external coupling, which in this case is the position  of the wall(s).

This result can be found using the language of gauge potentials. As we showed earlier for the global translations $X_L=X_R$ it is the momentum operator: $\mathcal A_+=\hat p$. So the renormalization can be also found from \eref{eq:kappa_RR_gs}
\be
\kappa_{++}=2\sum_{n\neq 0} {|\langle 0|\hat p|n\rangle|^2\over E_n-E_0}=m.
\label{eq:kappa_++_momentum}
\ee

\hrulefill

\exercise{Verify that \eref{eq:kappa_++_momentum} gives the correct expression for the mass [\eref{eq:kappa_++_potential}]. }

\exerciseshere
\hrulefill

Notice that the expression for the mass, \eref{eq:kappa_++_potential}, is expected to hold not only for a square well but any translationally invariant non-relativistic system. As we already discussed, it can be viewed as a sum rule or a quantum generalization of the equipartition theorem.

\begin{figure}[h]
\includegraphics[width=0.4\columnwidth]{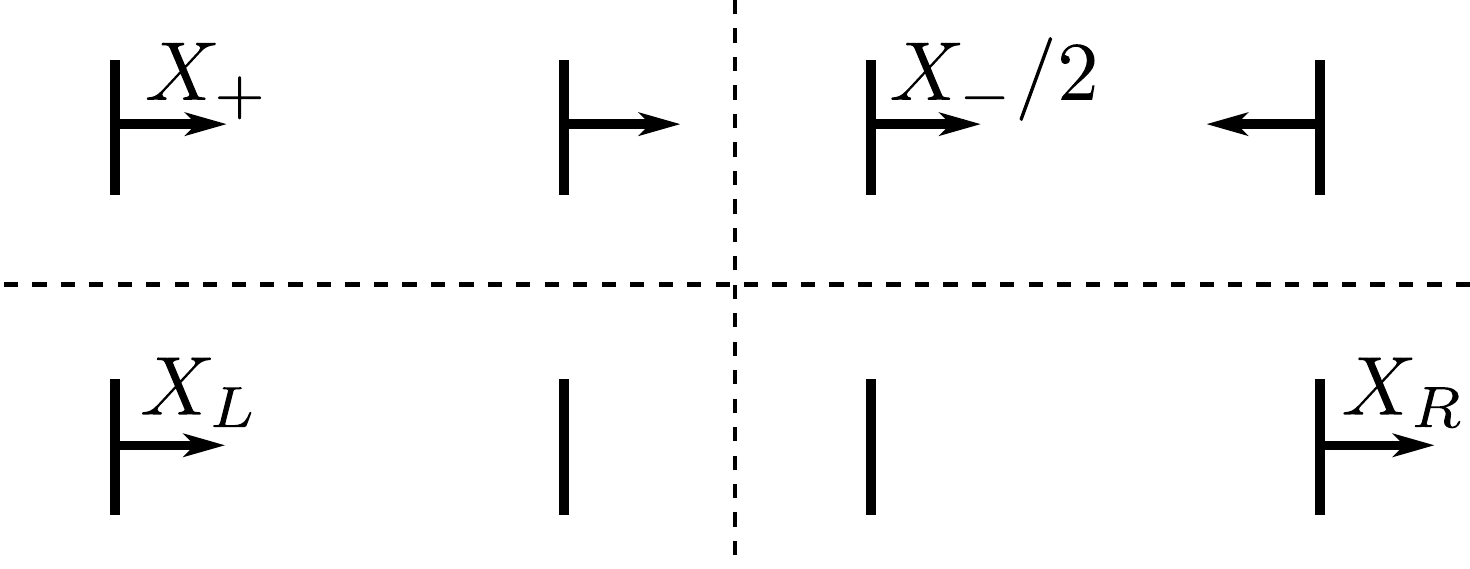}
\caption{Four possible modes of motion for the walls of the cavity.}
\label{fig:asymmetric_mass}
\end{figure}
 
\subsubsection{Dilations}

If instead we analyze the setup where the two walls are connected to a spring 
and move towards each other so that 
$\lambda=X_-$ is the (instantaneous) change in the length of the potential well (see \fref{fig:asymmetric_mass}) we find
\begin{equation}
\begin{split}
\kappa_{--}&=2 \hbar^2 \sum_{n\neq 0} \frac{V^2(\psi_0(X_L)\psi_n(X_L)+\psi_0(X_R) \psi_n(X_R))^2}{4 (E_n-E_0)^3}\\
&=m \frac{16}{\pi^2} \sum_{\substack{n=even \\ n\neq 0}} \frac{(n+1)^2}{[(n+1)^2-1]^3} = m \frac{\pi^2-6}{12\pi^2}\approx 0.033 m.
\label{eq:kappa-}
\end{split}
\end{equation}

Let us note a peculiar property of the dressed mass. Clearly 
$\kappa_{LL}+\kappa_{RR}\approx 0.56 m\neq \kappa_{++},\kappa_{--}$, i.e., the mass 
renormalization of the two walls is not the same as the sum of the mass 
renormalization of each wall measured separately. This is a result of 
interference, which is apparent in Eqs.~(\ref{eq:kappa+}) and (\ref{eq:kappa-}). 
Note that $(\kappa_{++} + 4 \kappa_{--})/2=\kappa_{LL}+\kappa_{RR}$. Thus the mass behaves 
similarly to the intensity in the double pass interferometer, where the sum of 
intensities in the symmetric and antisymmetric channels is conserved. This analogy will become more clear after \exref{ex:mass_tensor}, where you will compute the full mass renormalization tensor, whose normal modes correspond to $\kappa_{++}$ and $\kappa_{--}$, which interfere to give $\kappa_{LL}$ and $\kappa_{RR}$.

It is straightforward to generalize this calculation for a particle in the box prepared in an excited state $|n\rangle$:
\be
\kappa_{++}^n = 2 \sum_{n'\neq n} {|\langle n'| p| n\rangle|^2\over E_{n'}-E_n} =m \frac{64}{\pi^2} \sum_{n'=n+\mathrm{odd}} \frac{(n'+1)^2(n+1)^2}{[(n'+1)^2-(n+1)^2]^3}=m ~.
\label{eq:mass_piston_n}
\ee
Similar expressions hold for $\kappa_{--}^n$ and $\kappa_{RR}^n$, but unlike the Galilean mass $\kappa_{++}$ the latter two depend on $n$. In particular (cf. \exref{ex:dilation_mass_gauge}),
\be
\kappa_{RR}^n=2\hbar^2 \sum_{m\neq n} {|\langle m|\partial_{X_R}| n\rangle|^2\over (E_m-E_n)}=2\sum_{m\neq  n}{|\langle m|\mathcal D|n\rangle|^2\over E_m-E_n}={m\over 3}\left(1-{3\over 2\pi^2 n^2}\right),
\label{eq:mass_piston_dilation_n}
\ee
where
\[
\mathcal D={x \hat p+\hat px\over 2L}
\]
is the dilation operator introduced earlier [cf. \eref{eq:def_dilation_D}]. In the classical limit the renormalized mass approaches $m/3$. This result can be also easily recovered from the equipartition theorem. Indeed according to \eref{kappa_high} the high temperature asymptotic of the metric tensor is given by the variance of the gauge potential, which is the dilation operator in this case:
\be
\kappa_{RR}^n \stackrel{n \gg 1}{\xrightarrow{\hspace*{0.7cm}}}
{1\over k_B T}\langle \mathcal D\rangle^2\approx {1\over k_B T} {\langle x^2\rangle \langle p^2 \rangle\over L^2}={m\over 3},
\label{eq:kappa_RR_cl}
\ee
where we used that in the classical limit, according to the Gibbs statistics, probability distributions for the coordinate and the momentum factorize.

\hrulefill

\exercise{Complete the missing steps in deriving Eqs.~(\ref{eq:mass_piston_n}) and (\ref{eq:mass_piston_dilation_n}). }

\exerciseshere
\hrulefill

The fact that the mass $\kappa_{RR}$ or in short the {\em dilation mass}, since it corresponds to the dilations of the system, is equal to one third of the usual translational mass might look a bit strange. One would naively expect that the effect of interference terms appearing, e.g., in \eref{eq:kappa+} will disappear in the classical limit as usually happens. Indeed it is easy to see that such terms appear with opposite signs depending on whether the parity of the state $n$ is even or odd (for the excited state the equivalent expression will involve double summation over $n$
 and $n'$ and the sign of the interference term will depend on the parity difference between $n$ and $n'$). Because $E_{n'}-E_n$ is a smooth function of $n$ and $n'$ one would expect that these oscillations will cancel each other. However, this is not the case because the mass is always, even in the classical limit $n\gg 1$, dominated by the nearest excitations $n'=n\pm 1, n\pm 2$ so $E_{n'}-E_n$ can not be considered as a smooth continuous function of $n-n'$.

\begin{figure}[ht]
\includegraphics[width=0.8\columnwidth]{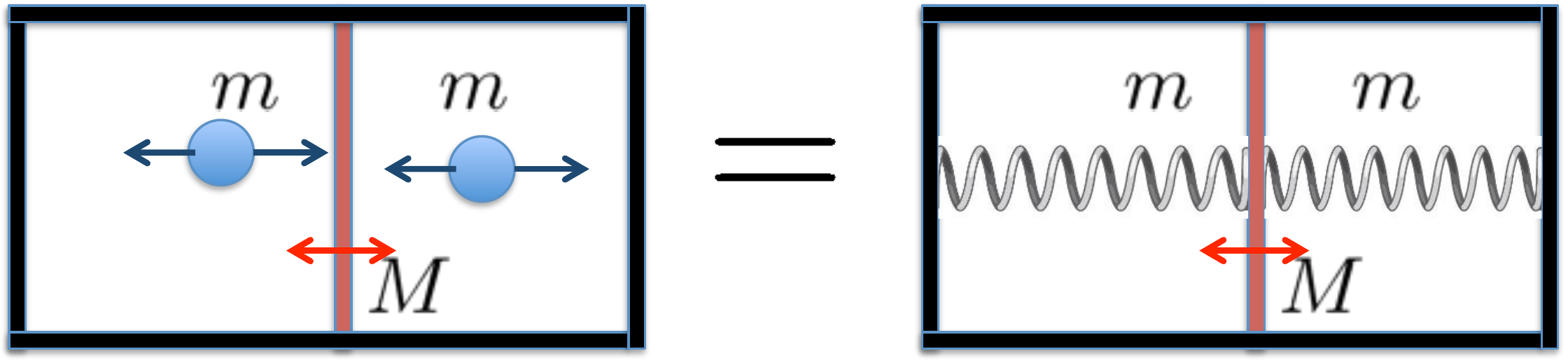}
\caption{A figure illustrating equivalence of a piston confined between two cavities with an ideal gas to the piston connected to two springs of mass $m$, where $m$ is the mass of the gas in each cavity. Note that analogy extends not only to forces (as it is usually discussed) but also to masses.}
\label{fig:two_pistons_springs}
\end{figure}

Instead this mass dressing can be qualitatively understood by noting that upon compression of the box (a.k.a. dilations, the generator of the $\kappa_{--}$ term), the mass $m$ pushes back against the walls much like a massive spring or a rubber band. Then if we push on the right end of the massive spring to give it a velocity $v_R$ with the other end held clamped at $x=0$, the velocity of the spring will be a linear function of the position, $v(x)=v_R x / L$. The kinetic energy of the massive spring in this case is 
\be
T = \int_0^L \frac{1}{2} \mu v^2(x) dx = \frac{\mu}{2} \int_0^L \frac{x^2 v_R^2}{L^2} = \frac{1}{6} m v_R^2 = {\kappa_{RR}\over 2} v_R^2~,
\ee
where $\mu=m / L$ is the mass density of the spring.  The corollary of this interpretation is that one can extend the analogy of the freely moving piston confined between two ideal gases (see \fref{fig:two_pistons_springs}). As is discussed in many textbooks, in this setup near equilibrium the two gases exert effective elastic forces on the piston from effective massless springs and the spring constant is proportional to the pressure. Our result shows that this analogy extends beyond this, at least in the non-interacting limit, giving equivalence of this setup to the piston coupled to two massive springs with the mass of each spring being the same as the mass of the gas on each side of the piston.

\hrulefill

\exercise{\label{ex:mass_tensor} Show that for an arbitrary eigenstate $n$, the mass tensor is diagonal in the $X_+, X_-$ basis, i.e., $\kappa_{+-}=\kappa_{-+}=0$. Then find the full mass tensor in the $X_R, X_L$ basis by using the Jacobian matrix $J=[\partial X_{+,-} / \partial X_{R,L}]$: $\kappa_{\{R,L\}} = J^T \kappa_{\{+,-\}} J$. Confirm that this gives the correct value of $\kappa_{RR}=\kappa_{LL}$ for the ground state [\eref{eq:kappa_RR_gs}].}

\exercise{Derive the gauge potential for the compression with respect to $X_-$. Using the equipartition theorem evaluate the mass $\kappa_{--}$ in the classical limit (corresponding to the highly excited state of the particle) and prove it is equal to $m/12$. Argue that in the high temperature limit the off-diagonal components of the metric tensor $g_{+-}=0$ and hence the mass tensor is also diagonal in the $+-$ space.}

\exercise{For the harmonic oscillator presented in \sref{sec:invitation}, translations and dilations correspond to changing $x_0$ and $k=m \omega^2$ respectively. Find the diagonal components of the mass tensor $\kappa_{x_0 x_0}$ and $\kappa_{kk}$ for an arbitrary eigenstate $|n\rangle$. Show that these connect to the metric tensor, which was derived for $x_0$ and $k$ in \exref{ex:geom_tens_sho} and \eref{eq:g_kk_n_sho}, respectively.}

\exercise{\label{ex:pendulum_earth} Consider the setup illustrated in \fref{fig:mass_pendulum_earth} in which a pendulum with mass $m$ attached to a box of mass $M$ is pulled away from the surface of the earth. Using the results of the previous problem, find the effective mass $M_\mathrm{eff}$ that setup will appear to have when lifted away from the earth as a function of its temperature $T$. For simplicity, you may assume that $T$ is large enough that the problem may be treated classically.}

\exerciseshere

\begin{figure}
\includegraphics[width=0.3\columnwidth]{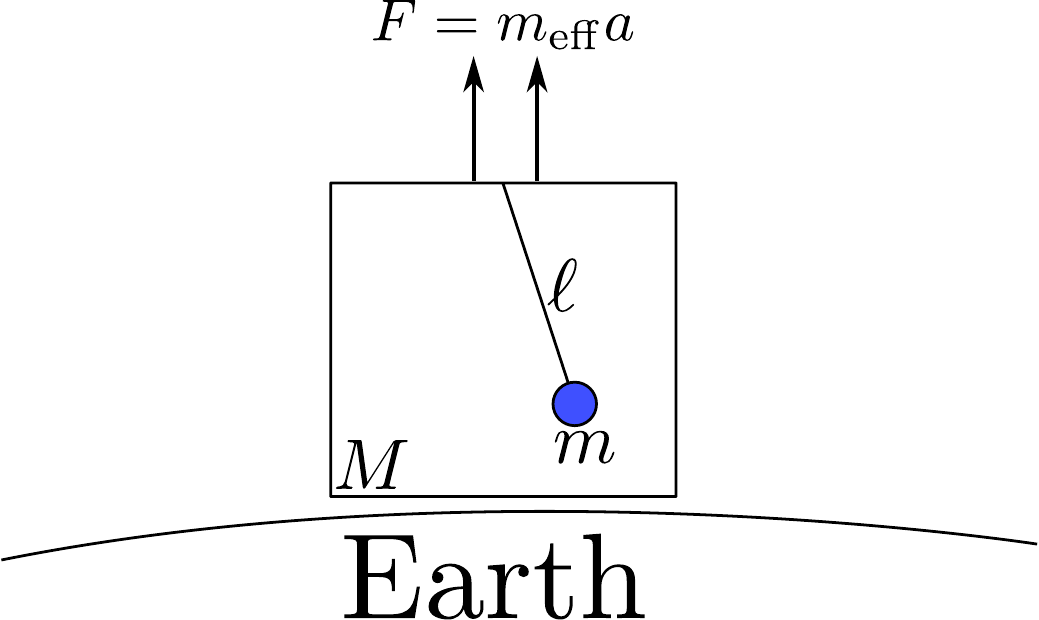}
\caption{Illustration of pendulum in a box being pulled away from the earth [\exref{ex:pendulum_earth}].}
\label{fig:mass_pendulum_earth}
\end{figure}

\hrulefill

\subsubsection{Classical derivation of the mass}\label{sec:classical_mass}

The example of the piston shows how the formalism of adiabatic perturbation theory can be used to find both the anticipated translational mass of the box with a particle inside and the less obvious dilation mass. These examples are sufficiently simple that they can be recovered from more elementary methods, although as we will see the actual derivations are more complicated and harder to extend to more complex setups. It is nevertheless instructive to see how the mass renormalization can be found from simple kinematics.

Let us start by computing the translational mass. Namely let us imagine a classical slow box of mass $M$, initially at rest, with a fast particle of mass $m$ inside it. At time $t=0$ we start accelerating the box with, for simplicity, constant acceleration $a$. Let us compute the force exerted on the box by the particle. We will find the force by computing the average momentum transferred to the particle during one cycle and divide by the period. Note that we are interested in the force averaged over the period. One can do the averaging in two equivalent ways: time averaging and space averaging. The second way, i.e., space averaging, is actually somewhat simpler because the wall is accelerating and time averaging should be done with some care. In quantum language this space averaging of the force is equivalent to the averaging of $\mathcal M$ over the stationary probability distribution.

\begin{figure}[ht]
\includegraphics[width=0.8\columnwidth]{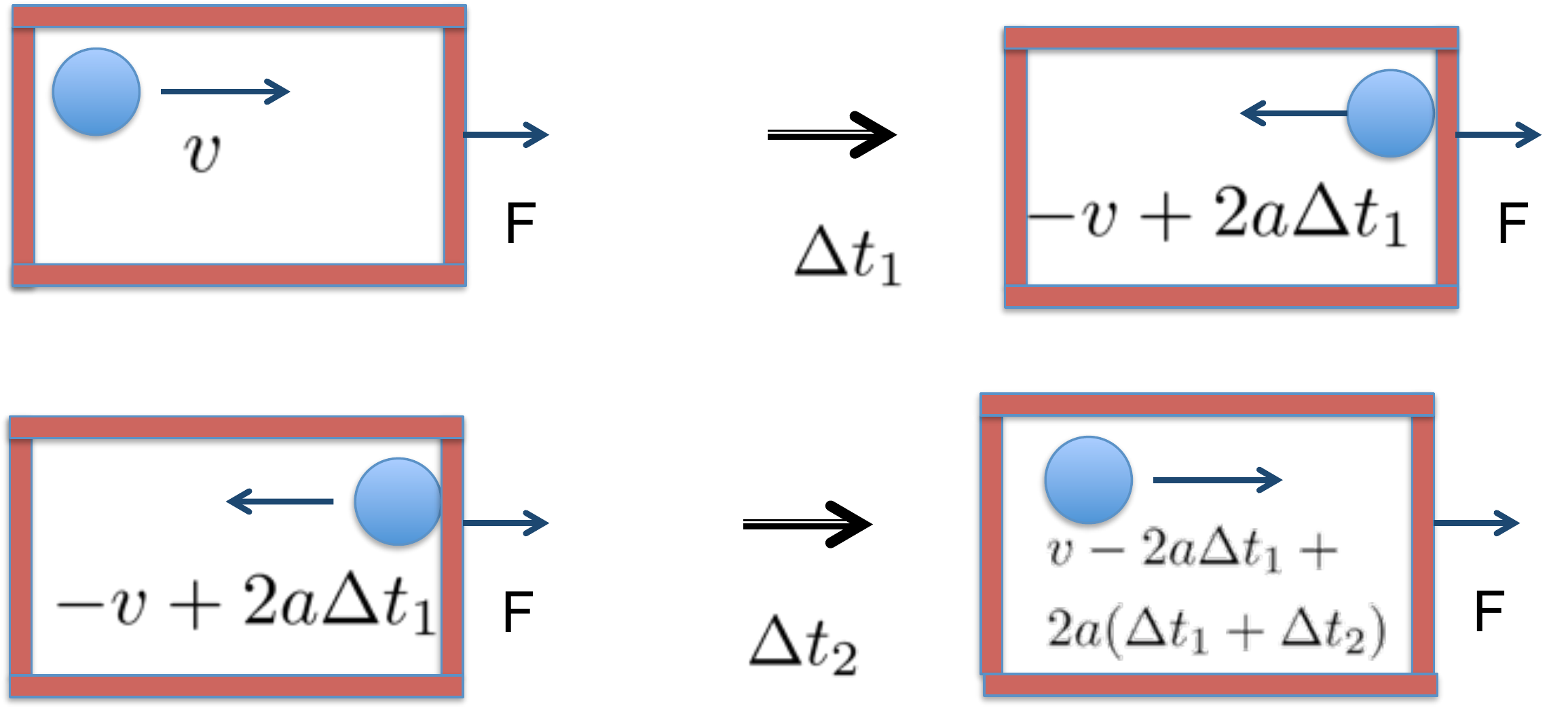}
\caption{Figure illustrating an elementary classical calculation of the translational mass. See text for details.}
\label{fig:piston_mass}
\end{figure}

Let us imagine that starting at time $t=0$ the box of length $L$ is pulled with a constant acceleration $a$ by some external force $F$ (see \fref{fig:piston_mass}). Let us also assume that the particle starts near the left wall (the position of the particle can be chosen arbitrarily) and moves in positive direction with initial velocity $v$. After time $\Delta t_1$ this particle collides with the right wall. By that time the wall moves distance $\Delta x_{1}=L+a\Delta t_1^2/2$ and acquires the velocity $V_1=a \Delta t_1$. Then it reflects back with velocity
\[
v_{1}=-v+2 V_1=-v+2 a\Delta t_1.
\] 
The total transferred momentum to the particle during this collision is 
\[
\Delta p_1=m (v_1-v)=2mv +2m a\Delta t_1.
\]
Then the particle moves backwards and collides with the left wall after time $\Delta t_2$. By that time the left wall moved by the distance $\Delta x_2=a (\Delta t_1+\Delta t_2)^2/2$ and acquired the velocity $v_2=a (\Delta t_1+\Delta t_2)$. The particle now reflects with the velocity
\[
v_2=-v_1+2a (\Delta t_1+\Delta t_2)=v+2a\Delta t_2.
\]
Hence the total transferred momentum to the particle is 
\[
\Delta p_2=m(v_2-v_1)=2m v +2ma(\Delta t_2-\Delta t_1).
\]
Now we can compute the force exerted by the wall on this particle as
\be
f={\Delta p_1+\Delta p_2\over \Delta t_1+\Delta t_2}={2m a\Delta t_2\over \Delta t_1+\Delta t_2}.
\label{eq:force_x}
\ee
This expression is rather complicated as we yet have to compute $\Delta t_1$ and $\Delta t_2,$ as functions of $v, a, L$. However, to find the mass we are interested only in the leading non-adiabatic response linear in acceleration. The numerator of \eref{eq:force_x} is already linear in $a$, which means that we can safely compute all time intervals only to zeroth order in $a$, which is trivial:
\[
\Delta t_1\approx \Delta t_2 \approx {L/v}.
\]
Combining all this together we find
\be
f\approx {2m a\over 2}=m a
\ee
as expected. So the total force required to accelerate the box and the particle is thus 
\be
f_{\rm tot}=(m+M) a,
\ee
which is precisely Newton's second law with the mass equal to the sum of the two masses. It is of course not surprising that we were able to reproduce this simple and expected result from more elementary methods. However, it is very instructive to see that we again relied in time scale separation and found this result only in the leading order adiabatic expansion with the small parameter $a\Delta t_1/v=a L/v^2$.

\begin{figure}[ht]
\includegraphics[width=0.8\columnwidth]{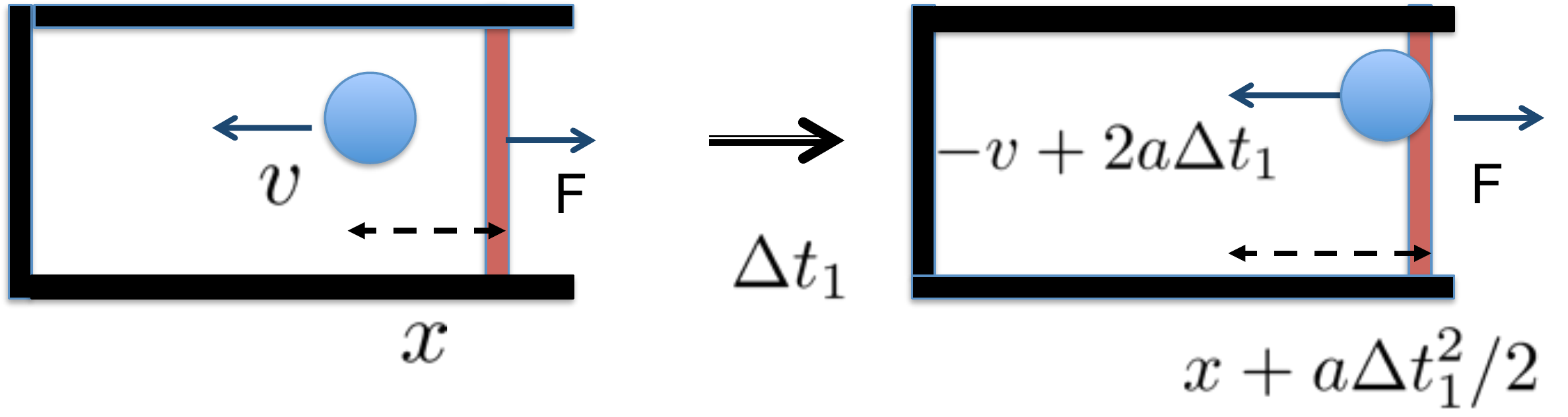}
\caption{Figure illustrating an elementary classical calculation of the dilation mass of the piston. See text for details.}
\label{fig:piston_mass_dilation}
\end{figure}

Now let us analyze another setup where the force $F$ is only applied to the right wall such that it moves with the acceleration $a$ while the left wall remains static (see \fref{fig:piston_mass_dilation}) . As we will see, the classical elementary derivation not involving gauge potentials becomes much more delicate as the force now depends on the initial position of the particle $x$. It is convenient to define $x$ measured from the left static wall in the interval $[-L,L]$ such that the subinterval $[-L,0]$ corresponds to the particle moving to the left (as shown in the figure) and the subinterval $[0,L]$ corresponding to the particle moving to the right, i.e., towards the moving wall. We assume that we start from a stationary probability distribution described by a uniform distribution of $x$. As in the previous example the particle hits the right wall after the time $\Delta t_1$, which can be found from 
\[
-L+x+v\Delta t_1= {a\Delta t_1^2\over 2}
\]
Instead of solving this quadratic equation in general we will only find $\Delta t_1$ to the order in acceleration:
\be
\Delta t_1\approx {L-x\over v}+{a\over 2} {(L-x)^2\over v^3}
\ee
The transferred momentum to the particle is thus
\[
\Delta p_1=-2mv+2m a \Delta t_1
\]
The particle will return to the original position (and thus will complete the cycle) after time $\Delta t_2$ which can be found from\footnote{More accurately the cycle is complete when the particle returns to a slightly shifted dilated position $x'$. However, it is easy to see that this effect is canceled in the linear order in $a$ as this shift has opposite effects on particles with opposite initial values of $x$.}
\[
\Delta t_2={L+x+ a\Delta t_1^2/2\over v-2a\Delta t_1},
\]
where we took into account that (i) the particle has to travel a longer distance because of the displacement of the wall and (ii) that it moves back with a reduced velocity. The force can be found as before by computing the ratio of the total momentum transfer over the period. To the leading linear order in acceleration it is 
\be
f(x)={\Delta p_1\over \Delta t_1+\Delta t_2}\approx -{m v^2\over L}+m a\left[ {5\over 2}-{2x\over L}-{x^2\over 2L^2}\right].
\ee
The first term is nothing but the usual generalized force proportional to pressure (which resists compression of the piston). The second term is proportional to the acceleration and thus should define the mass. Unlike the previous case of the translationally invariant motion, this term explicitly depends on the initial coordinate of the particle. Taking the average over these coordinates, which is equivalent to the average over the density matrix in the quantum case, we find the average force
\be
\overline f={1\over 2L}\int_{-L}^L f(x) dx=-{mv^2\over L}+2ma+{ma\over 3}\approx -{m (v-a L/v)^2\over L}+{ma\over 3}
\ee
The first term here is now the standard force due to the pressure averaged over the cycle, with $\overline v =v - aL/v$ being the average velocity of the particles. The second term is the non-adiabatic correction due to the acceleration, which gives the correct result from \eref{eq:kappa_RR_cl}. As we see, even in this simple example the ``elementary'' classical derivation of the dilation mass is very delicate. It requires careful analysis of several contributions to the force of the same order and the correct identification of different terms.

\subsection{Mass of a massless relativistic scalar field in a cavity}
\label{sec:photon_mass}

For a massive particle in a box, we have seen that the box acquires an extra mass due to translations or dilations that derives from the bare mass of the particle. We now ask what happens for massless particles in a box, such as a phonon, photon or some other excitation with a linear dispersion. For example one can imagine a vibrating string confined between two clamps (see \fref{fig:guitar_string}). The effective mass of photons in a cavity has been investigated since the early days of relativity, and the current theoretical understanding is that they appear to have a mass $E/c^2$ proportional to their energy (cf. Refs. ~\cite{Kolbenstvedt1995_1} and ~\cite{Wilhelm2015_1} for a recent discussion). Here we will compute the renormalization of the mass of the cavity containing particles with relativistic dispersion inside as before through the non-adiabatic correction to the generalized force. This will allow us to identify both the classical (thermal) and quantum (zero point) contributions to the photon mass.

\begin{figure}[ht]
\includegraphics[width=0.7\columnwidth]{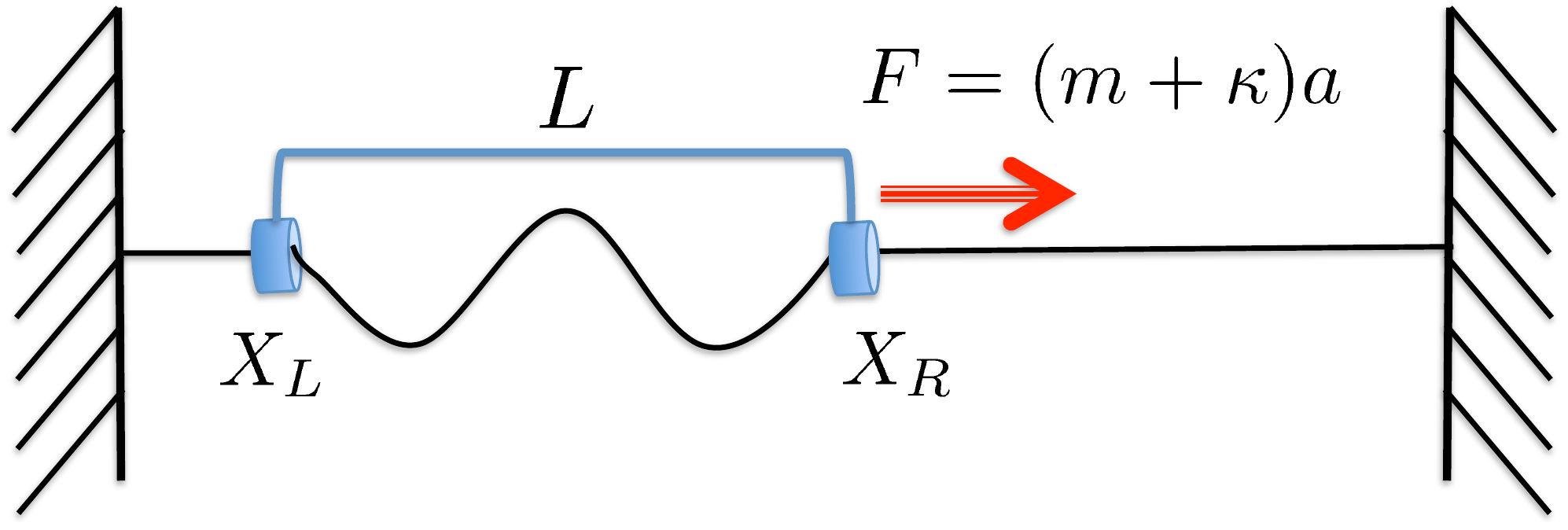}
\caption{Example of a system with a linear dispersion (guitar string) confined between two clamps. It is intuitively clear that moving the clamps is harder in the presence of vibrations as they should be dragged together with the clamps. This amounts to dressing the mass of the clamps analyzed here.}
\label{fig:guitar_string}
\end{figure}

A simple example realizing such a setup would be a vibrating string confined between two clamps (see \fref{fig:guitar_string}). Instead of the string one can imagine a Luttinger liquid confined between two impurities or a phonon (photon) gas confined between two reflecting mirrors. In our analysis we will ignore potential retardation effects on the confining potential. Specifically as before we will assume that $X_L(t)$ and $X_R(t)$ are given functions of time and, e.g., the symmetric mode $X_R(t)-X_L(t)={\rm const}(t)$ in the lab reference frame. This is perfectly justified in the case of a non-relativistic guitar string but might play an effect in the case of photons. For example, if we pull the right clamp with some force, there will be some delay before the left clamp starts moving. This implies that the bouncing photon will feel slightly different accelerations from the two walls and this may have some effect on the mass renormalization.  

We will consider the Klein-Gordon Hamiltonian describing a massless scalar harmonic field confined to the cavity:
\be
\mathcal H = \frac{1}{2} \int_{X_L}^{X_R} dx \left[ \Pi(x)^2 + c^2 \left( \frac{\partial \Phi}{\partial x} \right)^2 \right]~,
\label{eq:H_photon_1d_walls1}
\ee
where $\Phi$ is the field (describing displacement of atoms from equilibrium positions in the case of the string)
and $\Pi$ is the momentum canonically conjugate to $\Phi$. In the case of photons, $\Phi$ represents the vector potential $\Lambda$. Then the momentum $\Pi(x)$ and the gradient of $\Phi(x)$ appearing in the Hamiltonian represent the electric field $E = \partial_t \Lambda \propto \Pi$ and magnetic field $B \propto \partial_x \Lambda$ respectively. For computing the translational mass we will assume that the cavity of length $L$ extends from $X_L = -L/2 + X_+ $ to $X_R = L/2 + X_+$. We consider a simple choice of vanishing Dirichlet boundary conditions: $\Phi(X_L)=\Phi(X_R)=0$. For the string this implies that vibrations vanish at the boundary. For electromagnetic waves such boundary conditions can be realized by using a superconducting cavity such that the photons acquire a mass $\mu$ outside the cavity due to the Anderson-Higgs mechanism \cite{Anderson1963_1}. In the context of Klein-Gordon theory, this is represented by the Hamiltonian.
\be
\mathcal H = \frac{1}{2} \int_{-\infty}^\infty dx \left[ \Pi(x)^2 + c^2 \left( \frac{\partial \Phi}{\partial x} \right)^2 + \mu^2 \left( \Theta(X_R - x) + \Theta(x-X_L) \right) \Phi^2(x) \right]~.
\label{eq:H_photon_1d_with_walls}
\ee
The two Hamiltonians in \eref{eq:H_photon_1d_walls1} and \eref{eq:H_photon_1d_with_walls} are equivalent to each other in the limit $\mu\to\infty$. They can be used for two equivalent derivations of the mass renormalization as we show below: the first one is based on gauge potentials and the second one is based on generalized forces.

The Hamiltonian in \eref{eq:H_photon_1d_walls1} [similarly \eref{eq:H_photon_1d_with_walls}] is harmonic and thus can be diagonalized expanding the fields $\Phi(x)$ and $\Pi(x)$ in normal modes
\be
\Phi(x) = \sum_j f_j(x) Q_j ~,~ \Pi(x) = \sum_j f_j(x) P_j~,
\label{eq:normal_modes_cavity}
\ee
where the (real-valued) mode functions $f_j$ satisfy the usual orthonormality and completeness relations
\be
\int_{-\infty}^\infty f_j(x) f_k(x) dx = \delta_{jk},\quad \sum_j f_j(x)f_j(x')=\delta(x-x').
\label{eq:orthonormality_completeness}
\ee
The completeness relation ensures that the mode operators $Q_j$ and $P_j$ are canonically conjugate:
\be
[Q_j, P_k] = i \hbar \delta_{jk} ~.
\ee
The mode functions, diagonalizing the Hamiltonian in \eref{eq:H_photon_1d_with_walls} must satisfy the wave equation:
\be
-c^2 \partial_x^2 f_j = \omega_j^2 f_j
\ee
with vanishing boundary conditions for the Hamiltonian in \eref{eq:H_photon_1d_walls1}, and the Klein-Gordon equation with spatially dependent mass for the Hamiltonian in \eref{eq:H_photon_1d_with_walls}:
\be
-c^2 \partial_x^2 f_j + \mu^2\left( \Theta(X_R - x) + \Theta(x-X_L) \right)  f_j = \omega_j^2 f_j ~.
\ee 
It is straightforward to verify that in the limit $\mu\to\infty$ the mode functions are identical:
\be
f_j(x)=\sqrt{2\over L}\sin \left(k_j (x-X_L)\right),
\label{eq:mode_functions}
\ee
where $k_j=\pi j/L$, $j=1,2,\dots$ and the mode frequencies are $\omega_j=k_j c$. Then the Hamiltonian can be diagonalized in terms of usual ladder operators 
\[
a_j \equiv  \sqrt{\omega_j \over 2\hbar} Q_j + i \sqrt{1\over 2 \hbar\omega_j} P_j,\quad a_j^\dagger \equiv  \sqrt{\omega_j \over 2\hbar} Q_j - i \sqrt{1\over 2 \hbar\omega_j} P_j
\]
to give
\be
\mathcal H=\hbar \sum_j \omega_j (a_j^\dagger a_j + 1/2)~.
\label{eq:Hamiltonian_photons_normal_modes}
\ee 
The eigenstates of this Hamiltonian are clearly harmonic oscillator eigenstates $|n\rangle = |n_1, n_2, \ldots \rangle$, where $n_j = 0, 1, 2, \ldots$ denotes the number of photons in the mode $j$.

For this system it is possible to explicitly find the gauge potentials by writing the eigenstates of the Hamiltonian in the first quantized notation extending the derivation of \eref{eq:derivative_k_harmonic_state} to multiple modes. Each normal mode the wave function is given by the single-particle eigenstates of the harmonic oscillator:
\be
\phi_{n_j}(Q_j)=\sqrt{1\over \ell_j}\, \psi_{n_j}(Q_j/\ell_j),
\ee
where $\ell_j=\sqrt{\hbar/2\omega_j}$ and $\psi_{n_j}$ is expressed through the Hermite polynomials~\cite{Landau_Lifshitz_QM}. As will become clear shortly we will not need to explicitly know this function. The full photon many-body wave-function is just the properly normalized symmetrized product of the single-mode wave function:
\be
\Psi_{n_1,n_2\dots}(Q_1,Q_2,\dots)=C\sum_{\{\sigma\}} \prod_j \phi_{n_j} (Q_{\sigma_j}),
\ee
where $\{\sigma\}$ denotes all possible permutations of the mode indexes and $C$ is the normalization constant. Using that the normal coordinates $Q_j$ and the oscillator lengths $\ell_j$ can depend on $\lambda$ through both the mode functions and the mode frequencies, we can write
\be
\partial_\lambda \Psi=\sum_j  \left[{\partial Q_j\over \partial \lambda}{ \partial \Psi\over \partial Q_j}+{\partial \ell_j\over \partial \lambda}{\partial \Psi\over \partial \ell_j}\right].
\ee
Let us observe that
\[
{\partial Q_j\over \partial\lambda}=\int_{X_L}^{X_R} dx\, \partial_\lambda f_j(x)\, \Phi(x)+{\partial X_R\over \partial \lambda} f_j (X_R) \Phi(X_R)-{\partial X_L\over \partial \lambda} f_j (X_L) \Phi(X_L).
\]
For the vanishing boundary conditions that we are considering, the last two terms are equal to zero. In the first term we can re-express $\Phi(x)$ back through the mode functions (cf. \eref{eq:normal_modes_cavity}). Then we find 
\[
{\partial Q_j\over \partial\lambda}=\int_{X_L}^{X_R} dx\, \partial_\lambda f_j(x)\sum_i f_i(x') Q_i=\sum_i \zeta_{ji}^\lambda Q_i,
\]
where 
\[
\zeta_{ji}^\lambda=\int_{X_L}^{X_R} dx\, f_i(x) \partial_{\lambda} f_j(x).
\]
From differentiating the orthonormality relation of the mode functions, \eref{eq:orthonormality_completeness}, with respect to $\lambda$, we see that $\zeta_{ij}^\lambda=-\zeta_{ji}^\lambda$ and thus $\zeta^\lambda_{jj}=0$. As in \eref{eq:derivative_k_harmonic_state} for the harmonic oscillator we find 
\[
{\partial \phi_{n}(Q)\over \partial \ell}=-i {P Q+Q P\over 2\hbar \ell} \phi_{n}
\]
The last identity we need is
\[
{\partial \ell_j\over \partial \lambda}={\partial \ell_j\over \partial \omega_j}{\partial \omega_j\over \partial \lambda}=-{1\over 2}{\ell_j\over \omega_j} {\partial \omega_j\over \partial \lambda}.
\]
Combining all these results together we find
\be
i\hbar \partial_\lambda \Psi\equiv \mathcal A_{\lambda} \Psi=\left[\sum_{i\neq j} Q_i \zeta_{ij}^\lambda P_j-{1\over 2}\sum_j {1\over \omega_j}{\partial \omega_j\over \partial\lambda}{P_j Q_j+Q_j P_j\over 2}\right] \Psi.
\ee
Therefore the gauge potential is
\be
\mathcal A_\lambda=\sum_{i\neq j} Q_i \zeta_{ij}^\lambda P_j-{1\over 2}\sum_j {\partial \log\omega_j\over \partial \lambda}{P_j Q_j+Q_j P_j\over 2}.
\ee

It is convenient to rewrite this gauge potential in terms of the ladder operators:
\begin{multline}
\mathcal A_\lambda={i\hbar \over 4}\sum_{i\neq j} \zeta_{ij}^\lambda \left(\left[\sqrt{\omega_j\over \omega_i}-\sqrt{\omega_i\over \omega_j}\right](a_i^\dagger a_j^\dagger-a_i a_j)+ \left[\sqrt{\omega_j\over \omega_i}+\sqrt{\omega_i\over \omega_j}\right](a_j^\dagger a_i-a_i^\dagger a_j)\right)\\
-{i\hbar \over 4}\sum_j {\partial \log\omega_j\over \partial \lambda}(a_j^\dagger a_j^\dagger- a_j a_j)
\end{multline}
Clearly the only non-zero matrix elements of the gauge potential correspond either to scattering one particle or simultaneous creation or annihilation of two particles. In particular,
\beq
\langle \dots n_i-1\dots n_j+1\dots|\mathcal A_\lambda|\dots n_i\dots n_j\dots\rangle&=& i\hbar \sqrt{n_i (n_j+1)} {\omega_i+\omega_j\over 2\sqrt{\omega_i\omega_j}}\zeta^\lambda_{ij};\nonumber\\
\langle \dots n_i+1\dots n_j+1\dots|\mathcal A_\lambda|\dots n_i\dots n_j\dots\rangle
&=&-i\hbar \sqrt{(n_i+1) (n_j+1)} {\omega_i-\omega_j\over 2\sqrt{\omega_i\omega_j}}\zeta^\lambda_{ij};\nonumber\\
\langle \dots n_j+2\dots|\mathcal A_\lambda|\dots n_j\dots\rangle&=&-{i\hbar \over 4}{\partial {\log \omega_j}\over \partial\lambda}\sqrt{(n_j+1)(n_j+2)};\\
\langle \dots n_i-1\dots n_j-1,\dots|\mathcal A_\lambda|\dots n_i\dots n_j\dots\rangle
&=&i\hbar  \sqrt{n_i n_j} {\omega_i-\omega_j\over 2\sqrt{\omega_i\omega_j}}\zeta^\lambda_{ij};\nonumber\\
\langle \dots n_j-2\dots|\mathcal A_\lambda|\dots n_j\dots\rangle
&=&{i\hbar \over 4}{\partial {\log \omega_j}\over \partial\lambda}\sqrt{n_j(n_j-1)}.\nonumber
\eeq
Substituting these gauge potentials into the general expression for the mass, \eref{main_kappa}, and noting that the energy differences between the connected states are $\pm (\omega_i-\omega_j)$ for the scattering terms conserving the number of particles or $\pm (\omega_i+\omega_j)$ for non-conserving terms we find
\be
\kappa_\lambda={\hbar \over 4}\sum_{i\neq j} \bigg[{n_i-n_j\over \omega_j-\omega_i} {(\omega_i+\omega_j)^2\over \omega_i\omega_j} +{n_i+n_j+1\over \omega_i+\omega_j}{(\omega_i-\omega_j)^2\over \omega_i\omega_j} \bigg](\zeta^\lambda_{ij})^2\\
+{\hbar \over 8}\sum_j {2n_j+1\over \omega_j}\left({\partial \log \omega_j\over \partial\lambda}\right)^2
\label{eq:kappa_photons}
\ee
This expression splits into the two parts, namely $\kappa_\lambda^{\rm ph}$ which is linear in the occupation numbers, and 
$\kappa_\lambda^{\rm vac}$, which is the vacuum contribution:
\beq
\kappa_\lambda^{\rm ph}&=&\hbar\sum_{i\neq j} n_j  {\omega_i^2+3\omega_j^2\over \omega_j (\omega_i^2-\omega_j^2)}(\zeta_{ij}^\lambda)^2+{\hbar \over 4}\sum_j n_j {(\partial_\lambda \omega_j)^2\over \omega_j^3}\nonumber\\
\kappa_\lambda^{\rm vac}&=&{\hbar\over 4}\sum_{i\neq j} {(\omega_i-\omega_j)^2\over (\omega_i+\omega_j)\omega_i\omega_j}(\zeta_{ij}^\lambda)^2+{\hbar \over 8}\sum_j {(\partial_\lambda \omega_j)^2\over \omega_j^3}.
\label{eq:kappa_photons1}
\eeq

The expression above applies to any choice of the parameter $\lambda$. Moreover in this derivation we never used any specific dispersion relation so it applies both to massive and massless harmonic systems. In particular, it will apply to the massive Klein-Gordon theory confined to a cavity. And finally we never explicitly used the fact that the cavity is one dimensional. So if we extend the integrals defining the mode function overlaps $\zeta_{ij}^\lambda$ to $d$-dimensions, \eref{eq:kappa_photons1} will describe the mass renormalization of an arbitrary $d$-dimensional cavity with vanishing boundary conditions.

As with the single-particle case we focus on two possible motions: translations and dilations. For the translational motion $\lambda=X_+$ such that $\partial_+ X_R=\partial_+ X_L=1$, obviously $\partial_+\omega_j=0$. Using the explicit expressions for the mode functions, \eref{eq:mode_functions}, we find 
\be
\zeta_{ij}^+={1\over L} {2ij\over i^2-j^2} (1-(-1)^{i-j}).
\label{eq:overlap_+}
\ee
For the dilations $\lambda=X_R$, we find $\partial_\lambda X_R=1,\;\partial_\lambda X_L=0$, and
\be
\zeta_{ij}^R={1\over L} {2ij\over i^2-j^2} (1-\delta_{ij}).
\label{eq:overlap_R}
\ee
In addition 
\[
{d\log\omega_j\over d X_R}=-{1\over L}.
\]

Before proceeding with further analysis of the mass for translations and dilations of the cavity let us briefly show an alternative derivation of the same result based on generalized forces and \eref{eq:H_photon_1d_with_walls}. While the result will be equivalent, this derivation has its own advantages as it allows one to overcome the additional step of finding gauge potentials, which might prove difficult in more complicated setups, and therefore can be more amenable to numerical methods. Differentiating the Hamiltonian in \eref{eq:H_photon_1d_with_walls} with respect to $\lambda$ we find the generalized force operator:
\be
\mathcal M_\lambda\equiv -\partial_\lambda \mathcal H=-\mu^2   \left( \frac{\partial X_R}{\partial \lambda} \Phi^2(X_R)- \frac{\partial X_L}{\partial \lambda} \Phi^2(X_L)\right).
\label{eq:gen_force_photon_cavity}
\ee
Substituting the mode expansion of the fields and taking the large $\mu$ limit one finds
\be
\Phi^2(X_L) = \sum_{ij} f_i(X_L) f_j(X_L) Q_i Q_j = \frac{2 c^2}{\mu^2 L} \sum_{ij} k_i k_j Q_i Q_j,\quad \Phi^2(X_R) = \frac{2c^2}{\mu^2 L} \sum_{ij} (-1)^{i+j} k_i k_j Q_i Q_j.
\ee

\hrulefill

\exercise{Prove that using the generalized forces [\eref{eq:gen_force_photon_cavity}] and the general expression for the mass [\eref{main_kappa}] you can reproduce the identical expression for the mass as using the language of gauge potentials in \eref{eq:kappa_photons}.}
\exerciseshere

\hrulefill

\subsubsection{Translations}

Let us now analyze in detail the translational motion of the cavity. Substituting the expression for the overlap, \eref{eq:overlap_+}, into \eref{eq:kappa_photons1} we find
\be
\kappa_{+}^{\rm ph}={16\over L^2}\sum_{i-j~\mathrm{odd} } \hbar \omega_j n_j {\omega_i^2 (\omega_i^2+3\omega_j^2)\over (\omega_i^2-\omega_j^2)^3}={2\over c^2}\sum_j \hbar n_j \omega_j={2E\over c^2},
\ee
where $E=\sum_j \hbar n_j \omega_j$ is the total thermal energy of the photon gas inside the cavity. It is interesting that the result is completely universal, i.e., it does not depend on the energy distribution among the modes. Except for the prefactor of $2$, this result is fully consistent with expectations from special relativity. One possible origin for the discrepancy is that we assumed that the walls move with identical velocities in the lab frame, i.e., that we ignored any potential effects of retardation of the interaction keeping the walls of the cavity together. While this assumption might not be justified for real photons or other particles propagating with the speed of light, it is perfectly justified for slower excitations like phonons as in the setup shown in \fref{fig:guitar_string}. 

Next let us evaluate the vacuum contribution to the mass:
\be
\kappa_{+}^{\rm vac}={4\hbar\over L^2}\sum_{i+j~{\rm odd}} {\omega_i\omega_j\over (\omega_i+\omega_j)^3}={4\hbar \over \pi L c}\sum_{i+j~{\rm odd}} {ij\over (i+j)^3}.
\ee
This sum is formally divergent. The reason for this divergence comes from the assumption that the cavity is perfectly reflecting at all wavelengths. In reality this is never the case. For instance, if we are considering photons reflected from a metal, then the cutoff will be given by the plasma frequency, beyond which the metal becomes transparent. For the situation of the string shown in \fref{fig:guitar_string} the short distance cutoff would be given by the clamp radius: waves with very short wavelength would freely pass through the clamps, while longer wavelengths will be stopped. In the Klein-Gordon theory with a variable mass [\eref{eq:H_photon_1d_with_walls}] the cutoff is formally given by $\mu$. The easiest way to introduce cutoff to the problem is to add smooth cutoff function (e.g., a Gaussian) to the sum:
\be
\kappa_{+}^{\rm vac}\to {4\hbar \over \pi L c}\sum_{i+j~{\rm odd}} {ij\over (i+j)^3} \mathrm{e}^{-(\omega_i+\omega_j)^2/\omega_\Lambda^2}={4\hbar \over \pi L c}\sum_{i+j~{\rm odd}} {ij\over (i+j)^3} \mathrm{e}^{-\epsilon^2(i+j)^2},
\ee
where $\epsilon=\pi c/(L\omega_\Lambda)$ and $\omega_\Lambda$ is the cutoff frequency. We can evaluate this sum in two steps. First let us make the substitution $i+j=m$ and $i-j=n$ where $m$ and $n$ are integers: $m=3,5,\dots$ and $n=-m+2, -m+4,\dots m-2$. Then it is straightforward to evaluate the sum over $n$:
\be
\kappa_{+}^{\rm vac}={2\hbar \over 3\pi Lc}\sum_{m=3,5,\dots} {m^2-1\over m^2} \exp[-\epsilon^2 m^2]=C {\hbar \omega_\Lambda\over c^2}-{\pi \hbar \over 12 L c}=C{\hbar \omega_\Lambda\over c^2}+{2 E_c\over c^2},
\ee
where $C$ is a non-universal constant depending on the cutoff details and
\be
E_c=-{\pi \hbar c\over 24 L }
\ee
is the Casimir energy of the one-dimensional cavity~\cite{Bordag2001_1}, i.e., the universal (cutoff independent) contribution of the zero point fluctuations to the ground state energy of the cavity. It is interesting that as with the thermal energy there is an additional factor of two in the Casimir energy contribution to the cavity mass. The first, cutoff-dependent, correction to the mass does not depend on $L$ and thus can be interpreted as the renormalization of the mass of the cavity walls and absorbed into the definition of $M$. Apart from this correction we see that
\be
\kappa_{+}=2{E+E_c(L)\over c^2}
\ee

In a similar manner we can compute the snap modulus $\zeta$ representing the leading correction to Newtonian equations of motion (see \sref{sec:snap}). Using \eref{eq:snap_modulus} and repeating the same steps as deriving the mass we find
\be
\zeta_+=\zeta_+^{\rm ph}+\zeta_+^{\rm vac},
\label{eq:snap_photon}
\ee
where 
\be
\zeta_+^{\rm ph}={16 L^2\over \pi^4 c^4}\sum_{i+j~{\rm odd}} \hbar \omega_j n_j {i^2(i^4+10 i^2 j^2+5 j^4)\over  (i^2-j^2)^5}={L^2\over \pi^4 c^4}{\pi^4\over 6}\sum_i \hbar \omega_j n_j={E L^2\over 6 c^4},
\label{eq:zeta_photon_ph}
\ee
\be
\zeta_+^{\rm vac}={4\hbar L\over \pi^3 c^3}\sum_{i+j~{\rm odd}}  {i j\over (i+j)^5}={12-\pi^2\over 144 \pi}{\hbar L\over c^3}.
\label{eq:zeta_photon_vac}
\ee
Interestingly the first ``thermal'' contribution to the snap modulus also depends only on the total energy of the system. It gives a small correction to the Newtonian dynamics of the cavity as long as the round trip time of the photon $L/c$ is short compared to the characteristic time scales characterizing the motion of the cavity, e.g., the period of its oscillation. The second vacuum term has a very interesting interpretation related to the Unruh effect~\cite{Crispino2008_1}. Indeed this term is responsible for an energy correction proportional to the acceleration squared. On the other hand according to the Unruh effect an accelerated cavity acquires temperature proportional to the acceleration: $k_B T\sim \hbar \ddot \lambda /c$ and as a result the thermal energy $E_U\sim (k_B T)^2 L/(\hbar c)\sim \hbar L/c^2 \ddot\lambda^2$. So we see that
\[
{\zeta \ddot \lambda ^2\over 2} \sim {E_U\over c^2}.
\]
This contribution to the energy of the cavity, which goes beyond the standard paradigm of the Hamiltonian dynamics, can be interpreted as the result of vacuum heating by acceleration. We note that this interpretation is not precise as in order for the adiabatic perturbation theory to be valid the acceleration should be small such that the Unruh temperature should be smaller than the photon mode splitting. The Unruh effect is usually discussed in the continuum limit, when the cavity modes are not quantized. Nevertheless such an interpretation is very appealing and requires deeper investigation.

\hrulefill

\exercise{Derive the expressions for the snap modulus, Eqs.~(\ref{eq:zeta_photon_ph}) and (\ref{eq:zeta_photon_vac}). }

\exercise{Assume that the cavity with photons inside is connected to a spring and undergoes small oscillations. Using perturbation theory, analyze the leading effect of the snap modulus on the motion of the cavity. You can assume that at time $t=0$ the cavity is suddenly displaced by distance $\lambda_0$ from the equilibrium position and then released.}

\exerciseshere
\hrulefill

\subsubsection{Dilations}

Now let us analyze the second setup corresponding to dilations, where the left wall of the cavity is fixed and the right is free to move, i.e., $\lambda=X_R$. The derivations are very similar to the case of translations, so we will only quote the final results. As before it is convenient to split the mass into the thermal and vacuum contributions
\be
\kappa_R=\kappa_R^{\rm ph}+\kappa_R^{\rm vac},
\ee
where
\be
\kappa_R^{\rm ph}={\hbar\over L^2}\sum_{i\neq j} {n_i-n_j\over \omega_j-\omega_i} {ij\over (i-j)^2}+{\hbar\over L^2}\sum_{i,j} {n_i+n_j\over \omega_i+\omega_j} {ij\over (i+j)^2}={2E\over 3c^2}
\label{eq:kappa_R_ph}
\ee
and
\begin{eqnarray}
\kappa_R^{\rm vac}&=&{\hbar \over \pi L c}\sum_{ij}{ij\over (i+j)^3} \mathrm{e}^{-\epsilon^2(i+j)^2}={\hbar \over 6\pi L c}\sum_{m=2,3,\dots} {m^2-1\over m^2} \mathrm{e}^{-\epsilon^2 m^2} \nonumber
\\ &=&C' {\hbar \omega_\Lambda\over c^2}-{\hbar \over 12 \pi L c}\left(\frac{\pi^2}{3} + 1 \right)=C' {\hbar \omega_\Lambda\over c^2}+{2E_c\over 3c^2}\left( 1 + \frac{3}{\pi^2} \right).
\end{eqnarray}
The thermal contribution to the dilation mass is again, as in the single-particle case, one third of the thermal translational mass. Therefore the equivalence of the gas to the massive spring (cf. \fref{fig:two_pistons_springs}) extends to the relativistic gas. On the other hand, the quantum contribution to the dilation mass, as in the non-relativistic case [cf. \eref{eq:mass_piston_dilation_n}],  contains an additional correction.

\hrulefill

\exercise{Consider the massive Klein-Gordon Hamiltonian with vanishing boundary conditions:
\be
\mathcal H = \frac{1}{2} \int_{-X_L}^{X_R} dx \left[ \Pi(x)^2 + c^2 \left( \frac{\partial \Phi}{\partial x} \right)^2 +\mu_0^2 \Phi^2(x) \right],
\ee
where $\mu_0$ is now finite. 
\begin{itemize}
\item By repeating the arguments above, prove that mass is still given by \eref{eq:kappa_photons1} with same overlaps $\zeta_{ij}$ as in the massless case and the massive dispersion: $\omega_j=\sqrt{\mu_0^2+k_j^2}$, $k_j=\pi j/L$.
\item Evaluate the thermal contribution to the translational mass. In particular, prove that 
\be
\kappa_+^{\rm ph}=\sum_j \hbar \omega_j n_j \left(1+{c^2 k_j^2\over \omega_j^2}\right).
\label{eq:photon_mass_massive}
\ee
From this expression recover the non-relativistic limit as $\mu_0$ becomes large.
\item Show that for the dilation mass
\[
\kappa_R^{\rm ph}={\kappa_+^{\rm ph}\over 3}
\]
irrespective of $\mu_0$.
\item Analyze the Casimir vacuum contribution for the translational mass. Show that it vanishes as $\mu_0$ becomes large.
\end{itemize}
}

\exercise{Argue that, as was the case for the massive particle in a box, the photon mass tensor is diagonal in the $\{X_+,X_-\}$ basis, i.e., $\kappa_{+-}=0$. Using this compute the mass $\kappa_{--}\equiv \kappa_-$ from $\kappa_+$ and $\kappa_R$. }

\exercise{Consider a three-dimensional rectangular cavity with the Hamiltonian described by 
\be
\mathcal H = \frac{1}{2} \int d^3 r \left[ \Pi(r)^2 + c^2 \left( \nabla \Phi  \right)^2  \right]~,
\label{eq:H_photon_3d_with_walls}
\ee
where the integration goes over the interval $x\in[-X_L, X_R]$, $y\in [-L_x/2,L_x/2]$, $z\in[-L_z/2, L_z/2]$ with vanishing boundary conditions. Find the thermal contributions to the translational and dilation mass along the $x$-direction. For this observe that the $y$ and $z$ components of the momentum are conserved and therefore $c^2 (k_y^2+k_z^2)$ plays the role of the mass $\mu_0$ analyzed in the previous problem. Use the results of the previous problem to show that
\be
\kappa_{+,x}^{\rm ph}=\sum_{\vect j}\hbar \omega_{\vect j} n_{\vect j} \left(1+{k_{x_j}^2\over \vect k_{\vect j}^2}\right).
\label{eq:photon_mass_3d}
\ee
Using this result prove that
\be
\overline \kappa_+\equiv {\kappa_{+,x}+\kappa_{+,y}+\kappa_{+,z}\over 3}={4\over 3} {E\over c^2}.
\ee
}

\exerciseshere
\hrulefill

\subsubsection{Classical derivation of the mass}

Similar to the example of a massive particle in a box discussed in \sref{sec:classical_mass}, let us finally consider taking a semi-classical limit for the photon problem. First we note that the dominant contribution to the mass in \eref{eq:kappa_photons} in the semi-classical limit ($k_B T\gg \hbar c/L$)  is given by the first term in the sum, which corresponds to number conserving processes. This follows from observing that the first term is dominated by neighboring modes $|\omega_i-\omega_j|\sim c/L$ as it is singular when $|\omega_i-\omega_j|\to 0$ while the second term is regular. This immediately translates  to the suppression of the second contribution by a dimensionless factor $\hbar c/(L k_B T)$, which vanishes in the semi-classical limit.  Therefore, instead of photons it suffices to consider classical number-conserving particles with relativistic dispersion confined to a box, as illustrated in \fref{fig:mass_photon}a. Consider for simplicity the one-dimensional case in which we start with a microcanonical ensemble with particles with energy $E_0$ uniformly distributed within the box. We then gradually begin to accelerate the box until the final velocity $v$ is reached. During and after the acceleration, when the particle hits a wall moving away from it with velocity $v$, it is red-shifted from original frequency $E_0$ to the new energy $E_1 = E_0 (1-2 v /c+v^2/c^2) / \sqrt{1-v^2/c^2}$ (this energy shift is equivalent to the frequency shift for photons). It simply follows from the energy and momentum conservation. A similar blue shift occurs upon hitting a wall moving towards the particle. The combination of these processes causes particles to equilibrate in the lab frame such that forward-moving particles are blue shifted compared to the backwards moving particles (\fref{fig:mass_photon}). Numerically calculating the total energy of particles in the box, we can verify that the total energy after slowly accelerating to velocity $v$, averaged over initial conditions, is $E_\mathrm{tot} \approx E_0 (1 + v^2 / c^2) = E_0 + \kappa v^2 / 2$. Thus, as in the quantum case, we find that $\kappa=2 E / c^2$ in these semi-classical simulations. Similar simulations can be done for the case of massive relativistic particles or three-dimensional photons, all of which confirm the quantum predictions of \eref{eq:photon_mass_3d} (\fref{fig:mass_photon}c).

Le us comment that this mass renormalization $\kappa\sim E/c^2$ is typically tiny for real photons but can be observable for other types of systems. For example, in the guitar string setup illustrated in \fref{fig:guitar_string} one can easily show that $\kappa \propto \mu A_\mathrm{osc}^2 / L$, where $\mu$ is the mass density of the string and $A_\mathrm{osc}$ is the amplitude of the oscillations, so by either using a heavier string or plucking it more strongly, one can readily enhance this effect to the point that it might be observable.

\begin{figure}
\includegraphics[width=.7\columnwidth]{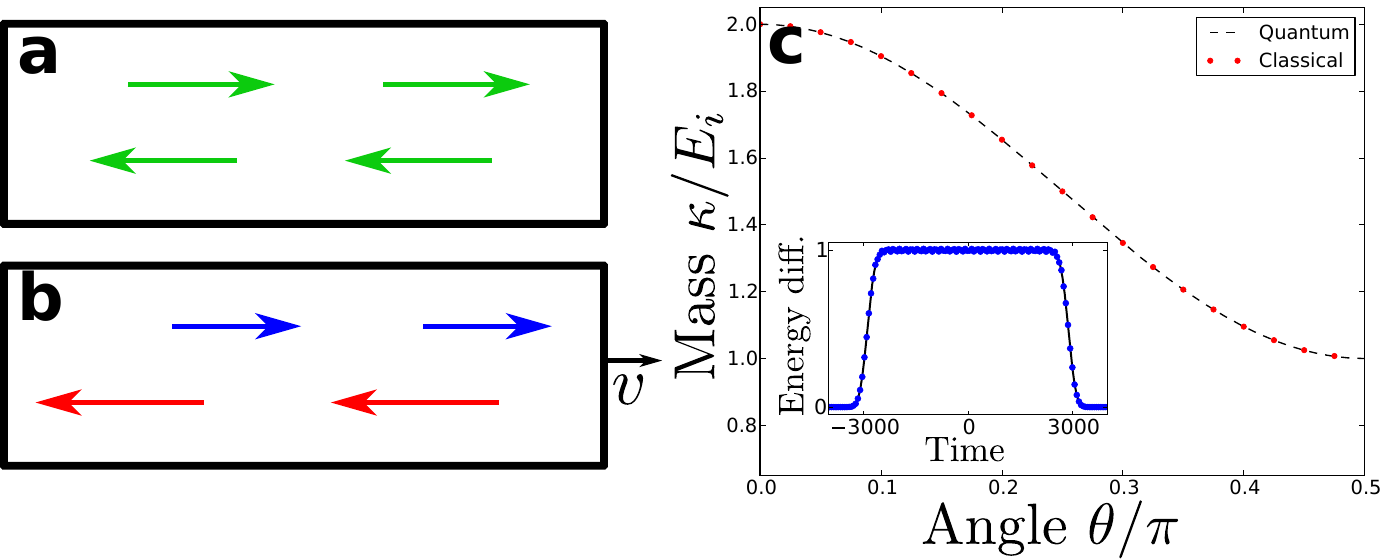}
\caption{(Semi)classical mass of relativistic particles. (a) Initial setup. Particles are prepared at initial energy $E_0$ and uniform probability distribution in space. (b) Cavity is slowly accelerated to velocity $v \ll c$. Doppler shifts of the particles upon hitting the cavity walls yield an equilibrium distribution in which forward-moving particles are blue-shifted and backward-moving particles are red-shifted. (c) The results of the simulation in a 3D cavity show that the semi-classical mass matches the quantum prediction in \eref{eq:photon_mass_3d}. Here $\theta= \tan^{-1} \big( k_\perp / k_x \big)$ parameterizes the initial direction of the particle in the cavity. The inset shows excess particle energy $(E-E_0)/v_0^2$ (blue dots) and $v^2 / v_0^2$ during a slow ramp from velocity $0$ to $v_0$ with $\theta=0$. The particles first heat as the box accelerates and then cool back down as the box decelerates, indicating that the dynamics is reversible as the mass correction should be.}
\label{fig:mass_photon}
\end{figure}

\hrulefill

\exercise{Confirm that the photon contribution to the dilation mass of the three-dimensional case satisfies $\kappa_{--}^\mathrm{3d-ph}({\vect p}) = \frac{1}{12}\kappa_{++}^\mathrm{3d-ph}({\vect p})$, where $\kappa_{++}^\mathrm{3d-ph} ({\vect p}) = 2 [1- (p_y^2 + p_z^2) / 2 p^2] E_{\vect p} / c^2$ (see \eref{eq:photon_mass_3d}).}

\exerciseshere

\hrulefill

\subsection{Classical central spin (rigid rotor) problem}

As another example let us consider a macroscopic rotor interacting with a bath of $N$ independent spin-1/2 particles (\fref{fig:rigid_rotor}a). This discussion closely follows that of Ref.~\cite{DAlessio2014_1}. We consider an interaction where the rotor with orientation $\hat n=( \sin\theta\cos\phi,\sin\theta\sin\phi,\cos\theta)$ produces a magnetic field parallel to $\hat n$ that interacts with the magnetic moments in the spin bath via Zeeman coupling of random magnitude. If instead of the rotor we use the quantum spin operator, this model is known as the central 
spin model.

The Hamiltonian describing the system of the form in \eref{eq:general} is
\be
\mathcal H_0=\frac{\vect L^2}{2I}+V(\vect{n}),\quad \mathcal H=-\vect n \cdot \sum_{i=1}^{N} 
\Delta_i\,\vect \sigma_i,
\ee
where $I$ is the momentum of inertia, which for simplicity we take to be isotropic, $V(\vect{n})$ is a time-dependent external 
potential, and $\vect L$ is the angular momentum of the rotor.
This example is similar to the one considered earlier except that the effective magnetic 
field is no longer confined to the $xz$ plane and we no longer assume that it is 
given by an external protocol. Rather, the time evolution of this system needs to be found self-consistently. Each spin evolves according to the von Neumann equation with the time-dependent Hamiltonian $\mathcal H(\vect{n}(t))$:
\begin{equation}
i\hbar \frac{d \rho}{dt}=\left[\mathcal H(\vect{n}(t)),\rho\right]~.
\label{vonN}
\end{equation} 
The rotor evolves according to the Hamilton equations of motion
\begin{equation}
I \dot{\vect{n}}=\vect L \times \vect n,\,\quad \dot{\vect{L}}=\vect n \times \left( 
\vect{M}_{ext} + \left< -\frac{\partial \mathcal H}{\partial \vect n} \right> \right)=\vect 
n \times \left( \vect{M}_{ext}+\sum_i \Delta_i \langle \vect \sigma_i 
\rangle\right)
\label{classical_rotor}
\end{equation} 
where $\vect{ M}_{ext}=-\frac{\partial V(\vect n)}{\partial \vect n}$ is the external
force on the rotor, such as the torque generated by an external magnetic field, and $\langle \dots \rangle$ indicates the quantum average over the density matrix $\rho(t)$. We assume that initially $\vect n_0=(0,0,1)$ and the spins are in thermal 
equilibrium with respect the Hamiltonian $\mathcal H(\vect n_0)$, giving $\langle \sigma^x_i 
\rangle_0=\langle \sigma^y_i \rangle_0=0$ and $\langle \sigma^z_i\rangle_0= \tanh(\beta \Delta_i)$.

For the toy model proposed here, these coupled equations can be easily solved 
numerically. In fact, according to the Ehrenfest theorem, the evolution of the 
expectation values follow the classical equation of motion and \eref{vonN} can be replaced with the much simpler equation $\hbar \dot{\vect 
m}_i=2\Delta_i \,\vect m_i \times \vect n$ where $\vect{m_i}=\langle \vect \sigma_i 
\rangle$. Therefore the exact dynamics of the system consists of the vectors 
($\vect L, \vect n, \{\vect m_i\})$ precessing around each other. 

We now compare the exact dynamics with the emergent Newtonian dynamics.
First, we note that the form of \eref{classical_rotor} 
immediately implies
\begin{equation}
\begin{split}
&\dot{\vect n} \cdot \vect L =0, \,\, \vect n \cdot \dot{ \vect L}=0 \implies
\vect n \cdot \vect L=\text{const} \\
&|\vect n|^2=\text{const} \implies \dot{\vect{n}}\cdot \vect n=0 \implies \ddot{\vect{n}}\cdot \vect n=-|\dot{\vect n}|^2
\end{split}
\label{conservation}
\end{equation}
We wish to compute the approximate generalized force $\langle \vect{\mathcal{M}} 
\rangle=-\langle \partial_{\vect{n}} \mathcal H\rangle$ in terms of the tensors 
$\kappa$ and $F$. The dissipative terms $\eta$ and $F^\prime$ 
are zero since there are no gapless excitations. Therefore \eref{off_diag} 
reduces to:
\begin{equation*}
 \langle \vect{\mathcal{M}} \rangle \approx \vect M_0+\hbar F_{\nu\mu}\dot{n}_{\mu}-\kappa_{\nu\mu}\ddot{n}_{\mu}
\end{equation*}
where $\nu,\mu \in \{x,y,z\}$. Using the expression for the spin-1/2 ground and excited states from earlier [\eref{eq:gs_es_qubit}], it follows that 
\begin{align*}
&\vect M_0\equiv \langle \vect{\mathcal{M}} \rangle_0= \hat n \sum_i \Delta_i \tanh(\beta \Delta_i) \\
&F_{\mu\nu}=F_0 \left(\begin{array}{ccc}
0 & -n_{z} & n_{y}\\
n_{z} & 0 & -n_{x}\\
-n_{y} & n_{x} & 0
\end{array}\right),\\
&\kappa_{\mu\nu}=\kappa_0 \left(\begin{array}{ccc}
1-n_{x}^{2} & -n_{x}n_{y} & -n_{x}n_{z}\\
-n_{y}n_{x} & 1-n_{y}^{2} & -n_{y}n_{z}\\
-n_{z}n_{x} & -n_{z}n_{y} & 1-n_{z}^{2}
\end{array}\right).
\end{align*}
where $F_0\equiv\frac{1}{2}\sum_i \tanh(\beta\Delta_i)$ and 
$\kappa_0\equiv\hbar^2\sum_i \frac{\tanh(\beta\Delta_i)}{4\Delta_i}$.
Substituting these expressions in \eref{classical_rotor} we find
\begin{equation*}
I \dot{\vect{n}}=\vect L \times \vect n,\,\, \dot{\vect{L}}=\vect n \times 
\vect{M}_{ext}-\hbar F_0\,\dot{\vect n}-\kappa_0\,(\vect n \times \ddot{\vect n})~.
\end{equation*}
To compute $I \ddot{\vect{n}}=\dot{\vect L}_\perp \times
\vect n+\vect L_\perp \times \dot{\vect n}$, it is now useful to split up $\vect L$ as $\vect L = \vect L_\perp + \hat n L_\parallel$, where $L_\parallel = \hat n \cdot \vect L$ is a constant of motion [see \eref{conservation}]. Then, using \eref{conservation} and the fact that $\vect{L}_\perp=I\,(\vect n \times \dot{\vect 
n})$ we arrive at:
\begin{equation}
I_{eff} \ddot{\vect{n}}=\left(\vect{n}\times\vect{ M}_{ext}\right)\times \vect n - L^\parallel_{eff} (\dot{\vect n} \times \vect n)-I_{eff} |\dot{\vect n}|^2 \vect n ~,
\label{rotor}
\end{equation}
where the renormalized moment of inertia is $I_{eff}=I+\kappa_0$ and the renormalized angular momentum is $L^\parallel_{eff} = L_\parallel + \hbar F_0$.

From this equation we see that the motion of the rotor is strongly 
renormalized by the interaction with the spin-$\frac{1}{2}$ particles.
Moreover we see that, even when the external force is absent 
($\vect{M}_{ext}=0$) and $L_\parallel=0$, the Berry curvature ($F_0$) causes a Coriolis-type force 
that tilts the rotation plane of the rotor. Indeed if we start with uniform 
rotations of the rotor in the $xz$ plane, i.e., $\vect n$ and $\dot{\vect n}$ lie in 
the $xz$ plane, we immediately see that the Berry curvature causes acceleration 
orthogonal to the rotation plane. The physics behind the Coriolis force is intuitively simple. At any finite 
angular velocity of the rotor, the spins will not be able to adiabatically follow the 
rotor and thus will be somewhat behind. As a result there will be a finite 
angle between the instantaneous direction of the spins and the rotor so the 
spins will start precessing around the rotor, and the rotor will in turn start precessing around the spins. \fref{fig:rigid_rotor}c shows an example where this Coriolis-induced precession can be observed. 

\begin{figure}
\includegraphics[width=.8\columnwidth]{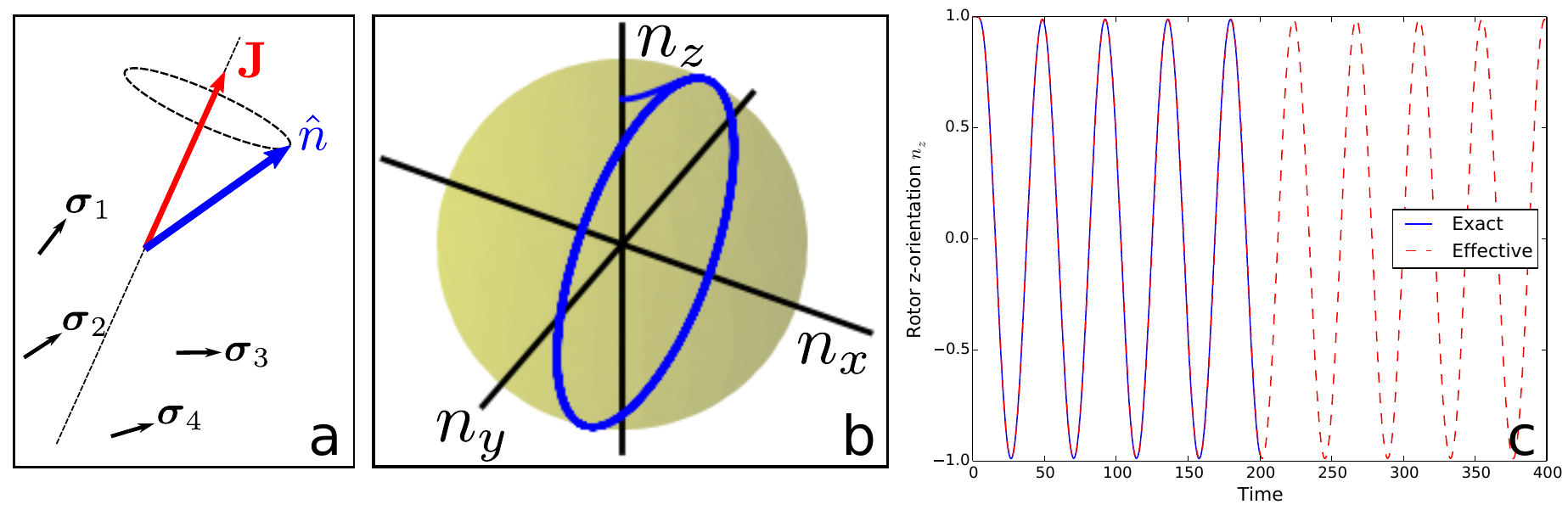}
\caption{Dynamics of the rigid rotor coupled to $N=20$ spins-1/2. (a) Illustration of the setup described in detail in the text. (b, c) Solution of the exact dynamics (blue line) and effective dynamics (dashed red line) to the model with $\beta=0.1$, $I=1$, and $\Delta_i$ randomly 
distributed in $(1,2)$. The initial conditions are $\vect{n}_0=(0,0,1)$ and $\vect{L}_0=(0,0,0)$ and initially the spins are in thermal equilibrium (see main text). The external force is ramped from its initial value of zero to final value $M^\ast=0.1$ in time $t_c=10$ according to the protocol $\vect{ M}_{ext}(t)=M^\ast \sin^2 \left( \frac{\pi t}{2 t_c} \right) \hat{x}$ for $0\le t\le t_c$, after which it is held fixed at $\vect{ M}_{ext} (t > t_c) = M^\ast \hat x$.}
\label{fig:rigid_rotor}
\end{figure}

\subsection{Quenched BCS superconductor}

In the previous sections, we have coupled the internal dynamics of our systems to external parameters such as the position of a box. A natural question that arises is whether these effective Newtonian dynamics can occur in situations where the classical dynamical degree of freedom is emergent, such as a macroscopic order parameter. We will now show that this possible in the case of a quenched BCS superfluid, which has been studied extensively since its realization in ultracold fermionic gases \cite{Barankov2004_1, Andreev2004_1, Barankov2006_1,Yuzbashyan2006_1}. It has been shown that the resulting equations of motion are integrable, but generally involve keeping track of every mode in the BCS theory. We will see how adiabatic perturbation theory gives new insight into this problem, allowing us to reduce the coupled equations of motion of the different momentum modes to a single integral equation in which the emergent slow mode -- the superconducting gap $\Delta$ -- is treated with the preceding formalism. In such a setup the equations of motion entirely emerge from the interactions with microscopic degrees of freedom and, for example, the mass is entirely determined by these interactions.

The system that we consider is a BCS superconductor with short-range interactions in which the interaction strength $g$ can be tuned as a function of time. This is a natural situation in, for example, ultracold atoms, where the interaction strength can be tuned by a Feshbach resonance \cite{Chin2010_1}. We start from the pairing Hamiltonian in $d$ dimensions
\be
\mathcal H = \sum_{\vect k \sigma} \epsilon_{\vect k} c^\dagger_{\vect k \sigma} c_{\vect k \sigma} - g \sum_{\vect k, \vect q} c_{\vect k \uparrow}^\dagger c_{-\vect k \downarrow}^\dagger c_{\vect q \uparrow} c_{-\vect q \downarrow}~,
\ee
which is exactly solvable within mean field theory. Here the single-particle energy of mode $\vect k$ is $E_{\vect k}$, from which the chemical potential is subtracted to get $\epsilon_{\vect k} = E_{\vect k} - \mu$. The mean-field decoupling consists of defining a gap $\Delta = g \sum_{\vect k} \langle c_{\vect k \uparrow} c_{-\vect k \downarrow} \rangle$, where the expectation value is taken over an arbitrary time-dependent wave function $|\psi(t)\rangle$ self-consistently. Making this replacement and switching to Anderson pseudospin notation  $\sigma$, where $\sigma^z_{\vect k} = 1$ ($-1$) corresponds to an unfilled (filled) pair, we get 
\be
\mathcal H= -\sum_{\epsilon} \big( \epsilon \sigma_{\epsilon}^z + \Delta \sigma^x_{\epsilon} \big) + \frac{\Delta^2}{g} \equiv \sum_\epsilon \mathcal H_\epsilon ~,
\label{eq:H_bcs_mft}
\ee
where without loss of generality we assumed that the gap starts real and remains real due to particle-hole symmetry. Note that we have switch from summing over the mode momentum $\vect k$ to their energies $\epsilon(\vect k)$. The last term in \eref{eq:H_bcs_mft} is often neglected, as it has no effect on the dynamics of the pseudospins. However, since we are interested in the dynamics of $\Delta$, it is convenient to write the Hamiltonian in this expressly energy-conserving form.

To better connect with our previous discussion let us introduce the momentum conjugate to the gap: $P_\Delta$ and the bare mass $m_0$, which we later send to zero. This gives the Hamiltonian in \eref{eq:H_bcs_mft} an additional term:
\be
\mathcal H\to \mathcal H +{P_\Delta^2\over 2 m_0}~,
\ee
such that the equations of motion for the gap read:
\be
m_0\ddot \Delta=\langle -\partial_\Delta \mathcal H\rangle=-{2\Delta \over g}+\sum_\epsilon \langle \sigma_\epsilon^x\rangle
\ee
In the limit of zero bare mass $m_0\to 0$ this equation simply reduces to the self-consistency equation:
\be
\Delta = \frac{g}{2} \sum_\epsilon \langle \sigma_\epsilon^x \rangle ~,
\ee
Note that the average is taken over the non-equilibrium density matrix, which is the solution of the von Neumann equation:
\be
i {d\rho\over dt}=[\mathcal H(\Delta(t)),\rho].
\ee

Starting in the ground state at some interaction strength $g_i$, we can ramp the interactions through some arbitrary protocol $g(t)$. For slow enough changes, we expect that the pseudospins $\sigma$ will be weakly excited above their ground state yielding leading Newtonian correction $\langle \mathcal M_\Delta \rangle \approx \langle \mathcal M_\Delta \rangle_0 - \kappa \ddot \Delta$, where 
\be
\langle \mathcal M_\Delta \rangle_0 = -\frac{2 \Delta}{g} + \sum_\epsilon \langle \sigma^x_\epsilon \rangle_0 = -\frac{2 \Delta}{g} + \sum_\epsilon \frac{\Delta}{\sqrt{\Delta^2 + \epsilon^2}} ~.
\ee
Similarly, using \eref{kappa_low}, the effective mass in the ground state will be 
\be
\kappa = 2 \sum_\epsilon \frac{| \langle e | \sigma_x | g \rangle|^2}{(E_e - E_g)^3} = \frac{1}{4} \sum_\epsilon \frac{\epsilon^2}{(\epsilon^2 + \Delta^2)^{5/2}} ~.
\ee
We can easily simulate both the exact and approximate equations of motion for this theory. More explicitly, we adopt the conventions of \rcite{Barankov2006_1} and expand near the Fermi surface by considering a uniform density of states $\nu$ extending in a band from $\epsilon = -W/2$ to $W/2$, with $W \gg \Delta$ playing the role of the UV cutoff. This band is then broken up into $N=\nu W$ discrete modes and the physical limit is achieved by taking $W, N \to \infty$. The ramp is specified in a UV-independent way as a function $\Delta_\mathrm{eq}(t)$, where from the gap equation in equilibrium (the ground state), 
\beq
\nn
\Delta_\mathrm{eq}(t) &=& \frac{g(t)}{2} \sum_\epsilon \frac{\Delta_\mathrm{eq}(t)}{\sqrt{\Delta_\mathrm{eq}(t)^2 + \epsilon^2}} 
\\ \implies g(t)^{-1} &=& \frac{1}{2} \sum_\epsilon \frac{1}{\sqrt{\Delta_\mathrm{eq}(t)^2 + \epsilon^2}} 
\approx \frac{\nu}{2} \int_{-W/2}^{W/2} d \epsilon \frac{1}{\sqrt{\Delta_\mathrm{eq}(t)^2 + \epsilon^2}} = \nu \ln \left( \frac{W}{\Delta_\mathrm{eq}(t)} \right) ~.
\eeq
Note that microscopic parameters such as $g$ can explicitly depend on the cutoff, while emergent objects such as the mass and the generalized force do not:
\beq
\nn \kappa &\approx& \frac{\nu}{4} \int_{-W/2}^{W/2} d\epsilon \frac{\epsilon^2}{(\epsilon^2 + \Delta^2)^{5/2}} \stackrel{W \to \infty}{\longrightarrow} \frac{\nu}{6 \Delta^2} ~.
\\ \langle \mathcal M_\Delta \rangle_0 &=& 2 \nu \Delta \ln \left( \frac{\Delta_\mathrm{eq}}{\Delta}\right) ~.
\eeq
Note also that both the mass and the generalized force are proportional to the density of states, i.e., they are extensive.

\begin{figure}
\includegraphics[width=\columnwidth]{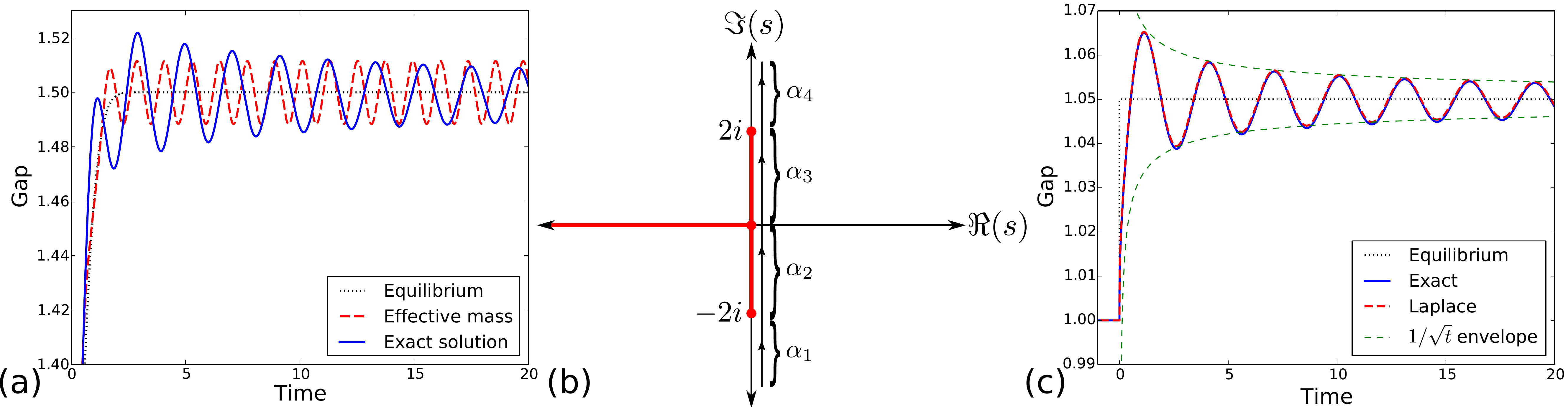}
\caption{Dynamics of the gap in a quenched BCS superfluid. (a) Gap vs. time for a ramp from $\Delta_i=1$ to $\Delta_f=1.5$ in time $T=1$. The exact solution (blue) is compared to the dynamics within the effective mass approximation (dashed red). This approximation is clearly insufficient, so we must instead solve the full integral equation, e.g., via Laplace transform. (b) The branch cut structure of $\tilde \alpha$, the Laplace transform of $\alpha \propto \Delta-\Delta_f$ [see \eref{eq:laplace_transform_alpha}]. (c) Dynamics of the gap after a small instantaneous quench from $\Delta_i=1$ to $\Delta_f=1.05$. The exact dynamics agree well with the solution of the integral equation. The dashed lines shows that $1/\sqrt t$ envelope that is analytically found at late times.}
\label{fig:dyn_bcs_gap}
\end{figure}

The simulations described above are plotted in \fref{fig:dyn_bcs_gap}a for a particular protocol in which $\Delta_\mathrm{eq}$ is slowly ramped from $\Delta_i$ to $\Delta_f$: $\Delta_\mathrm{eq}(t) = \Delta_i + (\Delta_f - \Delta_i) \mathrm{erf}(t/T)$. Quite surprisingly, the effective mass does not accurately described the dynamics of the ramped or quenched BCS superconductor. Other simulations confirm that this is true independent of the initial and final values of the gap or the time scale $T$ of the ramp. A particularly noticeable difference is that the simulations of the full model show damping of the oscillations, while the effective model with mass $\kappa$ undergoes infinitely long-lived oscillations about the minimum of the potential at $\Delta=\Delta_f$. 

To see where this comes from, let us consider small oscillations about the point at the end of the ramp. Linearizing about the final point, there is only one energy/time scale in the system, so the frequency of oscillations must scale as $\omega \sim \Delta_f$. We have seen that the effective mass gives a contribution to the generalized force $\kappa \ddot \Delta \sim \kappa A \omega^2 \sim \nu A$, where  $A$ is the amplitude of the oscillations. The next correction, which is non-Newtonian, is given by $\zeta \Delta^{(4)}$, where $\Delta^{(n)}$ denotes the $n$th time derivative. From the expansion that gave us the effective Newtonian dynamics, the coefficient $\zeta$ for the ground state is given by
\be
\zeta = 2 \sum_{n \neq 0} \frac{|\langle n | \partial_\Delta \mathcal H | 0 \rangle| ^2}{(E_n - E_0)^5} = \frac{1}{16} \sum_{\epsilon} \frac{\epsilon^2}{(\epsilon^2 + \Delta^2)^{7/2}} = \frac{\nu}{60 \Delta^4} ~.
\ee
So the correction is $\zeta \Delta^{(4)} \sim (\nu \Delta^{-4})(A \Delta^4) \sim \nu A$, i.e., it scales exactly the same way as the effective mass contribution. Indeed, if we consider an arbitrary term in the series $\chi_{n} \Delta^{(n)}$ for arbitrary positive even integer $n$, we will again find that its contribution scales as $\nu A$. Therefore, it is not okay to truncate at second order by considering just the effective mass - indeed, there is no limit where it will be fully correct to truncate at any finite order. This statement that we need to know not just $\Delta$ and its acceleration, but rather all of its higher-order (even) derivatives, is tantamount to saying that the local-in-time expansion about the time $t$ is not correct. Therefore, to solve this problem correctly, we must resort to the full integral equation from first order adiabatic perturbation theory [\eref{eq:gen_force_int}]:
\beq
\nn
\langle \mathcal M_\Delta \rangle_0 &=& 2 \int_{t_0}^t d t' \Dot{\Delta}(t') \sum_{m \neq n} \frac{\rho_n^0 \langle m | \mathcal M(t) | n  \rangle \langle n | \mathcal M(t') | m \rangle }{E_m(t') - E_n(t')} \mathrm{e}^{i \int_{t'}^t d\tau (E_m(\tau)-E_n(\tau))} + O(\dot{\Delta}^2)
\\ &=& \int_{t_0}^t dt' \Dot{\Delta}(t') \sum_\epsilon \frac{\epsilon^2}{(\epsilon^2 + \Delta(t')^2)\sqrt{\epsilon^2 + \Delta(t)^2}} \cos \left( 2 \int_{t'}^t d\tau \sqrt{\epsilon^2 + \Delta(\tau)^2} \right) ~,
\eeq
where we have taken the real part of the exponential because all the matrix elements are real. With a bit more effort, this integral equation can be solved numerically, and for slow ramps or small quenches, the integral equation agrees with the exact numerics (\fref{fig:dyn_bcs_gap}c).

We can gain a bit more understanding of the integral equation by consider the case of a small quench or equivalently the late-time behavior of a slow ramp. Assuming that the deviation $\alpha = (\Delta - \Delta_f)/\Delta_f$ of the gap from equilibrium is small, we can expand the integral equation about $\alpha=0$. The first order contribution is then 
\be
-2 \nu \Delta_f \alpha(t) = \int_{t_0}^t \dot{\alpha}(t') \sum_\epsilon \frac{\epsilon^2}{(\epsilon^2 + \Delta_f^2)^{3/2}} \cos \left( 2 \sqrt{\epsilon^2 + \Delta_f^2} (t - t') \right) ~.
\label{eq:integral_eqn_linearized}
\ee
This restores some degree of locality - the integral equation now only depends on the history of $\dot \alpha$ and not directly on $\alpha(t')$. \eref{eq:integral_eqn_linearized} can be Laplace transformed to get 
\be
-2 \nu \Delta_f \tilde{\alpha}(s) = (s {\tilde \alpha} - \alpha_0) \sum_\epsilon \frac{\epsilon^2}{(\epsilon^2 + \Delta_f^2)^{3/2}} \frac{s}{s^2 + 4(\epsilon^2 + \Delta_f^2)} ~, 
\label{eq:laplace_before_sum}
\ee
where $\tilde \alpha(s) = \int_0^\infty \mathrm{e}^{-s t} \alpha(t)$ is the Laplace transform of $\alpha$ and $\alpha_0 = (\Delta_i - \Delta_f)/\Delta_f$ is the initial condition. The equilibrium correlation function of the pseudospins $\sigma$ is encoded in the $\epsilon$-dependent terms, so the fermions can be integrated out to give
\be
\sum_\epsilon \frac{\epsilon^2}{(\epsilon^2 + \Delta_f^2)^{3/2}} \frac{s}{s^2 + 4( \epsilon^2 + \Delta_f^2)} = \frac{2 \nu}{s} \left[ -1 + \frac{\sqrt{4 \Delta_f^2 + s^2}}{s} \cosh^{-1} \left( \frac{\sqrt{4 + s^2/\Delta_f^2}}{2} \right) \right] ~.
\ee
Substituting this into \eref{eq:laplace_before_sum} and rearranging, we find that
\be
\tilde \alpha(s) = \alpha_0 \frac{-s + \sqrt{\Delta_f^2 + s^2} \cosh^{-1} \left( \frac{\sqrt{4 + s^2/\Delta_f^2}}{2} \right)}{s \sqrt{\Delta_f^2 + s^2} \cosh^{-1} \left( \frac{\sqrt{4 + s^2/\Delta_f^2}}{2} \right)} ~.
\label{eq:laplace_transform_alpha}
\ee
Note that, as expected, the only time scale in the problem is set by $\Delta$ and $\alpha$ scales linearly with $\alpha_0$. Therefore, rescaling $\alpha \to \alpha / \alpha_0$, $t \to t \Delta$, and $s \to s / \Delta$, we can attempt to invert the Laplace transform and solve for $\alpha(t)$.

These small quenches of the order parameter were studied in \rcite{Volkov1974_1}. Using very different methods, they nevertheless arrived at an equation of motion for the order parameters quite similar to \eref{eq:laplace_transform_alpha}. The inverse Laplace transform is given by
\be
\alpha(t) = \frac{1}{2\pi i} \int_{\gamma - i \infty}^{\gamma + i \infty} \mathrm{e}^{s t} \tilde{\alpha} (s) ds 
\ee
for any $\gamma > 0$. $\tilde \alpha$ has branch cuts associated with the square roots at $s=\pm 2 i$. We make the  the branch cut shown in \fref{fig:dyn_bcs_gap}(b), which does not cross the contour for any $\gamma > 0$. The inverse hyperbolic cosine also has a branch cut on the negative real axis, which again does not affect us. We simply make the obvious branch choice such that $\sqrt{4 + s^2}$ and $\cosh^{-1} \left( \sqrt{4 + s^2}/2 \right)$ are positive and real on the positive real axis, which uniquely defines the function on the chosen contour. Then taking the limit $\gamma \to 0^+$, we split the contour into four pieces, as shown in \fref{fig:dyn_bcs_gap}(b). Taking first $\alpha_3$, for which $s=ir$ with $0<r<2$, the branch choices gives $\sqrt{4 + s^2} = \sqrt{4 - r^2}$ and $\cosh^{-1} \left( \sqrt{4 + s^2}/2 \right) = i \cos^{-1} \left( \sqrt{4 - r^2}/2 \right)$. Thus
\beq
\nn \alpha_3(t) & = & \Re\left[ \frac{1}{2 \pi i} \int_0^2 i dr \mathrm{e}^{i r t} \frac{-ir + \sqrt{4-r^2} \left( i \cos^{-1} \left( \frac{\sqrt{4-r^2}}{2} \right) \right)}{i r \sqrt{4-r^2} \left( i \cos^{-1} \left( \frac{\sqrt{4-r^2}}{2} \right) \right) } \right]
\\ &=&  \frac{1}{2 \pi} \int_0^2 dr \sin(r t) \frac{-r + \sqrt{4-r^2} \cos^{-1} \left( \frac{\sqrt{4-r^2}}{2} \right)}{r \sqrt{4-r^2} \cos^{-1} \left( \frac{\sqrt{4-r^2}}{2}   \right) } ~.
\eeq
Similarly for the integral $\alpha_4$ ($s=ir$ for $r > 2$), the branch choices are $\sqrt{4 + s^2} = i \sqrt{r^2-4}$ and $\cosh^{-1} \left( \sqrt{4 + s^2}/2 \right) = i \pi / 2 +  \ln \left( (\sqrt{r^2 - 4} + r)/2 \right)$. So
\beq
\nn \alpha_4(t) & = & \Re\left[ \frac{1}{2 \pi i} \int_2^\infty i dr \mathrm{e}^{i r t} \frac{-ir + i \sqrt{r^2-4} \left(\frac{i \pi}{2} + \ln \left( \frac{r + \sqrt{r^2-4}}{2} \right) \right)}{i r (i \sqrt{r^2-4}) \left(\frac{i \pi}{2} + \ln \left( \frac{r + \sqrt{r^2-4}}{2} \right) \right) } \right]
\\ \nn &=& \frac{1}{2\pi} \int_2^\infty \frac{dr}{r \sqrt{r^2-4} \left(\frac{\pi^2}{4} + \ln \left( \frac{r + \sqrt{r^2-4}}{2} \right)^2 \right)} \Bigg[ \cos(rt) \left( \frac{\pi r}{2} \right) +
\\ &&~~ \sin(rt) \left( r \ln \left( \frac{r + \sqrt{r^2-4}}{2} \right) - \sqrt{r^2 - 4} \left( \frac{\pi^2}{4} + \ln \left( \frac{r + \sqrt{r^2-4}}{2} \right)^2 \right) \right) \Bigg] ~.
\eeq
One can easily show that $\alpha_1=\alpha_4$ and $\alpha_2=\alpha_3$, so $\alpha(t)=2(\alpha_3(t) + \alpha_4(t))$. These integrals can be evaluated numerically, the results of which are plotted in \fref{fig:dyn_bcs_gap}c. Clearly they match well with the exact dynamics.

We can also analyze the late time limit of these equations. In this limit, the cosines and sines yield fast-oscillatory integrals, which are then dominated by the stationary points of their integrands. Both $\alpha_3$ and $\alpha_4$ have a singularity at $r=2$ ($s=2i$). Therefore, the integrals are dominated by this point and we can simply expand the remainder of the integrand about $r=2$. Thus,
\beq
\nn
\alpha_3(t \gg 1/\Delta_f) & \approx&  \frac{1}{2\pi} \int_0^2 \frac{-\sin r t}{\pi \sqrt{2-r}} dr \approx - \frac{1}{2\pi^2} \int_{-\infty}^2 \frac{\sin r t}{\sqrt{2-r}} dr = (2\pi^3)^{-1/2} \frac{\cos 2t - \sin 2 t}{\sqrt t} ~.
\\ \alpha_4(t \gg 1/\Delta_f) & \approx&  \frac{1}{2 \pi^2} \int_2^\infty \frac{\cos rt}{\sqrt{r-2}} dr = (2\pi^3)^{-1/2} \frac{\cos 2t - \sin 2 t}{\sqrt t} ~.
\eeq
Thus, as seen in \rcite{Volkov1974_1}, the late time behavior of the gap is described by power law relaxation $\Delta \sim \cos(2 \Delta_f t + \varphi) / \sqrt t$, unlike the exponential relaxation expected in non-integrable (thermalizing) systems. This behavior can be traced back to the fact that the underlying BCS dynamics is integrable and has been termed collisionless relaxation \cite{Volkov1974_1}.

It is interesting to extend these results to the finite temperature case. Unlike the previous case where $\Delta$ was the only energy scale in the problem, the temperature now introduces a new energy scale that we might expect to cut off the correlation functions such that locality in time is restored. However, a quick calculation similar to that above (not shown) demonstrates that starting from a finite temperature ensemble yields qualitatively similar dynamics as those starting from the ground state. The reason for this is simple: as an integrable model, the adiabatically transported state from the thermal ensemble at $\Delta_i$ to the final value $\Delta_f$ \emph{is not thermal}. Such a non-thermal ensemble is referred to as a generalized Gibbs ensemble \cite{Rigol2007_1} and has been well-understood to occur in generic integrable systems. Here it manifests as an absence of thermalization that yields similar dynamics at finite energy density as those in the ground state. It is worth pointing out that previous works have shown that large quenches \cite{Barankov2006_1} and/or non-trivial initial states \cite{Yuzbashyan2006_1} can result in long-lived oscillations that do not relax. Whether or not such dynamics can be captured within the effective Newtonian framework is a fascinating open question which is beyond the scope of these lectures.

\section{Summary and outlook}

Over the course of these lectures, we have introduced the concept of gauge potentials and seen how they are connected to a wide variety of ideas from geometry and topology of quantum systems to the emergence of Newtonian dynamics. An important aspect of this perspective is its generality, allowing the derivation of effective dynamics in systems as different as photons in a cavity and quenched BCS superconductors. These ideas are therefore quite amenable to being used in many important experimental systems as a method for understanding the dynamics of slow variables. With numerical methods, these can even be used to understand dynamics in complicated interacting many-body systems using only equilibrium simulations, and therefore have the potential to solve dynamics in complicated systems above one dimension, where exact well-behaved numerical methods are scarce \cite{Zhou2008_1,Albuquerque2010_1,DeGrandi2011_1,KolodrubetzPRB2014_1,Wang2015_1}.

An interesting open topic is how these ideas can be utilized in ever more complicated systems, particularly towards understanding the gauge potentials for non-equilibrium systems. For instance, we have seen how the gauge potentials for excited states are ill-defined if the system is ergodic due to the problem of small denominators. We have provided two methods for regulating this problem, but connecting these ideas to adiabatic evolution in conventional thermodynamic systems remains an important open question. Furthermore, one may be able to generalize these ideas to truly non-equilibrium systems where equilibrium statistical mechanics does not apply.  A fascinating class of non-equilibrium Hamiltonians is periodically-driven systems, where one must differentiate between the effect of the parameters on the slow motion that can be written in terms of an effective time-independent Hamiltonian and the fast micromotion that it periodic with the same period as the drive~\cite{Shirley1965_1, Bukov2015_2}. Finally, all these questions become even more interesting the presence of coupling to an environment, which is usually the situation we are given in realistic experimental systems. These are all fascinating questions, and understanding them will prove very valuable in solving the dynamics of complicated quantum and classical systems.

\acknowledgements

These notes were partially based on works done jointly with L. D'Alessio, V. Gritsev, and Y. Kafri, with whom we acknowledge many useful discussions. We also acknowledge useful discussions with C. Jarzynski, E. Mueller, and E. Katz. We thank S. Davidson for preparing the solution manual for the problems. This manual is available upon request. Work of A. P. was supported by AFOSR FA9550-16-1-0334, NSF DMR-1506340 and ARO W911NF1410540. M.~K.~was supported by Laboratory Directed Research and Development (LDRD) funding from Berkeley Lab, provided by the Director, Office of Science, of the U.S. Department of Energy under Contract No. DE-AC02-05CH11231 as well as the U.S. DOE, Office of Science, Basic Energy Sciences as part of the TIMES initiative. D.S. acknowledges support of the FWO under grant No. 12M1518N as post-doctoral fellow of the Research Foundation - Flanders.

\appendix

\section{Metric tensor from Kubo response at finite temperature}
\label{sec:kubo_metric_finite_temp}

Consider a generic Hamiltonian $\mathcal H$ with eigenstates $|n\rangle$. We define the metric tensor with respect to single parameter $\lambda$, alternatively known as the fidelity susceptibility, at finite temperature by
\be
g_{\lambda \lambda} (T) = \sum_n \rho_n \sum_{m \neq n} \frac{|M_{nm}|^2}{(E_n - E_m)^2} ~,
\ee
where $\rho_n = \mathrm{e}^{-\beta E_n} / Z$ and $M_{nm}=\langle n | \partial_\lambda \mathcal H | m \rangle$. Define the (non-symmetrized) spectral function as
\be
S(\omega) =2\pi \sum_n \rho_n \sum_{m \neq n} |M_{nm}|^2 \delta(E_n - E_m + \omega) ~.
\label{eq:s_omega}
\ee
Then it is clear that 
\be
g = \int_{-\infty}^{\infty} {d \omega\over 2\pi} ~ \frac{S(\omega)}{\omega^2}=\int_0^\infty {d \omega\over 2\pi} {S(\omega)+S(-\omega)\over \omega^2}  ~.
\ee
We will now see that this expression can be connected to the out-of-phase susceptibility, which is measurable via linear response.

From standard Kubo response, the response function $\epsilon(\omega)$ of the magnetization $M$ to a small periodic perturbation of $\lambda=\lambda_0 \mathrm{e}^{-i \omega t}$ is given by \cite{Mahan2000_1}
\be
\epsilon(\omega) = i \int_0^\infty dt\, \mathrm{e}^{i \omega t-\delta t} \langle [ M(t), M(0) ] \rangle ~,
\label{eq:chi_omega}
\ee
where $\delta$ is an infinitesimal positive number added for convergence, $M(\omega) = \epsilon(\omega) \lambda(\omega)$, and the expectation value is over the thermal density matrix $\hat \rho = \mathrm{e}^{-\beta \mathcal H} / Z$. Note that, unlike the two-time correlation function used in defining the metric, the correlation function in \eref{eq:chi_omega} need not be connected; this is because the commutator makes $\langle [ A, B] \rangle = \langle [ A, B] \rangle_c$. Let us next use the Lehmann representation:
\beq
\nn \epsilon(\omega) & = & i \int_0^\infty dt\, \mathrm{e}^{i \omega t-\delta t} \sum_n \rho_n \big( \langle n | \mathrm{e}^{i \mathcal H t} M \mathrm{e}^{-i \mathcal H t} M | n \rangle - h.c. \big)
\\ \nn & = & i \int_0^\infty dt\, \mathrm{e}^{i \omega t-\delta t} \sum_n \rho_n \sum_{m} \big( \langle n | \mathrm{e}^{i \mathcal H t} M | m \rangle \langle m | \mathrm{e}^{-i \mathcal H t} M | n \rangle - h.c. \big)
\\ \nn & = & i \int_0^\infty dt\, \mathrm{e}^{i \omega t -\delta t} \sum_n \rho_n \sum_{m} \big( \langle n | \mathrm{e}^{i E_n t} M | m \rangle \langle m | \mathrm{e}^{-i E_m t} M | n \rangle - h.c. \big)
\\ \nn & = & i \int_0^\infty dt\, \mathrm{e}^{i \omega t-\delta t} \sum_n \rho_n \sum_{m} \big( \mathrm{e}^{i (E_n - E_m) t} - \mathrm{e}^{i (E_m - E_n) t} \big) \big| M_{nm} \big|^2
\\ \nn & = & i  \sum_n \rho_n \sum_{m \neq n}  \big| M_{nm} \big|^2 \left({1\over \delta-i (E_n-E_m+\omega)}-{1\over \delta-i (E_m-E_n+\omega)}\right).
\eeq
The imaginary part of the susceptibility $\epsilon''(\omega)  = \mathrm{Im} [\epsilon(\omega)]$ is thus
\begin{multline}
\epsilon''(\omega)= \sum_n \rho_n \sum_{m \neq n}  \big| M_{nm} \big|^2 \left({\delta\over \delta^2+ (E_n-E_m+\omega)^2}-{\delta \over \delta^2+(E_m-E_n+\omega)^2}\right) \\
=  \pi \sum_n \rho_n \sum_{m \neq n} \big| M_{nm} \big|^2 \big[ \delta(E_n - E_m + \omega) - \delta(E_m - E_n + \omega) \big]={S(\omega)-S(-\omega)\over 2},
\end{multline}
where we used the identity
\[
\lim_{\delta\to 0^+} {\delta\over \delta^2+x^2}=\pi \delta(x)
\]
and \eref{eq:s_omega} to get the last equality. In thermal equilibrium $S(\omega)$ and $S(-\omega)$ satisfy the fluctuation dissipation relation~\cite{Mahan2000_1}, which we derive for completeness from \eref{eq:s_omega}:
\begin{multline}
S(-\omega) =2\pi \sum_{m\neq n} {1\over Z} \mathrm{e}^{-\beta E_n} |M_{nm}|^2 \delta(E_m - E_n + \omega)=
2\pi \sum_{m\neq n} {1\over Z} \mathrm{e}^{-\beta E_m} |M_{nm}|^2  \delta(E_n-E_m+\omega)\\
=2\pi \mathrm{e}^{-\beta\omega}\sum_{m\neq n} {1\over Z} \mathrm{e}^{-\beta E_n} |M_{nm}|^2  \delta(E_n-E_m+\omega)= \mathrm{e}^{-\beta\omega} S(\omega),
\end{multline}
where in the first equality we changed summation indexes $n\leftrightarrow m$ and in the second equality used that $E_m=E_n-\omega$. Therefore  
\[
\epsilon''(\omega)={1\over 2} S(\omega) \left(1- \mathrm{e}^{-\beta\omega}\right)\;\Leftrightarrow \; S(\omega)={2\epsilon''(\omega)\over 1-\exp[-\beta\omega]}.
\]
and 
\be
g=\int_0^\infty {d \omega\over 2\pi} {S(\omega)+S(-\omega)\over \omega^2}=\int_0^\infty {d \omega\over \pi} {\epsilon''(\omega)\over \omega^2} {\exp[\beta \omega]+1\over \exp[\beta \omega]-1 }=\int_0^\infty {d \omega\over \pi} {\epsilon''(\omega)\over \omega^2} \coth(\beta\omega/2).
\ee

\bibliography{References}

\end{document}